%% file: main.tex
\documentclass[12pt,a4paper,twoside]{ipb}
\usepackage{csquotes}
\usepackage{amssymb}
\usepackage[portuguese,USenglish]{babel}
\graphicspath{{./images/}}
\usepackage{listings}
\usepackage{xcolor}
\usepackage{realboxes}
\usepackage{url}
\usepackage{enumitem}
\usepackage{amsmath}
\usepackage{bigints}
\usepackage{mathtools}
\usepackage{relsize, exscale}
\usepackage{bbold}
\usepackage[rightcaption]{sidecap}
\usepackage{empheq}
\usepackage{extarrows}
\usepackage[most]{tcolorbox}
\tcbset{myformula/.style={colback=white, %yellow!10!white,
    colframe=black, %red!50!black,
    top=0pt,bottom=0pt,left=0pt,right=0pt,
    boxsep=0pt,
    arc=0pt,
    outer arc=0pt,
}}
\usepackage{graphicx} %package to manage images
\graphicspath{ {images/} }
\usepackage[colorlinks=true, urlcolor=black, linkcolor=black, citecolor=black, bookmarks=true, pdfstartview=FitH]{hyperref}

\usepackage[style=ieee,backend=biber]{biblatex}
\usepackage{lipsum}
\usepackage{tabulary}
\usepackage{graphicx}
\usepackage{scalerel}
\usepackage{subfig}
\usepackage{float}  % Positioning of the figures
\usepackage[binary-units=true]{siunitx}
\usepackage[section]{placeins}
\usepackage{indentfirst}  % Indents all paragraphs
\usepackage{physics}
\usepackage{tensor}
\usepackage{multirow}
\usepackage{booktabs}
\usepackage{cellspace} % For spacing in cells
\usepackage{tabularray}
\usepackage{array}
\usepackage{comment}
\usepackage{mathtools}  % Allows adding multi line equations
\newcolumntype{P}[1]{>{\centering\arraybackslash}p{#1}}  % Centre col with width
\usepackage{enumitem}
\setlist{nosep}  % By default the separation is large

\usepackage{fancyref}  % This allows to use \fref{} command. This is very useful, it allows to easily refer to any \label{}. % See documentation: https://mirror.las.iastate.edu/tex-archive/macros/latex/contrib/fancyref/fancyref.pdf.

\title{$AdS_4\,/\,CFT_3$ : ABJM Theory, Brane Geometry, Correlators and Mellin Space}  % Title of your thesis
\author{Manikantt Mummalaneni}  % Student name
\supervisor{Prof. dr. Jan de Boer}
\cosupervisor{Dr. Lorenz Eberhardt}
\courseyear{2023-24}
\makeglossaries
\loadglsentries{chapters/acronyms}

\begin{document}
\beforepreface  % Places the cover page

%\clearpage
% Navigate to: chapters/acknowledgements.tex
%\input{chapters/acknowledgements} 

%\clearpage
% Navigate to: chapters/declaration.tex
%\input{chapters/declaration}

\clearpage
% Navigate to: chapters/abstract.tex 
\input{chapters/abstract}
\input{chapters/Thanks}
%\cleardoublepage  % Makes sure contents start on correct page
\afterpreface  % Places contents, list of tables and list of figures
\bodystart  % This sets stylistic parameters for the following content

% Here starts the actual content of your thesis
\input{chapters/Introduction}

\addcontentsline{toc}{chapter}{Introduction}
\input{chapters/research_design}
\input{chapters/research_results}

\input{chapters/situational_theoretical_analysis}
\input{chapters/rationale}
\input{chapters/conclusion_recommendations}

\input{chapters/conceptual_model}
\input{chapters/additional_chapters}
\addcontentsline{toc}{chapter}{Synthesis and Future Directions}
\input{chapters/appendix}

%\printglossary[type=\acronymtype,title={Definitions and Abbreviations}]

\clearpage
\addcontentsline{toc}{chapter}{References}
\printbibliography[title={References}]

 % Note, appendix must be last

\end{document}

%% file: chapters/acronyms.tex
\newacronym{API}{API}{Application Programming Interface}
\newacronym{TI}{TI}{Tecnologias da Informação}
\newacronym{ESTiG}{ESTiG}{Escola Superior de Tecnologia e Gestão}
\newacronym{IPB}{IPB}{Instituto Politécnico de Bragança}
\newacronym{CET}{CET}{Cursos de Especialização Tecnológica}
\newacronym{EI}{EI}{Engenharia Informática}
\newacronym{IG}{IG}{Informática de Gestão}
\newacronym{GSR}{GSR}{Gestão de Sistemas e de Redes}
\newacronym[longplural={Instituições de Ensino Superior}]{IES}{IES}{Instituição de Ensino Superior}
\newacronym{FPS}{FPS}{First Person Shooter}
\newacronym{RPG}{RPG}{Role-Playing Game}
\newacronym{OT}{OT}{Orientação Tutorial}
\newacronym{TP}{TP}{Trabalho Prático}
\newacronym{JT}{JT}{Jogo de Tabuleiro}
\newacronym{JC}{JC}{Jogo de Cartas}
\newacronym{GM}{GM}{Game Master}
\newacronym{ECTS}{ECTS}{European Credit Transfer System}
\newacronym{CESE}{CESE}{Curso de Estudos Superiores Especializados}
\newacronym{ANET}{ANET}{ Academia Nacional dos Engenheiros Técnicos}
\newacronym{FEANI}{FEANI}{Federação Europeia das Associações Nacionais de Engenheiros}
\newacronym{EHEA}{EHEA}{Área Europeia de Ensino Superior}
\newacronym{EQF}{EQF}{Quadro Europeu de Qualificações}
\glsunset{ECTS}
\glsunset{API}
\glsunset{FEANI}

%%--------------- INCLUDE your acronyms below ----------------------------------------

\newacronym{ML}{ML}{Machine Learning}
\newacronym{RL}{RL}{Reinforcement Learning}
\newacronym{GG-CNN}{GG-CNN}{Generative Grasping Convolutional Neural Network}
\newacronym{CV}{CV}{Computer Vision}
\newacronym{AI}{AI}{Artificial Intelligence}
\newacronym{MDP}{MDP}{Markov Decision Process}
\newacronym{LTL}{LTL}{Linear Temporal Logic}
\newacronym[longplural=Collaborative Robots]{Cobot}{Cobot}{Collaborative Robot}
%\newacronym{Cobots}{Cobots}{Collaborative Robots}
\newacronym{DAPG}{DAPG}{Demonstration Augmented Policy Gradient}
\newacronym{CNN}{CNN}{Convolutional Neural Network}
\newacronym{PUnS}{PUnS}{Planning with Uncertain Specifications}
\newacronym{CPU}{CPU}{Central Processing Unit}
\newacronym{GPU}{GPU}{Graphics Processing Unit}
\newacronym{RCNN}{R-CNN}{Regions with Convolutional Neural Networks}
\newacronym{YOLO}{YOLO}{You Only Look Once}
\newacronym{ROS}{ROS}{Robot Operating System}
\newacronym{MC}{MC}{Monte Carlo}
\newacronym{TD}{TD}{Temporal Difference}
\newacronym{ANN}{ANN}{Artificial Neural Network}
\newacronym{CCD}{CCD}{Charge Coupled Device}
\newacronym{CMOS}{CMOS}{Complementary Metal Oxide Semiconductor}
\newacronym{DRL}{deep RL}{Deep Reinforcement Learning}
\newacronym{DQN}{DQN}{Deep Q Learning Network}
\newacronym{ResNet}{ResNet}{Residual Neural Network}
\newacronym[longplural=Deep Convolutional Neural Networks]{DCNN}{DCNN}{Deep Convolutional Neural Network}
\newacronym{RGBD}{RGBD}{Red Green Blue Depth}
\newacronym{LQR}{LQR}{Linear–quadratic regulator}
\newacronym{TCP}{TCP}{Tool Center Point}

%% file: chapters/abstract.tex
\begin{center}
    ABSTRACT
\vspace{5mm} %5mm vertical space
\end{center}

%%%%%%%%%%%%%%%%%%%%%%%%%%%%%%
This thesis delves into the $AdS_4/CFT_3$ correspondence (M-theory on $AdS_4 \times S^7 / \mathbb{Z}_k$ $\leftrightarrow$ ABJM theory) in a comprehensive and detailed manner. The ABJM theory i.e. $\mathcal{N} = 6$ $U(N)_k \times U(N)_{-k}$ Chern-Simons matter theory, is presented in $\mathcal{N} = 2$ superspace, with explicit demonstration of R-symmetry invariance. The associated brane configuration in type IIB string theory is examined, and the low-energy effective field content is shown to align with that of the ABJM theory, by considering string modes and boundary conditions in a new intuitive perspective on brane intersection boundaries and staged compactification. The consequent lift to M-theory, along with the resulting geometry and its Killing spinors in 11D supergravity, is thoroughly analyzed. The $AdS_4/CFT_3$ statement is then articulated, and its implications are explored, with a focus on the computational machinery of Witten diagrams, first in position space and subsequently in the more convenient Mellin space. The thesis culminates with the computation of the four-point function of the stress-tensor multiplet superconformal primary, incorporating contributions from additional short multiplets involving Spin-3 and Spin-4 exchanges not previously considered in the literature. These computations hold significant practical implications for the M-theory S-matrix. As an overall remark, the streamlined nature of this comprehensive work also aims to inspire fresh insights from $AdS_4/CFT_3$, a duality that probes the non-perturbative aspects of M-theory; a theory that, if it exists, is a promising candidate for the theory of everything, yet remains shrouded in Mystery with much still unknown.
%%%%%%%%%%%%%%%%%%%%%%%%%%%%%%

\vspace{5mm} %5mm vertical space

%% file: chapters/Thanks.tex
\chapter*{}
\begin{center}
    \textit{{\larger[2] Thank you to my parents and my younger self}}
\end{center}

%% file: chapters/Introduction.tex
\chapter*{Introduction}\label{chap:Introduction}
The Theory of Everything, a theoretical physicist's ultimate dream, an apogee of philosophical direction, but a beacon so near yet so far. In an attempt to trace the beacon, Physicists unavoidably faced a reconciliation roadblock between the two giants of modern physics i.e. General Relativity and Quantum Mechanics. A mathematically elegant traversal was then provided by String theory, which to this day still remains the most comprehensive candidate for a theory of Quantum Gravity. But while incorporating fermionic degrees of freedom into the theory via supersymmetry, there emerged five different types of superstring theories during the first revolution (1984-1994) namely, Type I, Type IIA, Type IIB, $SO(32)$ Heterotic and $E_8 \times E_8$ Heterotic. This served as a momentary pit stop in the quest for `a' theory, until Edward Witten conjectured the existence of an M-theory in 1995 that unifies all the five superstring theories via a host of dualities (see fig in p.8). 

This sparked the second superstring revolution and was hailed as one of the most promising candidates for a theory of everything. However, nearly two decades later, a self-contained description that fully captures the quantum nature of M-theory remains elusive. Our understanding is still largely confined to a web of dualities, with the theory's low-energy effective limit approximated by classical eleven-dimensional supergravity. Beyond these aspects that probe special sectors, fundamental insights into M-theory are still sparse and have mostly been obtained via holographically dual quantum field theories. One such dual to M-theory in flat 11D spacetime is the \textit{BFSS matrix model}, which describes the quantum mechanics of $N \times N$ matrices in the large $N$ limit, and becomes an $n$-dimensional quantum field theory when $n$ spatial dimensions are toroidally compactified \cite{Becker_Becker_Schwarz_2006}. This conjecture was strongly motivated by the fact that this Matrix theory possesses the correct 11D supergravity (SUGRA) low-energy limit \cite{Banks_1997}. Although there is evidence supporting its validity for small values of $n$, it remains an incomplete description of M-theory, as its generalization to non-toroidal compactifications has yet to be established. While there are many others, two of the theories are dual to M-theory in AdS backgrounds, as conjectured by the $AdS_{d+1}/CFT_d$ (Anti-de Sitter/Conformal Field Theory) correspondence: the 6D (2,0) superconformal field theory in $AdS_7/CFT_6$ \cite{Maldacena_1999} and the 3D $U(N)_k \times U(N)_{-k}$ Chern-Simons matter theory (ABJM) in $AdS_4/CFT_3$ \cite{Aharony_2008}. Both theories probe the full non-perturbative structure of M-theory. However, while the former lacks a Lagrangian formulation, the latter does have one, making it particularly significant for the computability of exact results and the formulation of non-trivial tests. 

Therefore, the goal of this thesis is to conduct an extensive study of the $AdS_4/CFT_3$ correspondence. This involves a meticulous analysis of the components leading to the duality statement, followed by systematically laying the groundwork for the practical computations that the duality entails. The thesis then culminates in a potent example with significant practical implications in the determination of the M-theory S-matrix.
\begin{figure}[h]
\centering
\includegraphics[width=0.8\textwidth]{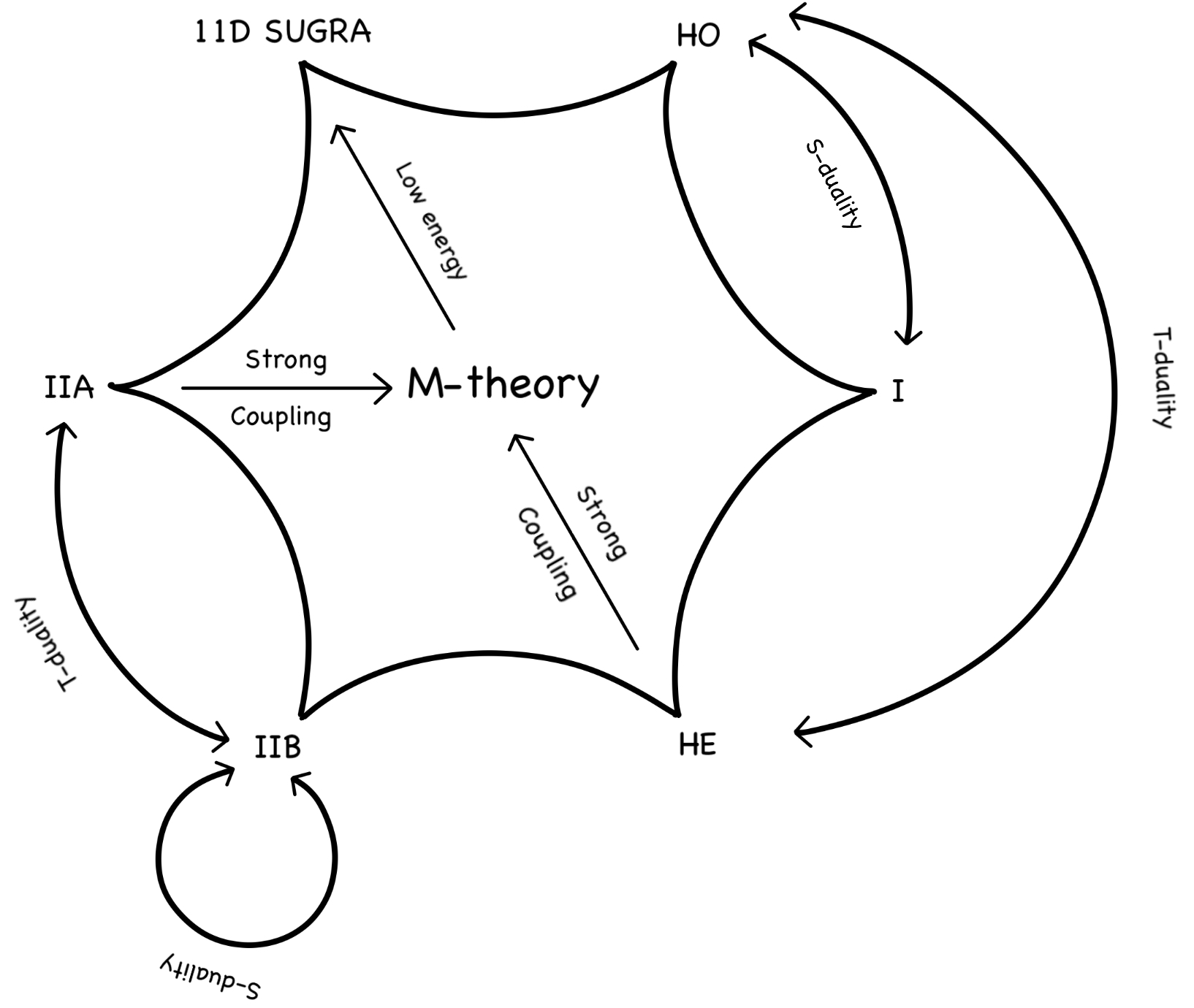}
\end{figure}

\noindent To be more specific for the convenience of the reader, the thesis is structured as follows.
\begin{itemize}[topsep = 5pt]
    \setlength\itemsep{0.5em}
    \item \textbf{Chapter 1}: This chapter delves into supersymmetric Chern-Simons matter theories, beginning with $\mathcal{N} = 2$ and progressing to $\mathcal{N} = 6$ (ABJM). Each theory is concisely described in $\mathcal{N} = 2$ superspace and also explicitly presented in a manifestly R-symmetry invariant form.
    \item \textbf{Chapter 2}: With the field content and action of the ABJM theory established, this chapter provides a detailed analysis of the corresponding brane configuration in type IIB string theory, as introduced in \cite{Aharony_2008}. A new intuitive perspective on brane intersection boundaries and staged compactification is offered. It is explicitly verified through consideration of string modes and boundary conditions that the low-energy effective theory residing on this brane configuration has the same field content as the ABJM theory.
    \item \textbf{Chapter 3}: Building on the previous chapter, this chapter lifts the brane configuration to a setup involving M-branes and Kaluza-Klein monopoles in M-theory, using the dualities illustrated in the figure above. The resulting geometric solution in 11D supergravity (SUGRA) is carefully analyzed, including its Killing spinors, demonstrating that it preserves $\mathcal{N} = 6$ supersymmetry in a special limiting region, consistent with the ABJM theory.
    \item \textbf{Chapter 4}:  Motivated by the groundwork laid in the preceding chapters, this chapter conjectures the $AdS_4/CFT_3$ correspondence statement in it's weak and strong forms. Additionally, other sections then detail the associated superconformal symmetry group $OSp(\mathcal{N} | 4)$ and briefly review three exact tests performed over the last decade, facilitated by the Lagrangian description of the ABJM theory.
    \item \textbf{Chapter 5}: In its weak form, the precise statement relates the generating functional of correlators in CFT to the on-shell supergravity action in AdS. This chapter explores the technical machinery required for such computations on the gravity side, known as Witten diagrams. Additionally, the latter parts of the chapter introduce modern concepts of CFT correlators, such as conformal blocks.
    \item \textbf{Chapter 6}: Considering the challenges in deriving closed-form expressions for Witten diagrams in position space, the first half of this final chapter delves into Mellin space, introduced by Mack in \cite{mack2009dindependent}, which serves as the analogue of momentum space for flat space scattering amplitudes. For the diagrams of interest, known as exchange diagrams, it was demonstrated in \cite{Hijano_2016} that Geodesic Witten diagrams correspond to conformal blocks and elegantly separate single trace and double trace operators in the OPE. This aligns well with the definition of Mellin amplitudes, which include only single trace poles, and thus, the derivation for arbitrary spin exchange is outlined. The second half of the chapter focuses on a specific example in $AdS_4/CFT_3$: the four-point function of the stress tensor multiplet. While previous literature \cite{Zhou_2018} considered only the exchange of operators within the stress-tensor multiplet, this thesis extends the computation to include potential contributions from other short multiplets, resulting in Spin-3 and Spin-4 exchanges. Finally, these computations are contextualized by a conjecture from Penedones \cite{Penedones_2011} that relates them to the M-theory S-matrix, and the thesis is concluded by setting the stage for future research.
\end{itemize}
While the motivations for the topics considered in this thesis have hopefully been made clear thus far, the reader may wonder about the purpose of this specific compositional structure. The answer is twofold. Firstly, to my knowledge, there is no single work that is as self-contained, detailed, introductory, and methodical with regards to the body of research concerning $AdS_4/CFT_3$. The effort to be as comprehensive as possible regarding the foundational topics ensures that readers of this thesis can approach almost any work within $AdS_4/CFT_3$ with familiarity. Secondly, this structure streamlines the progression from conceptualizing a duality to performing practical computations within it. This is because, I firmly believe that even seemingly mundane topics can reveal profound insights when approached with the right perspective and structured thought. Consequently, a few potentially novel insights into M-theory occurred to me, which are not included here due to being beyond the scope of this thesis, but which I intend to pursue in future research. I hope this work similarly inspires new insights, intuitions, or perspectives for the reader, because for a physicist, understanding how things work ultimately takes great precedence.

%% file: chapters/research_design.tex
\chapter{Chern-Simons matter theories}\label{chap:research}

The pure Chern-Simons theory in 2+1 dimensions is a topological field theory, characterized by a gauge group $\mathcal{G}$ and a level \textit{k}. The action in the language of differential forms is given by 
\begin{equation} \label{eq:CSA}
    S_\mathcal{CS} \equiv \frac{\textit{k}}{4\pi}\int_\mathcal{M} \,\, \text{Tr}\left(A \wedge dA + \frac{2}{3}A\wedge A\wedge A\right)
\end{equation}
where $\mathcal{M}$ is a topological 3-manifold, \textit{A} is the 1-form associated with the gauge field and transforms in the adjoint representation of $\mathcal{G}$. It can be seen that the metric doesn't appear anywhere in the action making the theory topological, and it can also trivially be noted that \textit{k} is dimension-less. Now under gauge transformations, the gauge field and correspondingly the action transform (up to a total divergence) as 
\begin{equation} \label{eq:CSG}
\begin{split}
    &A \rightarrow A' = g\,A\,g^{-1} + g \, dg^{-1} \,\,\,;\,\,\, g \in \mathcal{G} \\
    &S_\mathcal{CS} \rightarrow S'_\mathcal{CS} = S_\mathcal{CS} + 2\pi\textit{k}\cdot\delta S_\mathcal{CS} \\ &\delta S_\mathcal{CS} = \frac{1}{24\pi^2}\int \,\, \text{Tr}\left(g\,dg^{-1} \wedge g \, dg^{-1} \wedge g \, dg^{-1}\right)
\end{split}
\end{equation}
As can be seen from the proof provided in \cite{coleman_1985} under the study of instantons, referred to as the winding number, $\delta S_\mathcal{CS} \propto n \in \mathbb{Z}$. This then implies that, for $\delta S_\mathcal{CS} \neq 0$, the classical theory is not uniquely defined. However, the quantum theory can still be uniquely defined as long as the amplitude under the path integral is single-valued under the transformations above i.e., $e^{iS'_\mathcal{CS}} = e^{iS_\mathcal{CS}}$. This forces the condition, $\textit{k} \in \mathbb{Z}$. We can also vary the action (\ref{eq:CSA}) and figure out it's classical equations of motion, which are
\begin{equation}\label{eq:eomp}
    F = dA + A\wedge A = 0
\end{equation}
Under the gauge transformations (\ref{eq:CSG}), $F = 0 \implies F' = dA'+A'\wedge A' = 0$, which then allows us to find an apposite $g \in \mathcal{G}$ s.t $A\equiv 0$ locally. This implies that the theory has no propagating on-shell degrees of freedom (not true for higher dimensional theories \cite{Banados:1995mq}).

\section{Supersymmetric Chern-Simons-matter theories}

The natural progression from the pure theory is to add fermionic d.o.f and thereby make it supersymmetric. Since the gauge field \textit{A} (bosonic) in 2+1 dimensions has two off-shell d.o.f and zero
on-shell d.o.f, closure of supersymmetry algebra off-shell requires the completion of the supermultiplet, while the closure on-shell requires the additional use of eqn.(\ref{eq:eomp}). The off-shell supersymmetric CS theories coupled to matter, that are still conformal, were constructed in (\cite{doi:10.1142/S0217751X93001363, Schwarz_2004}) for the cases of $\mathcal{N}$ = 1, 2. Given the pivotal role of the $\mathcal{N}$ = 2 construction in the upcoming discussion, let's briefly introduce its superspace formulation below.

\subsection{$\mathcal{N}$ = 2 CSM Theory}
The Lorentz group in 2+1 dimensions is $SO(2,1)$, and it's covering group while talking about fermions is $Spin(2,1) \cong SL(2, \mathbb{R})$. The spinor representations therefore correspond to 2-component Majorana (real) spinors, and consequently $\mathcal{N}$ = 2 superspace consists of four fermionic d.o.f. Since the construction involves matter coupling, two $\mathcal{N}$ = 2 supermultiplets i.e., Vector (gauge)[\textit{V}] and Chiral (matter)[\textit{$\Phi$}] are considered \cite{Schwarz_2004}
\begin{itemize}[topsep = 5pt]
    \setlength\itemsep{0.3em}
    \item $\textit{\textbf{V}} : \{A_\mu, \chi, \sigma, \textit{D}\} \in \textit{\textbf{V}}$ \,;\, $A_\mu$ is the gauge field, $\chi$ is the two Majorana spinors combined into one complex spinor, $\sigma$ is a real scalar, \textit{D} is a real auxiliary scalar.
    
    \item $\mathbf{\Phi} : \{\phi, \psi, \textit{F}\} \in \mathbf{\Phi}$ \,;\, $\phi$ is a complex scalar, $\psi$ is the two Majorana spinors combined into a complex spinor, \textit{F} is a complex auxiliary scalar.
\end{itemize}
It can be seen that each supermultiplet has four bosonic and four fermionic off-shell d.o.f, thereby ensuring off-shell closure of supersymmetry. The four real supercharges can then be combined into a complex spinor $Q_\alpha(\alpha = 1, 2)$, and the corresponding supersymmetry algebra can then be written in the basis of $Q_\alpha$ and it's conjugate $\bar{Q}_\alpha$ (\cite{Aharony_1997, McKeon:2001su})
\begin{equation} \label{eq:ccalg}
    \{Q_\alpha, \bar{Q}_\beta\} = 2\gamma^\mu_{\alpha\beta}P_\mu + 2i\epsilon_{\alpha\beta}Z
\end{equation}
where $P_\mu$ is the 3-momentum in 2+1 dimensions, $Z$ is the central charge, and $\{\gamma^0, \gamma^1, \gamma^2\} = \{i\sigma_2, \sigma_1, \sigma_3\}$; $\sigma_i$ are the Pauli matrices. In the language of superspace, spanned by $\{x^\mu, \theta^\alpha, \bar{\theta}_\alpha\}$, the aforementioned supercharges and the corresponding supercovariant derivatives are given by
\begin{equation}\label{eq:superder}
    \begin{split}
        & Q_\alpha = \frac{\partial}{\partial \theta^\alpha} - \left(\gamma^\mu_{\alpha \beta} \bar{\theta}^\beta \right)\frac{\partial}{\partial x^\mu} \,\,,\,\, \bar{Q}_\alpha = -\frac{\partial}{\partial \bar{\theta}^\alpha} + \left(\theta^\beta \gamma^\mu_{\beta \alpha}\right)\frac{\partial}{\partial x^\mu} \\
        & D_\alpha = \frac{\partial}{\partial \theta^\alpha} + \left(\gamma^\mu_{\alpha \beta} \bar{\theta}^\beta \right)\frac{\partial}{\partial x^\mu} \,\,,\,\, \bar{D}_\alpha = -\frac{\partial}{\partial \bar{\theta}^\alpha} - \left(\theta^\beta \gamma^\mu_{\beta \alpha}\right)\frac{\partial}{\partial x^\mu}
    \end{split}
\end{equation}
where the raising and lowering of spinor index $\alpha$ is done using $\epsilon^{\alpha \beta}, \epsilon_{\alpha \beta} : \epsilon^{12} = \epsilon_{21} = 1$; This is because the finite-dimensional representation theory of $SL(2, \mathbb{R})$ $\equiv$ representation theory of $SU(2)$, which is the covering group of the group of rotations in the Wick-rotated theory i.e., $SO(3)$.  Now coming to the superfields, the vector-superfield $V$ in the Wess-Zumino gauge, the chiral superfield $\Phi$ : $\bar{D}_\alpha \Phi = 0$ and the anti-chiral superfield $\bar{\Phi}$ : $D_\alpha \bar{\Phi} = 0$  are written as \cite{SUSYDavid}
\begin{align}
        &V(x, \theta, \bar{\theta}) = -\theta \gamma^\mu \bar{\theta} A_\mu(x) - \theta \bar{\theta} \sigma(x) + i \theta^2 \bar{\theta}\bar{\chi}(x) - i {\bar{\theta}}^2 \theta \chi(x) + \frac{1}{2}\theta^2\bar{\theta}^2 D(x)\label{eq:fields}\\
        &\Phi(x, \theta, \bar{\theta}) = \phi(x) + \sqrt{2} \theta \psi(x) + \theta^2 F(x) + i\theta \gamma^\mu \bar{\theta}\partial_\mu \phi(x) - \frac{i}{\sqrt{2}}\theta^2 \partial_\mu \psi(x) \gamma^\mu \bar{\theta} - \frac{1}{4}\theta^2 \bar{\theta}^2 \partial^2 \phi(x) \nonumber\\
        &\bar{\Phi}(x, \theta, \bar{\theta}) = \bar{\phi}(x) + \sqrt{2} \bar{\theta} \bar{\psi}(x) + \bar{\theta}^2 \bar{F}(x) - i\theta \gamma^\mu \bar{\theta}\partial_\mu \bar{\phi}(x) + \frac{i}{\sqrt{2}}\bar{\theta}^2\theta \gamma^\mu \partial_\mu \bar{\psi}(x) - \frac{1}{4}\theta^2 \bar{\theta}^2 \partial^2 \bar{\phi}(x)\nonumber
\end{align}
where $\theta^2 = \theta_\alpha \theta^\alpha$, $\bar{\theta}^2 = \bar{\theta}_\alpha \bar{\theta}^\alpha$, $\partial^2 = \partial_\mu \partial^\mu$. The above mentioned $d = 3, \, \mathcal{N} = 2$ vector-superfield $V$ is obtained by dimensional reduction of the $d = 4, \,\mathcal{N} = 1$ vector-superfield; where $\sigma$ corresponds to $A_3$, which is the dimensionally-reduced component of the gauge field $A_\mu$ in $d = 4$ \cite{Schwarz_2004}. The central charge $Z$ in (\ref{eq:ccalg}) also corresponds to $P_3$, which is the dimensionally reduced component of the 4-momentum $P_\mu$ in $d = 4$. Consequently, the $d = 3, \, \mathcal{N} = 2$ Chern-Simons-matter theory can be obtained by the dimensional reduction of $d = 4, \; \mathcal{N} = 1$ Yang-Mills-matter theory, with the exception that the kinetic part of the vector supermultiplet is replaced by the supersymmetric version of (\ref{eq:CSA}). The relevant superspace action that does this is given by the sum of a non-trivial superspace kinetic term and the usual gauge invariant K\"{a}hler potential, as follows \cite{Gaiotto_2007}
\begin{equation}\label{eq:susycsm}
    S^{\mathcal{N} = 2}_{CSM} = \mathlarger{\mathlarger{\int}} d^3x \, \mathlarger{\mathlarger{\int}} d^4\theta \,\, \left\{\frac{k}{2\pi}\mathlarger{\int}_0^1 dt \,\, \text{Tr}\left[V\bar{D}^\alpha\left(e^{-tV}D_\alpha e^{tV}\right)\right] + \sum_{i = 1}^{N_f} \bar{\Phi}^i e^V \Phi^i \right\}
\end{equation}
where $i$ is the flavor index corresponding to a global $U(N_f)$ flavor symmetry acting on $\Phi$. Tr is normalized to be the trace in the fundamental representation when the gauge group is $U(N)$ or $SU(N)$, with $T^a$ as generators, so that $\text{Tr}(T^a T^b) = \frac{1}{2}\delta^{ab}$. Also, $\Phi^i$ is now a vector acted on by the representation $R_i$ of the gauge group. Now using this representation and the relations for spinors $\lambda, \epsilon$ : $\lambda\epsilon = \lambda_\alpha \epsilon^\alpha = \epsilon_\alpha \lambda^\alpha = \epsilon\lambda$\,\,,\,\,$\lambda\gamma^\mu\epsilon = \lambda_\alpha {\gamma^\alpha}_\beta \epsilon^\beta = -\epsilon_\alpha {\gamma^\alpha}_\beta \lambda^\beta = -\epsilon\gamma^\mu\lambda$\,\,,\,\,$(\gamma^\mu \lambda)\epsilon = -\lambda\gamma^\mu\epsilon$\,\,,\,\,$(\lambda \bar{\lambda})(\epsilon \bar{\epsilon}) = -(\lambda \epsilon)(\bar{\lambda}\bar{\epsilon}) -(\lambda\bar{\epsilon})(\epsilon\bar{\lambda})$; Substituting (\ref{eq:superder}) and (\ref{eq:fields}) into the kinetic part of (\ref{eq:susycsm}) and computing the integral over superspace yields
\begin{align}
        S^{\mathcal{N} = 2}_{CS} &= \mathlarger{\int} d^3x \, \mathlarger{\int} d^4\theta \,\, \frac{k}{2\pi}\mathlarger{\int}_0^1 dt \,\, \text{Tr}\left[V\bar{D}^\alpha\left(e^{-tV}D_\alpha e^{tV}\right)\right] \label{eq:susycs}\\
        &= \frac{k}{4\pi}\mathlarger{\int}\,\, \text{Tr}\left(A^aT^a\wedge dA^bT^b + \frac{2}{3}A^aT^a\wedge A^bT^b\wedge A^cT^c\right) - \bar{\chi}^a\chi^b\,\delta^{ab} + D^a\sigma^b\,\delta^{ab} \nonumber
\end{align}
where a = 1, 2,..., $dim(\mathcal{G})$, and \{$A_\mu = A_\mu^aT^a, \chi = \chi^aT^a, T^a\bar{\chi}^a = \bar{\chi}, \sigma^aT^a = \sigma, D^aT^a = D$\} $\in$ Lie algebra of the gauge group in the fundamental representation. Now using the definition of covariant derivative $D_\mu$ as follows 
\begin{equation*}
    D_\mu \{\phi^i, \psi^i\} = \partial_\mu \{\phi^i, \psi^i\} + \frac{i}{2}A_\mu^aT^a_{R_i}\{\phi^i, \psi^i\} \,\,\,\,;\,\,\,\, D_\mu \{\bar{\phi}^i, \bar{\psi}^i\} = \partial_\mu \{\bar{\phi}^i, \bar{\psi}^i\} - \frac{i}{2}\{\phi^i, \psi^i\}A_\mu^aT^a_{R_i}\nonumber
\end{equation*}
Substituting (\ref{eq:fields}) into the matter part of (\ref{eq:susycsm}) and integrating over superspace yields  
\begin{align}
        \mathlarger{\int} d^4\theta \,\, \sum_{i = 1}^{N_f} \bar{\Phi}^ie^V\Phi^i = &\sum_{i = 1}^{N_f}\biggl(D_\mu\bar{\phi}^i\,D^\mu\phi^i - i\bar{\psi}^i\gamma^\mu D_\mu \psi^i - \frac{1}{4}\bar{\phi}^i\sigma^a\sigma^b \, T^a_{R_i}T^b_{R_i}\phi^i
        + \frac{1}{2}\bar{\phi}^iD^aT^a_{R_i}\phi^i \nonumber\\
        &- \frac{1}{2}\bar{\psi}^i\sigma^aT^a_{R_i}\psi^i + \frac{i}{\sqrt{2}}\bar{\phi}^i\chi^aT^a_{R_i}\psi^i - \frac{i}{\sqrt{2}}\bar{\psi}^iT^a_{R_i}\bar{\chi}^a\phi^i + \bar{F}^iF^i\biggr) \label{eq:mattercs2}
\end{align}
where $T^a_{R_i}$, a = 1, 2,..., $dim(\mathcal{G})$, are generators of the gauge group $\mathcal{G}$ in representation $R_i$. Also, \{$A^i_\mu = A_\mu^aT^a_{R_i}, \chi^i = \chi^aT^a_{R_i}, T^a_{R_i}\bar{\chi}^a = \bar{\chi}^i, \sigma^aT^a_{R_i} = \sigma^i, D^aT^a_{R_i} = D^i$\} $\in$ Lie algebra of the gauge group in the representation $R_i$. Solving the equations of motion for $D, \chi, \bar{\chi}, F, \bar{F}$ in $S_{CSM}^{\mathcal{N} = 2}$ using (\ref{eq:susycs}) and (\ref{eq:mattercs2}) yields
\begin{equation}
    \begin{alignedat}{3}
        &D^a : \sigma^a = -\frac{2\pi}{k}\bar{\phi}^iT^a_{R_i}\phi^i \,\,\,\,\,\,\,\,\,\,\, &&F, \bar{F} : \bar{F} = 0, F = 0\\[0.5em]
        &\chi^a : \bar{\chi}^a = \frac{4\pi i}{k\sqrt{2}}\bar{\phi}^iT^a_{R_i}\psi^i  &&\bar{\chi}^a : \chi^a = -\frac{4\pi i}{k\sqrt{2}}\bar{\psi}^iT^a_{R_i}\phi^i  
    \end{alignedat} 
\end{equation}
Substituting these in $S_{CSM}^{\mathcal{N} = 2}$ and integrating out the fields gives the following action
\begin{align}
        S_{CSM}^{\mathcal{N} = 2} = S_{CS} + \mathlarger{\int} d^3x \,\, &D_\mu \bar{\phi}^i D^\mu \phi^i - i\bar{\psi}^i \gamma^\mu D_\mu \psi^i - \frac{\pi^2}{k^2}\left(\bar{\phi}^iT^a_{R_i}\phi^i\right)\left(\bar{\phi}^jT^b_{R_j}\phi^j\right)\left(\bar{\phi}^kT^a_{R_k}T^b_{R_k}\phi^k\right)\nonumber \\
        &+ \frac{\pi}{k}\left(\bar{\phi}^iT^a_{R_i}\phi^i\right)\left(\bar{\psi}^jT^a_{R_j}\psi^j\right) + \frac{2\pi}{k}\left(\bar{\psi}^iT^a_{R_i}\phi^i\right)\left(\bar{\phi}^jT^a_{R_j}\psi^j\right)
\end{align}
It can easily be seen from $S_{CS}$ that the dimension of $A_\mu$ is 1, and simultaneously from the gauge covariant kinetic terms for $\phi$ and $\psi$ that their dimensions are $\frac{1}{2}$ and 1 respectively. Consequently, it can clearly be noted that the couplings $\{\frac{\pi^2}{k^2}, \frac{\pi}{k}, \frac{2\pi}{k}\}$ are all dimension-less, hence justifying classical conformal invariance. It has been shown in \cite{KAPUSTIN_1994} that $S_{CS}$ is not renormalized beyond a possible finite 1-loop shift; And in \cite{Gaiotto_2007}, it has been argued that no IR relevant quantum corrections with dimension-less couplings can be added to $S_M^{\mathcal{N} = 2}$. So in conclusion, $\mathcal{N} = 2$ Chern-Simons-matter theory in 2+1 dimensions is conformally invariant even at the quantum level. For a more detailed description of the theory, the reader may refer to \{\cite{doi:10.1142/S0217751X93001363,Schwarz_2004,McKeon:2001su,Ivanov:1991fn}\}.

\subsection{$\mathcal{N} = 3$ CSM Theory}
The $\mathcal{N} = 4$ Yang-Mills-matter theory, whose kinetic term for the vector supermultiplet when replaced by the supersymmetric Chern-Simons term, breaks the $\mathcal{N} = 4$ supersymmetry to $\mathcal{N} = 3$ \cite{Kao:1992ig}. This method of obtaining the $\mathcal{N} = 3$ Chern-Simons-matter theory is in contrast with the $\mathcal{N} = 2$ case, where the supersymmetry was preserved. The three-dimensional $\mathcal{N} = 4$ supermultiplet itself is obtained by the dimensional reduction of either the corresponding six-dimensional $\mathcal{N} = 1$ supermultiplet or the four-dimensional $\mathcal{N}=2$ one \cite{seiberg1996gauge}. Therefore, the three-dimensional vector and matter multiplets for the $\mathcal{N} = 4$ Yang-Mills-matter theory and correspondingly the $\mathcal{N} = 3$ Chern-Simons-matter theory, organized in the $\mathcal{N} = 2$ superspace are as follows
\begin{itemize}[topsep = 2pt]
    \setlength\itemsep{0.3em}
    \item \textbf{Vector multiplet} $\equiv \textbf{V} \,\, | \,\,\textbf{Q} : \{A_\mu, \chi, \sigma, \textit{D}\} \in \textbf{V}\,, \,\{q, \lambda, S\} \in \textbf{Q}$ \,;\, \textbf{V} and \textbf{Q} are $\mathcal{N} = 2$ vector and chiral multiplets respectively, in the adjoint representation of $\mathcal{G}$
    
    \item \textbf{Matter multiplet} $\equiv (\mathbf{\Phi^i}, \mathbf{\tilde{\Phi}^i}) : \{\phi^i, \psi^i, \textit{F}^i\} \in \mathbf{\Phi^i}\,,\, \{\tilde{\phi}^i, \tilde{\psi}^i, \tilde{\textit{F}}^i\} \in \mathbf{\tilde{\Phi}^i}$ \,;\, $\mathbf{\Phi^i}$ and $\mathbf{\tilde{\Phi}^i}$ are $\mathcal{N} = 2$ chiral multiplets, transforming under representation $R_i$ and conjugate representation $\bar{R}_i$ of $\mathcal{G}$ respectively.
\end{itemize}
It can be seen that an auxiliary chiral multiplet $Q$ has been augmented to the $\mathcal{N} = 2$ vector multiplet $V$, whose scalars $\sigma$ and q (complex) correspond to the three components dimensionally reduced from the six-dimensional gauge field. The relevant action for $\mathcal{N} = 3$ Chern-Simons-matter theory in terms of the superfields above, in $\mathcal{N} = 2$ superspace is \cite{Gaiotto_2007}
\begin{equation}\label{eq:n3susycsm}
    \begin{split}
        S^{\mathcal{N} = 3}_{CSM} = S^{\mathcal{N} = 2}_{CS} &+ \mathlarger{\int} d^3x \mathlarger{\int} d^4\theta \,\, \sum_{i = 1}^{N_f} \left(\bar{\Phi}^ie^V\Phi^i + \tilde{\Phi}^ie^{-V}\bar{\tilde{\Phi}}^i\right)\\
        &+ \left[\mathlarger{\int} d^3x \mathlarger{\int} d^2\theta \,\, \left(\frac{k}{2\pi} \text{Tr}Q^2 - \sum_{i = 1}^{N_f} \tilde{\Phi}^i Q \Phi^i \right) + c.c\right]
    \end{split}
\end{equation}
where c.c stands for complex conjugate. In order to supersymmetrize the $V \,\,| \,\,Q$ multiplet with a Chern-Simons term, the $\{q, \lambda, S\} \in Q$ have to introduce terms in the action similar to how $\{\sigma, \chi, D\} \in V$ did respectively in (\ref{eq:susycs}). Tr$Q^2 + c.c$ does this job when introduced as an F-term in (\ref{eq:n3susycsm}), with it's coefficient fixed by supersymmetry and it's form fixed by the requirement of holomorphicity of a superpotential. This can be seen by writing the superspace expansion for $Q$ similar to (\ref{eq:fields}) and substituting it
\begin{equation}\label{eq:rsymmsee}
    \left[\mathlarger{\int} d^2\theta \,\, \left(\frac{k}{2\pi} \text{Tr}Q^2\right)\right] + c.c = \frac{k}{4\pi} \left(-\lambda^a_\alpha \lambda^{b\alpha} - \bar{\lambda}^a_\alpha \bar{\lambda}^{b\alpha}  + S^aq^b + \bar{S}^a\bar{q}^b\right)\delta^{ab}
\end{equation}
This term is also what breaks the $\mathcal{N} = 4$ supersymmetry to $\mathcal{N} = 3$ as mentioned in an earlier paragraph \cite{Kao:1992ig}. Among the remaining terms, $\tilde{\Phi}^i Q \Phi^i$ is the F-term that entails the $\mathcal{N} = 4$ supersymmetric completion of the usual gauge-invariant K\"{a}hler potential in $\mathcal{N} = 2$ \cite{MatteoBert}. Now $Q$ can be integrated out since it is auxiliary and has no dynamical d.o.f.
\begin{equation}\label{eq:supot}
\begin{split}
    Q \,\, : \,\, &Q^a = \frac{2\pi}{k} \left(\tilde{\Phi}^i T_{R_i}^a \Phi^i\right)\\
    \implies &W = \left(\frac{k}{2\pi} \text{Tr}Q^2 - \sum_{i = 1}^{N_f} \tilde{\Phi}^i Q \Phi^i \right) = -\frac{\pi}{k}\left(\tilde{\Phi}^i T_{R_i}^a \Phi^i\right)\left(\tilde{\Phi}^j T_{R_j}^a \Phi^j\right)
\end{split}   
\end{equation}
Therefore, the $\mathcal{N} = 3$ Chern-Simons-matter theory is just the superpotential $W$ added to the $\mathcal{N} = 2$ theory with matter as a hypermultiplet $(\Phi, \tilde{\Phi})$. Now taking the superspace expansions for the fields $\Phi, \tilde{\Phi}, Q$ similar to (\ref{eq:fields}), and substituting in $S^{\mathcal{N} = 3}_{CSM}$ yields
\small
\begin{flalign}
        &\mathlarger{\int} d^4\theta \,\, \sum_{i = 1}^{N_f} \left(\bar{\Phi}^ie^V\Phi^i + \tilde{\Phi}^ie^{-V}\bar{\tilde{\Phi}}^i\right)\nonumber = D_\mu \bar{\phi}^i D^\mu\phi^i  + D_\mu \tilde{\phi}^iD^\mu \bar{\tilde{\phi}}^i - i\bar{\psi}^i\gamma^\mu D_\mu\psi^i - i\tilde{\psi}^i\gamma^\mu D_\mu \bar{\tilde{\psi}}^i + \bar{F}^iF^i &&\\
        &- \frac{1}{4}\bar{\phi}^i\sigma^a\sigma^bT^a_{R_i}T^b_{R_i}\phi^i
        - \frac{1}{4}\tilde{\phi}^i \sigma^a\sigma^b T^a_{R_i}T^b_{R_i} \bar{\tilde{\phi}}^i
        + \frac{1}{2}\bar{\phi}^iD^aT^a_{R_i}\phi^i - \frac{1}{2}\tilde{\phi}^iD^aT^a_{R_i}\bar{\tilde{\phi}}^i - \frac{1}{2}\bar{\psi}^i\sigma^aT^a_{R_i}\psi^i\label{eq:n3matter}&&\\
        &+ \frac{1}{2}\tilde{\psi}^i\sigma^aT^a_{R_i}\bar{\tilde{\psi}}^i +\frac{i}{\sqrt{2}}\bar{\phi}^i\chi^aT^a_{R_i}\psi^i - \frac{i}{\sqrt{2}}\tilde{\psi}^i\chi^aT^a_{R_i}\bar{\tilde{\phi}}^i - \frac{i}{\sqrt{2}}\bar{\psi}^i\bar{\chi}^aT^a_{R_i}\phi^i + \frac{i}{\sqrt{2}}\tilde{\phi}^i\bar{\chi}^aT_{R_i}^a\bar{\tilde{\psi}}^i + \tilde{F}^i\bar{\tilde{F}}^i\nonumber&&
\end{flalign}
\normalsize
Similarly, substituting the superspace expansions of fields into the superpotential (\ref{eq:supot}) 
\small
\begin{flalign}
        \mathlarger{\int} d^2\theta \,\, W &= \frac{\pi}{k}\biggl[-2\left(\tilde{\phi}^i T_{R_i}^a F^i\right)\left(\tilde{\phi}^jT_{R_j}\phi^j\right) -2\left(\tilde{F}^iT_{R_i}^a\phi^i\right)\left(\tilde{\phi}^jT_{R_j}^a\phi^j\right) + \left(\tilde{\psi}^iT_{R_i}^a\phi^i\right)\left(\tilde{\psi}^jT_{R_j}^a\phi^j\right) \nonumber&&\\
        & +2\left(\tilde{\psi}^i T_{R_i}^a\phi^i\right)\left(\tilde{\phi}^j T_{R_j}^a\psi^j\right) + \left(\tilde{\phi}^iT_{R_i}^a\psi^i\right)\left(\tilde{\phi}^jT_{R_j}^a\psi^j\right) + 2\left(\tilde{\psi}^i T_{R_i}^a \psi^i\right)\left(\tilde{\phi}^jT_{R_j}^a\phi^j\right)\biggr]\label{eq:n3supot}&&
\end{flalign}
\normalsize
As mentioned earlier, it may be noted that the fields of the form $a^i, \bar{\tilde{a}}^i$ transform in the representation $R_i$ of $\mathcal{G}$, while fields of the form $\bar{a}^i, \tilde{a}^i$ transform in the conjugate representation $\bar{R}_i$ of $\mathcal{G}$. Now solving for the equations of motion for the auxiliary fields $F, \bar{F}, \tilde{F}, \bar{\tilde{F}}, \sigma, D, \chi, \bar{\chi}$ in (\ref{eq:n3susycsm}) yields
\small
\begin{equation}
    \begin{alignedat}{3}
        &F^i : \bar{F}^i = \frac{2\pi}{k}\tilde{\phi}^iT_{R_i}^a\left(\tilde{\phi}^jT_{R_j}^a\phi^j\right) \,\,\,\,\,\,\,\,\,\,\,\, &&\bar{F}^i : F^i = \frac{2\pi}{k}T_{R_i}^a\bar{\tilde{\phi}}^i\left(\bar{\phi}^jT_{R_j}^b\bar{\tilde{\phi}}^j\right)\\&\tilde{F}^i : \bar{\tilde{F}}^i = \frac{2\pi}{k}T_{R_i}^a\phi^i\left(\tilde{\phi}^jT_{R_j}^a\phi^j\right) \,\,\,\,\,\,\,\,\,\,\,\, &&\bar{\tilde{F}}^i : \tilde{F}^i = \frac{2\pi}{k}\bar{\phi}^iT_{R_i}^a\left(\bar{\phi}^jT_{R_j}^a\bar{\tilde{\phi}}^j\right)\\
        &\chi^a : \bar{\chi}^a = \frac{4\pi i}{k\sqrt{2}}\left(\bar{\phi}^iT_{R_i}^a\psi^i - \tilde{\psi}^iT_{R_i}^a\bar{\tilde{\phi}}^i\right) \,\,\,\,\,\,\,\,\,\,\,\, &&\bar{\chi}^a : \chi^a = \frac{4\pi i}{k\sqrt{2}}\left(\tilde{\phi}^iT_{R_i}^a\bar{\tilde{\psi}}^i - \bar{\psi}^iT_{R_i}^a\phi^i\right)\\
        &D^a : \sigma^a = \frac{2\pi}{k}\left(\tilde{\phi}^iT_{R_i}^a\bar{\tilde{\phi}}^i - \bar{\phi}^iT_{R_i}^a\phi^i\right)
    \end{alignedat}
\end{equation}
Substituting these back into (\ref{eq:n3matter}) and (\ref{eq:n3supot}), thereby integrating out the auxiliary fields, and writing the $S_{CSM}^{\mathcal{N} = 3}$ action in a manifestly $SU(2)$ invariant form, gives rise to
\vspace{1.25em}
\hrule width \textwidth
\vspace{-2.45em}
\small
\begin{flalign}
    S_{CS}^{\mathcal{N} = 3} &= S_{CS} + \mathlarger{\int} d^3x \,\,\biggl[D_\mu \bar{M}^i_\phi \mathbb{1}_2 D^\mu M^i_\phi -i\bar{M}_\psi^i\mathbb{1}_2\gamma^\mu D_\mu M_\psi^i - \frac{\pi^2}{k^2}\left(\bar{M}_\phi^iT_{R_i}^a T_{R_i}^b \mathbb{1}_2 M_\phi^i\right)\vec{S}_{\phi\phi}^a \cdot \vec{S}_{\phi\phi}^b \nonumber&&\\ &- \frac{\pi}{k}\vec{S}_{\psi\psi}^a \cdot \vec{S}_{\phi\phi}^a + \frac{\pi}{k}\left(S^a_{\phi\psi}\right)^\mu \left(S^a_{\psi\phi}\right)_\mu
    + \frac{\pi}{2k}\left(\bar{S}_{\phi\psi}^a\right)^\mu\left(S^a_{\phi\psi}\right)_\mu + \frac{\pi}{2k}\left(\bar{S}_{\psi\phi}^a\right)^\mu \left(S^a_{\psi\phi}\right)_\mu\biggr]&&\label{eq:n3fullaction}
\end{flalign}
\vspace{-0.55em}
\hrule width \textwidth
\vspace{-0.8em}
\normalsize
\noindent where $M^i_\phi, M^i_\psi$ are $SU(2)$ doublets, and $S^j_{\phi\phi}, S^j_{\psi\psi}, S^j_{\phi\psi}, S^j_{\psi\phi}\,;\, j = 1, 2, 3$ are $SU(2)$ triplets. 
\small
\begin{equation}\label{eq:doublets}
    \left(M_\phi^i\right)^A = \begin{pmatrix}\phi^i \\ \bar{\tilde{\phi}}^i\end{pmatrix} \,\,\,\,\,\,\,\, \left(M_\psi^i\right)_A = \begin{pmatrix}\psi^i \\ \bar{\tilde{\psi}}^i\end{pmatrix}
\end{equation}
\normalsize
\noindent$A = 1, 2$ is the $SU(2)$ index, whose raising and lowering is performed by $\epsilon^{AB}, \epsilon_{AB}$; $\epsilon^{12} = \epsilon_{21} = 1$. The triplets in the adjoint representation are then given by the following
\pagebreak
\begin{equation}\label{eq:Rsymmref}
    \begin{alignedat}{4}
        &\vec{S}_{\phi\phi}^a = \bar{M}_\phi^i T_{R_i}^a \vec{\sigma}M_\phi^i \,\,\,\,\,\,\,\, &&\left(S^a_{\phi\psi}\right)^\mu = \bar{M}_\phi^iT_{R_i}^a\sigma^\mu M_\psi^i \,\,\,\,\,\,\,\, &&&\left(\bar{S}_{\phi\psi}^a\right)^\mu = \bar{M}_\phi^i T_{R_i}^a \bar{\sigma}^\mu M_\psi^i\\
        &\vec{S}_{\psi\psi}^a = \bar{M}_\psi^i T_{R_i}^a \vec{\sigma}M_\psi^i \,\,\,\,\,\,\,\, &&\left(S^a_{\psi\phi}\right)^\mu = \bar{M}_\psi^iT_{R_i}^a\sigma^\mu M_\phi^i \,\,\,\,\,\,\,\, &&&\left(\bar{S}_{\psi\phi}^a\right)^\mu = \bar{M}_\psi^i T_{R_i}^a \bar{\sigma}^\mu M_\phi^i
    \end{alignedat}
\end{equation}
\vspace{-1em}
\begin{align*}
    \sigma^\mu = \left(\mathbb{1}_2, \vec{\sigma}\right) \,\,\,\,\,\, \bar{\sigma}^\mu = \left(-\mathbb{1}_2, \vec{\sigma}\right)
\end{align*}
where $\vec{\sigma}$ is the vector of Pauli matrices, and the contraction $\bar{M}\vec{\sigma}M \equiv \bar{M}^A\vec{\sigma}_{AB}M^B$. This $SU(2)$ invariance is an artifact of the R-symmetry group of the theory, which in the case of odd no. of dimensions and the minimal spinor being Majorana, is $SO(\mathcal{N})$ \cite{vanproeyen2016tools}; where $\mathcal{N}$ is the no. of supersymmetries. While it is clear from the action above that the matter transforms as doublets under this $SU(2)_R$, it can further be noted from (\ref{eq:rsymmsee}) that the four Majorana fermions packaged as $\chi, \lambda$ in the vector multiplet transform as a triplet and a singlet, and the three real scalars packaged as $q, \sigma$ transform as a triplet (as do the auxiliaries $D, S$). This $SU(2)_R$ symmetry also places a stronger constraint on renormalizability in contrast to the $\mathcal{N} = 2$ theory, as the $SU(2)_R$ charge of the fields cannot be renormalized. As a result, while the classical conformal invariance is manifest in (\ref{eq:n3fullaction}) with dimension-less couplings, invariance at the quantum level is argued in \cite{Gaiotto_2007}. For a more detailed description of the theory, the reader may refer to \{\cite{Kao:1992ig}, \cite{Kao_1996}, \cite{PhysRevD.50.2881}\}.

\subsection{$\mathcal{N} = 3$ CSM with Product Gauge group}\label{section:n3product}
As will be seen further down the line in the thesis, the theory of interest will have to be parity invariant, let's say under $x^2 \rightarrow -x^2$. It can clearly be noted that $S_{CS} \rightarrow -S_{CS}$ under this transformation, and therefore the theory will have to be gauged under a product gauge group with opposite Chern-Simons levels i.e., $\mathcal{G}_k \times \mathcal{G}_{-k}$, in order to respect parity invariance. This subsection thereby focuses on extending the prior $\mathcal{N} = 3$ theory to the case of the product gauge group $U(N)_k \times U(N)_{-k} \,\,(\text{or} \,\, SU(N)_k \times SU(N)_{-k})$, and the potential supersymmetry enhancement that comes with it. The multiplet content of this product gauge group theory can easily be foreseen from the multiplet structure of the prior $\mathcal{N} = 3$ theory with a single gauge group. Meaning, there will be an $\mathcal{N} = 3$ vector multiplet in the adjoint representation of each of the gauge group \{$V_r \,\, | \,\, Q_r$; r = 1, 2\} and similarly, there will now be two matter hypermultiplets \{($\Phi_r, \tilde{\Phi}_r$), r = 1, 2\} ($A_i, B_i$ in \cite{Aharony_2008}). 
\begin{itemize}[topsep = 10pt]
    \setlength\itemsep{0.5em}
    \item \textbf{Vector multiplet} $\in \mathcal{G}_k$ $\equiv \mathbf{V_1} \,\, | \,\,\mathbf{Q_1} : \{A_{1\mu}, \chi_1, \sigma_1, \textit{D}_1\} \in \mathbf{V_1}\,, \,\{q_1, \lambda_1, S_1\} \in \mathbf{Q_1}$ \,;\, $\mathbf{V_1}$ and $\mathbf{Q_1}$ are $\mathcal{N} = 2$ vector and auxiliary chiral multiplets respectively, in the adjoint representation of $\mathcal{G}_k$

    \item \textbf{Vector multiplet} $\in \mathcal{G}_{-k}$ $\equiv \mathbf{V_2} \,\, | \,\,\mathbf{Q_2} : \{A_{2\mu}, \chi_2, \sigma_2, \textit{D}_2\} \in \mathbf{V_2}\,, \,\{q_2, \lambda_2, S_2\} \in \mathbf{Q_2}$ \,;\, $\mathbf{V_2}$ and $\mathbf{Q_2}$ are $\mathcal{N} = 2$ vector and auxiliary chiral multiplets respectively, in the adjoint representation of $\mathcal{G}_{-k}$ 
    
    \item \textbf{Matter multiplet} $\equiv (\mathbf{\Phi_1}, \mathbf{\tilde{\Phi}_1}) \,\,\& \,\,(\mathbf{\Phi_2}, \mathbf{\tilde{\Phi}_2}) : \{\phi_r, \psi_r, \textit{F}_r\} \in \mathbf{\Phi_r}\,,\, \{\tilde{\phi}_r, \tilde{\psi}_r, \tilde{\textit{F}}_r\} \in \mathbf{\tilde{\Phi}_r}$ for r = 1, 2\,;\, $\mathbf{\Phi_r}$ and $\mathbf{\tilde{\Phi}_r}$ are $\mathcal{N} = 2$ chiral multiplets, transforming under the bifundamental ($N, \bar{N}$) and anti-bifundamental ($\bar{N}, N$) representations of $\mathcal{G}_k \times \mathcal{G}_{-k}$ respectively, where $\mathcal{G}$ is now $U(N) \,\, (\text{or} \,\,SU(N))$
\end{itemize}
Incase the reader is wondering as to why $\Phi$ is chosen to lie in the $(N, \bar{N})$ representation rather than $(N, N)$, you are free to choose the latter instead and it is perfectly fine. This is because the form of the lagrangian remains unchanged, as the gauge group is unitary and as there is also $\tilde{\Phi}$ in the conjugate representation anyways. The corresponding action in $\mathcal{N} = 2$ superspace can simply be written as an extension of (\ref{eq:n3susycsm}), by now imposing gauge invariance under the product group on the extended set of fields. 
\small
\begin{align}
    S_{ABJM} &= \left(S_{CS}^{\mathcal{N} = 2}\right)_k + \left(S_{CS}^{\mathcal{N} = 2}\right)_{-k} + \mathlarger{\int} d^3x \mathlarger{\int} d^4\theta \,\, \text{Tr}_2\left(\bar{\Phi}_r e^{V_1} \Phi_r e^{-V_2}\right) + \text{Tr}_1\left(\bar{\tilde{\Phi}}_r e^{V_2} \tilde{\Phi}_r e^{-V_1}\right)\nonumber\\
    &+ \left[\mathlarger{\int} d^3x \mathlarger{\int} d^2\theta \,\, \text{Tr}_1\left(\frac{k}{2\pi}Q_1^2 - \Phi_r Q_2 \tilde{\Phi}_r\right) - \text{Tr}_2\left(\frac{k}{2\pi}Q_2^2 + \tilde{\Phi}_r Q_1 \Phi_r\right)\right] + c.c  
\end{align}
\normalsize
where $\text{Tr}_1$ represents trace over the first gauge group ($\mathcal{G}_k$) with $V_1, Q_1$ belonging to it's Lie Algebra, and $\text{Tr}_2$ represents trace over the second gauge group ($\mathcal{G}_{-k}$) with $V_2, Q_2$ belonging to it's Lie Algebra. Also $\Phi, \tilde{\Phi}$ are now two-indexed objects due to the product nature of the gauge group, and thus can be represented as $N \times N$ matrices. Since $Q_1, Q_2$ are auxiliary fields, they can be integrated out as follows 
\small
\begin{align*}
        Q_1 : Q_1 = \frac{\pi}{k} \Phi_r \tilde{\Phi}_r \,\,\,\,\,\,\,\,\,\,\,\,\,\,\,\,\,\,\,\,\,\,\, Q_2 : Q_2 = -\frac{\pi}{k}\tilde{\Phi}_r\Phi_r
\end{align*}
\vspace{-2em}
\begin{align}
    W &= \text{Tr}_1\left(\frac{k}{2\pi}Q_1^2 - \Phi_r Q_2 \tilde{\Phi}_r\right) - \text{Tr}_2\left(\frac{k}{2\pi}Q_2^2 + \tilde{\Phi}_r Q_1 \Phi_r\right) = \frac{\pi}{2k}\text{Tr}_2\left(\tilde{\Phi}_r\Phi_r\tilde{\Phi}_s\Phi_s - \tilde{\Phi}_r\Phi_s\tilde{\Phi}_s\Phi_r\right)\nonumber\\
        &= \frac{\pi}{k}\text{Tr}_2\left(\tilde{\Phi}_1\Phi_1\tilde{\Phi}_2\Phi_2 - \tilde{\Phi}_1\Phi_2\tilde{\Phi}_2\Phi_1\right) = \frac{\pi}{2k}\epsilon^{ab}\epsilon^{\dot{a}\dot{b}}\text{Tr}_2\left(\tilde{\Phi}_{\dot{a}} \Phi_a \tilde{\Phi}_{\dot{b}} \Phi_b \right)
\end{align}
\normalsize
It can clearly be seen that after integrating out $Q_1, Q_2$, a new symmetry $SU(2)_\Phi \times SU(2)_{\tilde{\Phi}}$ is manifest in the superpotential, with $(\Phi_1, \Phi_2)$ and $(\tilde{\Phi}_1, \tilde{\Phi}_2)$ acting as the respective doublets. Additionally, it has been shown earlier in (\ref{eq:n3fullaction}) and (\ref{eq:doublets}) that there is an $SU(2)_R$ symmetry between $\Phi$ and $\bar{\tilde{\Phi}}$. Now if the generators $\in SU(2)_\Phi \times SU(2)_{\tilde{\Phi}}$ and $SU(2)_R$ are commuted, it can atleast intuitively be seen that this gives rise to generators that mix cross terms like $\Phi_1$ and $\bar{\tilde{\Phi}}_2$ for e.g, and thereby don't commute. This hints at the fact that maybe there is a larger symmetry group at play, and that these two are subgroups of it; Or even further with a bit of foresight, that $\Phi_1, \Phi_2, \bar{\tilde{\Phi}}_1, \bar{\tilde{\Phi}}_2$ form a quartet of $SU(4)$ \cite{Aharony_2008}. This is indeed the case, as is will be evident after integrating out even more auxiliary d.o.f at low energies. But first of all, taking the superspace expansions of $\Phi_r, \tilde{\Phi}_r$ similar to (\ref{eq:fields}) and substituting in $S_{ABJM}$ yields
\small
\begin{flalign}
     &\mathlarger{\int} d^4\theta \,\, \text{Tr}_2\left(\bar{\Phi}_r e^{V_1} \Phi_r e^{-V_2}\right) + \text{Tr}_1\left(\bar{\tilde{\Phi}}_r e^{V_2} \tilde{\Phi}_r e^{-V_1}\right) = \text{Tr}_2 \biggl[D_\mu \bar{\phi}_r D^\mu \phi_r + D_\mu\tilde{\phi}_r D^\mu \bar{\tilde{\phi}}_r - i\bar{\psi}_r\gamma^\mu D_\mu \psi_r\nonumber&&\\
        &- i\tilde{\psi}_r\gamma^\mu D_\mu \bar{\tilde{\psi}}_r - \frac{1}{4}(\bar{\phi}_r\sigma_1^2\phi_r + \bar{\phi}_r\phi_r\sigma_2^2 + \tilde{\phi}_r\sigma_1^2\bar{\tilde{\phi}}_r + \tilde{\phi}_r\bar{\tilde{\phi}}_r\sigma_2^2) + \frac{1}{2}(\bar{\phi}_r\sigma_1\phi_r\sigma_2 + \tilde{\phi}_r\sigma_1\bar{\tilde{\phi}}_r\sigma_2 +\bar{\phi}_rD_1\phi_r \nonumber&&\\
        &- \bar{\phi}_r\phi_rD_2 -\tilde{\phi}_rD_1\bar{\tilde{\phi}}_r + \tilde{\phi}_r\bar{\tilde{\phi}}_rD_2 - \bar{\psi}_r\sigma_1\psi_r + \bar{\psi}_r\psi_r\sigma_2 + \tilde{\psi}_r\sigma_1\bar{\tilde{\psi}}_r - \tilde{\psi}_r\bar{\tilde{\psi}}_r\sigma_2) + \frac{i}{\sqrt{2}}(\bar{\phi}_r\chi_1\psi_r\nonumber&&\\
        &- \bar{\phi}_r\psi_r\chi_2 - \bar{\psi}_r\bar{\chi}_1\phi_r + \bar{\psi}_r\phi_r\bar{\chi}_2 - \tilde{\psi}_r\chi_1\bar{\tilde{\phi}}_r + \tilde{\psi}_r\bar{\tilde{\phi}}_r\chi_2 + \tilde{\phi}_r\bar{\chi}_1\bar{\tilde{\psi}}_r - \tilde{\phi}_r\bar{\tilde{\psi}}_r\bar{\chi}_2) + \bar{F}_rF_r + \tilde{F}_r\bar{\tilde{F}}_r\biggr]\nonumber&&
\end{flalign}
\vspace{-2em}
\begin{flalign*}
    \bigl(&\textbf{\text{where}} \,\,\,\, D^\mu H_r = \partial^\mu H_r + \frac{i}{2}A_1^\mu H_r - \frac{i}{2}H_rA_2^\mu \,\,\,\,\textbf{\text{and}}\,\,\,\, D^\mu \bar{H}_r = \partial^\mu \bar{H}_r - \frac{i}{2}\bar{H}_r A_1^\mu + \frac{i}{2}A_2^\mu \bar{H}_r&&\\
    &\textbf{\text{for}} \,\,\, H_r = \{\phi_r, \bar{\tilde{\phi}}_r, \psi_r, \bar{\tilde{\psi}}_r\} \,\,\,\,, \,\,\,\, \bar{H}_r = \{\bar{\phi}_r, \tilde{\phi}_r, \bar{\psi}_r, \tilde{\psi}_r\}\bigr)&&
\end{flalign*}
\vspace{-1em}
\begin{flalign}\label{eq:abjm_2}
        \mathlarger{\int} &d^2\theta\,\, W  = \frac{\pi}{k}\text{Tr}_2\biggl[\tilde{F}_1\phi_1\tilde{\phi}_2\phi_2 + \tilde{\phi}_1F_1\tilde{\phi}_2\phi_2 + \tilde{\phi}_1\phi_1\tilde{F}_2\phi_2 + \tilde{\phi}_1\phi_1\tilde{\phi}_2F_2 - \tilde{F}_1\phi_2\tilde{\phi}_2\phi_1 - \tilde{\phi}_1F_2\tilde{\phi}_2\phi_1 \nonumber&&\\
        &- \tilde{\phi}_1\phi_2\tilde{F}_2\phi_1 - \tilde{\psi}_1\psi_1\tilde{\phi}_2\phi_2 - \tilde{\psi}_1\phi_1\tilde{\psi}_2\phi_2 - \tilde{\psi}_1\phi_1\tilde{\phi}_2\psi_2 - \tilde{\phi}_1\psi_1\tilde{\psi}_2\phi_2 - \tilde{\phi}_1\psi_1\tilde{\phi}_2\psi_2 - \tilde{\phi}_1\phi_1\tilde{\psi}_2\psi_2\nonumber&&\\
        &+ \tilde{\psi}_1\psi_2\tilde{\phi}_2\phi_1 + \tilde{\psi}_1\phi_2\tilde{\psi}_2\phi_1 + \tilde{\psi}_1\phi_2\tilde{\phi}_2\psi_1 + \tilde{\phi}_1\psi_2\tilde{\psi}_2\phi_1 + \tilde{\phi}_1\psi_2\tilde{\phi}_2\psi_1 + \tilde{\phi}_1\phi_2\tilde{\psi}_2\psi_1\biggr]&& 
\end{flalign}
\normalsize
Using these expressions in $S_{ABJM}$ and integrating out the auxiliary fields $F_r, \bar{F}_r, \tilde{F}_r, \bar{\tilde{F}}_r, \chi_r$ $\bar{\chi}_r$, $D_r, \sigma_r$ yields their equations of motion as follows
\small
\begin{equation}\label{eq:abjm_3}
    \begin{alignedat}{4}
    &\bar{F}_1 : F_1 = \frac{\pi}{k}\left(\bar{\tilde{\phi}}_2\bar{\phi}_2\bar{\tilde{\phi}}_1 - \bar{\tilde{\phi}}_1\bar{\phi}_2\bar{\tilde{\phi}}_2\right) \,\,\,\,\,\,&&;\,\,\,\,\,\, &&&F_1 : \bar{F}_1 = \frac{\pi}{k}\left(\tilde{\phi}_1\phi_2\tilde{\phi}_2 - \tilde{\phi}_2\phi_2\tilde{\phi}_1\right)\\
    &\bar{F}_2 : F_2 = \frac{\pi}{k}\left(\bar{\tilde{\phi}}_1\bar{\phi}_1\bar{\tilde{\phi}}_2 - \bar{\tilde{\phi}}_2\bar{\phi}_1\bar{\tilde{\phi}}_1\right) \,\,\,\,\,\,&&;\,\,\,\,\,\, &&&F_2 : \bar{F}_2 = \frac{\pi}{k}\left(\tilde{\phi}_2\phi_1\tilde{\phi}_1 - \tilde{\phi}_1\phi_1\tilde{\phi}_2\right) \\
    &\bar{\tilde{F}}_1 : \tilde{F}_1 = \frac{\pi}{k}\left(\bar{\phi}_1\bar{\tilde{\phi}}_2\bar{\phi}_2 - \bar{\phi}_2\bar{\tilde{\phi}}_2\bar{\phi}_1\right) \,\,\,\,\,\,&&;\,\,\,\,\,\, &&&\tilde{F}_1 : \bar{\tilde{F}}_1 = \frac{\pi}{k}\left(\phi_2\tilde{\phi}_2\phi_1 - \phi_1\tilde{\phi}_2\phi_2\right)\\
    &\bar{\tilde{F}}_2 : \tilde{F}_2 = \frac{\pi}{k}\left(\bar{\phi}_2\bar{\tilde{\phi}}_1\bar{\phi}_1 - \bar{\phi}_1\bar{\tilde{\phi}}_1\bar{\phi}_2\right) \,\,\,\,\,\,&&;\,\,\,\,\,\, &&&\tilde{F}_2 : \bar{\tilde{F}}_2 = \frac{\pi}{k}\left(\phi_1\tilde{\phi}_1\phi_2 - \phi_2\tilde{\phi}_1\phi_1\right)\\
    &D_1\,,\, \sigma_1 : \sigma_1 = \frac{\pi}{k}\left(\bar{\tilde{\phi}}_r\tilde{\phi}_r - \phi_r\bar{\phi}_r\right) &&; &&&D_2\,,\, \sigma_2 : \sigma_2 = \frac{\pi}{k}\left(\tilde{\phi}_r\bar{\tilde{\phi}}_r - \bar{\phi}_r\phi_r\right)\\
    &\bar{\chi}_1 : \chi_1 = \frac{+2\pi i}{k\sqrt{2}}\left(\bar{\tilde{\psi}}_r\tilde{\phi}_r - \phi_r\bar{\psi}_r\right) &&; &&& \chi_1 : \bar{\chi}_1 = \frac{-2\pi i}{k\sqrt{2}}\left(\bar{\tilde{\phi}}_r\tilde{\psi}_r - \psi_r\bar{\phi}_r\right)\\
    &\bar{\chi}_2 : \chi_2 = \frac{+2\pi i}{k\sqrt{2}}\left(\tilde{\phi}_r\bar{\tilde{\psi}}_r - \bar{\psi}_r\phi_r\right) &&; &&& \chi_2 : \bar{\chi}_2 = \frac{-2\pi i}{k\sqrt{2}}\left(\tilde{\psi}_r\bar{\tilde{\phi}}_r - \bar{\phi}_r\psi_r\right)
    \end{alignedat}
\end{equation}
\normalsize
Substituting these back in $S_{ABJM}$ gives rise to bosonic and fermionic potentials of sextic ($\phi^6$) and quartic ($\psi^2\phi^2$) forms. Owing to the insight in an earlier paragraph regarding the larger symmetry group being $SU(4)$, and taking inspiration from (\ref{eq:doublets}), the bosonic and fermionic quartets can be written as
\small
\begin{equation}\label{eq:quartets}
    \left(M_\phi\right)^A = \begin{pmatrix}
        \phi_1 \\ \phi_2 \\ \bar{\tilde{\phi}}_1 \\ \bar{\tilde{\phi}}_2
    \end{pmatrix} \,\,\,\,\,\,\,\, \left(M_\psi\right)_{\dot{A}} = \begin{pmatrix}
        \,\,\,\,\psi_2 \\ -\psi_1 \\ -\bar{\tilde{\psi}}_2 \\ \,\,\,\,\bar{\tilde{\psi}}_1
    \end{pmatrix}
\end{equation}
\normalsize
where an undotted upper index ($A$) corresponds to the fundamental \textbf{4} representation of $SU(4)$, and a dotted lower index ($\dot{A}$) corresponds to the anti-fundamental $\mathbf{\bar{4}}$ representation of $SU(4)$. If the reader is wondering as to why the quartet of $(M_\psi)_{\dot{A}}$ is not just $(\psi_1, \psi_2, \bar{\tilde{\psi}}_1, \bar{\tilde{\psi}}_2)$ similiar to $(M_\phi)^A$, it is because the the choice in (\ref{eq:quartets}) is what allows us to write all the terms in $S_{ABJM}$ coming from (\ref{eq:abjm_2}) \& (\ref{eq:abjm_3}) in a manifestly $SU(4)$ invariant form. Also, reading off the components in $(M_\psi)_{\dot{A}}$, \{$(\psi_2, -\psi_1), (-\bar{\tilde{\psi}}_2, \bar{\tilde{\psi}}_1)\}$ and $(\psi_2, -\bar{\tilde{\psi}}_2)$ [or $(-\psi_1, \bar{\tilde{\psi}}_1)$] are still doublets under $SU(2)_\Phi \times SU(2)_{\tilde{\Phi}}$ and $SU(2)_R$ respectively; This is because we can go from the fundamental representation to the anti-fundamental representation of $SU(2)$ using the $\epsilon^{ab}$ tensor. Using this representation, the only $SU(4)$ invariant sextic ($\phi^6$) and quartic ($\psi^2\phi^2$) terms that can be written are
\small
\begin{flalign}
    \mathcal{L}_{\phi^6} \,\,\,\,\,&= \text{Tr}_2\biggl[\bar{M}_{\phi\dot{A}}\,M_\phi^A\,\bar{M}_{\phi\dot{B}}\,M_\phi^B\,\bar{M}_{\phi\dot{C}}\,M_\phi^C\left(\textbf{a}\,\delta^{\dot{A}}_B \delta^{\dot{B}}_A \delta^{\dot{C}}_C + \textbf{b}\,\delta^{\dot{A}}_A \delta^{\dot{B}}_B \delta^{\dot{C}}_C + \textbf{c}\,\delta^{\dot{A}}_C \delta^{\dot{B}}_A \delta^{\dot{C}}_B + \textbf{d}\,\delta^{\dot{A}}_B \delta^{\dot{B}}_C \delta^{\dot{C}}_A\right)\biggr]&&\nonumber\\
    \mathcal{L}_{\psi^2\phi^2} &= \text{Tr}_2\biggl[\bar{M}_\psi^A\,M_{\psi\dot{A}}\,\bar{M}_{\phi\dot{B}}\,M_\phi^B\left(\textbf{e}\,\delta^{\dot{A}}_A\delta^{\dot{B}}_B  + \textbf{f}\,\delta^{\dot{B}}_A\delta^{\dot{A}}_B\right) + \bar{M}_\psi^A\,M_\phi^B\,\bar{M}_{\phi\dot{B}}\,M_{\psi\dot{A}}\left(\textbf{g}\,\delta^{\dot{A}}_A\delta^{\dot{B}}_B  + \textbf{h}\,\delta^{\dot{B}}_A\delta^{\dot{A}}_B\right)&&\nonumber\\
&\,\,\,\,\,\,\,\,\,\,\,\,\,\,\,\,\,\,\,+\textbf{i}\,\epsilon_{ABCD}\,\bar{M}_\psi^A\,M_\phi^B\,\bar{M}_\psi^C\,M_\phi^D + \textbf{j}\,\epsilon^{\dot{A}\dot{B}\dot{C}\dot{D}}\,\bar{M}_{\phi\dot{A}}\,M_{\psi\dot{B}}\,\bar{M}_{\phi\dot{C}}\,M_{\psi\dot{D}}\biggr] \label{eq:potentialenergyabjm}&&
\end{flalign}
\normalsize
where $\epsilon_{ABCD}$ \& $\epsilon^{\dot{A}\dot{B}\dot{C}\dot{D}}$ are the $SU(4)$ invariant tensors with $\epsilon^{1234} = -\epsilon_{1234}$ = 1. Now by simply comparing with the terms in $S_{ABJM}$, the coefficients can be determined as follows
\small
\begin{equation}\label{eq:constantcondabjm}
    \begin{alignedat}{5}
        &\bar{\phi}_1\phi_1\bar{\phi}_1\phi_1\bar{\phi}_1\phi_1 \,\,:\,\, \textbf{a} + \textbf{b} + \textbf{c} + \textbf{d} = 0 \,\,\,\,\,\,\,\,;\,\,\,\,\,\,\,\,&&\tilde{\phi}_1\bar{\tilde{\phi}}_1\bar{\phi}_1\phi_1\tilde{\phi}_1\bar{\tilde{\phi}}_1 \,\,:\,\, \textbf{a} + 3\textbf{b} = -\frac{\pi^2}{4k^2}\\ &\tilde{\phi}_1\phi_1\bar{\phi}_1\bar{\tilde{\phi}}_1\bar{\phi}_2\phi_2 \,\,:\,\, \textbf{a} = -\frac{\pi^2}{2k^2} \,\,\,\,\,\,\,\,\,\,\,\,\,\,\,\,\,\,\,\,\,\,\,\,\,\,\,\,\,;\,\,\,\,\,\,\,\,&&\tilde{\phi}_1\phi_1\bar{\phi}_2\bar{\tilde{\phi}}_1\bar{\phi}_1\phi_2 \,\,:\,\, \textbf{d} = \frac{\pi^2}{3k^2}\\
        &\bar{\psi}_1\psi_1\bar{\phi}_2\phi_2 \,\,\,\,\,\,\,\,\,\,\,\,\,:\,\, \textbf{e} + \textbf{f} = -\frac{\pi}{2k} \,\,\,\,\,\,\,\,\,\,\,\,\,\,\,\,\,\,\,\,\,\,\,; &&\bar{\psi}_1\psi_1\bar{\phi}_1\phi_1 \,\,\,\,\,\,\,\,\,\,\,\,\,:\,\, \textbf{e} = \frac{\pi}{2k}\\
        &\bar{\psi}_1\phi_2\bar{\phi}_2\psi_1 \,\,\,\,\,\,\,\,\,\,\,\,\,:\,\, \textbf{g} + \textbf{h} = \frac{\pi}{2k} \,\,\,\,\,\,\,\,\,\,\,\,\,\,\,\,\,\,\,\,\,\,\,\,\,\,; &&\bar{\psi}_1\phi_1\bar{\phi}_1\psi_1 \,\,\,\,\,\,\,\,\,\,\,\,\,:\,\, \textbf{g} = -\frac{\pi}{2k}\\
        &\tilde{\psi}_1\phi_1\tilde{\psi}_2\phi_2 \,\,\,\,\,\,\,\,\,\,\,\,\,:\,\, \textbf{i} = \frac{\pi}{2k} \,\,\,\,\,\,\,\,\,\,\,\,\,\,\,\,\,\,\,\,\,\,\,\,\,\,\,\,\,\,\,\,\,\,\,\,\,\,; &&\tilde{\phi}_1\psi_1\tilde{\phi}_2\psi_2 \,\,\,\,\,\,\,\,\,\,\,\,\,:\,\, \textbf{j} = -\frac{\pi}{2k}
    \end{alignedat}
\end{equation}
\normalsize
Solving these gives $\textbf{a} = -\frac{\pi^2}{2k^2}$, $\textbf{b} = \textbf{c} = \frac{\pi^2}{12k^2}$, $\textbf{d} = \frac{\pi^2}{3k^2}$, $\textbf{e} =  -\textbf{g} = \frac{\pi}{2k}$, $-\textbf{f} = \textbf{h} = \frac{\pi}{k}$, $\textbf{i} = -\textbf{j} = \frac{\pi}{2k}$. Now with these coefficients, $S_{ABJM}$ in a manifestly $SU(4)$ invariant form is
\vspace{1em}
\hrule width \textwidth
\vspace{-0.3em}
\small
\begin{equation*}\label{eq:ABJMTheory}
    S_{ABJM} = \left(S_{CS}\right)_k + \left(S_{CS}\right)_{-k} + \mathlarger{\int} d^3x\,\, \text{Tr}_2\biggl[D_\mu\bar{M}_{\phi\dot{A}}\,D^\mu M_\phi^A - i\bar{M}_\psi^A\gamma^\mu D_\mu M_{\psi\dot{A}}\biggr] + \mathcal{L}_{\phi^6} + \mathcal{L}_{\psi^2\phi^2}\tag{$\boldsymbol{\star}$}
\end{equation*}
\normalsize
\vspace{-0.35em}
\hrule width \textwidth
\vspace{1em}
This is the ABJM theory, named after Ofer Aharony, Oren Bergman, Daniel Louis Jafferis and Juan Maldacena, first introduced in \cite{Aharony_2008}, 2008. It has an $SU(4)_R \cong SO(6)_R$ symmetry as can be seen in (\ref{eq:ABJMTheory}), and as mentioned earlier in the paragraph below (\ref{eq:Rsymmref}), $SO(6)_R$ symmetry in 2+1 dimensions corresponds to an $\mathcal{N} = 6$ supersymmetry. If this reverse implication $SO(6)_R \implies \mathcal{N} = 6$ seems inconclusive, the reader may refer to \cite{Bandres_2008} where the supersymmetry is explicitly verified by writing down the supersymmetric transformations of fields. In addition to the $SU(4)_R$, it is also easy to see that there is a global $U(1)_b$ symmetry, with $(M_\phi, \bar{M}_\psi)$ and $(\bar{M}_\phi, M_\psi)$ charged $b = 1$ and $b = -1$ respectively, when the gauge group is $SU(N)_k \times SU(N)_{-k}$. But when the gauge group is $U(N)_k \times U(N)_{-k}$, this global symmetry can be gauged since $U(1)$ is a subgroup of $U(N)$. However by adding an additional $U(1)_{\bar{b}}$ gauge field $A_{\bar{b}}$, and considering it's coupling to the $U(1)_{b}$ gauge field $A_b$ via the Chern-Simons terms $kA_b \wedge dA_{\bar{b}}$; An additional $U(1)_{b}$ global symmetry can be realized in addition to the $U(1)$ gauge symmetry \cite{Aharony_2008}. Therefore, in both the $SU(N)_k \times SU(N)_{-k}$ and $U(N)_k \times U(N)_{-k}$ cases, the global symmetry group of the theory is $SU(4)_R \times U(1)_b$. This group also arises as the isometry group of the gravitational background dual to the ABJM theory, as we will see in Chapter \ref{chap:situational_theoretical_analysis}.

The ground state manifold or the moduli space of vacua can also be determined for the $U(N)_k \times U(N)_{-k}$ case. This corresponds to the vanishing potential energy (excluding the coupling to $A_1, A_2$ coming from $D_\mu$) of (\ref{eq:ABJMTheory}) i.e., $\mathcal{L}_{\phi^6} + \mathcal{L}_{\psi^2\phi^2} = 0$. From (\ref{eq:constantcondabjm}), since \textbf{a} + \textbf{b} + \textbf{c} + \textbf{d} = 0, it can easily be seen from (\ref{eq:potentialenergyabjm}) that $\mathcal{L}_{\phi^6} = 0$ if $(M_{\phi})^A$ are the maximal subset of commuting matrices i.e., $N \times N$ diagonal matrices. Similar argument holds for $\mathcal{L}_{\psi^2\phi^2} = 0$ and $(M_{\psi})_{\dot{A}}$ since \textbf{e} + \textbf{g} = \textbf{f} + \textbf{h} = \textbf{i} + \textbf{j} = 0. It can also be checked that for any non-diagonal matrix, the off-diagonal modes become massive \cite{berenstein2009aspects}, thereby implying that the subspace of diagonal matrices is infact the entire moduli space of the theory. These diagonal matrices correspond to a reduced gauge symmetry $(U(1)_k \times U(1)_{-k})^N$ upto a permutation of the diagonal elements.

Let the $N$ pairs of $U(1)_{k} \times U(1)_{-k}$ gauge fields be $(A_1^j, A_2^j)$, j = 1, 2,..., $N$, with their corresponding gauge transformations $(A_1^j - d\Lambda_1^j, A_2^j - d\Lambda_2^j)$. Since $(M_\phi)^A$ and $(M_\psi)_{\dot{A}}$ are in the $(N, \bar{N})$ and $(\bar{N}, N)$ representations respectively under $U(N)_{k} \times U(N)_{-k}$, their corresponding gauge transformations under $(U(1)_{k} \times U(1)_{-k})^N$ are $e^{i(\Lambda^j_1 - \Lambda^j_2)}(M_\phi)_{jj}^A$ and $e^{i(\Lambda^j_2 - \Lambda^j_1)}(M_\psi)^{jj}_{\dot{A}}$ respectively. Now $A_1^j, A_2^j$ can be gauge fixed to be zero thereby decoupling them from matter in $D_\mu$, and since $\mathcal{L}_{\phi^6} + \mathcal{L}_{\psi^2\phi^2} = 0$, the theory of diagonal matrix-valued matter is just a free theory for matter with CS terms for the gauge fields.
\begin{equation}
    S_{Moduli} = (S_{CS})_k + (S_{CS})_{-k} + \mathlarger{\int d^3x \,\,\text{Tr}_2\left[\partial_\mu \bar{M}_{\phi \dot{A}}\partial^\mu M_\phi^A - i\bar{M}_\psi^A\gamma^\mu \partial_\mu M_{\psi \dot{A}}\right]}
\end{equation}
If the story ended here, the moduli space would simply be $(\mathbb{C}^4)^N$ upto permutations, with $\mathbb{C}^4$ corresponding to the four complex valued fields in the quartet of $SU(4)_R$. However, there is a residual gauge symmetry associated to constant $\Lambda_1^j, \Lambda_2^j$. Such gauge transformations are anomalous without the right conditions imposed, since they induce a finite change to $S_{Moduli}$ via $S_{CS}$ as follows
\begin{equation}\label{eq:modulitrans}
    \delta S_{Moduli} = \frac{k}{2\pi}\mathlarger{\int} \, \Lambda_1^j \wedge dA_1^j - \Lambda_2^j \wedge dA_2^j
\end{equation}
As mentioned near (\ref{eq:CSG}), such transformations $\delta S_{Moduli} = 2\pi n$, $n \in \mathbb{Z}$ in order to have a single-valued amplitude under the path integral. Also, $\int dA_i = \int F_i$ is just the magnetic charge under the field $A_i$ since $\int$ is over a 2-manifold, which by Dirac quantization is equal to $2\pi m$, $m \in \mathbb{Z}$ for a unit electric charge. Therefore using these two conditions, it can be seen from (\ref{eq:modulitrans}) that $\Lambda_1^j, \Lambda_2^j = \frac{2\pi n}{k}$; $k \in \mathbb{Z}$. Now since $e^{i(\Lambda^j_1 - \Lambda^j_2)}(M_\phi)_{jj}^A$ and $e^{i(\Lambda^j_2 - \Lambda^j_1)}(M_\psi)^{jj}_{\dot{A}}$ are the gauge  transformations of matter, the moduli space has an additional identification of states under this residual gauge symmetry i.e., $(C^4 /\mathbb{Z}_k)^N/ S_N$; where $S^N$ just denotes the permutation of diagonal elements. This orbifold action is also realized in the geometry of a specific configuration of branes in Chapter \ref{chap:situational_theoretical_analysis}, whose low-energy world volume theory is the same as the ABJM theory. The construction of such a configuration of branes in String (M) - theory will be the subject of exploration starting from the next chapter.

%% file: chapters/research_results.tex
\chapter{Brane construction for the ABJM theory}\label{chap:research_results}

D-branes are non-perturbative higher dimensional objects in string theory, that are dynamical. However, in the perturbative regime of string theory ($g_s \ll$ 1), their description is limited to being objects on which open strings can end. Now, similar to how D-branes impose Dirichlet boundary conditions on the bosonic fields of the open string, they impose appropriate boundary conditions on the fermionic fields as well that relate the left and the right moving supercharges, thereby preserving only half of the supersymmetries \cite{Becker_Becker_Schwarz_2006}. Also, since D-branes are non-pertrubative objects, their tension is expected to be safely extrapolated to the strong coupling regime. With these two facts in mind, the stable D-branes (the ones that couple to the R-R fields) of a type of string theory are taken to be \textit{half-BPS} states, while the unstable ones break all of the supersymmetries. To be more explicit, let the two Majorana-Weyl spinors representing the supercharges of the string theory be $Q_L \,\,\& \,\,Q_R$. Since a D\textit{p}-brane breaks the full $SO(9, 1)$ Lorentz group to $SO(p, 1) \times SO(9-p)$, we change the basis $(Q_L, Q_R) \rightarrow (\tilde{Q}_L = aQ_L + bQ_R, \tilde{Q}_R = cQ_L + dQ_R)$ s.t the new basis is a Lorentz covariant spinor under $Spin(p, 1)$
\begin{equation}
\begin{split}
    &(\tilde{Q}_L, \tilde{Q}_R) \rightarrow (\mathcal{S}\tilde{Q}_L, \mathcal{S}\tilde{Q}_R) \,\,\, \text{for} \,\,\, (Q_L, Q_R) \rightarrow (\mathcal{S}Q_L, \mathcal{S}Q_R) \,\,\,;\,\,\, \mathcal{S} \in Spin(p, 1)\\
    \implies &[a, \mathcal{S}] = [b, \mathcal{S}] = [c, \mathcal{S}] = [d, \mathcal{S}] = 0\\
    \implies & a, b, c, d = \pm \mathbb{1} \,\,\, \text{or} \,\,\,\pm \prod_{\mu = 0}^p \Gamma^\mu
\end{split}
\end{equation}
where $\Gamma^\mu$ are the $9+1$ dimensional gamma matrices, any subset of which i.e., $\mu = 0, 1, 2, 3,..., p$ can be considered as the $p+1$ dimensional gamma matrices. One particular choice of $a, b, c, d$ s.t there are equal number of supercharges in $\tilde{Q}_L$ and $\tilde{Q}_R$ is
\begin{equation}
    \tilde{Q}_L = Q_R - \left(\prod_{\mu = 0}^p \Gamma^\mu\right)Q_L \,\,\,\,\,;\,\,\,\,\, \tilde{Q}_R = Q_R + \left(\prod_{\mu = 0}^p \Gamma^\mu\right)Q_L
\end{equation}
The supersymmetry transformations in the new basis then become $\bar{\tilde{\epsilon}}_L \tilde{Q}_L + \bar{\tilde{\epsilon}}_R \tilde{Q}_R$, upon which the boundary conditions imposed on the fermionic fields lead to $\tilde{\epsilon}_L = 0$. Therefore stable \textit{half-BPS} D\textit{p}-branes preserve supersymmetries of the form
\begin{equation}\label{eq:dpsusyrel}
    \bar{\epsilon}_L Q_L + \bar{\epsilon}_R Q_R \,\,\,\,\,\,\textbf{\text{s.t}}\,\,\,\,\,\, \epsilon_L =(-1)^{p-1}\left( \prod_{\mu = 0}^p \Gamma^\mu\right)\epsilon_R
\end{equation}
Further in this description, at low energies (E $\ll \alpha'^{-1/2}$) where only the massless string modes are considered relevant, the dynamics of the open string is effectively described by a supersymmetric gauge theory living in the world volume of D-branes. To be more specific, a stack of $\textbf{N}$ coincident D-branes gives rise to the $U(N)$ gauge group for oriented strings ($SO(N)$ or $USp(N)$ for unoriented), with the adjoint representation labeled by the multi-indexed Chan-Paton factor; The index of one end of the string is associated to the fundamental $\textbf{N}$ representation, while the other end is associated to the anti-fundamental $\mathbf{\bar{N}}$ representation. Combining this gauge group information with the supersymmetry information from (\ref{eq:dpsusyrel}), specific gauge theories in $d+1$ dimensions with $\mathcal{N}$ supersymmetries can be realized in the low energy world volume theory of configurations involving stacks of different types of stable branes.

\section{Brane construction for the ABJM theory}
The ABJM theory described in the previous chapter can also be constructed by taking the $\mathcal{N} = 4$ Yang-Mills theory in 2+1 dimensions with a product gauge group, deforming it to the $\mathcal{N} = 3$ Yang-Mills Chern-Simons theory, and then flowing to the IR \cite{Aharony_2008}. This roadmap is what will be used to construct the configuration of branes in type IIB superstring theory, whose low energy world volume theory corresponds to the ABJM theory.

\subsection{$\mathcal{N} = 4$ $U(N) \times U(N)$ Yang-Mills theory in 2+1 dimensions}\label{sec:n4brane}
\begin{figure}[h]
\includegraphics[width=\textwidth]{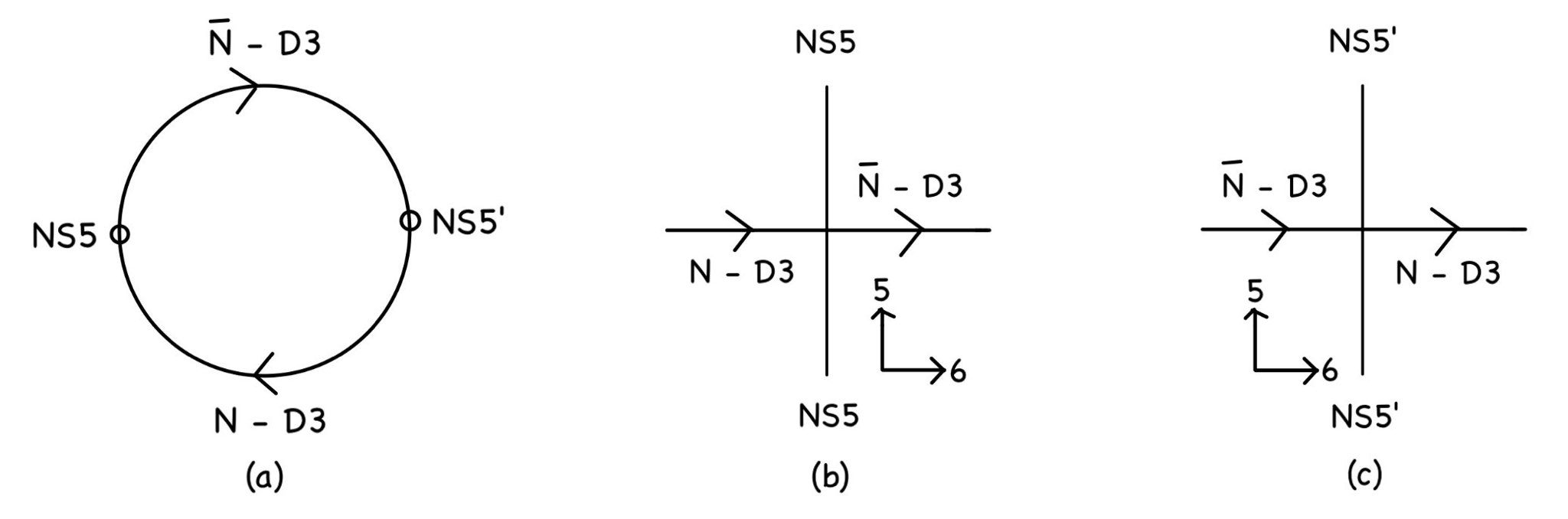}
\caption{(a) N D3-branes wrapping the compact 6 direction, with NS5 and NS5' branes separated along it. (b) Zoomed in near NS5-brane in the 56 plane. (c) Zoomed in near NS5'-brane in the 56 plane.}
\label{fig:brane_1}
\end{figure}

This is an $\mathcal{N} = 4$ theory in odd no. of dimensions with the minimal spinor being Majorana, and therefore has to have $SO(4)_R$ symmetry \cite{vanproeyen2016tools}. Although R-symmetries are only approximate symmetries of the string theory, it is still really helpful to have a brane configuration that manifestly showcases this R-symmetry, which then helps us to identify the low-energy multiplet structure explicitly. Since topologically $SO(4)$ is the double cover of $SO(3) \times SO(3)$, let's say the two triplets of coordinates under this $SO(3) \times SO(3)$ are $(x^3, x^4, x^5)$ denoted by 345 and $(x^7, x^8, x^9)$ denoted by 789. The 2+1 dimensions for the field theory can then be 012 with the separation along $6$ possibly leading to the product gauge group structure. One such configuration considered in \cite{Aharony_2008} is two parallel NS5-branes (NS5 \& NS5') along 012345 separated along the compact direction 6, and a stack of $N$ coincident D3-branes along 0126, see figure \ref{fig:brane_1} (a). \textbf{N}, $\mathbf{\bar{N}}$ in the figure denote the representation in which the Chan-Paton index of the $\sigma = 0$ end lies in, for an open oriented string starting from that brane. Further, NS5-brane is a stable \textit{half-BPS} magnetic dual of the fundamental string, and is magnetically charged under the Kalb-Ramond two-form $B_2$. It is then trivial to note that this D3-NS5-NS5' system preserves only 1/4 of the supersymmetries i.e., 8 supercharges, which corresponds to $\mathcal{N} = 4$ supersymmetry in 012. However, if there is a lingering concern as to whether D3-branes can have their boundary on NS5-branes, it can indeed be shown using various dualities that D3 branes can end on NS5-branes (or even D5-branes) \cite{Strominger_1996}. 

Now let's analyze the string modes in this configuration and hence make a note of the multiplets that are generated at low energies, at the scale of which, the relative size of the compact 6 dimension gets extremely smaller. Let this compactification be done in two stages, first by bringing the upper half of the circle ($\mathbf{\bar{N}}$ - D3) closer to the lower half (\textbf{N} - D3), and then by bringing NS5 and NS5' closer to each other. If we zoom in near NS5 (figure \ref{fig:brane_1} (b)), the circular D3 branes can be treated as \textbf{N} - D3 branes ending on NS5-brane from the left and $\mathbf{\bar{N}}$ - D3 branes ending on NS5-brane from the right, and vice-versa for the NS5'-brane (figure \ref{fig:brane_1} (c)). In both the cases, there are two types of strings, one that starts on the left brane and ends on the right brane  and the other that does the opposite. During the first stage of compactification where the left and the right branes meet, the aforementioned strings give rise to massless modes, of which the bosonic modes in the lightcone gauge are of the form $\alpha_{-1}^\mu$ ($\mu = 1, 3, 4, 5, 6, 7, 8, 9)$. However, boundary conditions imply that only the $\mu = 6, 7, 8, 9$ modes survive after compactification \cite{Hanany_1997}. Firstly near the NS5-brane, let us call these a, b, a', b' (\textbf{N}$\mathbf{\bar{N}}$ strings) representation and c, d, c', d' ($\mathbf{\bar{N}}$\textbf{N} strings) representation s.t a + ib = $\phi_1$, a' + ib' = $F_1$, c + id = $\tilde{\phi}_1$ and c' + id' = $\tilde{F}_1$, with their complex superpartners $\psi_1, \tilde{\psi}_1$. These eight bosonic d.o.f correspond to the fluctuations along 6789 of the 012 D3-brane subspace in NS5-brane, relative to the NS5 and NS5' branes. Similarly, there are $\phi_2, F_2, \tilde{\phi}_2, \tilde{F}_2, \psi_2, \tilde{\psi}_2$ corresponding to the modes of the $\mathbf{\bar{N}}$\textbf{N} and \textbf{N}$\mathbf{\bar{N}}$ strings near the NS5' brane. Again, the eight bosonic d.o.f correspond to the fluctuations along 6789 of the 012 D3-brane subspace in NS5'-brane, relative to the NS5 and NS5' branes.

\noindent Additionally, there are two other types of strings, ones which start and end on the \textbf{N} - D3 branes and ones which start and end on the $\mathbf{\bar{N}}$ - D3 branes. Once again, each of them have bosonic string modes of the form $\alpha_{-1}^\mu$ ($\mu = 1, 3, 4, 5, 6, 7, 8, 9)$, of which only the $\mu = 1, 3, 4, 5$ survive after imposing boundary conditions \cite{Hanany_1997}. These give rise to eight bosonic d.o.f :  e, f, g, h, e', f', g', h' (each in the adjoint representation) s.t f + ig = $q_1$, f' + ig' = $q_2$, h = $\sigma_1$, h' = $\sigma_2$, and their superpartners $\chi_1$, $\chi_2$, $\lambda_1$, $\lambda_2$. More specifically, after compactification, e and e' are the propagating gauge d.o.f in the 012 subspace of NS5 and NS5' respectively and, \{(f, f'), (g, g'), (h, h')\} correspond to the motion of the 012 subspace in the \{$x^3, x^4, x^5$\} directions of (NS5, NS5') branes. Finally, the second stage of compactification is also done, rendering these modes massless, and effectively dimensionally reducing the $x^6$ direction.

At the end of it all, we are now left with NS5 and NS5' branes with the D3-branes breaking into a 012 subspace in each of them. Coming to the bosonic and fermionic d.o.f, there are \{e, $\sigma_1$, $\chi_1$, $q_1$, $\lambda_1$\} $\in V_1 \, | \, Q_1$ in the $(N^2, 1)$ representation; \{e', $\sigma_2$, $\chi_2$, $q_2$, $\lambda_2$\} $\in V_2 \, | \, Q_2$  in the $(1, N^2)$ representation; \{$\phi_1, \psi_1, \phi_2, \psi_2$\} $\in (\Phi_1, \Phi_2)$ in the (\textbf{N}, $\mathbf{\bar{N}}$) representation and \{$\tilde{\phi}_1, \tilde{\psi}_1, \tilde{\phi}_2, \tilde{\psi}_2$\} $\in (\tilde{\Phi}_1, \tilde{\Phi}_2)$ in the ($\mathbf{\bar{N}}$, \textbf{N}) representation; where each representation is under the $U(N) \times U(N)$ group. These d.o.f can easily be identified with the $\mathcal{N} = 4$ (non-deformed) product gauge group multiplets, already mentioned under section \ref{section:n3product}. Therefore, the brane construction desribed thus far gives rise to an $\mathcal{N} = 4$ $U(N) \times U(N)$ Yang-Mills theory in 2+1 dimensions.

\subsection{$\mathcal{N} = 2$ $U(N) \times U(N)$ Yang-Mills theory in 2+1 dimensions}\label{sec:n2ymbrane}
\begin{figure}[h]
\centering
\includegraphics[width=0.7\textwidth]{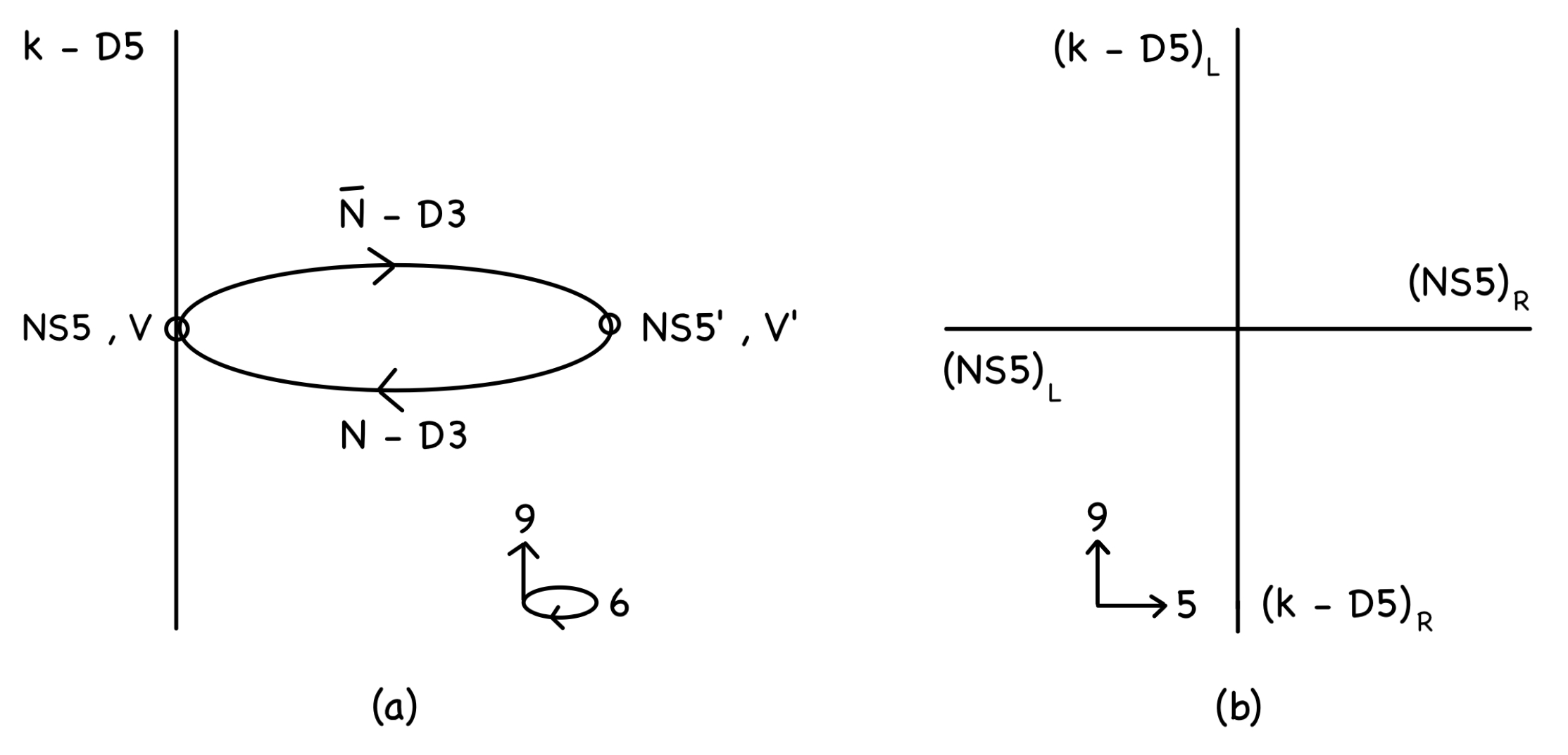}
\caption{(a) \textit{k} D5-branes along 012349 intersecting the NS5-brane in 01234, N D3-branes in 012; V and V' are the 012 subspaces of D3-branes in NS5 and NS5' branes respectively. (b) \textit{k} D5-branes and the NS5-brane breaking into two pieces each $\{(\textit{k} - \text{D5})_\text{L}$, $(\textit{k} - \text{D5})_\text{R}\}$ \& $\{(\text{NS5})_\text{L}$, $(\text{NS5})_\text{R}\}$  at the intersection in the 59 plane.}
\label{fig:brane_2}
\end{figure}

Now that an $\mathcal{N} = 4$ theory has been described as a configuration of branes, the natural next step is to deform this configuration and break the supersymmetry to $\mathcal{N} = 3$. However since each brane breaks half the supersymmetries, the manifestly brane constructive way to do it would be to first break the supersymmetry to $\mathcal{N} = 2$ and then find a way to enhance it to $\mathcal{N} = 3$. But before that, the form of supersymmetries preserved by the NS5-brane would have to be mentioned, similar to the ones mentioned in (\ref{eq:dpsusyrel}) for the case of D\textit{p}-branes. The result can be arrived at via the $SL(2, \mathbb{Z})$ symmetry in type IIB superstring theory, which acts on the doublet of charges as follows \cite{Sen_1994}
\begin{equation*}
    \begin{pmatrix}
        p \\ q
    \end{pmatrix} \rightarrow \begin{pmatrix}
        p'\\ q'
    \end{pmatrix} = \textbf{A} \begin{pmatrix}
        p \\ q
    \end{pmatrix} \,\,;\,\, \textbf{A} = \begin{pmatrix}
        a && b \\ c && d
    \end{pmatrix} \in SL(2, \mathbb{Z}) \,\,\,\,\,\, \textbf{\text{i.e.}} \,\,\,\,\,\, a, b, c, d \in \mathbb{Z} \,\,\,\, \boldsymbol{\&} \,\,\,\, ad - bc = 1
\end{equation*}
where $p, q$ are either the electric or magnetic charges of the state, corresponding to the Kalb-Ramond two-form $B_2$ and the R-R two-form $C_2$ respectively. Since NS5-brane couples magnetically to $B_2$, the state vector transforming under $SL(2, \mathbb{Z})$ is given by $\begin{pmatrix}
    1 \\ 0 
\end{pmatrix}$, while for the D5-brane it is $\begin{pmatrix}
    0 \\ 1
\end{pmatrix}$ since it magnetically couples to $C_2$. The corresponding $SL(2, \mathbb{Z})$ matrix \textbf{A} that transforms an NS5-state to a D5-state is $\begin{pmatrix}
    0 && 1 \\ -1 && 0
\end{pmatrix}$. Now since the left and right supercharges ($Q_L, Q_R$) of the type IIB theory also transform as a doublet under this $SL(2, \mathbb{Z})$ \cite{Bars_1997}, the preserved supersymmetries in (\ref{eq:dpsusyrel}) thereby become
\begin{equation*}
    \bar{\epsilon}_L Q_L + \bar{\epsilon}_R Q_R \,\,\rightarrow\,\, \bar{\epsilon}_L Q_R - \bar{\epsilon}_R Q_L = \bar{\epsilon'}_L Q_L  + \bar{\epsilon'}_R Q_R \,\,\,\,\,\, ; \,\,\,\,\,\,  \epsilon_L =(-1)^{p-1}\left( \prod_{\mu = 0}^p \Gamma^\mu\right)\epsilon_R
\end{equation*}
Therefore an NS5-brane in the 012345 directions preserves supersymmetries of the form
\begin{equation}\label{eq:ns5susy}
    \bar{\epsilon}_L Q_L + \bar{\epsilon}_R Q_R \,\,\,\,\,\,;\,\,\,\,\,\, \epsilon_L = -\Gamma^0\Gamma^1\Gamma^2\Gamma^3\Gamma^4\Gamma^5 \epsilon_L \,\,\,\, \boldsymbol{\&} \,\,\,\, \epsilon_R = \Gamma^0\Gamma^1\Gamma^2\Gamma^3\Gamma^4\Gamma^5 \epsilon_R
\end{equation}
Therefore for the brane configuration mentioned thus far, using (\ref{eq:ns5susy}) for the NS5-branes and (\ref{eq:dpsusyrel}) for the D3-branes, combined with the fact that the left and right supercharges are of the same chiraity in type IIB theory i.e. $\Gamma^{11} \{Q_L, Q_R\} = \{Q_L, Q_R\}$; where $\Gamma^{11} = \prod_{\mu = 0}^{9} \Gamma^\mu$, the following relation is derived
\begin{equation}
    \epsilon_L = \Gamma^0\Gamma^1\Gamma^2\Gamma^7\Gamma^8\Gamma^9 \epsilon_R
\end{equation}
Therefore D5-branes added to the configuration along 012789 preserve the $\mathcal{N} = 4$ supersymmetry, and hence any other type of \textit{half-BPS} addition breaks the supersymmetry to $\mathcal{N} = 2$. Since an $\mathcal{N} = 2$ theory in 2+1 dimensions has $SO(2)_R$ symmetry, it is preferable that this R-symmetry is manifest in the brane configuration. Let the directions 78 transform under this $SO(2)_R$, and let \textit{k} D5-branes be added along 012349, such that they interesect the NS5-brane along 01234 and the D3-branes along 012 (see figure \ref{fig:brane_2} (a)) \cite{Aharony_2008}. 

In this new configuration, there are new types of strings in addition to the ones discussed previously i.e. oriented open strings between D5 and D3 branes. Similar discussion to the one made in the paragraph below figure \ref{fig:brane_1} can be made here as well; Let us pick one of the \textit{k} D5-branes, and consider it to be much heavier than the D3-brane. Then the non-vanishing bosonic modes under boundary (D3-D5) conditions ($\alpha_{-1}^\mu$; $\mu$ = 5, 6, 7, 8) of the strings between (D5, $\mathbf{\bar{N}}$-D3) and (D5, N-D3) combined near the NS5-brane give rise to four bosonic d.o.f. These correspond to the fluctuations of \textbf{V} in the 5678 directions relative to the D5-brane after compactification. The four bosonic d.o.f along with their superpartners (2 Majorana fermions) belong to two chiral multiplets, one in the \textbf{N} representation ($\varphi_1$) and the other in the $\mathbf{\bar{N}}$ representation ($\tilde{\varphi}_1$). The same argument can be made for the vicinity of NS5'-brane as well, and since there are \textit{k} D5-branes, this finally leads to the addition of \textit{k} massless chiral multiplets in the fundamental (\textbf{N}) and \textit{k} massless chiral multiplets in the anti-fundamental ($\mathbf{\bar{N}}$) representations of each of the $U(N)$ factors.

To summarize the multiplet structure so far, we have vector multiplet $V_1 \,|\, Q_1$ in the adjoint representation of the first $U(N)$ factor and the vector multiplet $V_2 \,|\, Q_2$ in the adjoint representation of the second $U(N)$ factor under the $U(N) \times U(N)$ gauge group. Coming to matter, there are bifundamental matter multiplets $\Phi_i, \tilde{\Phi}_i$ (i = 1, 2) in the $(\textbf{N}, \mathbf{\bar{N}})$ and $(\mathbf{\bar{N}}, \textbf{N})$ representations respectively. Additionally, there are \textit{k} chiral multiplets in the \textbf{N} representation ($\varphi_1$) and \textit{k} chiral multiplets in the $\mathbf{\bar{N}}$ representation ($\tilde{\varphi}_1$) of the first $U(N)$ factor, and similarly $\varphi_2, \tilde{\varphi}_2$ for the second $U(N)$ factor. Currently, all of these fields are massless and the overarching theory respects $\mathcal{N} = 2$ supersymmetry. 

\subsection{Chern-Simons deformation}
The next goal is to find a way to deform the brane configuration such that the Chern-Simons term arises naturally in the low energy theory. The key observation that helps us in doing that is to note from the discussion in the previous sections, as to how the chiral multiplets $\varphi_i, \tilde{\varphi}_i$ each have a single Majorana fermion, whereas the bifundamental matter multiplets $\Phi_i, \tilde{\Phi}_i$ each have two Majorana fermions (or one complex fermion). The mass term of a spinor $\omega$ : $m\bar{\omega}\omega$ in a theory with odd no. of fermions and in odd no. of dimensions is parity violating \cite{Delbourgo_1994}; where parity transformation in odd no. of dimensions is the reflection of one of the spatial coordinates. It has a parity violating spin $\frac{|m|}{2m}$ and such a mass term clearly doesn't respect gauge invariance as well. However, it has been shown in \cite{PhysRevD.29.2366} that these fermions can be integrated out by regulating the UV divergences in a certain way, to recover either the parity symmetry or the gauge symmetry but not both. If we choose to recover the gauge symmetry, the outline of the method as described in \cite{PhysRevD.29.2366} is as follows
\begin{itemize}[topsep = 5pt]
    \setlength\itemsep{0.3em}
    \item Let the action of the theory with odd no. of fermions (n) in odd no. of dimensions, with a parity violating and gauge non-invariant mass ($m$) term be $S[A, \omega; m]$; where A is the gauge field and $\omega$ is the massive fermion.
    
    \item Integrate out the massive fermionic d.o.f and introduce a heavy parity-violating Pauli-Villars regulator field with a mass parameter $\Lambda$ to obtain $S_{eff}[A; m, \Lambda]$

    \item Define a regulated effective action $S_{eff}^R = \lim_{\Lambda \to \infty}S[A; m = 0, \Lambda] - S[A; m \rightarrow \infty, \Lambda]$ during the computation of which, the $\Lambda$ divergent term in $S[A; m = 0, \Lambda]$ cancels the $\Lambda$ divergent term in $S[A; m \rightarrow \infty]$

    \item $S[A; m \rightarrow \infty, \Lambda]$ contains the Chern-Simons term $S_{CS}$ with a coefficient \[\sum_{i = 1}^n\frac{|m_i|}{8\pi m_i}\] coming from the parity violating spin of each of the massive fermions that have been integrated out. This Chern-Simons term, being parity-violating, cancels the gauge non-invariance of $S[A; m = 0, \Lambda]$.  
\end{itemize}
Therefore, this method outlined above takes a gauge non-invariant parity-violating theory with odd no. of massive fermions in odd no. of dimensions, and regularizes it to a gauge-invariant theory with a Chern-Simons term but without the massive fermions. Back to our current brane configuration, the multiplets with odd no. of fermions are $\{\varphi_i^N, \tilde{\varphi}_i^N \,\,;\,\, N = 1, 2, 3,..., k \,\,\& \,\, i = 1, 2\}$ with each having a single massless Majorana fermion. So if we can deform the brane configuration such that half of these fermions acquire a positive real mass term, while the other half acquire a negative real mass term, then by the prescription mentioned above, this gives rise to Chern-Simons terms of levels \textit{k} and -\textit{k} post regularization, which is what is required for the ABJM theory. But before doing that, there has been an important inconsistency that hasn't been addressed yet. 

\begin{figure}[h]
\setlength{\belowcaptionskip}{-16pt}
\centering
\includegraphics[width=0.9\textwidth]{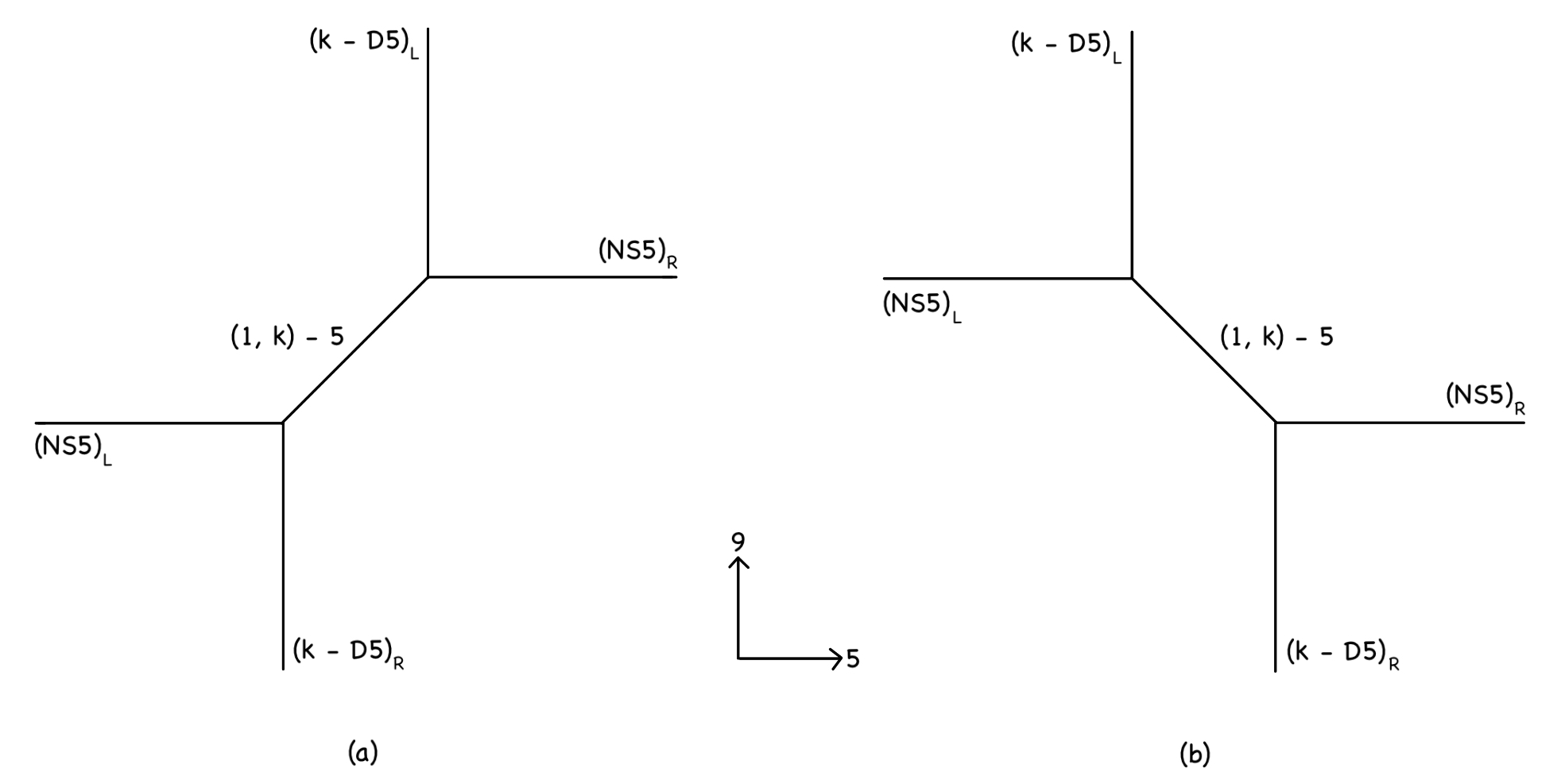}
\caption{(a) $\left[(\text{NS5})_{\text{L}}, (\textit{k} - \text{D5})_{\text{R}}\right]$ and $\left[(\text{NS5})_{\text{R}}, (\textit{k} - \text{D5})_{\text{L}}\right]$ pieces merging into an intermediate (1, \textit{k}) 5-brane at the intersection. (b) $\left[(\text{NS5})_{\text{L}}, (\textit{k} - \text{D5})_{\text{L}}\right]$ and $\left[(\text{NS5})_{\text{R}}, (\textit{k} - \text{D5})_{\text{R}}\right]$ pieces merging into an intermediate (1, \textit{k}) 5-brane at the intersection.}
\label{fig:brane_3}
\end{figure}

The \textit{k} D5-branes intersect the NS5-brane in 01234, and hence naively break into two pieces $(k-D5)_L$ and $(k-D5)_R$ ending on the NS5 brane on either side in the $x^9$ direction (see figure \ref{fig:brane_2} (b)). The boundary of each piece in 01234 lying in the NS5-brane worldvolume, has a flux associated with it's magnetic charge under the R-R two-form $C_2$, which is not a field in the NS5-brane worldvolume. This clearly leads to a violation of charge conservation, thereby rendering the current view of intersection inaccurate. This is where the $SL(2, \mathbb{Z})$ symmetry of the type IIB theory, mentioned at the beginning of section \ref{sec:n2ymbrane}, comes to the rescue. This symmetry allows the existence of bound states of \textit{p} NS5-branes and \textit{q} D5-branes as intermediate $(p, q)$ 5-branes, and hence the current configuration of NS5-brane and \textit{k} D5-branes merge into an intermediate $(1, k)$ 5-brane at the intersection (as shown in figure \ref{fig:brane_3}), thereby respecting charge conservation \cite{Aharony_1997_b}.

The orientation of breaking in figure \ref{fig:brane_3} (a) (NS5-(1, \textit{k})-D5) clockwise) is opposite to the orientation of breaking in figure \ref{fig:brane_3} (b) (NS5-D5-(1, \textit{k}) clockwise). So to place them on the same footing, perform an $SL(2, \mathbb{Z})$ transformation that changes the (1, \textit{k}) 5-brane $\begin{pmatrix}
    1 \\ k
\end{pmatrix}$ to an NS5-brane $\begin{pmatrix}
    1 \\ 0
\end{pmatrix}$, by $A \in SL(2, \mathbb{Z})$ = $\begin{pmatrix}
    1 && 0 \\ -k && 1
\end{pmatrix}$. This transformation then changes the NS5-brane to a (1, -\textit{k}) 5-brane. So with the orientations now matched, we will refer to the intermediate brane in the two ways of breaking in figures \ref{fig:brane_3} (a) and (b) as (1, \textit{k}) and (1, -\textit{k}) respectively from now on. The next step is to determine the angle of the intermediate brane w.r.t the NS5-brane in the 59 plane. Let us consider the lower half of figure \ref{fig:brane_3} (a) i.e., the $(\text{NS5})_{\text{L}}$ and $(k - \text{D5})_{\text{R}}$ pieces merging to form the intermediate (1, \textit{k}) 5-brane in the 59 plane, at the intersection point let's say (0, 0). The 01234 boundary of $(k - \text{D5})_{\text{R}}$ is \textit{k} 4-branes propagating in 5+1 dimensions (012345), and can thus be treated as \textit{k} point particles in 1+1 dimensions (05). Since $(k - \text{D5})_{\text{R}}$ spans the $x^9$ direction which is transverse to $(\text{NS5})_{\text{L}}$, the $x^9$ position of $(\text{NS5})_{\text{L}}$ can therefore be treated as the potential due to the aforementioned point particles in 1+1 dimensions (05). The Poisson's equation in 05 then becomes \cite{Aharony_1997_b}
\begin{equation}
    \nabla^2 \, x^9 = k\,\delta(x^5) \implies x^9 = \frac{k|x^5|}{2} + \frac{kx^5}{2} 
\end{equation}
where the constants of integration are fixed by requiring that $x^9 = 0$ for $x^5 \rightarrow -\infty$. It can be seen from the solution to the Poisson's equation that for $x^5 > 0$ : $x^9 = kx^5$, which is the region of the (1, \textit{k}) 5-brane. Therefore, the intermediate brane makes an angle $\theta$ with the NS5-brane s.t $\text{tan}\,\theta = k$. 

Now that the issue of charge conservation has been properly addressed, back to the subject of providing masses to the chiral multiplets $\varphi_1, \tilde{\varphi}_1, \varphi_2, \tilde{\varphi}_2$, which are in the (\textbf{N}, 1), ($\mathbf{\bar{N}}$, 1), (1, \textbf{N}) and (1, $\mathbf{\bar{N}}$) representations of $U(N) \times U(N)$ respectively. The gauge invariant terms in the superspace action that can potentially generate mass terms like $m\bar{\omega}\omega$ ($\omega$ represents a spinor in these chiral multiplets) are
\begin{equation}\label{eq:posstermsmass}
\begin{split}
    &\mathlarger{\int} d^4\theta \,\, \text{Tr}_2\left[\bar{\varphi}_1 e^{V_1}\varphi_1 + \tilde{\varphi}_1e^{-V_1}\bar{\tilde{\varphi}}_1\right] + \text{Tr}_1\left[\bar{\varphi}_2 e^{V_2}\varphi_2 + \tilde{\varphi}_2e^{-V_2}\bar{\tilde{\varphi}}_2\right]\\
    &\implies \bar{\omega}\omega : \frac{1}{2}Tr_2\left[\tilde{\omega}_1 \sigma_1\bar{\tilde{\omega}}_1 - \bar{\omega}_1\sigma_1\omega_1\right] + \frac{1}{2}Tr_1\left[\tilde{\omega}_2 \sigma_2\bar{\tilde{\omega}}_2 - \bar{\omega}_2\sigma_2\omega_2\right]
\end{split}
\end{equation}
where $V_1$ and $V_2$ are the vector multiplets valued in the $(N^2, 1)$ and $(1, N^2)$ representations respectively. Such terms were already seen in the superspace expansions of gauge theories in the previous chapter. It can clearly be noted that the mass terms can be generated for non-zero vacuum expectation values $\langle \sigma_1 \rangle$ and $\langle \sigma_2 \rangle$. As mentioned earlier, the bosonic d.o.f $\sigma_1$, $\sigma_2$ correspond to the motion of the 012 subspace of D3-branes in the $x^5$ direction of NS5 and NS5' branes respectively; Meaning $\langle \sigma_1 \rangle$ and $\langle \sigma_2 \rangle$ specify their respective average $x^5$ positions. In the presence of heavy D5-branes, this position will have to be mentioned relative to them, which in the intersecting case implies $\langle \sigma_1 \rangle = \langle \sigma_2 \rangle = 0$, reassuring that they are currently massless. However, if $(k - \text{D5})_L$ and $(k - \text{D5})_R$ in figure \ref{fig:brane_3} move by a distance of $\tilde{m}$ and $m$ respectively on either side of the origin in the $x^5$ direction, then $\langle \sigma_i \rangle$ can take on four values : position of the D3-branes relative to $(k - \text{D5})_L$ [$x^5 - \tilde{m}$] or relative to $(k - \text{D5})_R$ [$x^5 + m$] in the (1, \textit{k}) breaking, position of the D3-branes relative to $(k - \text{D5})_L$ [$x^5 + \tilde{m}$] or relative to $(k - \text{D5})_R$ [$x^5 - m$] in the (1, -\textit{k}) breaking. In (\ref{eq:posstermsmass}), let one of the choices be such that \cite{Bergman_1999}
\begin{equation*}\label{eq:fermmassbrane}
    \begin{split}
        &-\bar{\omega}_1\sigma_1\omega_1 : \,\,\, \textbf{\text{mass}} \,\, = -\langle \sigma_1 \rangle = -(x^5_{D3} - \tilde{m}) \,\,\,\,\,\,\,\,\,\, \tilde{\omega}_1 \sigma_1\bar{\tilde{\omega}}_1 : \,\,\, \textbf{\text{mass}} \,\, = \langle \sigma_1 \rangle = x^5_{D3} + m \\
        &-\bar{\omega}_2\sigma_2\omega_2 : \,\,\, \textbf{\text{mass}} \,\, = -\langle \sigma_2 \rangle = -(x^{5'}_{D3} + \tilde{m}) \,\,\,\,\,\,\,\,\,\, \tilde{\omega}_2 \sigma_2\bar{\tilde{\omega}}_2 : \,\,\, \textbf{\text{mass}} \,\,= \langle \sigma_2 \rangle = x^{5'}_{D3} - m
    \end{split}
\end{equation*}
where $x^5_{D3}$ and $x^{5'}_{D3}$ are the $x^5$ positions of the 012 subspace of D3-branes in NS5 and NS5' respectively. We now have massive fermions $\omega_1, \tilde{\omega}_1, \omega_2, \tilde{\omega}_2$ each with \textit{k} copies, with the masses given above. Using the regularization procedure described at the beginning of this section, integrating out these massive fermions gives rise to Chern-Simons terms for each $U(N)$ factor in the $U(N) \times U(N)$ gauge group, with the levels as follows
\begin{equation}
    \begin{split}
        &\mathbb{1} \times U(N) \,\, : \,\,\,\text{CS-level} \,\,\,= \,\, \frac{k}{2} \,sgn\,(\tilde{m} - x^5_{D3}) + \frac{k}{2} \,sgn\, (m + x^5_{D3})\\
        &U(N) \times \mathbb{1} \,\, : \,\,\,\text{CS-level} \,\,\,= \,\, \frac{k}{2} \,sgn\,(x^{5'}_{D3} - m) - \frac{k}{2} \,sgn\, (x^{5'}_{D3} + \tilde{m})
    \end{split}
\end{equation}
where $sgn(x) = \frac{|x|}{x}$. We can clearly see that for $-m \,<\, x^5_{D3} \,<\, \tilde{m}$ and $-\tilde{m} \,< x^{5'}_{D3} <\, m$, CS-levels for the first  and the second $U(N)$ factors are \textit{k} and -\textit{k} respectively. Since $-m \,<\, x^5_{D3} \,<\, \tilde{m}$ and $-\tilde{m} \,< x^{5'}_{D3} <\, m$ correspond to the region of $(1, k)$, it can be considered to be the only relevant part in the NS5-D5 brane system from now on. Therefore we now have the brane configuration as an NS5'-brane along 012345, a (1, \textit{k}) 5-brane along 01234$[5, 9]_\theta$ and N-D3 branes breaking into two 012 subspaces in the NS5' and (1, \textit{k})5 branes; where $[5, 9]_\theta = x^5 cos(\theta) + x^9 sin(\theta)$ s.t $tan(\theta) = k$. The corresponding low energy effective theory is an $\mathcal{N} = 2$ Yang-Mills Chern-Simons theory with the product gauge group $U(N)_k \times U(N)_{-k}$, with four massless bifundamental matter multiplets ($\Phi_1, \Phi_2, \tilde{\Phi}_1, \tilde{\Phi}_2$) and their complex conjugates, two massless adjoint chiral multiplets ($Q_1, Q_2$) and two vector multiplets $V_1, V_2$ whose fermions $\chi_1, \chi_2$ have masses $\frac{g_{YM}^2k}{2\pi}$ and -$\frac{g_{YM}^2k}{2\pi}$ respectively.
\begin{equation}\label{eq:vecmass}
    \frac{1}{g_{YM}^2}\left(-\frac{g_{YM}^2k}{4\pi}\,Tr_1[2\bar{\chi}_1\chi_1] + \frac{g_{YM}^2k}{4\pi}\,Tr_2[2\bar{\chi}_2\chi_2]\right)
\end{equation}
Now the next step is to find a way to enhance the supersymmetry from $\mathcal{N} = 2$ to $\mathcal{N} = 3$.

\subsection{$\mathcal{N} = 3$ enhancement and the lift to M-theory}
\begin{figure}[h]
\centering
\includegraphics[width=0.9\textwidth]{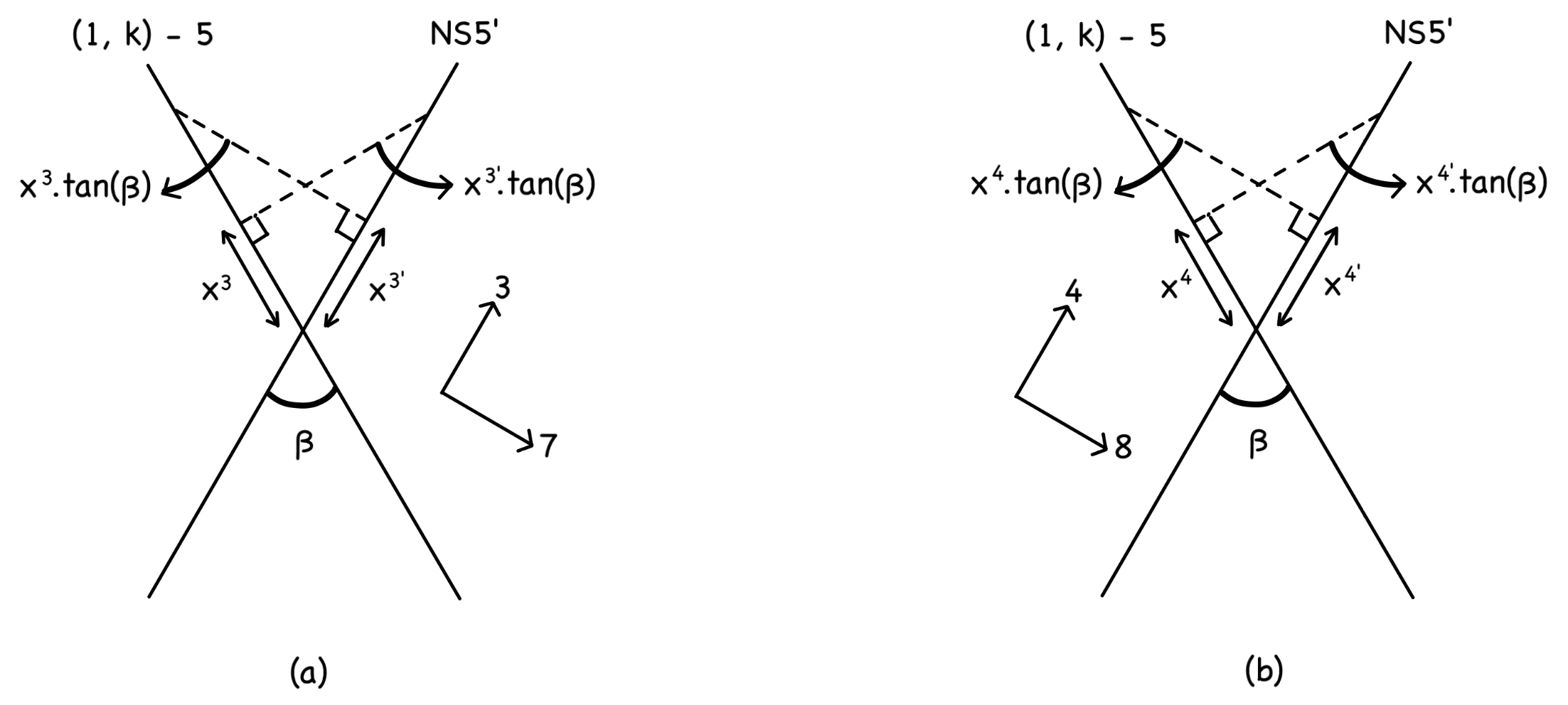}
\caption{(a) (1, \textit{k}) 5-brane rotated by an angle $\beta$ w.r.t the NS5'-brane in the 37 plane. (b) (1, \textit{k}) 5-brane rotated by an angle $\beta$ w.r.t the NS5'-brane in the 48 plane.}
\label{fig:brane_4}
\end{figure}

In the previous section it was seen that by having a (1, \textit{k}) 5-brane at an angle $\theta$ w.r.t the NS5' brane, the vector multiplets $V_1$ and $V_2$ whose bosonic d.o.f correspond to the motion of the D3-branes along $x^5$ in the NS5-branes, acquired masses $\frac{g_{YM}^2 tan(\theta)}{2\pi}$ and -$\frac{g_{YM}^2 tan(\theta)}{2\pi}$ respectively. The choice of the plane 59 can be generalized to planes having one direction longitudinal to the NS5'-brane and the other transverse to it. The two other such planes are 37 and 48, so rotate the (1, \textit{k}) 5-brane by angles $\alpha, \gamma$ w.r.t the NS5' brane in these two planes \cite{Aharony_2008}. As discussed earlier in section \ref{sec:n4brane}, the bosonic d.o.f of the adjoint chiral multiplets $Q_1, Q_2$ correspond to the motion of the D3-branes in the 34 directions of NS5 (now (1, \textit{k})) and NS5' branes respectively. Now by generalizing the case of the 59 plane, this means the multiplets $Q_1, Q_2$ acquire masses $\frac{g_{YM}^2 tan(\beta)}{2\pi}$ and -$\frac{g_{YM}^2 tan(\beta)}{2\pi}$ respectively; where note that $\alpha = \gamma = \beta$ since the d.o.f belonging to the same $\mathcal{N} = 2$ multiplet must have the same mass in order to preserve $\mathcal{N} = 2$ supersymmetry. See figure \ref{fig:brane_4} for a pictorial depiction of the factor $tan(\beta)$ using relative distances, similar to the discussion in the previous section.

\noindent The gauge invariant terms in the action that lead to such mass terms for $Q_1$ and $Q_2$ are
\begin{equation}\label{eq:adjmass}
\begin{split}
    &\frac{1}{g_{YM}^2}\mathlarger{\int}d^2\theta \,\, \left(\frac{g_{YM}^2 tan(\beta)}{2\pi}\text{Tr}_1\left[Q_1^2\right] - \frac{g_{YM}^2 tan(\beta)}{2\pi}\text{Tr}_2\left[Q_2^2\right]\right) + c.c\\
    &= \frac{1}{g_{YM}^2}\left(-\frac{g_{YM}^2 tan(\beta)}{4\pi}\text{Tr}_1\left[\lambda_1 \lambda_1 + \bar{\lambda}_1\bar{\lambda}_1\right] + \frac{g_{YM}^2 tan(\beta)}{4\pi}\text{Tr}_2\left[\lambda_2 \lambda_2 + \bar{\lambda}_2\bar{\lambda}_2\right]\right)
\end{split}  
\end{equation}
where $\lambda_1, \lambda_2$ are the complex spinors in $Q_1, Q_2$. It can be seen from (\ref{eq:vecmass}) that for the vector multiplets $V_j$, the two Majorana spinors $\chi_{jR}, \chi_{jI} : \chi_j = \chi_{jR} + i\chi_{jI}$ have masses of the same sign. On the contrary, for the adjoint chiral multiplets $Q_j$, the two Majorana spinors $\lambda_{jR}, \lambda_{jI} : \lambda_j = \lambda_{jR} + i\lambda_{jI}$ have masses of the opposite sign, which can be seen from (\ref{eq:adjmass}). For $\theta = \beta$ all of these fermions will have the same magnitude of mass, and the supersymmetry is enhanced to $\mathcal{N} = 3$ with the R-symmetry group $SO(3)_R \cong SU(2)_R$; where $\chi_{jR}, \chi_{jL}, \lambda_{jR}$ form the triplet under $SU(2)_R$ and $\lambda_{jI}$ forms the singlet. In section \ref{sec:n4brane}, it was mentioned that the R-symmetry group of $\mathcal{N} = 4$ i.e. $SO(3)_A \times SO(3)_B$ acts independently on the 345 and 789 subspaces respectively. For $\mathcal{N} = 3$, this $SO(3)_L \times SO(3)_R$ has broken down to it's diagonal subgroup $SO(3)_D$, which corresponds to rotation by the same $SO(3)$ element $\theta$ in the 37, 48 and 59 planes. 

Therefore we finally have $\mathcal{N} = 3$ $U(N)_k \times U(N)_{-k}$ Yang-Mills Chern-Simons theory as the low energy effective theory on the configuration : N-D3 branes along 0126, (1, \textit{k}) 5-brane along 012$[3, 7]_\theta [4, 8]_\theta [5, 9]_\theta$ and an NS5'-brane along 012345, with the D3-branes breaking into two 012 subspaces on the (1, \textit{k})-5 and NS5' branes which are separated along $x^6$. Also $[i, j]_\theta = x^i cos(\theta) + x^j sin(\theta)$; $tan(\theta) = k$. The Yang-Mills coupling in three dimensions, unlike the four-dimensional case, has an energy scale in it i.e. $g_{YM}^2 \propto E$, which can easily be seen by dimensional analysis. Therefore in the IR limit, the dimensionless coupling $\frac{E}{g_{YM}^2}$ is small and hence the Yang-Mills part of the theory becomes an irrelevant operator. On the field theory side, the auxiliary fields $Q_1, Q_2$ can then be integrated out in the resulting Chern-Simons matter theory to make the $\mathcal{N} = 6$ supersymmetry manifest, which was explicitly seen in the previous chapter. However, to see such an enhancement in the geometry of the brane configuration, further T-duality transformations and the lift to M-theory will have to be performed \cite{Aharony_2008}.

T-dualizing along a compact direction of radius $R$ maps Neumann boundary conditions into Dirichlet boundary conditions along a dual direction of radius $\tilde{R} = \frac{\alpha'}{R}$ (and vice-versa), thereby transforming a D(\textit{p}+1)-brane (wrapping the dualized direction) to a D\textit{p}-brane (and vice-versa) \cite{polchinski1996notes}. Similarly, T-dualizing along a direction transverse to the NS5-brane transforms it into a Kaluza-Klein monopole associated with the dual direction \cite{Eyras_1998}. Thereby in the current configuration, T-dualizing along the direction 6 transforms the D3-branes into D2-branes along 012, and the NS5'-brane into a KK-monopole along 012345 associated with the dual direction $\tilde{6}$. The bound state (1, \textit{k})5-brane transforms into a bound state of KK-monopole associated with $\tilde{6}$ (T-dual of the NS5) and \textit{k} D6-branes (T-dual of D5) acting as flux on the KK-monopole \cite{Sen_1997}, along 012$[3, 7]_\theta [4, 8]_\theta [5, 9]_\theta$. We now have a type IIA theory with the aforementioned dual brane-configuration, which when taken to the strong coupling limit $g_s \rightarrow \infty$ makes the eleventh dimension ($x^{10}$) of radius $l_s g_s$ non-compact, thereby lifting the theory to M-theory. In this lift, the D2-branes become M2-branes and the D6-branes become KK-monopoles associated with the 10 direction, and the other KK-monopoles associated with $\tilde{6}$ remain the same \cite{Townsend_1995}. We therefore have \textbf{N} M2-branes along 012, a KK-monopole along 012$[3, 7]_\theta [4, 8]_\theta [5, 9]_\theta$ associated with the linear combination of $\tilde{6}$ and 10, and a KK-monopole along 012345 associated with $\tilde{6}$. The geometry of this configuration preserves six supercharges ($\mathcal{N} = 3$ in 2+1 dimensions) \cite{Gauntlett_1997}, and the low energy limit corresponds to localizing to the intersection region of the two KK-monopoles, which has a $\mathbb{C}^4 \, / \, \mathbb{Z}_k$ singularity that preserves twelve supercharges, thereby finally leading to the $\mathcal{N} = 6$ ABJM theory. The upcoming chapter will delve into a more detailed exploration of this geometry.

%% file: chapters/situational_theoretical_analysis.tex
\chapter{Brane geometry for the ABJM theory}\label{chap:situational_theoretical_analysis}
In the previous chapter, D-branes were treated in the light of being higher-dimensional objects on which open strings can end in the perturbative regime of string theory. However, there is another perspective that sheds more light on their true non-perturbative nature, in which they appear as either elementary (singular) or solitonic (non-singular) solutions of supergravity that saturate the BPS bound of the corresponding supersymmetry algebra. Naturally for the supergravity approximation to be valid, we would have to be in the weak curvature regime i.e., $L \gg l_s$ ($l_p$ for M-theory); where L is the characteristic length scale of the spacetime geometry sourced by the brane. Similarly in the case of M-theory, the M2-brane is an elementary BPS solution of 11D supergravity w.r.t the three-form field $A_3$, and the M5-brane is it's magnetic solitonic dual. The other solutions of interest for us are the elementary (F1-string) and it's solitonic dual (NS5-brane) w.r.t the Kalb-Ramond two-form $B_2$, the elementary (Kaluza-Klein modes) and it's solitonic dual (KK-monopoles) w.r.t the KK-gauge field ($G_{\mu d}$; where $G$ is the metric and $d$ represents the dimension being reduced). Since these are exact non-trivial solutions to the highly non-linear supergravity approximation of string theory (and M-theory), they thereby probe the non-linear structure of the theory; which when combined with the fact that their tensions can be safely extrapolated to arbitrarily strong couplings by the virtue of a protective BPS bound, captures the trace of their non-perturbative nature. 

\section{M2-Brane solution}\label{sec:m2branesol}
Since the previous chapter ended in a cliffhanger with a configuration of M2-branes and two KK-monopoles, let us first look at the singular solution to 11D supergravity corresponding to M2-branes, which was constructed in \cite{Duff:1990xz}. The derivation is long and procedural, and since it is not the main objective of this thesis, only an outline of it will be presented with important intermediate steps highlighted following \cite{phdthesis}. The 012 worldvolume M2-branes has Poincare group invariance $ISO(2, 1)$, while the subspace transverse to isolated M2-branes has an $SO(8)$ symmetry group. Following these symmetries, let the ansatz for the line element be
\begin{equation}\label{eq:lineelement}
    ds^2_{M2} = e^{2A(r)}ds^2_{M_{1, 2}} + e^{2B(r)}ds^2_{X_8}
\end{equation}
where $ds^2_{M_{1, 2}}$ is the Minkowski metric in 2+1 dimensionsm $ds^2_{X_8}$ is the Euclidean metric for the transverse eight dimensions (the reason for it's notation to be $X_8$ rather than $E_8$ will be clear later), and \textit{r} is the radial distance in this transverse space $X_8$. The bosonic part of the 11-dimensional supergravity action is \cite{Becker_Becker_Schwarz_2006}
\begin{equation}\label{eq:sugra11action}
\begin{split}
    &2\kappa_{11}^2 S_{\text{bosonic}} = \mathlarger{\int}_{M, \,G} R \star \mathbb{1} - \frac{1}{2} F_4 \wedge \star F_4 - \frac{1}{6} A_3 \wedge F_4 \wedge F_4 \,\,\,;\,\,\, 2\kappa_{11}^2 = \frac{1}{2\pi}(2\pi l_p)^9\\
    &\star \,\,\coloneq\,\, \text{Hodge star operator}\,\,\,\,|\,\,\,\, G \coloneq \text{Metric tensor} \,\,:\,\, G_{MN} = \eta_{AB}E^A_M E^B_N 
\end{split}
\end{equation}
where $E^A_M$ is the elfbein (the 11D analogue of vierbein in the tetrad formalism of four dimensions), $F_4 = dA_3$ is the field strength associated with the three-form field ($A_3$) in M-theory and $R$ is the Ricci scalar. Also, only the bosonic part of the supergravity action is relevant for the classical solutions that we seek to construct, since a classical solution always has vanishing background fermionic fields. However, the supersymmetry transformations of the full theory are relevant to analyze, since our classical solutions preserve a certain number of supersymmetries (BPS). 
\begin{align}
    &\delta E^A_M = \bar{\epsilon}\,\Gamma^A \Psi_M \,\,\,;\,\,\, \delta A_{MNP} = -3\bar{\epsilon}\,\Gamma_{[MN}\Psi_{P]} \,\,\,;\,\,\, \delta \Psi_M = \nabla_M \epsilon + \frac{1}{12}\left(\Gamma_M{\textbf{F}}^{(4)} - 3\textbf{F}_M^{(4)}\right)\epsilon\nonumber\\
    &\textbf{\text{where}} \,\,\,\,\, {\textbf{F}}^{(4)} = \frac{1}{4!}F_{MNPQ}\Gamma^{MNPQ} \,\,\,;\,\,\,\textbf{F}^{(4)}_M = \frac{1}{2}\left[\Gamma_M, \textbf{F}^{(4)}\right] \,\,\,;\,\,\, \Gamma_M = E^A_M \Gamma_A \\
    &\nabla_M \epsilon = \partial_M \epsilon + \frac{1}{4}\omega_{MAB} \Gamma^{AB} \epsilon \,\,\,;\,\,\, \omega_{MAB} = \frac{1}{2}(-\Omega_{MAB} + \Omega_{ABM} - \Omega_{BMA}) \,\,\,;\,\,\, \Omega_{MN}^A = 2\partial_{[N}E^A_{M]}\nonumber   
\end{align}
where $M,N,P,...$ represent the coordinate basis and $A,B,C,...$ represent the local tangent-space basis. Consequently, $\Gamma_M$ and $\Gamma_A$ are the coordinate-dependent (curved space) and coordinate-independent (flat space) 11D gamma matrices; $\Gamma_{MN...}, \Gamma_{AB...}$ represent the anti-symmetric product of gamma matrices and so does the [.] bracket. The trick to construct classical solutions that saturate the BPS bound is to find \textit{Killing spinors} \cite{MatthiasBlau}; which are spinors ($\epsilon_K$) that parametrize the supersymmetry transformations ($\delta_{\epsilon}$) such that they leave a specific field ($E^A_M, A_{MNP}, \Psi_M$) configuration invariant i.e., $\delta E^A_M = \delta A_{MNP} = \delta\Psi_M = 0$. The idea is similar to \textit{Killing vectors} that characterize bosonic symmetries. Now note that for classical solutions $\delta E^A_M = \delta A_{MNP} = 0$, since the fermionic field on the RHS (Gravitino $\Psi_M$) vanishes anyways. Therefore the only equation left to solve in order to find $\epsilon_K$ is
\begin{equation}\label{eq:killingspinoreq}
    \delta \Psi_M = \nabla_M \epsilon_K + \frac{1}{12}\left(\Gamma_M{\textbf{F}}^{(4)} - 3\textbf{F}_M^{(4)}\right)\epsilon_K = 0
\end{equation}
In order to conveniently use the ansatz (\ref{eq:lineelement}), split the labels into $m, \hat{m}$ and similarly $a, \hat{a}$ 

\noindent where the labels without the \,\,$\hat{}$\,\, correspond to the longitudinal 012 directions and the 
labels with the \,\,$\hat{}$\,\, correspond to the transverse directions. The elfbein corresponding to the ansatz (\ref{eq:lineelement}) is then
\begin{equation}\label{eq:elfbeinansatz}
    E^A_M = \begin{cases}
        e^{A(r)}\,\delta^A_M & A = a\\
        e^{B(r)}\,\delta^A_M & A = \hat{a}
    \end{cases}
\end{equation}
The gamma matrices can be written in a new basis that makes them manifestly compatible with the splitting of labels mentioned earlier. Consequently the spinor can be decomposed accordingly, making it manifestly respect $ISO(2, 1) \times SO(8)$ symmetry
\begin{align}\label{eq:basischoicegamma}
    &\Gamma^A = (\Gamma^a, \Gamma^{\hat{a}}) = (\gamma^a \otimes \Sigma^9, \mathbb{1}_2 \otimes \Sigma^{\hat{a}}) \,\,\,;\,\,\, \Sigma^9 = \prod_{\hat{b}}\Gamma^{\hat{b}} \implies \left(\Sigma^9\right)^2 = \mathbb{1}_8\nonumber\\
    &\epsilon(x^m, x^{\hat{m}}) = \eta_0 \otimes \varepsilon(r) \,\,\,\,\,\textbf{\text{s.t}}\,\,\,\,\, \eta_0 \in Spin(2, 1) \,\,\,;\,\,\, \varepsilon(r) = \varepsilon_L(r) + \varepsilon_R(r)  \in Spin(8)
\end{align}
where $\gamma^a$, $\Sigma^{\hat{a}}$ are the gamma matrices of 2+1 dimensional Minkowski space and eight-dimensional Euclidean space respectively. Also, $\eta_0$ is a constant spinor. Now since we are looking for singular solutions (as opposed to solitonic), let the symmetry respecting ansatz for the three-form field $A_3$ be
\begin{equation}\label{eq:threeformansatz}
    A_3 = \pm e^{C(r)} dx^0 \wedge dx^1 \wedge dx^2
\end{equation}
Substituting the ansätze (\ref{eq:elfbeinansatz}) and (\ref{eq:threeformansatz}) in (\ref{eq:killingspinoreq}), with the choice of basis (\ref{eq:basischoicegamma}) for the gamma matrices results in the following
\begin{align}
    &(a) \,\,\delta \Psi_m = 0 \,\,:\,\, -\frac{1}{6}e^{-3A(r)}\Sigma^{\hat{a}}\gamma_m \partial_{\hat{a}} \left(e^{3A(r)}\Sigma^9 \mp e^{C(r)}\right) \, \epsilon = 0 \label{eq:kslong}\\
    &(b) \,\,\delta \Psi_{\hat{m}} = 0 \,\,:\,\, \left[\partial_{\hat{m}} \mp \frac{1}{6}e^{-3A(r)}\Sigma^9\partial_{\hat{m}}e^{C(r)} - \frac{1}{2}\Sigma^{\hat{n}}_{\,\,\,\hat{m}}\left(\partial_{\hat{n}}B(r) \pm \frac{1}{6}e^{-3A(r)}\Sigma^9\partial_{\hat{n}}e^{C(r)}\right)\right]\,\epsilon = 0 \nonumber
\end{align}
The trick to solving these equations is to remember that we are looking for 1/2-BPS solutions, and gamma matrices or more specifically chiral projection matrices do the job of decomposing a spinor into two halves of components. Thereby, it can be noted in (\ref{eq:kslong} (a)) that for the choice $C(r) = 3A(r)$, we get the chiral projector in the transverse space $\frac{1}{2}(1 - \Sigma^9)$. Similarly, it can be noted in (\ref{eq:kslong}(b)) that for the choices $C(r) = 3A(r)$ and $C(r) = -6B(r)$, we get $\frac{1}{2}(1 - \Sigma^9)$ in the equation. With these choices, the solutions to the above equations are
\begin{align}
    &(\ref{eq:kslong}) \,(a) \,\,:\,\, \epsilon = f(x^{\hat{m}})(1 \pm \Sigma^9)\epsilon_0 \,\,\implies\,\, (\ref{eq:kslong})\,(b) \,\,:\,\, f(x^{\hat{m}}) =  e^{\frac{C(r)}{6}} \nonumber\\
    &\text{Therefore} \,\,\,\,\, \epsilon = e^{\frac{C(r)}{6}}\eta_0 \otimes \varepsilon_0 \,\,\,;\,\,\, (1 \mp \Sigma^9)\varepsilon_0 = 0, \,\, C(r) = 3A(r) = -6B(r) \label{eq:kssol} 
\end{align}
where $\epsilon_0, \eta_0, \varepsilon_0$ are constant spinors. It can clearly be seen from (\ref{eq:kssol}) that only half the components of $\varepsilon_0$ and hence of $\epsilon$ are non-zero, thereby justifying the 1/2-BPS nature of solutions. Now to solve for the explicit form of the function $C(r)$, equations of motion for the three-form $A_3$ in the action (\ref{eq:sugra11action}) will have to be solved, resulting in the following 
\begin{align}
    &\text{EOM}\,\,:\,\,d \star F_4 + \frac{1}{2}F_4 \wedge F_4 = 0 \,\,\,\,\,\,\,\,| \,\,\,\,\,\,\,\, d \,\,\coloneq\,\,\text{Exterior derivative}\nonumber\\
    &\implies e^{-C(r)} \equiv H(r) = 1 + \frac{L^6}{r^6} \label{eq:functionthree}
\end{align}
Using the relations $C(r) = 3A(r) = -6B(r)$ derived earlier from the Killing spinor equation, the supergravity solution for a stack of \textit{N'} coincident M2-branes finally becomes
\begin{empheq}[box=\fbox]{align}\label{eq:m2branesol}
  \begin{split}
        ds^2_{M2} = H(r)^{-\frac{2}{3}}ds^2_{M_{1, 2}} + H(r)^{\frac{1}{3}}ds^2_{X_8} \,\,\,\,&\Big\vert\,\,\,\, H(r) = 1 + \frac{L^6}{r^6}\\
        \,\,A_3 = \pm \,H(r)^{-1} dx^0 \wedge dx^1 \wedge dx^2 \,\,\,\,\,\,\,\,\,\,\,&\Big\vert\,\,\,\, L^6 = 32\pi^2N'l_p^6
   \end{split}
\end{empheq}
where $L$ is the characteristic length scale of the geometry. The relation $L = 32\pi^2Nl_p^6$ can be derived by requiring the metric component $g_{00}$ to be related to the Newtonian potential in the asymptotic limit. In this limit the M2-brane can be treated as a point source in eight dimensions, and hence the relation for the Schwarzschild solution in higher dimensions can be generalised to this case \cite{Emparan_2008}
\begin{equation}\label{eq:charlengthscale}
    \lim_{r\to\infty} g_{00} \approx -1 + \frac{2}{3}\left(\frac{L^6}{r^6}\right) = -1 + \frac{16\pi G_{11}N'T_{M2}}{9\Omega_7 r^6}
\end{equation}
where $T_{M2} = \frac{1}{(2\pi)^2l_p^3}$ is the tension of the M2-brane, $G_{11} = \frac{\kappa_{11}^2}{8\pi}$ is the 11-dimensional Newton's constant from (\ref{eq:sugra11action}), and $\Omega_7 = \pi^4/3$ is the volume of a unit seven-sphere. Substituting all of these in (\ref{eq:charlengthscale}) gives us the relation $L^6 = 32\pi^2N'l_p^6$.

\section{Embedding of KK-Monopoles}
KK-monopoles in four dimensions are solitonic solutions in the five dimensional Kaluza-Klein theory, that are magnetically charged with respect to the KK $U(1)$ gauge field $g_{\mu d}$; where $d$ is the dimension being reduced. The solution in the spatial directions is characterized by the metric of a Taub-NUT instanton \cite{Gross:1983hb}
\begin{equation}\label{eq:taubnut}
    \begin{split}
        &ds^2_5 = -dt^2 + ds^2_{TN} \,\,\,\,\,;\,\,\,\,\, ds^2_{TN} = V(r)(dr^2 + r^2d\Omega^2_2) + \frac{1}{V(r)}{(dy + A)}^2\\
        &\text{where}\,\,\,\,V(r) = 1 + \frac{R}{2r}\,,\,\,\, A = R\,sin^2\left(\frac{\theta}{2}\right)d\phi\,,\,\,\,d\Omega_2^2 = d\theta^2 + sin^2\theta d\phi^2
    \end{split}
\end{equation}
where $A$ is the gauge field of the monopole, consistent with the fact that the field strength ($F_2 = \frac{R}{2}\,sin\theta\,d\theta \wedge d\phi$) is proportional to the $S^2$(two-sphere) volume form $\Omega_2 = sin\theta\,d\theta\wedge d\phi$. The consequent non-trivial magnetic charge of the monopole is $2\pi R$, and by making a geometric analogue of the Dirac quantization argument i.e., by requiring the solution to be non-singular as $r \rightarrow 0$, one concludes that the coordinate $y$ has period $2\pi R$. This bodes well with the fact that $y$ represents the compact dimension, whose physical radius is now $\tilde{R}(r) = {V(r)}^{-\frac{1}{2}} R$, which approaches $R$ for $r \rightarrow \infty$ and zero for $r \rightarrow 0$. Therefore a KK-monopole is specified by a circle that shrinks ($\tilde{R} \rightarrow 0$) at it's core ($r \rightarrow 0$), and three spatial dimensions transverse to the monopole. The generalization to the case of multiple monopoles comes in the form of a multi-center Taub-NUT metric, by making use of the property that these monopoles do not interact with each other (\cite{Gibbons:1978tef, Gross:1983hb})
\begin{equation}
    ds^2_5 = -dt^2 + V(\vec{x})d\vec{x} \cdot d\vec{x} + \frac{1}{V(\vec{x})}{(dy + \vec{A}\cdot d\vec{x})}^2 \,\,\,\,;\,\,\,\, V(\vec{x}) = 1 + \frac{R}{2}\sum_{\alpha = 1}^{M}\frac{1}{|\vec{x} - \vec{x}_\alpha|}
\end{equation}
where $\vec{x}_\alpha$ corresponds to the core of each of the $M$ monopoles, $\vec{A}$ is the component vector of the one-form gauge field $A$, and $\vec{x}$ is the vector of three spatial coordinates transverse to the monopole. The spatial part of this solution is the special case ($\epsilon = 1$, $R_\alpha = R$, $y \sim y + 2\pi R$) of a Gravitational Multi-Instanton solution called the Gibbons-Hawking metric which is \cite{Gibbons:1978tef}
\begin{equation}
    ds^2_{GH} = U d\vec{x} \cdot d\vec{x} + U^{-1}{(dy + \vec{\omega} \cdot d\vec{x})}^2 \,\,\,;\,\,\, \vec{\nabla}U = \vec{\nabla} \times \vec{\omega} \,,\,\,\,U = \epsilon + \frac{1}{2}\sum_{\alpha = 1}^M\frac{R_\alpha}{|\vec{x} - \vec{x}_\alpha|}
\end{equation}
where $\vec{\nabla}$ is the gradient operator. Continuing on the theme of generalization, Gibbons-Hawking metrics themselves are a subclass of metrics on four dimensional hyper-K\"{a}hler manifolds with a triholomorphic $S^1$ (compact dimension) isometry \cite{article}. A brief but a more formal introduction to hyper-K\"{a}hler manifolds will be given in section \ref{sec:preservedsusy}, but for now let us stick to the agenda of arriving at the metric for the monopole configuration of interest. For this, an important observation to make is the derivation of monopole solutions by embedding Gibbons-Hawking metrics in the space transverse to the monopole world volume ($x^0$). This can be generalized to eleven dimensions, where say the Gibbons-Hawking metric is embedded in the space transverse to the world volume of a D6-brane with $y$ being the M-theory circle $x^{10}$; This gives rise to a monopole solution with $x^{10}$ shrinking at it's core. But as mentioned towards the end of the previous chapter, one of the monopoles of interest is associated with the linear combination of two circles $\tilde{6}$ and $10$, rather than a single circle. To find a solution to this, it is only natural to consider a subclass of metrics on hyper-K\"{a}hler manifolds with triholomorphic $T^2 (S^1 \times S^1)$ isometry, which are called toric hyper-K\"{a}hler manifolds (\cite{published_papers, Goto1991pm}). The general 4\textit{n} dimensional toric hyper-K\"{a}hler metric has the local form \cite{Gauntlett_1997}
\begin{align}
    &ds^2 = U_{ij} d\vec{x}^i \cdot d\vec{x}^j + U^{ij}(dy_i + A_i)(dy_j + A_j) \,\,\,;\,\,\, A_i = d\vec{x}^j \cdot \vec{\omega}_{ji} = dx^j_r \omega^r_{ji}\label{eq:torhyperkahlermetric}\\
    &\text{where}\,\,\,\,\, \partial_{x^j_r} \omega^s_{ki} - \partial_{x^k_s}\omega^r_{ji} = \epsilon^{rst}\partial_{x^j_t}U_{ki} \,\,\,\,|\,\,\, \partial_{x^{[j}_t}U_{k]}i = 0\,,\,\,\,U^{ij}\partial_{\vec{x}^i}\cdot\partial_{\vec{x}^j}U = 0 \label{eq:kahlerconditions}
\end{align}
where $\vec{x}^i = \{x^i_r\,;\,r = 1,2,3\}$ are \textit{n} triplets of coordinates, $U_{ij}$ are the entries of a positive definite symmetric $n \times n$ matrix $U$ with $U^{ij}$ being the entries of $U^{-1}$, and $\vec{\omega}_{ji}$ is a triplet of $n \times n$ matrix functions. Also $y_i \sim y_i + 2\pi$ for $i = 1, 2,..., n$ represent the \textit{n} compact dimensions of $T^n$. Solving the linear constraints on $U$ in (\ref{eq:kahlerconditions}), gives rise to solutions 
\begin{equation}\label{eq:Usolhk}
    U_{ij} = U_{ij}^{(\infty)} + \sum_{\{p\}}\sum_{m = 1}^{N(\{p\})} U_{ij}[\{p\}, \vec{a}_m(\{p\})] \,\,\,\,\Big\vert\,\,\,\,U_{ij}[\{p\}, \vec{a}] = \frac{p_i p_j}{2|\sum_k p_k\vec{x}^k - \vec{a}|}
\end{equation}
where \{$p_k\,;\,k = 1, 2,..., n$\} is a vector of \textit{n} real numbers that are required to be co-prime integers, for the solution to be non-singular. $N(\{p\})$ denotes the number of solutions with the specific \textit{p}-vector \{\textit{p}\}, and \{$\vec{a}_m(\{p\})$\,;\,$m = 1, 2,..., N(\{p\})$\} are the set of arbitrary $N(\{p\})$ three-vectors associated with the \textit{p}-vector $\{p\}$. The $\sum_{\{p\}}$ then represents the sum over solutions with distinct \textit{p}-vectors. The solution (\ref{eq:torhyperkahlermetric}) is then specified by a host of 3(\textit{n} - 1)\textit{d}-planes, which for a \textit{p}-vector $\{p\}$ and an arbitrary vector $\vec{a}_m(\{p\})$ is given by
\begin{equation}\label{eq:planehk}
    \sum_{k = 1}^n p_k \vec{x}^k = \vec{a}_m
\end{equation}
The metric (\ref{eq:torhyperkahlermetric}) is also $SL(n, \mathbb{Z})$ invariant in the sense that, the action of $S \in SL(2, \mathbb{Z})$ on $U (U \rightarrow S^T U S)$ and on $\{p\}(\{p\} \rightarrow S\{p\})$, takes a solution (\ref{eq:Usolhk}) specified by the planes (\ref{eq:planehk}) parametrized by $\{p\}$, to another solution of the same form (\ref{eq:Usolhk}) now specified by planes (\ref{eq:planehk}) parametrized by $S\{p\}$. The requirement for the entries of \textit{p}-vectors to stay co-prime integers restricts the group to be $SL(n, \mathbb{Z})$ rather than $SL(n, \mathbb{R})$. The solutions of interest for us are the ones with $T^2$ isometry i.e., \textit{n} = 2. Let us call ($y_1, y_2$) as ($\tilde{x}^6, x^{10}$) respectively and [$\vec{x}^1, \vec{x}^2$] as [$(x^3, x^4, x^5), (x^7, x^8, x^9)$] respectively. Also, consider the case of two distinct \textit{p}-vectors $\{p\} : [p_{(1)}, p_{(2)}] = [(p_1, q_1), (p_2, q_2)]$ and let the arbitrary vectors be $\vec{a} = \vec{0}$. The angle $\theta$ between the two 3-planes corresponding to the two \textit{p}-vectors (\ref{eq:planehk}), subject to the metric (\ref{eq:torhyperkahlermetric}), is then given by  
\begin{equation}\label{eq:anglehk}
    cos\,\theta = \frac{p_{(1)}\cdot p_{(2)}}{\sqrt{p_{(1)}^2 p_{(2)}^2}} \,\,\,;\,\,\, p \cdot q = {\left(U^{(\infty)}\right)}^{ij}p_i q_j
\end{equation}
It has been shown in \cite{Gauntlett_1997} that the eight dimensional metric (\ref{eq:torhyperkahlermetric}) corresponding to this special case, when embedded in the transverse space $ds^2_{X_8}$ of the general M2-brane solution (\ref{eq:m2branesol}), dimensionally reduced along 10 and T-dualized along $\tilde{6}$ can be interpreted as : Two $(p_1, q_1)$ and $(p_2, q_2)$5-branes at an angle $\theta$ w.r.t each other in the 37, 48 and 59 planes, with D3-branes suspended between them along the separating $6$ direction in type IIB supergravity. This is precisely the configuration seen in the previous chapter for $p_{(1)} = (1, 0)$ and $p_{(2)} = (1, k)$, with $tan\,\theta = k$. Further restricting oneself to solutions s.t $det(U^{(\infty)}) = 1$, it can easily be seen from (\ref{eq:planehk}) that (\ref{eq:anglehk}) is invariant under $SL(2, \mathbb{R})$, with the coordinate pairs $(x^3, x^7), (x^4, x^8)$ and $(x^5, x^9)$ transforming as $SL(2, \mathbb{R})$ doublets. Therefore for the current case of two \textit{p}-vectors, an $SL(2, \mathbb{R})$ transformation can always be performed on the coordinates s.t the angle $\theta$ remains invariant, while the vacuum solution $U^{(\infty)}$ is brought to be equal to the identity matrix $\mathbb{1}_2$. Therefore finally, the eleven dimensional supergravity solution corresponding to \textbf{N} M2-branes along 012, a KK-monopole along 012$[3, 7]_\theta [4, 8]_\theta [5, 9]_\theta$ associated with the linear combination of $\tilde{6}$ and 10 circles, and a KK-monopole along 012345 associated with the $\tilde{6}$ circle is obtained by $\rightarrow$ Embedding (\ref{eq:torhyperkahlermetric}) in (\ref{eq:m2branesol}) by using (\ref{eq:Usolhk}) for the case of two \textit{p}-vectors $(1, 0)$ and $(1, k)$
\begin{empheq}[box=\fbox]{align}\label{eq:m2kkbranesol}
  \begin{split}
        &ds^2_{M2-KK} = H^{-\frac{2}{3}}ds^2_{M_{1, 2}} + H^{\frac{1}{3}}\left[U_{ij}d\vec{x}^i \cdot d\vec{x}^j + U^{ij}(dy_i + A_i)(dy_j + A_j)\right]\\
        &U = \mathbb{1}_2 + \begin{pmatrix}
            h_1 && 0 \\ 0 && 0 
        \end{pmatrix} + \begin{pmatrix}
            h_2 && kh_2 \\ kh_2 && k^2h_2
        \end{pmatrix} \,\,\,\,;\,\,\,\, h_1 = \frac{1}{2|\vec{x}_1|} \,,\,\,\, h_2 = \frac{1}{2|\vec{x}_1 + k\vec{x}_2|}
   \end{split}
\end{empheq}
where $H$ is now a harmonic function on the transverse toric hyper-K\"{a}hler 8-manifold, as opposed to the free M2-brane case (\ref{eq:m2branesol}) where it was a harmonic function on the $SO(8)$ invariant transverse euclidean space. 

\section{Unbroken supersymmetry}\label{sec:preservedsusy}
The next task is to find the no. of supersymmetries preserved by a solution of the form (\ref{eq:m2kkbranesol}). As mentioned in section \ref{sec:m2branesol}, it can be done by solving the killing spinor equation (\ref{eq:killingspinoreq}), which reads (reiterated to introduce the notation of supercovariant derivative $D_M$)
\begin{equation}
    D_{{M}}\epsilon_K = \left[\nabla_{M} + \frac{1}{12}\left(\Gamma_{M}{\textbf{F}}^{(4)} - 3\textbf{F}_{M}^{(4)}\right)\right]\epsilon_K = 0
\end{equation}
The solution and the steps to arrive at it are almost exactly the same as the isolated M2-brane case, except for the fact that the geometry of the hyper-K\"{a}hler manifold enters via $\nabla_{\hat{m}}$ into the terms containing $B(r)$ in (\ref{eq:kslong} (b)). Therefore the solution (\ref{eq:kssol}) is slightly generalized with regards to the conditions on $\varepsilon_0$ and $\eta_0$, as follows
\begin{equation}\label{eq:covconst}
    \epsilon = H^{-\frac{1}{6}}\eta_0 \otimes \varepsilon_0\,\,\,;\,\,\,(1 \mp \Sigma^9)\varepsilon_0 = 0 \,\,\,|\,\,\,\nabla_{\hat{m}}\varepsilon_0 = 0\,,\,\,\,\nabla_m\eta_0 = 0
\end{equation}
where the reader may be reminded that $m, \hat{m}$ correspond to labels representing the longitudinal and transverse directions respectively. $\eta_0$ is the covariantly constant ($\nabla_m \eta_0 = 0$) spinor $\in$ $SL(2, \mathbb{R})$ in the longitudinal Minkowski space; And $\varepsilon_0 = \varepsilon_{s} + \varepsilon_{c}$ is the covariantly constant ($\nabla_{\hat{m}} \varepsilon_0 = 0$) spinor in the transverse space, with either non-zero $\varepsilon_s \in$ $Spin(8)$ or non-zero $\varepsilon_c \in$ $Spin(8)$, depending on the sign in the chirality condition $(1 \mp \Sigma^9)\varepsilon_0 = 0$. Earlier in the case of isolated M2-branes, covariantly constant $\varepsilon_0$ turned out to be a trivial constant spinor as the transverse space was Euclidean, but it is not the case anymore as the transverse space is now a toric hyper-K\"{a}hler 8-manifold with a richer geometric structure. However, before examining covariantly constant spinors on this manifold, a brief but a more formal introduction to hyper-K\"{a}hler manifolds is in order \cite{Nigelhitchin}. 
\\

\noindent \textbf{Definition} : A hyper-K\"{a}hler manifold is a $4n$-dimensional Riemannian manifold ($M, g$) with three covariantly constant orthogonal automorphisms ($T_pM \rightarrow T_pM)$ \textit{I}, \textit{J} and \textit{K} of the tangent bundle which satisfy the quarternionic identities $I^2 = J^2 = K^2 = IJK = -1$; where $T_pM$ denotes the tangent space at point $p$ of the manifold $M$, and -1 denotes the negative of the identity automorphism $1 = id$ on the tangent bundle. The presence of these three complex structures induces three K\"{a}hler two-forms $\omega_I, \omega_J, \omega_K$ on $M$
\begin{equation*}
    \omega_I(X, Y) = g(IX, Y) \,\,\,;\,\,\, \omega_J(X, Y) = g(JX, Y) \,\,\,;\,\,\, \omega_K(X, Y) = g(KX, Y) \,\,\,|\,\,\, X, Y \in T_pM
\end{equation*}
such that $\Omega_I \coloneq \omega_J + i\omega_K$ is holomorphically symplectic with respect to I, and so on cyclically. Now given a connection ($\nabla$) on the tangent bundle of $M$, the parallel transport of $T_pM$ around a loop $\gamma : [0, 1] \rightarrow M$ s.t $\gamma(0) = \gamma(1) = p$ gives an automorphism element $hol_{\nabla}(\gamma) : T_pM \rightarrow T_pM$, called the holonomy. The collection of all such elements for all the loops $\gamma$ based at $p$ yields the holonomy group $Hol(p)$. Since the complex structures \textit{I}, \textit{J} and \textit{K} are covariantly constant, $Hol(p) \subset [O(4n) \cap GL(n, \mathbb{H})] \subset Sp(n)$; where $GL(n, \mathbb{H})$ is the group of invertible matrices over the field of quarternions, and $Sp(n) \coloneq Sp(2n, \mathbb{C}) \cap SU(2n)$ is the compact symplectic group. 

The isometries on $M$ generated by killing vectors $X$ are called triholomorphic, if they preserve the three complex structures and their corresponding K\"{a}hler forms \cite{Hitchin:1986ea}
\begin{equation}\label{eq:triholo}
    \mathcal{L}_X\omega_I = \mathcal{L}_X\omega_J = \mathcal{L}_X\omega_K = 0 \,\,\,;\,\,\, \mathcal{L}_XI = \mathcal{L}_XJ = \mathcal{L}_XK = 0
\end{equation}
where $\mathcal{L}_X$ is the Lie derivative w.r.t $X$, and the structures \textit{I}, \textit{J}, \textit{K} are assumed to not rotate under the isometries. Now back to the problem of analyzing covariantly constant spinors on the toric hyper-K\"{a}hler manifold with $T^2$ isometry. Let the killing vectors corresponding to the two compact dimensions $y_1$ and $y_2$ be
\begin{equation}\label{eq:killingantisymm}
    Y^i = \frac{\partial}{\partial y_{i}} \,\,\,;\,\,\, i = 1, \,2 \,\,\,\,\,\Big|\,\,\,\,\, (\mathcal{L}_{Y^i})G_{\hat{m}\hat{n}} = 0 \implies \nabla_{\hat{m}}(Y^i)_{\hat{n}} + \nabla_{\hat{n}}(Y^i)_{\hat{m}} = 0
\end{equation}
Since these are isometries, we expect the the preserved supersymmetries $\varepsilon_0$ to be independent of $y_i$, which can be written using the Lie derivative of spinors as \cite{lichnerowicz}
\begin{equation}\label{eq:liederspin}
    (\mathcal{L}_{Y^i})\varepsilon_0 = (Y^i)^{\hat{m}}\nabla_{\hat{m}}\,\varepsilon_0 - \frac{1}{4}\Gamma^{\hat{m}}\Gamma^{\hat{n}}\nabla_{\hat{m}}(Y^i)_{\hat{n}}\,\varepsilon_0 = -\frac{1}{4}\Gamma^{{\hat{m}}{\hat{n}}}R[\nabla Y^i]_{\hat{m}\hat{n}}\,\varepsilon_0 = 0
\end{equation}
where the covariant constancy of $\varepsilon_0 : \nabla_{\hat{m}}\varepsilon_0 = 0$ from (\ref{eq:covconst}) has been used. From (\ref{eq:killingantisymm}), it can be seen that $\nabla Y^i$ is an $8 \times 8$ anti-symmetric matrix, hence $\in$ Lie algebra $so(8)$, and $\epsilon_0$ lies in a representation $R$ of $so(8)$, which is what the $R[\nabla Y^i]$ represents in (\ref{eq:liederspin}). But this is not the end of the story as we will see now. Take the structure $K$, a vector $X$, and consider the lie derivative of the vector $KX$ w.r.t $Y^i$
\begin{align}
    &(\mathcal{L}_{Y^i})KX = [(\mathcal{L}_{Y^i})K]X + K[(\mathcal{L}_{Y^i})X]\, {=} K[(\mathcal{L}_{Y^i})X]\nonumber\\
    \implies &(\nabla_{Y^i})KX - (\nabla_{KX})Y^i = K[(\nabla_{Y^i})X] - K[(\nabla_X)Y^i]\nonumber\\
    \implies &K[(\nabla_X)Y^i] = (\nabla_{KX})Y^i \label{eq:commutecovar}
\end{align}
where the second line used the Lie derivative for a torsion free (necessary for hyper-K\"{a}hler manifolds) connection; $\mathcal{L}_X Y = [X, Y] = \nabla_X Y - \nabla_Y X$. The result (\ref{eq:commutecovar}) combined with the earlier deduction of $so(8)$ implies that, $DY^i$ actually belongs to the Lie algebra $so(8) \cap gl(n, \mathbb{H})$. This means (\ref{eq:liederspin}) holds only if $\varepsilon_0$ transforms as a singlet under the corresponding Lie group, which for the maximal intersection case is $Sp(2)$. A further result from the Bogomolov decomposition theorem \cite{FABogomolov_1974} may be noted that, the holonomy group of any Calabi-Yau metric on a simply connected holomorphically symplectic manifold of dimension 4\textit{n} with $h^{2,0}(M) = 1$ is exactly $Sp(n)$; where $h^{p, q}$ is the hodge number. This implies that the hyper-K\"{a}hler manifold of interest for $n = 2$, that has $h^{2,0} = 1$ (see the Hodge diamond for $n 
= 2$ case in \cite{Beri2022}), has the holonomy group as exactly $Sp(2)$ rather than a proper subgroup of it. Combining this note with the earlier deduction regarding singlets therefore implies that, the amount of unbroken supersymmetry arises as singlets in the decomposition from representations of $Spin(8)$ to representations of $Sp(2) \cong Spin(5)$. Such decompositions are given by branching rules $Spin(8) \downarrow Sp(2)$, which are (\cite{yamatsu2020finitedimensional, Gauntlett_1997})
\begin{equation}\label{eq:branchingrules}
    \mathbf{8}_s \rightarrow \mathbf{5} \oplus \mathbf{1} \oplus \mathbf{1} \oplus \mathbf{1} \,\,\,\,|\,\,\,\, \mathbf{8}_c \rightarrow \mathbf{4} \oplus \mathbf{4}
\end{equation}
where $\mathbf{8}_s$ and $\mathbf{8}_c$ are the two opposite chirality spinor representations of $Spin(8)$. From (\ref{eq:covconst}) it can be seen that depending on the sign of the condition $(1 \mp \Sigma^9)\varepsilon_0 = 0$, which in turn depends on the sign of the M2-brane charge under the three form field $A_3$ (\ref{eq:threeformansatz}), either $\varepsilon_s$ or $\varepsilon_c$ survives; where $\varepsilon_0 = \varepsilon_s + \varepsilon_c$. Therefore for an appropriate sign of the brane charge, $\epsilon_s \in \mathbf{8}_s$ survives, and (\ref{eq:branchingrules}) then implies that 6 ($\epsilon_0 = \eta_0 \times \varepsilon_s$ : $2 \times 3$) supersymmetries are preserved, which indeed is $\mathcal{N} = 3$ supersymmetry in the world volume of the M2-brane. It may also be noted that for the opposite sign of the charge all the supersymmetry is broken, since there are no singlets in the decomposition of $8_c$ in (\ref{eq:branchingrules}).

\section{Near-Horizon limit}
Now that the fact of $\mathcal{N} = 3$ supersymmetry has been established, it remains to be seen if any of the limits localizing to a region of the solution (\ref{eq:m2kkbranesol}), yields the enhancement to $\mathcal{N} = 6$ supersymmetry. As such, there are four notable limits to the transverse $X_8$ space : \textbf{(a)} $|\vec{x}_1| \rightarrow \infty, \,|\vec{x}_1 + k\vec{x}_2| \rightarrow \infty$, \textbf{(b)} $|\vec{x}_1| \rightarrow 0, \, |\vec{x}_1 + k\vec{x}_2| \rightarrow \infty$, \textbf{(c)} $|\vec{x}_1| \rightarrow \infty, \, |\vec{x}_1 + k\vec{x}_2| \rightarrow 0$, and \textbf{(d)} $|\vec{x}_1| \rightarrow 0, \, |\vec{x}_1 + k\vec{x}_2| \rightarrow 0$. The first one (a) is the usual asymptotically flat limit of the transverse space, and the second one (b) corresponds to (from \ref{eq:m2kkbranesol})
\begin{equation}\label{eq:limlocfirstmonopole}
\begin{split}
    &U_{(b)} \approx \begin{pmatrix}
        h_1 && 0 \\ 0 && 1
    \end{pmatrix} \,\,\,\,;\,\,\,\, h_1 = \frac{1}{2|\vec{x}_1|}\\
    &(b) : ds^2_{X_8} \approx H^{-\frac{2}{3}}ds^2_{M_{1, 2}} + H^{\frac{1}{3}}[\,ds^2_{TN}(\vec{x}_1 \rightarrow \infty, y_1) + d\vec{x}_2 \cdot d\vec{x}_2 + dy_2^2\,]\\
    &\text{where}\,\,\,\,\, ds^2_TN (\vec{x}_1, y_1) = (\ref{eq:taubnut})\,:\,V(|\vec{x}_1|) = 1 + \frac{1}{2|\vec{x}_1|} 
\end{split}
\end{equation}
which is the just the limit that localizes to the core of the first monopole, and hence the transverse space splits into Taub-NUT $\times \,\,\mathbb{R}^4$. As mentioned near (\ref{eq:taubnut}), $ds^2_{TN}(\vec{x}_1, y_1)$ is non-singular as $|\vec{x}_1| \rightarrow$ 0 iff $y_1$ has a period $2\pi R$. Since in the current case $R = 1$, and $y_1 \sim y_1 + 2\pi$ for a hyper-K\"{a}hler metric as mentioned above (\ref{eq:Usolhk}), the local region (b) is a non-singular space. This makes (b) not interesting as it preserves sixteen Poincaré supersymmetries upto a first order-approximation of the space i.e., $M_{1, 2} \times \mathbb{R}^8$. Similarly, for the limit (c), the corresponding $U_{(c)}$ is
\begin{equation}
    U_{(c)} \approx \begin{pmatrix}
        h_2 && kh_2 \\ kh_2 && k^2h_2
    \end{pmatrix} \,\,\,\,;\,\,\,\, h_2 = \frac{1}{2|\vec{x}_1 + k\vec{x}_2|}
\end{equation}
Since $\mathbb{1}_2 + U_{(c)}$ is also a solution to (\ref{eq:Usolhk}) just as $\mathbb{1}_2 + U_{(b)}$ is, due to the linear nature of the $U_{ij}$ in (\ref{eq:Usolhk}), the limit (c) can be arrived at from limit (b) via an $SL(2, \mathbb{Z})$ transformation (see below (\ref{eq:planehk})); where $U^{\infty}$ (with the constraint $det(U^{\infty}) = 1$) has been fixed to $\mathbb{1}_2$ due to the $SL(2, R)$ symmetry in (\ref{eq:anglehk}). Therefore the local region (c) is a non-singular space since (b) is non-singular, and hence is equally uninteresting. The limit (d) on the other hand is very interesting and corresponds to the intersection region of the two monopoles
\begin{equation}
    U_{(d)} \approx \begin{pmatrix}
        h_1 + h_2 && kh_2 \\ kh_2 && k^2h_2
    \end{pmatrix}\,\,\,\,;\,\,\,\,h_1 = \frac{1}{2|\vec{x}_1|}\,,\,\,\,h_2 = \frac{1}{2|\vec{x}_1 + k\vec{x}_2|}
\end{equation}
Define a new patch of coordinates local to (d), and transform the transverse space metric $ds^2_{X_8}$ accordingly as follows
\begin{align}
    &\vec{x}'_{1} \coloneq \vec{x}_1 \,\,\,;\,\,\, \vec{x}'_2 \coloneq \vec{x}_1 + k\vec{x}_2 \implies U_{ij}\,d\vec{x}^i\cdot d\vec{x}^j \rightarrow U'_{ij}\,d\vec{x}'^{i}\cdot d\vec{x}'^j\nonumber\\
    &\text{where}\,\,\,\,\,U' = L^T U L \,\,\,\Big|\,\,\,L^i_{\,\,j} = \frac{\partial \vec{x}^i}{\partial \vec{x}^{'j}} = \begin{pmatrix}
        1 && 0 \\ -\frac{1}{k} && \frac{1}{k}
    \end{pmatrix}\label{eq:transformedU}\\
    &\text{Also}\,\,\,\,\,U^{ij}(dy_i + A_i)(dy_j + A_j) \rightarrow U'^{ij}(dy'_i + A'_i)(dy'_j + A'_j)\nonumber\\
    &\text{Therefore}\,\,\,\,\,y'_j = L^i_{\,\,j}\,y_i \,\,\,\,\text{and}\,\,\,\,A'_j = L^i_{\,\,j}\,A_i\label{eq:transformedy}
\end{align}
Therefore the transverse space metric for the local region (d) using (\ref{eq:transformedU}) and (\ref{eq:transformedy}) is
\begin{equation}\label{eq:intersectionregion}
\begin{split}
    &ds^2_{X_8(d)} = U'_{(d)ij}\,d\vec{x}'^id\vec{x}'^j + U'^{ij}_{(d)}\,(dy'_i + A'_i)(dy'_j + A'_j)\,\,\,\,\Big|\,\,\,\, U'_{(d)} = \begin{pmatrix}
        h_1 && 0 \\ 0 && h_2
    \end{pmatrix}\\
    &\text{where}\,\,\,\,\,h_1 = \frac{1}{2|\vec{x}'_1|}\,,\,\,h_2 = \frac{1}{2|\vec{x}'_2|} \,\,\,\,;\,\,\,\,y'_1 = y_1 - \frac{1}{k}y_1\,,\,\,y'_2 = \frac{1}{k}y_2
\end{split} 
\end{equation}
Since $(y_1, y_2) \sim (y_1, y_2) + (2\pi, 2\pi)$, it is clear from (\ref{eq:intersectionregion}) that $y'_1 \sim y'_1 - \frac{2\pi}{k}$ and $y'_2 \sim y'_2 + \frac{2\pi}{k}$. This violates the periodicity condition required by a hyper-K\"{a}hler metric to be non-singular, and therefore (\ref{eq:intersectionregion}) has a singularity characterized by this angular deficit. From the form of $U'_{(d)}$, it is also clear that the transverse space localized to (d) is divided into two orthogonal subspaces labelled by $(\vec{x}_1, y_1) \times (\vec{x}_2, y_2)$. Now since any Riemannian manifold is locally Euclidean upto first order, this singularity can then be locally represented by the fixed point of the oribifold action $\mathcal{C}^2/\mathbb{Z}_k \times \mathcal{C}^2/\mathbb{Z}_k = \mathcal{C}^4/\mathbb{Z}_k$; where each $\mathcal{C}^2$ is one of the two orthgonal subspaces of the transverse space. The action of $\mathbb{Z}_k$ on $\mathcal{C}^4$ is given by
\begin{equation}
    \mathcal{C}^4/\mathbb{Z}_k \,\,:\,\, z^i \sim e^{\frac{2\pi i}{k}\cdot b_z} \,z^i \,\,\,|\,\,\, z^i \in \mathbb{C}^4\,\,;\,\,i = 1, 2, 3, 4
\end{equation}
where $b_z = 1$ is the $U(1)$ charge of the fundamental $\mathbf{4}$ representation in the decomposition of $8_v \in Spin(8)$ under $SU(4) \times U(1)$, which is the reduced isometry group of $\mathcal{C}^4$ under the action of $\mathbb{Z}_k$. The branching rules of this reduction of isometry $so(8) \supset su(4) \oplus u(1)$ can then be used to give the action of $\mathbb{Z}_k$ on the spinors $8_s, 8_c$ of $Spin(8)$, which helps determine the invariant spinors under $\mathbb{Z}_k$ and hence preserved supersymmetry \cite{HALYO_2000}.
\begin{align}
    &\mathbf{8}_v \rightarrow \mathbf{4}_1 \,\oplus\, \mathbf{\bar{4}}_{-1} \,\,\,\,|\,\,\,\, \mathbf{8}_s \rightarrow \mathbf{1}_2 \,\oplus\, \mathbf{1}_{-2} \,\oplus\, \mathbf{6}_0 \,\,\,\,|\,\,\,\,\mathbf{8}_c \rightarrow \mathbf{4}_{-1} \,\oplus\, \mathbf{\bar{4}}_{1}\\
    \implies &\mathbb{Z}_k \,\,:\,\, \varepsilon_s \rightarrow \left(\mathbb{1}_1\,e^{\frac{2\pi}{k}\cdot 2} \oplus  \mathbb{1}_1\,e^{\frac{2\pi}{k}\cdot (-2)} \oplus \mathbb{1}_6\,e^{\frac{2\pi}{k}\cdot 0}\right) \, \varepsilon_s \label{eq:branchingzk}
\end{align}
where $\varepsilon_s$ is the killing spinor as mentioned in (\ref{eq:covconst}), which is the chirality preserved given an appropriate sign of $A_3$ as mentioned near (\ref{eq:branchingrules}). It can then easily be seen from (\ref{eq:branchingzk}) that, for $k = 1, 2$ all of $1_{2}, 1_{-2}$ and $6_0$ remain invariant, while for $k > 2$ only the $6_0$ piece remains invariant. Therefore in the killing spinor $\epsilon_0 = H^{-\frac{1}{6}}\eta_0 \otimes \varepsilon_0$, 16 supersymmetries ($\mathcal{N} = 8$) are preserved for $k = 1, 2$, and 12 supersymmetries ($\mathcal{N} = 6$) are preserved for $k > 2$, by $\mathbb{C}^4/\mathbb{Z}_k$. This is precisely the supersymmetry desired by the ABJM theory, as seen in Chapter \ref{chap:research}. Finally, the limits $|\vec{x}_1| \rightarrow 0, |\vec{x}_1 + k\vec{x}_2|$ corresponding to the localization (d), translate to $|\vec{x}'_1|, |\vec{x}'_2| \rightarrow 0$ which is nothing but $r \rightarrow 0$ in $\mathbb{C}^4/\mathbb{Z}_k$; where $r$ is the radial distance in the transverse space. Therefore the line element local to this region is (\ref{eq:m2kkbranesol})
\begin{equation*}
    ds^2_{ABJM} = \lim_{r\to 0}\left[H^{-\frac{2}{3}}ds^2_{M_{1, 2}} + H^{\frac{1}{3}}ds^2_{\mathbb{C^4}/\mathbb{Z}_k}\right] = \lim_{r\to 0}\left[H^{-\frac{2}{3}}ds^2_{M_{1, 2}} + H^{\frac{1}{3}}\left(dr^2 + r^2ds^2_{S^7/\mathbb{Z}_k}\right)\right]
\end{equation*}
where $H$ is now the harmonic function on $\mathbb{C}^4/\mathbb{Z}_k$, which is $H(r) = 1 + \frac{L^6}{r^6}$. Substituting this in the above equation, and performing the change of coordinates $z = \frac{L^3}{2r^2}$ yields
\vspace{-0.6em}
\begin{equation}
    ds^2_{ABJM} = \frac{L^2}{4z^2}\left(dz^2 + ds^2_{M_{1, 2}}\right) + L^2ds^2_{S^7/\mathbb{Z}_k}
\end{equation}
\begin{empheq}[box=\fbox]{align}\label{eq:abjmbranesol}
  \begin{split}
        &ds^2_{ABJM} = ds^2_{AdS_4}(L/2) + ds^2_{S^7/\mathbb{Z}_k}(L) \,\,\,\,\Big\vert\,\,\,L^6 = 32\pi^2N'l_p^6\\
        &F_4  = \frac{3}{8}L^3 \,d\,\text{vol}(AdS_4) \,\,\,\,\Big\vert\,\,\,\, d\,\text{vol}(AdS_4) = -\frac{1}{z^4}dt \wedge dx^0 \wedge dx^1 \wedge dz 
   \end{split}
\end{empheq}

\vspace{0.5em}
\noindent where $ds^2_{AdS_4}(L/2)$ and $ds^2_{S^7/\mathbb{Z}_k}(L)$ are the line elements of the spaces in the subscript with radii $L/2$ and $L$ respectively. Also $N' = kN$ in order to have $N$ quantized flux on $S^7/\mathbb{Z}_k$, since the volume of $S^7/\mathbb{Z}_k$ is smaller than that of $S^7$ by a factor of $k$ \cite{Aharony_2008}. 

Therefore to summarize the discussions of Chapters \ref{chap:research_results} and \ref{chap:situational_theoretical_analysis}, the ABJM theory described in Chapter \ref{chap:research} is the same as the low-energy theory living on \textbf{N} M2-branes, placed at the $\mathbb{C}^4/\mathbb{Z}_k$ singularity in it's transverse space. Infact, this conclusion is just a launching pad to a much stronger conjectured duality, which is called the $AdS_4 \,/\,CFT_3$ correspondence, which claims that the ABJM theory is dual to M-theory in the near-horizon geometry background (\ref{eq:abjmbranesol}) i.e., $AdS_4 \times S^7/\mathbb{Z}_k$. This duality will be the subject of exploration starting from the next chapter.

%% file: chapters/rationale.tex
\chapter{$AdS_4\,/\,CFT_3$ Correspondence}\label{chap:rationale}
In the January of 1998, Juan Maldacena published a landmark paper \cite{Maldacena_1999} initiating a study into the conjectured duality between compactifications of M / String theory on $AdS \times X$ backgrounds and Conformal Field Theories (CFTs) living on the boundary of $AdS$; where $X$ is some compact manifold. The previous three chapters served to establish ABJM theory as the low energy dynamical theory living in the world volume of \textbf{N} M2-branes probing a $\mathbb{C}^4/\mathbb{Z}_k$ singularity. This world volume (012) is the boundary of $AdS_4$ in the near-horizon geometry with the bulk coordinate labelled by $z$ (\ref{eq:abjmbranesol}), which thereby prompts to conjecture a corresponding $AdS_4 \,/\, CFT_3$ duality.
\vspace{1em}
\hrule width \textwidth
\[
  \begin{gathered}
    \mathcal{N} = 6 \,\,\,\text{\text{superconformal Chern-Simons}}\\
    \text{\text{matter theory in 2+1 dimensions}}\\
    \text{\text{with gauge group}}\,\,\, U(N)_k \times U(N)_{-k}\\
    \text{\text{(also called ABJM theory)}}
  \end{gathered}
  \quad\mathlarger{\mathlarger{\mathlarger{\Longleftrightarrow}}}\quad
  \begin{gathered}
    \text{\text{M-theory on}} \,\,AdS_4 \times S^7/\mathbb{Z}_k \,\,\text{\text{with}}\\
    N \,\,\text{\text{units of flux}} \,\,\,\int_{S^7/\mathbb{Z}_k}\star F_4
  \end{gathered}
\]
\hrule width \textwidth
\vspace{1em}
\noindent It can be checked from (\ref{eq:abjmbranesol}) that $\int_{S^7/\mathbb{Z}_k} \star F_4 = \frac{2L^6\pi^4}{k} = N(2\kappa_{11}^2 T_{M2})$; where $T_{M2}$ is the M2-brane tension and $2\kappa_{11}^2$ is the eleven dimensional gravitational coupling, as mentioned near (\ref{eq:charlengthscale}). As a preliminary test to this duality, the moduli space of the ABJM theory which was shown to be $(\mathbb{C}^4/\mathbb{Z}_k)^N / S_N$ in Chapter \ref{chap:research}, is clearly the same for $N$ units of flux probing a $\mathbb{C}^4/\mathbb{Z}_k$ singularity. A more fundamental test however is the matching of symmetries on either side of the duality. The sphere $S^7$ when embedded into $\mathbb{R}^8$ or $\mathbb{C}^4$ has an isometry group $SO(8)$, which upon the action of $\mathbb{Z}_k$ breaks into a subgroup $SU(4) \times U(1)$ as seen in the previous chapter. This can be identified with the global symmetry group $SU(4)_R \times U(1)_b$ of ABJM theory discussed in Chapter \ref{chap:research}. Similarly the $AdS_4$ of signature $(3, 1)$ when embedded into $\mathbb{R}^{3, 2}$ has an isometry group $SO(3, 2)$. This can be identified with the conformal group on the field theory side, since the group of all globally defined conformal transformations in $\mathbb{R}^{p,q}$ is isomorphic to $SO(p+1, q+1)$ \cite{inbookschotten}. However, since we are dealing with a superconformal field theory on the CFT side and a supersymmetric background on the M-theory side, the algebra of the aforementioned bosonic symmetry group is extended to include supersymmetry generators. Since $su(4)_R \cong so(6)_R$ and $so(3, 2) \cong sp(4)$, the superconformal group that has it's maximal bosonic subalgebra as $sp(4) \oplus so(6)_R$ is the Orthosymplectic super Lie group $OSp(6|4)$ \cite{Bandres_2008}. For $k = 1, 2$, it is $OSp(8|4)$ since there is an $\mathcal{N} = 6\,(so(6)_R) \rightarrow \mathcal{N} = 8\,(so(8)_R)$ supersymmetry enhancement as mentioned near (\ref{eq:branchingzk}). Let us now explore this $OSp(\mathcal{N}|4)$ group in more detail

\section{$OSp(\mathcal{N}\,|\,4)$ : Algebra and Representations}\label{sec:ospn4algebra}
The generators of the conformal group \textit{Conf}$(\mathbb{R}^{3, 1}) \cong SO(3, 2)$ are the three Lorentz generators $\{\mathcal{M}_{\mu\nu} | \mathcal{M}_{\mu\nu} = -\mathcal{M}_{\nu\mu}\}$, the three generators of spacetime translations $\mathcal{P}_\mu$, the three generators of special conformal transformations $\mathcal{K}_\mu$, and the generator of dilations $\mathcal{D}$; where $\mu = 0, 1, 2$ is the spacetime index. Since the fermionic generators are due to be included later, it is convenient to convert the spacetime index to a spinor index, which can be done using the 2+1 dimensional gamma matrices as follows
\begin{equation}
\begin{split}
    &\mathcal{P}^{\,\alpha}_{\,\,\,\,\beta} = (\gamma^\mu)^{\alpha}_{\,\,\,\beta}\,\mathcal{P}_\mu\,\,\,\,\,\Big|\,\,\,\,\, \mathcal{K}^{\,\alpha}_{\,\,\,\,\beta} = -(\bar{\gamma}^{\mu})^{\alpha}_{\,\,\,\beta}\,\mathcal{K}_\mu\,\,\,\,\,\Big|\,\,\,\,\,\mathcal{M}^{\,\alpha}_{\,\,\,\,\beta} = \frac{1}{2}(\gamma^\mu)^{\alpha}_{\,\,\,\delta}\,(\bar{\gamma}^{\nu})^{\delta}_{\,\,\,\beta}\,\mathcal{M}_{\mu\nu}\\
    &\text{where}\,\,\,\,\,(\gamma^\mu)^{\alpha}_{\,\,\,\beta} = (i\sigma_2, \sigma_1, \sigma_3) \,\,\,\, \text{and} \,\,\,\, (\bar{\gamma}^{\mu})_{\alpha}^{\,\,\,\beta} = (-i\sigma_2, \sigma_1, \sigma_3)
\end{split}
\end{equation}
where $\alpha, \beta, \delta$ are the spinor indices and $\sigma_i$ are the Pauli matrices. The spinor indices are raised and lowered by $\epsilon^{\alpha\beta}, \,\epsilon_{\alpha\beta}$; $\epsilon^{12} = -\epsilon_{12} = 1$. Written out explicitly, the generators are
\begin{equation}\label{eq:spinorconfgen}
\begin{gathered}
        \mathcal{P}_{\alpha\beta} = \begin{pmatrix}
        \mathcal{P}_0 - \mathcal{P}_1 && \mathcal{P}_2\\ \mathcal{P}_2 && \mathcal{P}_0 + \mathcal{P}_1
    \end{pmatrix} \,\,\,\,;\,\,\,\, \mathcal{K}^{\alpha\beta} = \begin{pmatrix}
        -(\mathcal{K}_0 + \mathcal{K}_1) && \mathcal{K}_2 \\ \mathcal{K}_2 && \mathcal{K}_1 - \mathcal{K}_0
    \end{pmatrix} \\[1.5ex]
        \mathcal{M}_\alpha^{\,\,\,\beta} = \begin{pmatrix}
            -\mathcal{M}_{01} && \mathcal{M}_{02} - \mathcal{M}_{12} \\
            \mathcal{M}_{02} + \mathcal{M}_{12} && \mathcal{M}_{01}
        \end{pmatrix} 
\end{gathered}
\end{equation}
The well known conformal algebra \cite{eberhardt2020superconformal} rewritten in terms of spinor indices (\ref{eq:spinorconfgen}) is
\begin{align}
    &[\mathcal{M}_\alpha^{\,\,\,\beta}, \mathcal{P}_{\gamma\delta}] = 2\delta_{\{\gamma}^{\,\,\,\,\,\,\beta}\,\mathcal{P}_{{\textstyle\mathstrut}\delta\}\alpha} - \delta_{\alpha}^{\,\,\,\beta}\,\mathcal{P}_{\gamma\delta} \,\,\,\,\,\,\,\,;\,\,\, [\mathcal{M}_\alpha^{\,\,\,\beta}, \mathcal{K}^{\gamma\delta}] = -2\delta_\alpha^{\,\,\,\{\gamma}\mathcal{K}^{\delta\}\beta} + \delta_\alpha^{\,\,\,\beta}\mathcal{K}^{\gamma\delta}\\
    &[\mathcal{M}_\alpha^{\,\,\,\beta}, \mathcal{M}_\gamma^{\,\,\,\delta}] = -\delta_\alpha^{\,\,\,\delta}\mathcal{M}_{\gamma}^{\,\,\,\beta} + \delta_{\gamma}^{\,\,\,\beta}\mathcal{M}_{\alpha}^{\,\,\,\delta} \,\,\,\,\kern 0.13em;\,\,\, [\mathcal{K}^{\alpha\beta}, \mathcal{P}_{\gamma\delta}] = 4\delta_{\{\gamma}^{\,\,\,\{\alpha}\mathcal{M}_{\delta\}}^{\,\,\,\beta\}} + 4\delta_{\{\gamma}^{\,\,\,\{\alpha}\delta_{\delta\}}^{\,\,\,\beta\}}D\\[0.5ex]
    &[\mathcal{D}, \mathcal{P}_{\alpha\beta}] = \mathcal{P}_{\alpha\beta}\,\,\,;\,\,\, [\mathcal{D}, \mathcal{K}^{\alpha\beta}] = -\mathcal{K}^{\alpha\beta} \,\,\,;\,\,\, [\mathcal{D}, \mathcal{M}_\alpha^{\,\,\,\beta}] = 0\label{eq:cartanosp1}\\
\intertext{Now introducing the $2\mathcal{N}$ Poincaré supercharges $\mathcal{Q}_{\alpha r}$ of $\mathcal{N}$ supersymmetry in 2+1 dimensions; where $r$ is the fundamental representation index of the R-symmetry group $so(\mathcal{N})_R$}
    &\{\mathcal{Q}_{\alpha r}, \mathcal{Q}_{\beta s}\} = 2\delta_{rs}\mathcal{P}_{\alpha\beta}\,\,\,;\,\,\,[\mathcal{M}_\alpha^{\,\,\,\beta}, \mathcal{Q}_{\gamma r}] = \delta_\gamma^{\,\,\,\beta}\mathcal{Q}_{\alpha r} - \frac{1}{2}\delta_\alpha^{\,\,\,\beta}\mathcal{Q}_{\gamma r} \,\,\,;\,\,\, [D, \mathcal{Q}_{\alpha r}] = \frac{1}{2}\mathcal{Q}_{\alpha r}\label{eq:cartanosp2}
\end{align}
The additional $2\mathcal{N}$ conformal supercharges $\mathcal{S}^{\alpha}_{\,\,\,r}$ are then generated by commuting the Poincaré supercharges with the special conformal generators (\cite{Kac:1977qb, SCHWARZ1981221})
\begin{equation}\label{eq:anticommss}
    [\mathcal{K}^{\alpha\beta}, \mathcal{Q}_{\gamma r}] = -2i\delta_{\gamma}^{\,\,\,\{\alpha}\mathcal{S}^{\beta\}}_{\,\,\,r}\,\,\,;\,\,\,\{\mathcal{S}^{\alpha}_{\,\,\,r}, \mathcal{S}^{\beta}_{\,\,\,s}\} = -2\delta_{rs}\mathcal{K}^{\alpha\beta}
\end{equation}
These $\mathcal{S}^{\alpha}_{\,\,\,r}$ transform under the remaining conformal transformations as follows
\begin{align}
    &[\mathcal{P}_{\alpha\beta}, \mathcal{S}^{\gamma}_{\,\,\,r}] = -2i\delta_{\{\alpha}^{\,\,\,\,\,\gamma}\mathcal{Q}_{{\textstyle\mathstrut}\beta\}r}\,\,\,;\,\,\, [\mathcal{M}_\alpha^{\,\,\,\beta}, \mathcal{S}^{\gamma}_{\,\,\,r}] = -\delta_\alpha^{\,\,\,\gamma}\mathcal{S}^{\beta}_{\,\,\,r} + \frac{1}{2}\delta_\alpha^{\,\,\,\beta}\mathcal{S}^{\gamma}_{\,\,\,r}\,\,\,;\,\,\,[D, \mathcal{S}^{\alpha}_{\,\,\,r}] = -\frac{1}{2}\mathcal{S}^{\alpha}_{\,\,\,r}\label{eq:cartanosp3}
\intertext{The $\mathcal{N}(\mathcal{N} - 1)/2$ antisymmetric R-charges $\mathcal{R}_{rs}$ of the R-symmetry group $so(\mathcal{N})_R$, appear in the algebra in the anti-commutator between Poincaré and conformal supercharges}
&\{\mathcal{Q}_{\alpha r}, \mathcal{S}^{\beta}_{\,\,\,s}\} = 2i\delta_{rs}(\mathcal{M}_\alpha^{\,\,\,\beta} + \delta_\alpha^{\,\,\,\beta}\mathcal{D}) + 2\delta_\alpha^{\,\,\,\beta}\mathcal{R}_{rs}\,\,\,\,;\,\,\,\,[\mathcal{R}_{rs}, \mathcal{R}_{tu}] = 2i(\delta_{r[u}\mathcal{R}_{t]s} - \delta_{s[u}\mathcal{R}_{t]r})\label{eq:anticommqs}
\intertext{The supercharges transform in the fundamental vector representation of $so(\mathcal{N})_R$ as follows}
&\quad\quad\quad\quad[\mathcal{R}_{rs}, \mathcal{Q}_{\alpha t}] = i(\delta_{rt}\mathcal{Q}_{\alpha s} - \delta_{st}\mathcal{Q}_{\alpha r}) \,\,\,;\,\,\, [\mathcal{R}_{rs}, \mathcal{S}^{\alpha}_{\,\,\,t}] = i(\delta_{rt}\mathcal{S}^{\alpha}_{\,\,\,s} - \delta_{st}\mathcal{S}^{\alpha}_{\,\,\,r})\label{eq:cartanosp3}
\end{align}
Note that [\,\,.\,\,] and \{\,.\,\} over indices stand for weighted anti-symmetrization and symmetrization respectively in all the aforementioned relations. It may also be noted that the generators $\mathcal{R}$ commute with all the conformal generators $(\mathcal{P}, \mathcal{K}, \mathcal{M}, \mathcal{D})$, since the bosonic subalgebra is a direct sum $sp(4) \oplus so(N)_R$. Finally to finish up on the algebra, the hermitian conjugates are defined for the generators so that they form an anti-automorphism of the algebra, consistent with $[A, B\}^\dag = [B^\dag, A^\dag\}$, as follows
\begin{equation}\label{eq:conjgenosp}
\begin{gathered}
    (\mathcal{P}_{\alpha\beta})^\dag = \mathcal{K}^{\alpha\beta}\,\,,\,\,\,\,(\mathcal{K}_{\alpha\beta})^\dag = \mathcal{P}^{\alpha\beta} \,\,,\,\,\,\, (\mathcal{M}_\alpha^{\,\,\,\beta})^\dag = \mathcal{M}_\beta^{\,\,\,\alpha}\,\,,\,\,\,\,\mathcal{D}^\dag = \mathcal{D}\\
    (\mathcal{Q}_{\alpha r})^\dag = -i\mathcal{S}^\alpha_{\,\,\,r}\,\,,\,\,\,\,(\mathcal{S}^\alpha_{\,\,\,r})^\dag = -i\mathcal{Q}_{\alpha r}\,\,,\,\,\,\,(\mathcal{R}_{rs})^\dag = \mathcal{R}_{rs}
\end{gathered}
\end{equation}
Now to find the unitary irreducible representations (u-irreps), they are all infinite dimensional since $osp(\mathcal{N}|4)$ is non-compact. However, a finite dimensional subspace of states $V_B$ called the ($osp(\mathcal{N} | 4), B$) module (Harish-Chandra module), consisting of $B$-finite vectors w.r.t a maximal compact subgroup $B \subset OSp(\mathcal{N} | 4)$, is enough to recover the entire original state space $V$ \cite{inbookharishchand}. Infact going further, the highest weight states of the u-irreps of $osp(\mathcal{N} | 4)$ are completely specified by the highest weight states of the u-ireps of $B$, without needing the module structure of $V_{B}$. These will be the only relevant states of interest which are called superconformal primaries. The maximal compact subgroup of $SO(3, 2)$ is $SO(3) \times SO(2)$, generated by the Wick-rotated Lorentz generators $\mathcal{M}_W$ and the dilation generator $\mathcal{D}$. Therefore by including $SO(\mathcal{N})_R$ and taking the covering group of $SO(3)$, the maximal compact subgroup $B$ is $SU(2)_{M_W} \times U(1)_D \times SO(\mathcal{N})_R$. The product structure of this subgroup can also be seen from their corresponding generators commuting with each other, and thereby forming a Cartan subalgebra of $sp(4) \oplus so(\mathcal{N})_R$. The irreducible representations of $B$ are therefore labelled by 
\begin{equation}\label{eq:representationosp1}
    [j]^{R_n}_{\Delta} \,\,\,\Big|\,\,\,(\mathcal{M}_W)_{\alpha}^{\,\,\,\alpha} \ket{j, m} = m\ket{j, m}\,,\,\,\mathcal{D} \,:\, \Delta \,,\,\, \mathcal{H}(so(\mathcal{N})_R) \,:\, R_n
\end{equation}
where $\mathcal{H}(so(\mathcal{N})_R)$ is the $[\frac{\mathcal{N}}{2}]$ dimensional Cartan subalgebra of $so(\mathcal{N})$, under which the representations are labeled by the Dynkin labels $R_n$, n = 1, 2,..., $[\frac{\mathcal{N}}{2}]$. Also the index $\alpha$ is not summed over in $(\mathcal{M}_W)_{\alpha}^{\,\,\,\alpha}$, which represents one of the diagonal elements ($\equiv (\mathcal{M}_W)_{01}$) of the traceless $(\mathcal{M}_W)_\alpha^{\,\,\,\beta}$ (\ref{eq:spinorconfgen}). Consequently it is accompanied by an $SU(2)$ spin representation $\ket{j, m}$, the eigenvalue under which is $m$. The eigen value $\Delta$ of the state w.r.t $\mathcal{D}$ is called the scaling dimension. Now coming to the raising and lowering operators, it can be seen from the commutators with $\mathcal{D}$ (\ref{eq:cartanosp1}, \ref{eq:cartanosp2}, \ref{eq:cartanosp3}) that $\mathcal{P}_{\alpha\beta}$ \& $\mathcal{K}^{\alpha\beta}$ raise and lower $\Delta$ by one respectively, while $\mathcal{Q}_{\alpha r}$ \& $\mathcal{S}^{\alpha}_{\,\,\,r}$ raise and lower $\Delta$ by half respectively. Combining with the ones for $\ket{j, m}$ and $R_n$, the complete set for (\ref{eq:representationosp1}) is then as follows
\begin{align}
    &\text{Raising}\,\,\,\,\,\,\,:\,\,\,(\mathcal{M}_W)_{2}^{\,\,\,1}\,,\,\,\,\mathcal{R}^{+}_{rs}\,,\,\,\,\mathcal{P}_{\alpha\beta}\,,\,\,\mathcal{Q}_{\alpha r}\\
    &\text{Lowering}\,\,\,:\,\,\, (\mathcal{M}_W)_{1}^{\,\,\,2}\,,\,\,\,\mathcal{R}^{-}_{rs}\,,\,\,\,\mathcal{K}^{\alpha\beta}\,,\,\,\,\mathcal{S}^{\alpha}_{\,\,\,r}\label{eq:loweringreposp}
\end{align}
where $\mathcal{R}^{+}_{rs}$ \& $\mathcal{R}^{-}_{rs}$ are the raising and lowering operators of $so(\mathcal{N})_R$ respectively. Therefore finally, a \textit{superconformal primary} is the highest $\Delta$ weight state labelled by (\ref{eq:representationosp1}), annihilated by $\Delta$ lowering operator $\mathcal{S}^{\alpha}_{\,\,\,r}$ (and hence by $\mathcal{K}^{\alpha\beta}$ since $\{\mathcal{S}, \mathcal{S}\} \sim \mathcal{K}$ (\ref{eq:anticommss})). The states obtained by acting on the primary with $\Delta$ raising operator $\mathcal{Q}_{\alpha r}$ (or by $\mathcal{P}_{\alpha\beta}$ since $\{\mathcal{Q}, \mathcal{Q}\}\sim \mathcal{P}$ (\ref{eq:cartanosp2})) are called \textit{descendants} of the primary state. Let the primaries be denoted by the operator $\mathcal{V}$ which by state operator correspondence implies the state $\ket{\mathcal{V}}$ in radial quantization. Similarly, let the descendant of level $l \geq 1$ be denoted by the operator $\mathcal{O}_l$ ($\ket{\mathcal{O}_l}$), which denotes a state obtained by acting with $Q_{\alpha r}$ on $\ket{\mathcal{V}}$ \textit{l}-times in a radially ordered manner. Unitarity then implies that these primaries and descendants must have a non-negative norm ($\braket{\mathcal{V}}{\mathcal{V}}$, $\braket{\mathcal{O}_l}{\mathcal{O}_l} \geq 0$). Now $\braket{\mathcal{O}_l}{\mathcal{O}_l} \sim \bra{\mathcal{V}}Q_l^\dag...Q_1^\dag Q_1...Q_l\ket{\mathcal{V}}$, and since $Q^\dag = -iS$ (\ref{eq:conjgenosp}), by using $\{\mathcal{Q}, \mathcal{S}\}$ all the $\mathcal{S}$ can be brought to the right which then annihilate the primary ($\mathcal{S}\ket{\mathcal{V}} = 0$). The condition $\braket{\mathcal{O}_l}{\mathcal{O}_l} \geq 0$ then translates to a function $\bra{\mathcal{V}}\mathcal{F}(\{\mathcal{Q}, \mathcal{S}\})\ket{\mathcal{V}} \geq 0$, and since $\{\mathcal{Q}, \mathcal{S}\} \sim \mathcal{M}_W + \mathcal{D} - \mathcal{R}$ (\ref{eq:anticommqs}), their action on $\ket{\mathcal{V}}$ gives rise to a polynomial in $j, \Delta_{\mathcal{V}}, R_n$ for each $l$. The classes of unitarity bounds for the current case of three dimensions are therefore as follows
\begin{equation*}
    \begin{gathered}
        f(j) + g(R_n) + \delta_A \equiv \Delta_A\\
        \,\,\,\,f(j_B) + g(R_n) + \delta_B \equiv \Delta_B 
    \end{gathered}
    \quad;\quad \delta_A > \delta_B
\end{equation*}
\begin{equation}
           {\text{Multiplets}}\,\, \coloneq\begin{cases} \textit{\text{Long}}\quad\,\,\,\,\,\,\,\,\mathcal{L}[j]^{R_n^{\mathcal{V}}}_{\Delta_\mathcal{V}} & \Delta_{\mathcal{V}} > \Delta_A\\[1ex]
            \textit{\text{Short}}\quad\,\,\,\,\,\,\,\mathcal{A}_l[j]^{R_n^{\mathcal{V}}}_{\Delta_{\mathcal{V}}} & \Delta_{\mathcal{V}} = \Delta_A\\[1ex]
            \textit{\text{Isolated}}\quad\,\,\mathcal{B}_l[j_B]^{R_n^{\mathcal{V}}}_{\Delta_{\mathcal{V}}} & \Delta_{\mathcal{V}} = \Delta_B
        \end{cases}
\end{equation}
The multiplets $\mathcal{A}_l, \mathcal{B}_l$ contain null states with zero norm starting from level \textit{l} since they saturate the bound, and will have to be removed from the spectrum. Also, as the value of $\Delta_{\mathcal{V}}$ for the long multiplet $\mathcal{L}$ decreases, it reaches the saturation point from above, potentially leading to its fragmentation into short multiplets and isolated short multiplets. These transformations are governed by recombination rules. Similarly, when subjected to exactly marginal deformations in the superconformal theory, short or isolated short multiplets can vanish from the theory by recombining into long multiplets according to the same recombination rules. Notably, the isolated short multiplets that do not participate in any recombination rules remain absolutely protected under such deformations, suggesting their significance for further study. The cases of interest for us are $\mathcal{N} = 6, \,8$ supersymmetry, whose superconformal primaries, unitarity bounds and recombination rules will now be summarized in the tables that follow \cite{cordova2016multiplets}.
\subsection*{$osp\,(6 \,|\,4)$ \textit{Superconformal primaries and Recombination rules}}
\setlength{\tabcolsep}{12pt}
\renewcommand{\arraystretch}{2}
\begin{table}[h]
    \centering % Extra spacing between rows
    \begin{tabular}{|c|l|l|l|} % Add vertical lines between columns
        \hline
        \textbf{Name} & \quad\,\,\,\,\,\textbf{Primary} & \quad\,\,\textbf{Unitarity Bound} & \quad\textbf{Null State}\\
        \hline
        $\mathcal{L}$ & $[j]^{(R_1, R_2, R_3)}_{\Delta}$ & $\Delta > j + R_1 + \frac{(R_2 + R_3)}{2} + 1$  & \quad\quad\quad\,\,- \\
        \hline
        $\mathcal{A}_1$ & $[j]^{(R_1, R_2, R_3)}_\Delta$\,,\,\,$j \geq \frac{1}{2}$ & $\Delta = j + R_1 + \frac{(R_2 + R_3)}{2} + 1$ & $[j-1/2]^{(R_1 + 1, R_2, R_3)}_{\Delta + \frac{1}{2}}$ \\
        \hline
        $\mathcal{A}_2$ & $[0]^{(R_1, R_2, R_3)}_\Delta$ & $\Delta = R_1 + \frac{(R_2 + R_3)}{2} + 1$ & $[0]^{(R_1 + 2, R_2, R_3)}_{\Delta+1}$ \\
        \hline
        $\mathcal{B}_1$ & $[0]^{(R_1, R_2, R_3)}_\Delta$ & $\Delta = R_1 + \frac{(R_2 + R_3)}{2}$ & $[1/2]^{(R_1 + 1, R_2, R_3)}_{\Delta + \frac{1}{2}}$ \\
        \hline
    \end{tabular}
    \label{tab:n6scp}
\end{table}
\addcontentsline{lot}{table}{4.16\,\, $OSp(6 | 4)$ superconformal multiplets and unitarity bounds}
\vspace{-1.4em}
\begin{equation}\label{eq:recombrulesn6}
\begin{split}
    &\mathcal{L}[j \geq 1]^{(R_1, R_2, R_3)}_\Delta \,\,\longrightarrow\,\, \mathcal{A}_1[j]_{\Delta_A}^{(R_1, R_2, R_3)} \,\,\,\oplus\,\,\, \mathcal{A}_1[j-1/2]^{(R_1 + 1, R_2, R_3)}_{\Delta_A + \frac{1}{2}}\\
    &\mathcal{L}[j = 1/2]_{\Delta}^{(R_1, R_2, R_3)} \,\,\longrightarrow\,\, \mathcal{A}_1[1/2]^{(R_1, R_2, R_3)}_{\Delta_A} \,\,\,\oplus\,\,\, \mathcal{A}_2[0]^{(R_1 + 1, R_2, R_3)}_{\Delta_A + \frac{1}{2}}\\
    &\mathcal{L}[j = 0]^{(R_1, R_2, R_3)}_\Delta \,\,\longrightarrow\,\, \mathcal{A}_2[0]^{(R_1, R_2, R_3)}_{\Delta_A} \,\,\,\oplus\,\,\, \mathcal{B}_1[0]^{(R_1 + 2, R_2, R_3)}_{\Delta_A + 1}\\
    &\text{where}\,\,\,\,\, \Delta_A \equiv j + R_1 + \frac{1}{2}(R_2 + R_3) + 1
\end{split}
\end{equation}
It can be seen from (\ref{eq:recombrulesn6}) that multiplets of the form $\mathcal{B}_1[0]^{(R_1, R_2, R_3)}_{R_1 + \frac{(R_2 + R_3)}{2}}$ with $R_1 = 0, 1$ are absolutely protected. Also $(R_1, R_2, R_3)$ ; $R_1, R_2, R_3 \in \mathbb{Z}_{\geq 0}$ are the $so(6)_R$ Dynkin labels.
\subsection*{$osp\,(8 \,|\,4)$ \textit{Superconformal primaries and Recombination rules}}
\setlength{\tabcolsep}{5.3pt}
\renewcommand{\arraystretch}{2}
\begin{table}[h]
    \centering % Extra spacing between rows
    \begin{tabular}{|c|l|l|l|} % Add vertical lines between columns
        \hline
        \textbf{Name} & \quad\,\,\,\,\,\textbf{Primary} & \quad\,\,\textbf{Unitarity Bound} & \quad\textbf{Null State}\\
        \hline
        $\mathcal{L}$ & $[j]^{(R_1, R_2, R_3, R_4)}_{\Delta}$ & $\Delta > j + R_1 + R_2 + \frac{(R_3 + R_4)}{2} + 1$  & \quad\quad\quad\,\,- \\
        \hline
        $\mathcal{A}_1$ & $[j]^{(R_1, R_2, R_3, R_4)}_\Delta$\,,\,\,$j \geq \frac{1}{2}$ & $\Delta = j + R_1 + R_2 + \frac{(R_3 + R_4)}{2} + 1$ & $[j-1/2]^{(R_1 + 1, R_2, R_3, R_4)}_{\Delta + \frac{1}{2}}$ \\
        \hline
        $\mathcal{A}_2$ & $[0]^{(R_1, R_2, R_3, R_4)}_\Delta$ & $\Delta = R_1 + R_2 + \frac{(R_3 + R_4)}{2} + 1$ & $[0]^{(R_1 + 2, R_2, R_3, R_4)}_{\Delta+1}$ \\
        \hline
        $\mathcal{B}_1$ & $[0]^{(R_1, R_2, R_3, R_4)}_\Delta$ & $\Delta = R_1 + R_2 + \frac{(R_3 + R_4)}{2}$ & $[1/2]^{(R_1 + 1, R_2, R_3, R_4)}_{\Delta + \frac{1}{2}}$ \\
        \hline
    \end{tabular}
    \label{tab:n8scp}
\end{table}
\addcontentsline{lot}{table}{4.17\,\, $OSp(8 | 4)$ superconformal multiplets and unitarity bounds}
\pagebreak
\vspace*{-2.6em}
\begin{equation}\label{eq:recombrulesn8}
\begin{split}
    &\mathcal{L}[j \geq 1]^{(R_1, R_2, R_3, R_4)}_\Delta \,\,\longrightarrow\,\, \mathcal{A}_1[j]_{\Delta_A}^{(R_1, R_2, R_3, R_4)} \,\,\,\oplus\,\,\, \mathcal{A}_1[j-1/2]^{(R_1 + 1, R_2, R_3, R_4)}_{\Delta_A + \frac{1}{2}}\\
    &\mathcal{L}[j = 1/2]_{\Delta}^{(R_1, R_2, R_3, R_4)} \,\,\longrightarrow\,\, \mathcal{A}_1[1/2]^{(R_1, R_2, R_3, R_4)}_{\Delta_A} \,\,\,\oplus\,\,\, \mathcal{A}_2[0]^{(R_1 + 1, R_2, R_3, R_4)}_{\Delta_A + \frac{1}{2}}\\
    &\mathcal{L}[j = 0]^{(R_1, R_2, R_3, R_4)}_\Delta \,\,\longrightarrow\,\, \mathcal{A}_2[0]^{(R_1, R_2, R_3, R_4)}_{\Delta_A} \,\,\,\oplus\,\,\, \mathcal{B}_1[0]^{(R_1 + 2, R_2, R_3, R_4)}_{\Delta_A + 1}\\
    &\text{where}\,\,\,\,\, \Delta_A \equiv j + R_1 + R_2 + \frac{1}{2}(R_3 + R_4) + 1
\end{split}
\end{equation}
From (\ref{eq:recombrulesn8}), multiplets of the form $\mathcal{B}_1[0]^{(R_1, R_2, R_3, R_4)}_{R_1 + R_2 + \frac{(R_3 + R_3)}{2}}$ with $R_1 = 0, 1$ are absolutely protected. Also $(R_1, R_2, R_3, R_4)$ ; $R_1, R_2, R_3, R_4 \in \mathbb{Z}_{\geq 0}$ are the $so(8)_R$ Dynkin labels.

\section{Tests of $AdS_4\,/\,CFT_3$ Correspondence}
Since $AdS\,/\,CFT$ correspondence is not a rigorously proven statement, but rather a technically inspired yet beautifully consistent conjecture, it is often put under a microscope and subject to a lot of scrutiny in order to test the various hypotheses that are derived from the conjecture. Since the gravity side of the duality is a String / M-theory, it is more often than not a highly intractable problem to design tests that are non-perturbative and exact in nature. Infact the problem is even worse in the current case of $AdS_4\,/\,CFT_3$ because unlike String theory, there is no tunable coupling constant in M-theory as it is already the strong coupling limit of the Type IIA theory. To make matters worse, there is hardly anything concrete that is known about M-theory beyond the eleven dimensional supergravity limit and the information from it's various dualities with String theory. Now for the supergravity approximation to be valid, $L$ (curvature length scale) $\gg l_p$ (eleven dimensional planck length), which are related by (\ref{eq:abjmbranesol}) as $(\frac{L}{l_p})^6 = 32\pi^2Nk$. Now since the coupling constants of the ABJM theory (\ref{eq:ABJMTheory}) go as $\frac{1}{k}$, define the corresponding 't Hooft coupling \cite{tHooft:1973alw} as $\lambda \equiv \frac{N}{k}$, under which the aforementioned equation becomes $(\frac{L}{l_p})^6 = 32\pi^2k^2\lambda$. Therefore for a fixed $k$ of order $\sim 1$ and large coupling $\lambda$, $L \gg l_p$ and hence the supergravity approximation is valid. This further adds to the list of complications, as even a test with supergravity approximation on the gravity side of the duality, would require information from the strongly coupled non-planar limit of the ABJM theory, which is extremely difficult to obtain. However, all hope is not lost as some reliable extrapolations can be done from weak coupling to strong coupling in the presence of extended supersymmetry ($\mathcal{N} = 6,8$ in this case). Some of the important tests that have been performed on $AdS_4\,/\,CFT_3$ duality by making use of such supersymmetry protected information, will be briefly reviewed during the rest of this chapter. They are
\vspace{0.5em}
\begin{enumerate}[topsep = 2pt, label=\textbf{\arabic*})]
    \setlength\itemsep{0.5em}
    \item Superconformal index of the ABJM theory compared with the index over supergravitons in $AdS_4 \times S^7/\mathbb{Z}_k$ in the large $N$ limit, for different units of KK-momenta.
    \item Entropy of asymptotically $AdS$ magnetically (or electrically) charged $BPS$ black holes in four dimensions, compared with the degeneracy of states on the CFT side.
    \item Free energy of \textit{N} M2-branes from the ABJM partition function on $S^3$\,, \,at large \textit{N}.
\end{enumerate}

\subsection{Superconformal index of the ABJM theory}
The superconformal index of a superconformal field theory closely resembles the Witten index (SUSY partition function) \cite{WITTEN1982253}, and is defined to contain all the information about absolutely protected multiplets of the theory. This index, invariant under exactly marginal deformations, was first defined for four dimensions $d = 4$ in \cite{Kinney_2007}, and consequently later for $d = 3, 5$ and 6 in \cite{Bhattacharya_2008}. Given a choice of a pair of nilpotent ($A : A^2 = 0$) supercharges $\mathcal{Q}, \mathcal{S} \in osp(6 | 4)$, that commute with the Cartan generators $\mathcal{M}_D (\equiv \mathcal{D} + (\mathcal{M})_\alpha^{\,\,\,\alpha}), R_2, R_3$ without loss of generality, the index for ABJM theory is \cite{Bhattacharya_2009}
\begin{equation}\label{eq:superconformalindex}
    I(x, y, z) \coloneq \text{Tr}\left[(-1)^F\,e^{-\beta'\{Q, S\}}\,x^{\mathcal{M}_D}\,y^{R_2}\,z^{R_3}\right]
\end{equation}
where $F$ is the fermion number operator. Also, $I$ receives contribution only from states annihilated by $\mathcal{Q}, \mathcal{S}$ and hence is independent of $\beta'$. This index admits a path integral representation on $S^2 \times S^1$, with $S^1$ parametrized by Euclidean time $\tau \sim \tau + \beta + \beta'$; where $x \equiv e^{-\beta}$. It is then computed via SUSY localization i.e., by inserting a $\mathcal{Q}$-exact operator $e^{-t\{Q, V\}}$ into $I$ which leaves it invariant, and taking the limit $t \rightarrow \infty$; where $V$ is any gauge invariant operator. This insertion translates to the modification of the action $S \rightarrow S + t\{Q, V\}$ in the path integral representation, due to the nilpotency of $\mathcal{Q}$. In \cite{Kim_2009}, the $\mathcal{Q}$-exact term is chosen to be similar to the kinetic term of $d = 3, \mathcal{N} = 2$ Yang-Mills theory; Consequently for $t \rightarrow \infty$, the path integral reduces to a sum of contributions from saddle points which are characterized by monopole solutions for the ABJM gauge field strengths ($F_1, F_2$) (\ref{eq:ABJMTheory}). Their corresponding flux through $S^2$ and holonomy along $S^1$ are 
\small
\begin{equation}
    \mathlarger{\int}_{S^2}\frac{F_j}{2\pi} = \text{diag}\,(n_{r1},n_{r2},...,n_{rN}) \,\,\,\Bigg|\,\,\, \mathcal{P}\,\text{exp}\,(i\mathlarger{\int}_{S^1}A_j) = \text{diag}\,(e^{i\alpha_{r1}}, e^{i\alpha_{r2}}, ..., e^{i\alpha_{rN}})
\end{equation}
\normalsize
where $\mathcal{P}$ denotes the Polyakov loop. Finally by conducting a 1-loop analysis, the complete index has been computed as detailed in \cite{Kim_2009}, which for a saddle point labelled $\{n_{jk}\}$ is
\vspace{1.1em}
\hrule width \textwidth
\vspace{-0.4em}
\small
\begin{align}
    &I(x, y, z) = {\mathlarger{\mathlarger{\int}}}d\alpha_{ri}\,\,\frac{e^{-S}x^{\epsilon_0}}{4\pi^2S_\alpha}\prod_{s = 1}^2\Biggl[\,\,\prod_{j < k} 4\,\delta^{n_{sj}}_{n_{s_k}}\,\text{sin}^2\left(\frac{\alpha_{sj} - \alpha_{sk}}{2}\right) \times \prod_{j, k = 1}^{N}\text{exp}\left(h_{sjk}(x, y, z)\right)\,\Biggr]\nonumber\\[0.3ex]
    &h \coloneq f_{sjk}(x) \,+\, g_{jk}(x, y, z) \,+\, g_{jk}^*(x, \frac{1}{y}, z)\,\,\,\Bigg|\,\,\,f \coloneq \sum_{p = 1}^\infty \frac{\left(\delta^{n_{sj}}_{n_{sk}}-1\right)}{p}\kern 0.05em x^{\,p\,|n_{sj} - n_{sk}|}\,e^{-ip\,(\alpha_{sj} - \alpha_{sk})}\label{eq:abjmsuconfind}\\[0.7ex]
    &g \coloneq \sum_{p = 1}^\infty \frac{x^{p\,(|n_{1j} - n_{2k}| + 1/2)}\,(y^p + z^p - x^p - (xyz)^p)}{p\,(1 - x^{2p})(\sqrt{yz})^{p}}e^{-ip\,(\alpha_{1j} - \alpha_{2k})}\,\,\,\Big|\,\,\,e^{-S} =  e^{ik\sum_{i = 1}^N(n_{1i}\alpha_{1i} - n_{2i}\alpha_{2i})}\nonumber
\end{align}
\normalsize
\vspace{-0.6em}
\hrule width \textwidth
\vspace{0.9em}
\noindent where $S_\alpha$ is the symmetry factor of identical variables among $\{\alpha_{1j}\}$ or $\{\alpha_{2j}\}$, and $\epsilon_0$ is the Casimir energy $\sum_{i, j = 1}^N |n_{1i} - n_{2j}| - \sum_{r = 1}^2\sum_{i < j}|n_{ri} - n_{rj}|$. The reader may then refer to Section 3 of \cite{Kim_2009}, where the large $N$ limit of (\ref{eq:abjmsuconfind}) is numerically and analytically shown to perfectly agree with the index over super-gravitons in $AdS_4 \times S_7/\mathbb{Z}_k$ from \cite{Bhattacharya_2009}, for the cases of 1, 2 and 3 units of KK momenta along the M-theory circle (Hopf-fiber of $S^7/\mathbb{Z}_k$).

\subsection{Entropy of a charged $AdS_4$ Black hole}
The analytic solution for a charged $BPS$ Black Hole that is asymptotically $AdS_4$ in $\mathcal{N} = 2, d = 4$ gauged supergravity coupled to $n_V$ abelian vector multiplets, was constructed in \cite{Cacciatori_2010}. The bosonic field content of the theory is the vierbein $e^a_\mu$, $n_V + 1$ gauge fields $A_\mu^I$ (including the one from the graviton multiplet), $I = 0, 1,..., n_V$ and $n_V$ complex scalars $z^{\alpha}, \alpha = 1, 2,..., n_V$. These scalars parametrize an $n_V$-dimensional Hodge-K\"{a}hler manifold $\mathcal{M}$, which is a local Special K\"{a}hler manifold (torsion-free flat connection $\nabla$) equipped with a rank $2n + 2$ flat symplectic bundle $\mathcal{H} \rightarrow \mathcal{M}$ (structure group $Sp(2n + 2, \mathbb{R})$), whose sections are covariantly holomorphic (\ref{eq:sectionshodkah}). The tensor space $\mathcal{H} = \mathcal{L} \otimes \mathcal{S}\mathcal{V}$, where $\mathcal{L} \rightarrow \mathcal{M}$ is a line bundle with sections characterized by the possible superpotentials, and $\mathcal{S}\mathcal{V} \rightarrow \mathcal{M}$ is a flat holomorphic vector bundle with sections characterized by the superspace components of the electric and magnetic field strengths corresponding to $A_\mu^I$ \cite{Andrianopoli_1997}. The sections $\Omega$ of $\mathcal{H} \rightarrow \mathcal{M}$ have the following structure
\small
\begin{equation}\label{eq:sectionshodkah}
    \Omega = 
        \left(\mathcal{X}^{\Lambda},\,\mathcal{Y}_\Lambda\right) 
    \,\,\,;\,\,\, \Lambda = 0,1,...n_V \,\,\,\Big|\,\,\,\nabla_{\bar{\alpha}}\Omega = \partial_{\bar{\alpha}}\Omega - \frac{1}{2}(\partial_{\bar{\alpha}}\mathcal{K})\Omega = 0\,\,\,\Big|\,\,\,\mathcal{Y}_\Lambda = \frac{\partial \mathcal{Y}}{\partial \mathcal{X}^\Lambda}
\end{equation}
\normalsize
where $\mathcal{K} = -\,\text{log}\left[i\left(\bar{\mathcal{X}}^{\Lambda}\mathcal{Y}_\Lambda - \bar{\mathcal{Y}}_\Lambda\mathcal{X}^\Lambda\right)\right]$ is the K\"{a}hler potential, $\mathcal{Y} \equiv \mathcal{Y}(\mathcal{X})$ is the prepotential, and \,\,$\bar{}$ \,\,represents conjugates. Now consider the specific case of $n_V = 3$ and prepotential $\mathcal{Y} = -2i(\mathcal{X}^0\mathcal{X}^1\mathcal{X}^2\mathcal{X}^3)^{1/2}$ in the maximal $\mathcal{N} = 8$ gauged supergravity truncated to $\mathcal{N} = 2$. The spherically symmetric black hole solution for this specific case with four magnetic charges $\mathcal{Q}_M^\Lambda \equiv (n_a, n_4)$ and Bekenstein-Hawking entropy $\mathcal{S}_{BH}$ is (\cite{Hristov_2011}, \cite{Benini_2016}) 
\begin{equation}\label{eq:blackholeads4}
\begin{gathered}
    ds^2 = -e^{\mathcal{K}(\mathcal{X})}\left(gr - \frac{c}{2gr}\right)^2dt^2 + \frac{e^{-\mathcal{K}(\mathcal{X})} dr^2}{\left(gr - \frac{c}{2gr}\right)^2} + 2e^{-\mathcal{K}(\mathcal{X})}\,r^2(d\theta^2 + sin^2\theta \,d\phi^2)\\
    S_{BH} = -\frac{2\pi g^2}{G_{4D}}\sqrt{\mathcal{X}^0(r_h)\mathcal{X}^1(r_h)\mathcal{X}^2(r_h)\mathcal{X}^3(r_h)}\sum_\Lambda\frac{\mathcal{Q}_M^\Lambda}{\mathcal{X}^\Lambda(r_h)}
\end{gathered} 
\end{equation}
where $g, c$ are parameters and $r_h$ is the horizon radius. This black hole is dual to the $k = 1, \mathcal{N} = 8$ ABJM theory on $S^2 \times \mathbb{R}$ which is topologically twisted to $\mathcal{N} = 2$, by a choice of partial inclusion map $\iota : U(1)_R \subset SO(8)_R \rightarrow SO(3)$; where $SO(3)$ is the rotation group on $S^2$. Let the Cartan generators of $SO(8)_R$ be $J_1, J_2, J_3, J_4$, of which $J_4$ generates the $U(1)_R$ after inclusion $\iota$ while $J_a, a = 1, 2, 3$ are defined to commute with the supercharge $\mathcal{Q}$ and Hamiltonian $H$ of the $\mathcal{N} = 2$ theory. The partition function is then computed by the usual Witten index, now modified to include the operators that commute with $H$ similar to (\ref{eq:superconformalindex}), called the topologically twisted index \cite{Benini_2015}, as follows
\small
\begin{equation}\label{eq:twistedtopindex}
    Z(\{y_a\};\{n_a\}) = \text{Tr}\left[(-1)^Fe^{-\beta H} e^{i\Delta_aJ_a}\right]\,\,\,;\,\,\,y_a \equiv e^{i\Delta_a}
\end{equation}
\normalsize
This has a path integral representation on $S^2 \times S^1$ and has been computed in \cite{Benini_2016} via SUSY localization, where the correspondence with $S_{BH}$ is also shown in the large $N$ limit. 
\small
\begin{equation}
    \mathbb{R}e\,\text{log}\,Z \,\Big|_{\bar{\Delta}_a}(n_a) = S_{BH}(n_a) \,\,\,;\,\,\, \frac{\partial\, \mathbb{R}e\,\text{log}\,Z}{\partial \Delta_{1,2,3}}\Big|_{\sum_a \Delta_a = 2\pi} (\bar{\Delta}_a) = 0\,\,\,;\,\,\,\sum_a n_a + n_4 = 2
\end{equation}
\normalsize

\subsection{Type IIA on $AdS_4 \times \mathbb{C}\mathbb{P}^3$ and Free energy}\label{sec:freeenergy}
The compact $S^7/\mathbb{Z}_k$ manifold considered thus far can equivalently be viewed as an $S^1$ Hopf fibration over the complex projective plane $\mathbb{C}\mathbb{P}^3$, with the action of $\mathbb{Z}_k$ changing the periodicity of $S^1$ from $2\pi L \rightarrow \frac{2\pi}{k}L$. Therefore in the regime of supergravity validity (large $N$), the radius of this M-theory circle ($S^1$) is $\frac{L}{k}$, which for large $k$ becomes very small and hence the theory reduces to Type IIA supergravity on $AdS_4 \times \mathbb{C}\mathbb{P}^3$. Now further dimensionally reduce to four dimensions ($AdS_4$) in a $\mathbb{C}\mathbb{P}^3$ invariant way \cite{Bak_2010} and Euclideanize the theory. The resulting four dimensional geometry and the corresponding gravitational part of the supergravity action are as follows (\cite{Balasubramanian_1999}, \cite{Emparan_1999})
\small
\begin{equation*}
    \begin{gathered}
        ds^2_{AdS_4} = \frac{L^2}{4}\left(d\rho^2 + \text{sinh}^2\rho\,d\Omega_3^2\right) \,\,\,\Bigg|\,\,\, S_{grav} = S_{\text{bulk}} + S_{\text{surf}} + S_{ct}\,\,\,\Bigg|\,\,\,S_{\text{surf}} = -\frac{1}{8\pi G_4}\mathlarger{\int}_{S^3}d^3x \,\sqrt{h}\mathcal{K}\\S_{\text{bulk}} = -\frac{1}{16\pi G_4} \mathlarger{\int}_{AdS_4} d^4x\,\sqrt{g}\left(\mathcal{R}_g - 2\Lambda\right)
        \,\,\,\Bigg|\,\,\,S_{ct} = \frac{1}{8\pi G_4}\mathlarger{\int}_{S^3}d^3x \,\sqrt{h}F(\Lambda, \mathcal{R}_h, \nabla \mathcal{R}_h) 
    \end{gathered}
\end{equation*}
\normalsize
where $\Lambda = -\frac{12}{L^2}$ is the cosmological constant, $S^3$ is the boundary of Euclidean $AdS_4$ i.e. $\rho \rightarrow \infty$, $g$ is the metric on $AdS_4$ and $h$ is the induced metric on $S^3$; Correspondingly $\mathcal{R}_g$ and $\mathcal{R}_h$ are their respective Ricci scalars, while $\mathcal{K}$ is the trace of the extrinsic curvature of $S^3$ as embedded in $AdS_4$. $S_{\text{bulk}}$ is the usual Einstein-Hilbert term, while $S_{\text{surf}}$ is the Gibbons-Hawking-York boundary term, which is needed to obtain Einstein equations while varying $S_{\text{grav}}$ with metric fixed at the boundary. $S_{ct}$ is the counterterm added on the boundary in order to remove divergences arising from $S_{\text{bulk}}$ and $S_{\text{surf}}$ as $\rho \rightarrow \infty$. The scalar $\mathcal{R}_g$ for $AdS_4$ is $-\frac{48}{L^2}$, so evaluating $S_{\text{bulk}}$ on the hypersurface $\rho = \rho_0$
\small
\begin{equation}
    S_{\text{bulk}}(\rho_0) = \frac{3L^2}{32\pi G_4}\int d\Omega_3 \int_0^{\rho_0}d\rho\,\,\text{sinh}^3\rho = \frac{3\pi L^2}{16G_4}\left[\frac{1}{12}\text{cosh}(3\rho_0) - \frac{3}{4}\text{cosh}(\rho_0) + \frac{2}{3}\right]
\end{equation}
\normalsize
Now as $\rho_0 \rightarrow \infty$, the divergences from the $\text{cosh}$ terms are removed by $S_{\text{surf}}$ and $S_{\text{ct}}$ \cite{Emparan_1999}. By using $L^6 = 32\pi^2Nkl_p^6$ (\ref{eq:abjmbranesol}), $G_{11} = \text{Vol}(S^1 \times \mathbb{C}\mathbb{P}^3) G_4$; where $\text{Vol}(S^1 \times \mathbb{C}\mathbb{P}^3) = \frac{2\pi L}{k} \times \frac{\pi^3L^6}{3!}$, and $G_{11} = 16\pi^7l_p^9$, $S_{\text{grav}}$ on the boundary then becomes the following 
\begin{equation}\label{eq:freesugra}
    \lim_{\rho_0\to\infty}S_{\text{grav}}(\rho_0) = \frac{\pi L^2}{8G_4} = \frac{\sqrt{2}\pi}{3}k^{\frac{1}{2}}N^{\frac{3}{2}}
\end{equation}
On the CFT side, it was shown in \cite{Kapustin_2010} via SUSY localization of a SUSY Wilson loop, that the partition function for a SUSY Chern-Simons theory localizes to a non-Gaussian matrix model. Consequently, the partition function $Z$ was computed in \cite{Drukker_2011} for the ABJM theory on $S^3$. In the large $N$ limit, the free energy $F$ has a genus expansion as follows 
\small
\begin{equation}
    F = \text{log}\,Z = \sum_{g = 0}^{\infty} g_s^{2g - 2}F_g(\lambda)\,\,\,;\,\,\,\lambda \equiv \frac{N}{k}\,,\,\,g_s \equiv \frac{2\pi i}{k} 
\end{equation}
\normalsize
The reader may then refer to \cite{Drukker_2011}, where it is shown that the leading term of this expansion matches exactly the supergravity computation (\ref{eq:freesugra}) i.e., |\,\,$g_s^{-2}F_0 = -\lim_{\rho_0\to\infty}S_{\text{grav}}(\rho_0)$\,\,|

\section{A More Precise Statement of $AdS_4\,/\,CFT_3$ Duality}
The word duality is often used in various contexts implying a host of different equivalence relations, but is hardly anything concrete by itself in order to do actual calculations. Therefore a more formal and a more precise statement of the said duality $AdS_4/CFT_3$ is in order, which is most easily described in Euclidean signature. It's strongest form is \cite{witten1998anti}
\begin{equation}\label{eq:preciseads4cft3}
    \left\langle \text{exp}\left(\int_{S^3} d^3x \,\phi_0\mathcal{V}\right)\right\rangle_{ABJM} = Z_{M-theory}\,\Bigg|_{\,\lim_{z\to 0}\left(\phi(z, x)z^{\Delta - 3}\right) = \phi_0(x)}
\end{equation}
where $z$ is the $AdS_4$ (radius $L/2$) bulk coordinate, $\phi(z, x)$ is a field propagating in the bulk (non-boundary region of $AdS_4$), and $\Delta$ is the scaling dimension of the superconformal primary $\mathcal{V}$ (see Section \ref{sec:ospn4algebra}) which is dual to $\phi$. $Z_{M-theory}$ is the partition function of M-theory on $AdS_4$ compactified on $S^7 / \mathbb{Z}_k$, integrated over metrics that respect the conformal structure of the boundary $S^3$ ($z \rightarrow 0$). The LHS on the other hand is the $\mathcal{V}$ generating functional sourced by $\phi_0$ in ABJM theory. Also if $\phi$ is a scalar field, $\Delta \equiv \Delta_{+}$ is one of the solutions to the equation $\Delta(\Delta - 3) = m^2\left(\frac{L}{2}\right)^2$; where $m$ is the mass of the field $\phi$. Combined with the other solution $\Delta_{-}$, these two characterize the normalisable ($\phi_+$) and non-normalisable modes ($\phi_0$) respectively of $\phi$ near the boundary.
\begin{align*}
    &(\square_{AdS_4} - m^2)\phi = 0 \,\,\,\Big|\,\,\,\square_{AdS_4} = \frac{4}{L^2}\left(z^2\partial_z^2 - 2z\partial_z + z^2\eta_{\mu\nu}\partial^\mu\partial^\nu\right) \implies \phi(z, x) \sim e^{ip^\mu x_\mu}z^{\Delta}\\[1ex]
    &\Delta(\Delta - 3) = \frac{m^2L^2}{4} \implies \Delta_{\pm} = \frac{\left(3 \pm \sqrt{9 + m^2L^2}\right)}{2} \,\,\,\Bigg|\,\,\, \lim_{z\to 0}\phi(z, x) \sim \phi_0(x)z^{\Delta_{-}} + \phi_+(x)z^{\Delta_{+}}
\end{align*}
where the $z$ dependence of $\phi$ and the condition $\Delta(\Delta - 3) = \frac{m^2L^2}{4}$ are obtained from the Klein-Gordon equation on $AdS_4$ as shown above. As seen already in (\ref{eq:preciseads4cft3}), $\phi_0(x)$ sources the generating functional in the CFT, while $\phi_+ (x)$ may be identified with the VEV of operator $\mathcal{V}$ via dimensional analysis. However a small note is that, for operators with $\Delta < 3/2$ and satisfying the unitarity bounds mentioned in section \ref{sec:ospn4algebra}, $\Delta \equiv \Delta_{-}$ and therefore the roles of $\phi_{+}$ and $\phi_0$ are exchanged. The mass-dimension relations for a general fields that are not necessarily scalars can be derived similarly, and are summarized in table \ref{tab:massdimrel} \cite{Ammon_Erdmenger_2015}. Now the statement (\ref{eq:preciseads4cft3}) is not really useful for any calculations, as there isn't an existing formulation of M-theory. Therefore it's weaker version is obtained by performing a saddle point approximation in the path integral representation of $Z_{M-theory}$, around it's low energy effective theory i.e. eleven dimensional supergravity. 
\begin{equation}\label{eq:saddlepointadscft3}
    \left\langle \text{exp}\left(\int_{S^3} d^3x \,\phi_0\mathcal{V}\right)\right\rangle_{ABJM} \approx e^{-S_{sugra}}\,\Bigg|_{\,\lim_{z\to 0}\left(\tilde{\phi}(z, x)z^{\Delta - 3}\right) = \phi_0(x)}
\end{equation}
where $S_{sugra}$ is the on-shell action of eleven dimensional supergravity in the bulk $AdS_4$, reduced on $S^7/\mathbb{Z}_k$; $\tilde{\phi}$ is a solution with the leading asymptotic behavior $z^{3 - \Delta}\phi_0$ as $z \rightarrow 0$.

\setlength{\tabcolsep}{10pt}
\renewcommand{\arraystretch}{1.5}
\begin{table}[h]
    \centering % Extra spacing between rows
    \begin{tabular}{|l|l|} % Add vertical lines between columns
        \hline
        \quad\quad\quad\quad\textbf{Type of field} & \,\,\,\,\textbf{Relation between m and $\Delta$}\\
        \hline
        scalars, massive spin two fields & $m^2L^2/4 = \Delta(\Delta - 3)$\\
        \hline
        massless spin two fields & $m^2L^2/4 = 0, \Delta = 3$ \\
        \hline
        \textit{p}-form fields & $m^2L^2/4 = (\Delta - p)(\Delta + p - 3)$ \\
        \hline
        spin 1/2, spin 3/2 fields & $|m|L/2 = \Delta - 3/2$ \\
        \hline
        rank `\textit{s}' symmetric traceless tensor & $m^2L^2/4 = (\Delta + s - 2)(\Delta - s - 1)$\\
        \hline
    \end{tabular}
    \caption{Mass dimension relations}
    \label{tab:massdimrel}
    \vspace{-0.8em}
\end{table}
\noindent Let us call the LHS of (\ref{eq:preciseads4cft3}) and (\ref{eq:saddlepointadscft3}) $\mathcal{J}[\phi_0]$. Since $\mathcal{J}[\phi_0]$ is the generating functional for correlation functions of $\mathcal{V}$, consequently $\mathcal{W}[\phi_0] = \text{log}\,J[\phi_0]$ is the generating functional for connected correlation functions of $\mathcal{V}$. Generalizing to the case of multiple composite gauge invariant superconformal primaries $\mathcal{V}_i$, with their corresponding sources $\phi_0^i$, the general connected n-point correlation function is then given by
\begin{equation}\label{eq:correlationholographic}
    \left\langle\mathcal{V}_{i_1}(x_1)\mathcal{V}_{i_2}(x_2)...\mathcal{V}_{i_n}(x_n)\right\rangle_{ABJM, \,\text{conn.}} = -\frac{\delta^n \mathcal{W}}{\delta \phi_0^{i_1}(x_1)\,\delta \phi_0^{i_2}(x_2)...\delta\phi_0^{i_n}(x_n)}\Bigg|_{\,\phi_0^i = 0}
\end{equation}
where it goes without mention that both the equations (\ref{eq:preciseads4cft3}) and (\ref{eq:saddlepointadscft3}) now include the bulk fields (solutions) $\phi^i$ ($\tilde{\phi}^i$) and corresponding dual operators $\mathcal{V}_i$; where $\lim_{z\to 0}\left(\phi^i(z, x)z^{\Delta - 3}\right) = \phi_0^i(x)$, same goes for $\tilde{\phi}^i(z, x)$. In summary, computing the connected correlation functions as in (\ref{eq:correlationholographic}) corresponds to the following procedure
\vspace{1em}
\begin{itemize}
    \setlength\itemsep{0.5em}
    \item Find the bulk fields $\phi^i$ dual to operators $\mathcal{V}_i$ of scaling dimensions $\Delta_i$, via table \ref{tab:massdimrel}.
    \item Include the relevant $\phi^i$ terms in eleven dimensional supergravity and dimensionally reduce it on $S^7/\mathbb{Z}_k$ to four dimensions, to obtain the off-shell $S_{sugra}$.
    \item Find the solution $\tilde{\phi}^i$ corresponding to the equations of motion for $\phi^i$ in $S_{sugra}$ such that the solution is asymptotically ($z \rightarrow 0$) : $\tilde{\phi}^i(z, x) \sim z^{3 - \Delta_i}\phi_0^i(x)$.
    \item Compute the on-shell $S_{sugra}[\phi_0^i]$ by substituting $\tilde{\phi}^i$ into the off-shell $S_{sugra}$.
    \item Since from (\ref{eq:saddlepointadscft3}), $S_{sugra}[\phi_0^i] = -\mathcal{W}[\phi_0^i]$, vary it as in (\ref{eq:correlationholographic}) to obtain connected correlation functions in the holographically dual ABJM theory.
\end{itemize}
\vspace{1em}
The patterns among the terms that appear during the above mentioned computation, are pictorially captured by \textit{Witten diagrams}, which are the Feynman rules for the holographic computation of correlation functions in $AdS$ geometry. Infact the 1-loop diagrams are at the order (1/$N^4$) which can be ignored for supergravity (large $N$ fixed $k$) and therefore only the tree level diagrams contribute. These diagrams and their application to a specific example of operators belonging to the stress tensor multiplet, will be explored in the next chapter. As we will see, this leads to a highly non-trivial test of the \,$AdS_4\,/\,CFT_3$\, duality.  

%% file: chapters/conclusion_recommendations.tex
\chapter{Witten Diagrams and CFT Correlators}\label{cap:conclusions_recommendations}
\vspace{-0.5em}
In any quantum field theory, correlation functions are key quantities to compute, since they constitute fundamental information regarding the theory. Although this may not be obvious from their description as objects relevant to the analysis of interactions, they do in fact probe the quantum skeleton of the theory. This statement was made rigorous in Minkowski space by the axiomatic description of quantum field theory owing to Wightman \cite{wightmanqft}, whose reconstruction theorem implies that the Hilbert space and quantum field operators can be recovered from the knowledge of all the vacuum correlation functions of all the fields. For the Wick-rotated Euclidean theory, these Wightman correlation functions translate to the corresponding Schwinger functions that satisfy the Osterwalder–Schrader axioms \cite{ostschradqft}, accompanied by a similar reconstruction theorem. The geometric background of interest to us is the boundary of $AdS$ which is conformally flat, and therefore it is reasonable to expect the existence of a similar reconstruction theorem, albeit hopefully the importance of correlation functions has already been briefly impressed upon.
\[
  \begin{gathered}
    \text{Correlation functions}\\\text{(CFT)}
  \end{gathered}
  \quad\mathlarger{\mathlarger{\mathlarger{\Longleftrightarrow}}}\quad
  \begin{gathered}
    \text{Holographic amplitudes}\\\text{(AdS)}
  \end{gathered}
\]
As mentioned towards the end of the previous chapter, $S_{sugra}$ can be used to compute the connected correlation functions in the large $N$ fixed $k$ limit. These connected functions contain all the information that the full correlation functions contain, since any disconnected diagram is merely a product of connected diagrams. This in conjunction with the argument from the previous paragraph implies that, the tree-level amplitudes in $S_{sugra}$, computed via \textit{Witten diagrams}, probe the quantum structure of ABJM theory in the strongly coupled non-planar limit. Therefore a match between these holographic amplitudes and the correlation functions computed in ABJM is a highly non-trivial test of $AdS_4 /CFT_3$ duality. Now by using this duality in the other direction, one may be ambitious enough to hope to recover the supergravity footprint of the full M-theory S-matrix, or maybe even the partition function $Z_{M-theory}$ via (\ref{eq:preciseads4cft3}). But let us not get ahead of ourselves, and instead start by describing the `Witten diagrams' in the following section.  

\section{Witten Diagrams in Euclidean $AdS_{d+1}$}
Witten diagrams are the Feynman rules governing the patterns in the computation of holographic amplitudes in $AdS$ space, that are dual to correlation functions in the conformal field theory living on the boundary of $AdS$. As can be seen in figure \ref{fig:wd_1} below, they are represented by a circle denoting the conformal boundary of $AdS_{d + 1}$ i.e. $S^d$, with the interior denoting the bulk of $AdS_{d + 1}$. Since we are in the supergravity approximation, we will focus only on the tree-level diagrams [\,$O(1/N^2)$\,] in this section.
\begin{figure}[h]
\centering
\includegraphics[width=\textwidth]{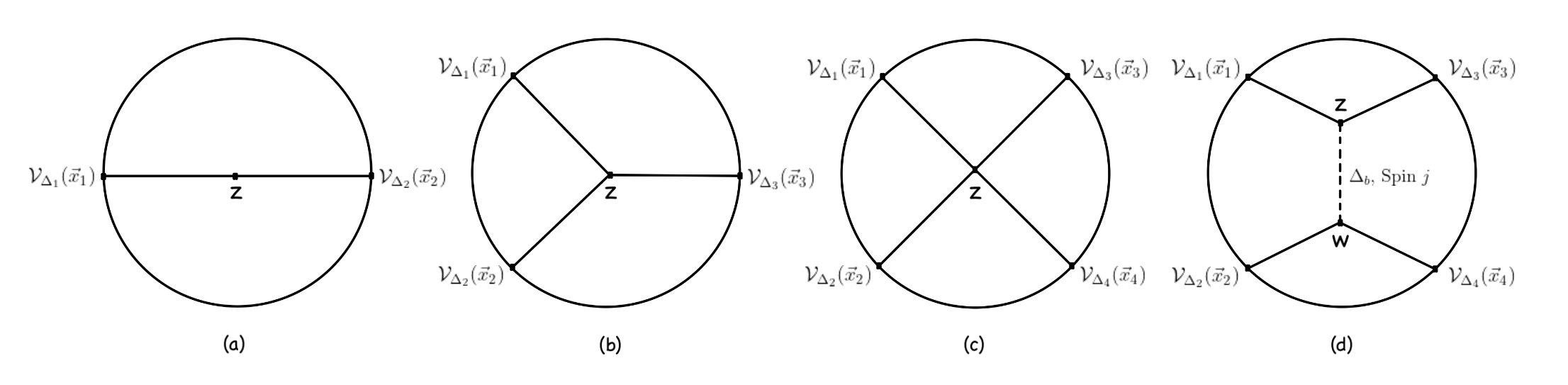}
\vspace{-1.5em}
\caption{Witten diagrams with boundary ($\partial AdS_{d+1}$) circle and interior ($AdS_{d+1}$) bulk. (a) 2-point contact (b) 3-point contact (c) 4-point contact (d) 4-point \textit{t}-channel exchange}
\label{fig:wd_1}
\end{figure}

The points on the boundary are labelled by superconformal primaries dual to the bulk field $\phi(z)$, sourced by $\phi_0(\vec{z})$. The solid line propagators that emanate from a source on the boundary to a point in the bulk are called \textit{bulk-boundary} propagators, and the dotted lines (as in figure \ref{fig:wd_1} (d)) between two bulk points ($z$ and $w$) are called \textit{bulk-bulk} propagators. The vertices in the bulk are governed by the interaction terms in $S_{sugra}$, for e.g. the vertices in (b) and (d) arise from cubic coupling terms. The notation to distinguish between bulk and boundary points is that boundary points have a \,$\vec{}$\, over them, while the bulk points do not. A bulk point $z$ is further sub-labelled as ($z_0$, $\vec{z}$) where $z_0$ is the radial coordinate of $AdS_{d+1}$; such that $(z - \vec{x})^2 \equiv z_0^2 + (\vec{z} - \vec{x})^2$. Also throughout this chapter from now on, the curvature radius of $AdS$ space will be set to one for convenience; Therefore the Wick rotated $AdS$ line element in this new notation in Poincar\'{e} coordinates is
\begin{equation}\label{eq:adsmetric}
    ds^2_{AdS_{d+1}} = \frac{1}{z_0^2}\left(\,dz_0^2 + d\vec{z}\cdot d\vec{z}\,\right)\,\,\,\,\,\Big|\,\,\,\,\,z \coloneq (z_0, \vec{z})\,,\,\,\vec{z} \in \mathbb{R}^d
\end{equation}
It is also convenient to define the chordal distance $\xi$ in terms of geodesic distance $d(z, w)$
\begin{equation*}
    d(z, w) \equiv \int_z^w ds_{AdS_{d+1}} = \text{log}\left(\frac{1 + \sqrt{1 - \xi^2}}{\xi}\right) \implies \xi\,:\,\xi(z, w) = \frac{2z_0w_0}{z_0^2 + w_0^2 + (\vec{z} - \vec{w})^2}
\end{equation*}
Symbolically, the general method to compute these propagators is to first find the terms containing the bulk field $\phi$ in the action $S_{sugra}$, derive the equations of motion say $L[\phi] = 0$, and then compute the propagator $G$ as a kernel for the EOM with appropriate source $S$ i.e. $L[G] = S$. Here on, $G_{b\partial}$ and $G_{bb}$ stand for bulk-boundary and bulk-bulk respectively.

\subsection{Massive Scalar field | Spin-0}\label{sec:scalarprop}
Let us consider a bulk scalar field $\phi$ with mass $m : m^2 = \Delta(\Delta - d)$ (table \ref{tab:massdimrel}), dual to the superconformal primary $\mathcal{V}_{\Delta}$. The relevant part of the action $S_{sugra}$ containing $\phi$ is 
\begin{equation}\label{eq:scalaraction}
    \begin{split}
    &S_{scalar} = \frac{1}{2}\int d^{d+1}z\,\sqrt{g}\left(g^{\mu\nu}\partial_\mu\phi\,\partial_\nu\phi + m^2\phi^2\right) \implies \text{EOM}:(\kern 0.05em\square_{AdS} - m^2\kern 0.05em)\,\phi = 0\\
    &\text{where}\,\,\,\,\,\square_{AdS}\phi = \nabla_\mu \nabla^\mu \phi = \frac{1}{\sqrt{g}}\partial_\mu(\sqrt{g}g^{\mu\nu}\partial_\nu\phi)= \left(z_0^2\partial_{z_0}^2 - 2z_0\partial_{z_0} + z_0^2\partial_{\vec{z}}\cdot\partial_{\vec{z}}\right)\phi 
    \end{split} 
\end{equation}
where $\nabla_\mu$ is the connection on $AdS_{d+1}$. Also note that $L_{scalar}$ can be chosen to be $g^{\mu\nu}\phi\nabla_\mu\nabla_\nu \phi$ as well. This is because both the choices give the same equations of motion and describe the same dynamics in the bulk, however the one in (\ref{eq:scalaraction}) is the right choice in order to have a non-vanishing generating functional for correlators on the conformal boundary \cite{Boschi_Filho_1999}. Now as mentioned before, the Bulk-Boundary propagator emanates from $\phi_{0, \Delta}$ on the boundary, while the Bulk-Bulk propagator emanates at a bulk point $w$ with a source let's say $J(w)$. The former, $\phi_{0, \Delta}$, doesn't enter the EOM as a source but rather as a boundary condition, since it's information is already embedded in the definition of $AdS/CFT$ (\ref{eq:saddlepointadscft3}); However the latter $J(w)$, does source the equations of motion. Therefore, introducing the propagators as the appropriate integral kernels, as follows
\begin{equation*}
\begin{alignedat}{3}
    &G_{b\partial} \,\,:\,\, \phi_\Delta(z) = \int_{\partial AdS} d^d\vec{x}\,\,G_{b\partial}^\Delta(z, \vec{x})\phi_{0, \Delta}(\vec{x}) \,\,\,\,\,\,\,\kern 0.05em;\,\,\,&&\left(\square_{AdS} - m^2\right)\phi_\Delta(z) = 0\,\Big|_{\,L_1}\\
    &G_{bb} \,\,:\,\, \tilde{\phi}_\Delta(z) = \int_{AdS}d^{d+1}w \,\sqrt{g}G_{bb}^{\Delta}(z, w) J(w) \,\,\,\kern 0.05em;\,\,\, &&\left(\square_{AdS} - m^2\right)\tilde{\phi}_\Delta(z) = J(z)\\
    &\,&&\left(\square_{AdS} - m^2\right)G^{\Delta}_{bb}(z, w) = \frac{\delta^{d+1}(z-w)}{\sqrt{g}}
\end{alignedat}
\end{equation*}
where $L_1$ is the limit, $\lim_{z_0\to 0}\left(\phi_\Delta(z_0, \vec{z})z_0^{\Delta - d}\right) = \phi_{0, \Delta}(\vec{z})$. The solutions are (\cite{witten1998anti}, \cite{dhoker2002supersymmetric})
\small
\begin{equation}\label{eq:scalarprop}
\begin{gathered}
    G^{\Delta}_{b\partial} = C_{\Delta}\left(\frac{z_0}{(z - \vec{x})^2}\right)^{\Delta}\,\,\,\Bigg|\,\,\,G^{\Delta}_{bb}(\xi) = \frac{C_\Delta}{2^\Delta(2\Delta - d)}\xi^\Delta\cdot\, \tensor[_2]{F}{_1}\left(\frac{\Delta}{2}, \frac{\Delta + 1}{2};\Delta - \frac{d}{2} + 1;\xi^2\right)\\
    \text{where}\,\,\,\,C_\Delta = \frac{\Gamma(\Delta)}{\pi^{d/2}\Gamma\left(\Delta - \frac{d}{2}\right)} \,\,\,\text{for}\,\,\,\Delta > \frac{d}{2}
\end{gathered}
\end{equation}
\normalsize
where \,$\tensor[_2]{F}{_1}$ is a hypergeometric function. Also, for $\Delta < \frac{d}{2}$, just replace $\Delta \rightarrow d - \Delta$ to go from $\Delta \equiv \Delta_{+}$ to $\Delta \equiv \Delta_{-}$ as mentioned in the previous chapter. It can also be seen that $G_{b\partial}$ has the necessary singular behavior $\lim_{z_0\to 0}\left(G_{b\partial}(z, \vec{x})\,z_0^{\Delta - d}\right) = \delta(\vec{z} - \vec{x})$. Since $0 \leq \xi \leq 1$ as defined earlier, the hypergeometric function with the argument $\xi$ has a convergent taylor series expansion for all values of $\xi$ except for $\xi(z, w) = 1$ i.e. ($z = w$).
\begin{equation}
    \tensor[_2]{F}{_1}(a,b;c;\xi) = \sum_{n = 0}^\infty \frac{(a)_n(b)_n}{(c)_n}\cdot\frac{\xi^n}{n!} \,\,\,\,\,\,\Bigg|\,\,\,\,\,\, (q)_n = \begin{cases}
        1 & n = 0 \\ q(q+1)...(q+n-1) & n > 0
    \end{cases}
\end{equation}

\subsection{Massive Vector field | Spin-1}
Now let us shift our focus onto propagators of fields with a spin. Starting with Spin-1 in this section and moving onto Spin-2 in the next, only the bulk-bulk propagators will be described in the main text, as they are the only ones required for computations later on in the next chapter. The interested reader may then refer to \cite{Costa_2014} for the description of the general bulk-boundary propagator for a Spin-\textit{j} field, within the embedding formalism. Consider a bulk gauge field $A_\mu$ with mass $m : m^2 = (\Delta - 1)(\Delta + 1 - d)$ (table \ref{tab:massdimrel}), dual to the superconformal primary $\mathcal{V}_\Delta$. The relevant part of the action $S_{sugra}$ with a classical source $J^\mu$ contains the usual well-known terms of the gauge field
\begin{align}
    &S_{vector} = \int d^{d+1}z \,\sqrt{g}\left[\frac{1}{2}(\nabla_\mu A_\nu)^2 - \frac{1}{2}(\nabla^\mu A_\mu)^2 + \frac{1}{2}m^2 A^\mu A_\mu - A_\mu J^\mu\right]\label{eq:vectoraction}\\ 
    &\text{EOM}\,\,:\,\,(\square_{AdS} - m^2)A_\mu - \nabla_\mu(\nabla^\nu A_\nu) = -J_\mu\,\,\,|\,\,\, \nabla_\mu J^\mu = 0 \implies \nabla^\mu A_\mu = 0
    \end{align}
As mentioned above, working in the restricted space of covariantly-conserved currents ($\nabla_\mu J^\mu = 0$), acting on the EOM with $\nabla^\mu$ from the left, leads to the gauge $\nabla^\mu A_\mu = 0$, which is what will be used for the massless case. Introducing the bulk-bulk propagator ($G_{\mu;\nu'}^\Delta$) for the gauge field as a kernel, similar to the scalar field, leads to the following \cite{D_Hoker_1999}
\begin{equation}\label{eq:greensfunctionvector}
    \begin{gathered}
        A_\mu(z) = \int d^{d+1}w \,\sqrt{g}\,G_{\mu;\nu'}^\Delta(z, w) J^{\nu'}(w)\\
        \implies \nabla^{\mu} \partial_{\,[\,\mu}\,G^\Delta_{\nu\,];\,\nu'} (z, w) - m^2G^{\Delta}_{\nu;\nu'}(z, w) = -g_{\nu\nu'}\delta^{d+1}(z-w) + \partial_{\nu'} \Lambda_\nu(z, w)
    \end{gathered}
\end{equation}
where $\partial_\mu = \frac{\partial}{\partial z^\mu}$ and $\partial_{\mu'} = \frac{\partial}{\partial w^{\mu'}}$. Also, $\Lambda_\nu$ is a vector function that reflects the gauge freedom, and the term $\partial_{\nu'}\Lambda_\nu(z, w)$ vanishes when the above equation is multiplied by the covariantly conserved current $J^{\nu'}$ and integrated over. The reader may also note that the $bb$ notation is dropped in the bulk-bulk propagator $G^\Delta_{\mu\nu'}$ for convenience, and that it's bi-vector form is indicative of it's belonging to the vector field from here on. Now defining $u \equiv 1/\xi$, the solution to the above equations is given by a linear combination of the tensor structures $\partial_\mu\partial_{\nu'}u$ and $\partial_\mu u\, \partial_{\nu'} u$ \cite{Costa_2014}
\begin{equation}
    G_{\mu;\nu'}^{\Delta}(z, w) = -\partial_\mu \partial_{\nu'}u \,g_0(u) + \partial_\mu u\, \partial_{\nu'}u\,g_1(u)
\end{equation}
\vspace{-1em}
\small
\begin{equation*}
    \begin{split}
        &g_0(u) = (d - \Delta) H_{a}(u) - \frac{1 + u}{u}H_{a+1}(u) \,\,\,\Big|\,\,\,g_1(u) = \frac{(1+u)(d - \Delta)}{u(2 + u)}H_a(u) - \frac{d + (1+u)^2}{u^2(2+u)}H_{a+1}\\
        &H_a(u) = \mathcal{N}_m(2u)^{-\Delta}\,\tensor[_2]{F}{_1}\left(a, \frac{1 - d + 2\Delta}{2}, 1 - d + 2\Delta, -\frac{2}{u}\right) \,\,\,\Big|\,\,\,\mathcal{N}_m = -\frac{\Gamma(\Delta + 1)}{2\pi^{d/2}m^2\Gamma(\Delta + 1 - d/2)}
    \end{split}
\end{equation*}
\normalsize
where $a \equiv \Delta$. In the massless limit, the gauge and physical components separate \cite{D_Hoker_1999}
\begin{equation}
    G^\Delta_{\mu;\nu'}(z, w) = -\partial_\mu \partial_{\nu'}u \, H(u) + \partial_\mu\, \partial_{\nu'}\,S(u)\,\,\,\Big|\,\,\, H(u) = \frac{\Gamma((d - 1)/2)}{4\pi^{(d+1)/2}}\frac{1}{(u(u+2))^{(d-1)/2}}
\end{equation}
where $S(u)$ is a gauge artifact that drops out of (\ref{eq:greensfunctionvector}) and can be discarded from solution.

\subsection{Massive Tensor field | Spin-2}\label{subsec:massivespin2}
In order to derive the relevant terms in the action for a massive spin-2 field ($m^2 = \Delta(\Delta - d)$; table \ref{tab:massdimrel}), one starts with the Einstein-Hilbert action with a negative cosmological constant $\Lambda = -\frac{d(d - 1)}{2}$. The boundary term and counter term contributions as seen in section \ref{sec:freeenergy} are not necessary, since we are interested in bulk-bulk propagators only.
\begin{equation}
    S_{AdS_{d+1}} = \int d^{d+1}z \,\sqrt{g}\left[\frac{1}{2\kappa_{d+1}^2}(R - 2\Lambda) + \mathcal{L}_M\right]
\end{equation}
where $\mathcal{L}_M$ is the matter lagrangian and $\kappa_{d+1} = 1$ for convenience. Now perturbing this action with a tensor field $h_{\mu\nu}$ i.e. $g_{\mu\nu} \rightarrow g_{\mu\nu} + h_{\mu\nu}$, and analyzing upto quadratic order in $h$, the action splits into $S_{AdS_{d+1}} + S_{tensor} + O(h^3)$. The $S_{tensor}$ is then given by \cite{Buchbinder_2000}
\begin{align}\label{eq:tensoraction}
    S_{tensor} = &\int d^{d+1}z\,\sqrt{g}\Biggl[\frac{1}{4}\nabla_\mu h \nabla^\mu h - \frac{1}{4}\nabla_\mu h_{\nu\alpha} \nabla^\mu h^{\nu\alpha} - \frac{1}{2}\nabla^\mu h_{\mu\nu} \nabla^\nu h + \frac{1}{2}\nabla_\mu h_{\nu\alpha} \nabla^\alpha h^{\nu\mu}\nonumber\\[-1.5ex]
     &+ \frac{1}{2}h_{\mu\nu}h^{\mu\nu} - \frac{(d - 1)}{2}h^2 - h_{\mu\nu}J^{\mu\nu} + \,\,\,\text{mass-term}\left\{\frac{m^2}{4}(h^2 - h_{\mu\nu}h^{\mu\nu})\right\}\Biggr]
\end{align}
where $h$ is the trace $g_{\mu\nu}h^{\mu\nu}$, and $J_{\mu\nu} \coloneq \frac{1}{\sqrt{g}}\frac{\delta(\sqrt{g}\mathcal{L}_M)}{\delta g^{\mu\nu}}$ is a classical source. It may also be noted that the Fierz–Pauli mass term \cite{Fierz:1939ix} has been added to the quadratic order action. The equation of motion for $h_{\mu\nu}$ from $S_{tensor}$ is the following \cite{Naqvi_1999}
\small
\begin{equation*}
    (\square_{AdS} - m^2 + 2)h_{\mu\nu} + \nabla_\mu\nabla_\nu h - \nabla_\mu \nabla^\sigma h_{\sigma\nu} - \nabla_\nu \nabla^\sigma h_{\mu\sigma} - \left(2 + \frac{m^2}{d-1}\right)g_{\mu\nu}h = -J_{\mu\nu} + \frac{J_\sigma^{\,\,\sigma}}{d - 1}g_{\mu\nu}
\end{equation*}
\normalsize
Introducing the bulk-bulk propagator ($G_{\mu\nu;\mu'\nu'}^\Delta$) for the tensor field  as a kernel leads to  
\begin{equation}\label{eq:tensorpropagator}
    h_{\mu\nu}(z) = \int d^{d+1}w \,\sqrt{g}G^\Delta_{\mu\nu;\mu'\nu'}(z, w)J^{\mu'\nu'}(w)
\end{equation}
\vspace{-1.7em}
\begin{align*}
    (\square_{AdS} - m^2 + 2)(G^\Delta)_{\mu\nu;\mu'\nu'} + \nabla_\mu\nabla_\nu (G^\Delta)_{\sigma\,\,;\mu'\nu'}^{\,\,\sigma} - \nabla_\mu \nabla^\sigma (G^\Delta)_{\sigma\nu;\mu'\nu'} - \nabla_\nu \nabla^\sigma (G^\Delta)_{\mu\sigma;\mu'\nu'}&\\ - \left(2 + \frac{m^2}{d-1}\right)g_{\mu\nu}(G^\Delta)_{\sigma\,\,;\mu'\nu'}^{\,\,\sigma}
    = \left(\frac{2}{d - 1}g_{\mu\nu}g_{\mu'\nu'} - g_{\mu\mu'}g_{\nu\nu'} - g_{\mu\nu'}g_{\nu\mu'}\right)\delta(z, w)&
\end{align*}
The solution to this equation is obtained by decomposing $G^\Delta_{\mu\nu;\mu'\nu'}$ onto a basis of five irreducible $SO(d, 1)$ tensors $T^{(i)}_{\mu\nu;\mu'\nu'}$. The full expression is in Appendix. In the massless limit ($\Delta = d$), three linear combinations of $T^{(3)}, T^{(4)}$ and $T^{(5)}$ correspond to diffeomorphisms,

\noindent and therefore the only physical components are that of $T^{(1)}$ and $T^{(2)}$, given by \cite{D_Hoker_1999}
\small
\begin{flalign}
    &G_{\mu\nu;\mu'\nu'} = (\partial_\mu \partial_{\mu'}u\,\kern 0.1em\partial_\nu\partial_{\nu'}u + \partial_\mu\partial_{\nu'}u\,\kern 0.1em\partial_\nu\partial_{\mu'}u)\,G(u) + g_{\mu\nu}g_{\mu'\nu'}H(u) + \nabla\tensor[_{(\mu}]{S}{_{\nu);\mu'\nu'}} + \nabla\tensor[_{(\mu'}]{S}{_{\mu\nu);\nu'}}\nonumber&&\\[0.5ex]
    &G(u) = G_{bb}^d(u) \,\,(\ref{eq:scalarprop})\, \,\,\,\Big|\,\,\,H(u) = -\frac{1}{d-1}\left[2(1+u)^2G(u) + 2(d-2)(1+u)\int_\infty^u dv\,G(v)\right]\label{eq:gravitonpositionprop}&&
\end{flalign}
\normalsize
\subsection{Amplitudes for diagrams}
To summarize the progress thus far, the fields considered have been rewritten in terms of an integral kernel (propagator), which dictates their behavior in Euclidean $AdS$ as a result of sources either in the bulk or on the boundary. The obvious advantage of this representation is that, the process of evaluating $S_{sugra}$ and hence the dual correlators (\ref{eq:saddlepointadscft3}, \ref{eq:correlationholographic}), with the right boundary conditions for the fields, is made systematic in a manifestly diagrammatic way via propagators. It is due to the fact that for example
\begin{equation}\label{eq:identityscalar}
    \frac{\delta\phi_{\Delta}(z)}{\delta\phi_{0, \Delta}(\vec{y})} = \frac{\delta}{\delta \phi_{0, \Delta}(\vec{y})}\int_{\partial AdS}d^d\vec{x}\,\,G^{\Delta}_{b\partial}(z, \vec{x})\phi_{0, \Delta}(\vec{x}) = G^{\Delta}_{b\partial}(z, \vec{y})
\end{equation}
However so far, only the free field terms have been considered in $S_{sugra}$, but by adding the interaction terms the equations of motion become coupled and highly non-linear making it difficult to solve them exactly. Therefore interactions are treated perturbatively in their coupling constants. For simplicity, and also because they are the only ones we need later on, we will only consider diagrams with external legs (bulk-boundary propagators, figure \ref{fig:wd_1}) belonging to the scalar field. To start off, consider the n-point contact diagram (n = 2, 3, 4 figure \ref{fig:wd_1} (a), (b), (c)). These correspond to the interaction term $\lambda_n\prod_{i = 1}^n \phi_{\Delta_i}$ in $S_{sugra}$. Therefore perturbatively expanding $\phi(z)$ in orders of $\lambda_n$
\small
\begin{equation}\label{eq:scalarperturb}
    \phi_{\Delta}(z) = \vec{\phi}_{\Delta}(z) + \lambda_n \int d^{d+1}w \, G_{bb}^\Delta(z, w)\vec{\phi}_{\Delta}(w) + ... \,\,\,\Bigg|\,\,\,\vec{\phi}_{\Delta}(z) = \int d^{d}\vec{x}\,\,G_{b\partial}^\Delta(z, \vec{x})\phi_{0, \Delta}(\vec{x})
\end{equation}
\normalsize
where ... stands for higher orders in $\lambda_n$ obtained by recursive substitution of $\vec{\phi}(w)$. Also note that in the expansion, the source $J(w)$ seen in section \ref{sec:scalarprop} has been taken to be the field generated at $w$ due to the external source $\phi_0(\vec{x})$ i.e. $\vec{\phi}(w)$. Now by substituting (\ref{eq:scalarperturb}) into the interaction term $\int\lambda_n\prod_{i = 1}^n \phi_{\Delta_i}$ and using (\ref{eq:identityscalar}), gives rise to the tree-level amplitude at order $\lambda_n$ for the n-point contact diagram as follows
\begin{equation}\label{eq:contactamplitude}
    \mathcal{A}_{contact}^n(\{\vec{x}_i\}) = \lambda_n \int \frac{d^{d+1}z}{z_0^{d+1}}\prod_{i = 1}^n G_{b\partial}^{\Delta_i}(z, \vec{x}_i)
\end{equation}
where $1/z_0^{d+1}$ is $\sqrt{g}$ for the metric (\ref{eq:adsmetric}). Now consider the diagram in figure \ref{fig:wd_1} (d), it has two cubic vertices with a Spin-\textit{j} bulk-bulk propagator. First consider $j = 0$, in which case the vertices are governed by the interaction term $\frac{\lambda_{\Delta_i\Delta_r\Delta_k}}{s}\phi_{\Delta_i}\phi_{\Delta_r}\phi_{\Delta_k}$; where $s$ is the symmetry factor. Now by substituting the expansion (\ref{eq:scalarperturb}) for $n = 3$ into this interaction term, expanding to quadratic order in $\lambda$, and using (\ref{eq:identityscalar}) gives rise to the amplitude for the four-point scalar exchange diagram as follows
\begin{equation}\label{eq:scalarexchamplitude}
\begin{split}
    \mathcal{A}^{scalar}_{t-exch.}(\vec{x}_1, \vec{x}_2, \vec{x}_3, \vec{x}_4) &= \lambda_{\Delta_1\Delta_3\Delta} \int \frac{d^{d+1}z}{z_0^{d+1}}\,G_{b\partial}^{\Delta_1}(z, \vec{x}_1)G_{b\partial}^{\Delta_3}(z, \vec{x}_3)A_\Delta(z, \vec{x}_2, \vec{x}_4)\\
    \text{where}\,\,\,\,A_\Delta(z, \vec{x}_2, \vec{x}_4) &= \lambda_{\Delta_2 \Delta_4\Delta} \int \frac{d^{d+1}w}{w_0^{d+1}}\,G_{bb}^{\Delta}(z, w) G_{b\partial}^{\Delta_2}(w, \vec{x}_2)G_{b\partial}^{\Delta_4}(w, \vec{x}_4)
\end{split}
\end{equation}
Next up, consider a vector ($j = 1$) exchange. As seen in (\ref{eq:vectoraction}), the field $A_\mu$ couples to matter via $A_\mu J^\mu$. Let us work in the restricted space of massless vectors coupled to covariantly conserved currents ($\nabla_\mu J^\mu = 0$), which is the Noether current corresponding to $U(1)$ global symmetry in the case of a complex scalar field; $J^\mu = i(\phi \nabla^\mu \phi^{*} - \phi^{*}\nabla^\mu \phi)$. Let $\phi_{\Delta_i\Delta_r} = (\phi_{\Delta_i} + i\phi_{\Delta_r})/\sqrt{2}$, and let the coupling constant be $\lambda_{\Delta\Delta_i\Delta_r}^{vec}$, so expanding the vector field and the scalar field perturbatively in $\lambda$ using (\ref{eq:greensfunctionvector})
\small
\begin{align}\label{eq:perturbvector}
    A_\mu^\Delta(z) = \lambda_{\Delta\Delta_i\Delta_r}^{vec}\int d^{d+1}w \sqrt{g} G^{\Delta}_{\mu;\nu'}J^{\nu'}_{\Delta_i\Delta_r}(w)\,\,\,\Bigg|\,\,\, J^{\nu'}_{\Delta_i\Delta_r} = (\phi_{\Delta_i}\nabla^{\nu'}\phi_{\Delta_r} - \phi_{\Delta_r}\nabla^{\nu'}\phi_{\Delta_i})
\end{align}
\normalsize
 Higher orders in $\lambda$ are then obtained by recursively expanding in $\phi$, using the expansion (\ref{eq:scalarperturb}) with $\lambda_n$ now replaced by $\lambda_{\Delta\Delta_i\Delta_r}^{vec}$. By substituting (\ref{eq:perturbvector}) into the coupling term $\int \lambda^{vec}_{\Delta\Delta_k\Delta_l}A^{\Delta}_\mu J^\mu_{\Delta_k\Delta_l}$, taking $i, r, k, l = 1, 3, 2, 4$ (for figure \ref{fig:wd_1}), expanding $\phi$, and performing the variation w.r.t $\phi_{0, \Delta_1}, \phi_{0, \Delta_3}, \phi_{0, \Delta_2}, \phi_{0, \Delta_4}$; One obtains the amplitude for the vector exchange diagram at tree level and quadratic order in $\lambda$ as follows
 \small
\begin{align}
    &\mathcal{A}^{vector}_{t-exch.}(\vec{x}_1, \vec{x}_2, \vec{x}_3, \vec{x}_4) = \lambda^2_{vec} \int \frac{d^{d+1}z}{z_0^{d+1}}\,\int \frac{d^{d+1}w}{w_0^{d+1}}\,T^{\mu}_{\Delta_1\Delta_3}(z, \vec{x}_1, \vec{x}_3)G_{\mu;\nu'}^{\Delta}(z, w)T^{\nu'}_{\Delta_2\Delta_4}(w, \vec{x}_2, \vec{x}_4)\nonumber\\
    &\text{where}\,\,\,\,T^{\rho}_{\Delta_i\Delta_r}(z, \vec{x}_i, \vec{x}_r) = \left[G_{b\partial}^{\Delta_i}(z, \vec{x}_i)\nabla^{\rho}G_{b\partial}^{\Delta_r}(z, \vec{x}_r) - G_{b\partial}^{\Delta_r}(z, \vec{x}_r)\nabla^{\rho}G_{b\partial}^{\Delta_i}(z, \vec{x}_i)\right] \label{eq:vectorexchamplitude}
\end{align}
\normalsize
where $\lambda^2_{vec} = \lambda^{vec}_{\Delta\Delta_1\Delta_3}\lambda^{vec}_{\Delta\Delta_2\Delta_4}$. Now moving onto a tensor (j = 2) exchange, as seen in (\ref{eq:tensoraction}), the field $h_{\mu\nu}$ couples to matter via $h_{\mu\nu}J^{\mu\nu}$. Once again, let us work in the restricted space of massless tensors (gravitons, $\Delta = d$) coupled to covariantly conserved currents ($\nabla_\mu J^{\mu\nu} = 0$); which is the stress-energy tensor of a scalar field; $J^{\mu\nu} = \nabla^\mu\phi\nabla^\nu\phi - \frac{1}{2}g^{\mu\nu}(\nabla^{\rho}\phi\nabla_\rho\phi + m^2\phi^2)$, $m$ is the mass of the scalar field. Let the coupling constant be $\lambda^{grav}_{\Delta\Delta_i}$, so expanding the graviton field and the scalar field perturbatively in $\lambda$ using (\ref{eq:tensorpropagator})
\begin{equation}\label{eq:tensorperturb}
    \begin{gathered}
        h_{\mu\nu}^{\Delta}(z) = \lambda^{grav}_{\Delta\Delta_i}\int d^{d+1}w \sqrt{g}G^{\Delta}_{\mu\nu;\mu'\nu'}J^{\mu'\nu'}_{\Delta_i}(w) + ...\\
        J^{\mu'\nu'}_{\Delta_i} =  \nabla^{\mu'}\phi_{\Delta_i}\nabla^{\nu'}\phi_{\Delta_i} - \frac{1}{2}g^{\mu'\nu'}(\nabla^{\rho}\phi_{\Delta_i}\nabla_\rho\phi_{\Delta_i} + m^2\phi^2_{\Delta_i})
    \end{gathered}
\end{equation}
where $...$ denotes higher orders in $\lambda$. By substituting (\ref{eq:tensorperturb}) into the coupling term $\int \lambda^{grav}_{\Delta\Delta_r}h_{\mu\nu}^{\Delta}J^{\mu\nu}_{\Delta_r}$, taking $i, r = 1, 2$ (for $\Delta_1 = \Delta_3$ and $\Delta_2 = \Delta_4$ in figure \ref{fig:wd_1} (d)), expanding $\phi$, performing the variation w.r.t $\phi_{0, \Delta_1}(\vec{x}_1), \phi_{0, \Delta_1}(\vec{x}_3), \phi_{0, \Delta_2}(\vec{x}_2), \phi_{0, \Delta_2}(\vec{x}_4)$; One obtains the amplitude for the graviton exchange diagram at tree level and quadratic order in $\lambda$ 
\small
\begin{equation*}
    \mathcal{A}^{grav}_{t-exch.}(\vec{x}_1, \vec{x}_2, \vec{x}_3, \vec{x}_4) = \frac{\lambda^2_{grav}}{4} \int \frac{d^{d+1}z}{z_0^{d+1}}\,\int \frac{d^{d+1}w}{w_0^{d+1}}\,T^{\mu\nu}_{\Delta_1}(z, \vec{x}_1, \vec{x}_3)G_{\mu\nu;\mu'\nu'}^{\Delta}(z, w)T^{\mu'\nu'}_{\Delta_2}(w, \vec{x}_2, \vec{x}_4)
\end{equation*}
\vspace{-1em}
\begin{equation}\label{eq:gravexchange}
    \begin{split}
        \text{where}\,\,\,\,T^{\mu\nu}_{\Delta_1}(z, \vec{x}_1, \vec{x}_3) = \nabla^{\mu}G_{b\partial}^{\Delta_1}(z, \vec{x}_1)\nabla^{\nu}G^{\Delta_1}_{b\partial}(z, \vec{x}_3) - \frac{1}{2}g^{\mu\nu}&\Bigl[\nabla^{\rho}G^{\Delta_1}_{b\partial}(z, \vec{x}_1)\nabla_\rho G^{\Delta_1}_{b\partial}(z, \vec{x}_3)\\
        &+ m^2G^{\Delta_1}_{b\partial}(z, \vec{x}_1)G^{\Delta_1}_{b\partial}(z, \vec{x}_3)\Bigr]
    \end{split}
\end{equation}
\normalsize
\vspace{-0.1em}
$T^{\mu'\nu'}_{\Delta_2}(w, \vec{x}_2, \vec{x}_4)$ is then obtained by \,\,$\mu, \nu, z, \Delta_1, \vec{x}_1, \vec{x}_3 \,\kern 0.047em\,\rightarrow\,\kern 0.047em\,\mu', \nu', w, \Delta_2, \vec{x}_2, \vec{x}_4$ \,\,in the above.
\newenvironment{LastLineToRight}%
  {\setlength{\parindent}{0pt}\setlength{\leftskip}{0pt plus 1fil}\setlength{\rightskip}{0pt plus -1fil}}{\par}
\begin{LastLineToRight}
    Also $\lambda^2_{grav} = \lambda^{grav}_{\Delta\Delta_1}\lambda^{grav}_{\Delta\Delta_2}$, \,and $\frac{1}{4} (\frac{1}{2!2!})$ is the symmetry factor due to $\Delta_1 = \Delta_3$ \& $\Delta_2 = \Delta_4$.
\end{LastLineToRight}

\noindent Although the amplitudes mentioned have been for the cases of massive scalars, massless Spin-1 and Spin-2, the procedure must be clear for the reader to extend this analysis to other diagrams or for even higher-spin exchanges, by considering more and more couplings. However, the cases mentioned are the only ones we will need for the four-point computations in the next chapter. It should also be mentioned that all the coupling constants ($\lambda$) in this section are obtained by Kaluza-Klein reduction of eleven dimensional supergravity (for d = 3). Next, we will address correlators within the CFT formalism.

\section{Correlators in CFT and conformal blocks}\label{sec:conformalcft}
The subject of conformal field theory is really vast with extensive research and literature present on numerous topics across the subject. Therefore, this section will only be a very brief listing of some important facts about Euclidean CFTs that will be of relevance to us, and will mostly be following \cite{Francesco1999ConformalFT}. To begin, for the special case of $d = 2$, the conformal algebra is locally well-defined and infinite dimensional. However, the conformal group is finite dimensional and globally well-defined for $d > 2$, and is generated by translations ($\mathcal{P}$), dilations ($\mathcal{D}$), rotations ($\mathcal{M}$), and special conformal transformations ($\mathcal{K}$). The algebra was described in section \ref{sec:ospn4algebra} as a subalgebra of superconformal $osp(\mathcal{N} | 4)$. Since we are ultimately interested in $d = 3$, we will only confine the discussion to the $d > 2$ case. Observables such as correlation functions are then covariant under the global conformal group, which thereby places strong constraints on the terms that appear in them. Consider four spinless primary operators $\Phi_i(\vec{x}_i)$ with scaling dimensions $\Delta_i$ respectively for $i = 1,2,3,4$. They transform under conformal transformations as the following
\begin{equation}
    \vec{x} \rightarrow \vec{x}{\,'} \,\,\,\Big|\,\,\,\Phi(\vec{x}_i) \rightarrow \Phi_i\kern 0.025em'(\vec{x}_i\kern 0.025em') = \left|\frac{\partial \vec{x}_i\kern 0.025em'}{\partial \vec{x}_i}\right|^{-\frac{\Delta_i}{d}} \phi(\vec{x}_i)
\end{equation}
where $|.|$ is the Jacobian of the conformal transformation. Starting with two-point functions, covariance under dilations ($\vec{x} \rightarrow \lambda\vec{x}$), rotations and translations constrains it to be of the form (with $\ket{0}$ annihilated by conformal generators)
\begin{equation*}
    \langle\Phi_i(\vec{x}_i)\Phi_j(\vec{x}_j)\rangle = \langle\Phi_i\kern 0.025em'(\vec{x}_i\kern 0.025em')\Phi_j\kern 0.025em'(\vec{x}_j\kern 0.025em')\rangle\,\,\,\Big|\,\,\,\langle\Phi_i(\vec{x}_i) \Phi_j(\vec{x}_j)\rangle = \frac{C_{ij}}{|\vec{x}_i - \vec{x}_j|^{\Delta_i + \Delta_j}} 
\end{equation*}
where $C_{ij}$ is a constant coefficient. However, covariance under special conformal transformations (SCT) restricts it further by demanding $\Delta_i = \Delta_j$, as follows
\begin{equation}\label{eq:2pointconf}
    \text{SCT}\,\,:\,\,\left|\frac{\partial \vec{x}\,'}{\partial \vec{x}}\right| = \frac{1}{(1 - 2\vec{b}.\vec{x} + {|b|}^2{|x|}^2)^d}\,\,\,\Bigg|\,\,\,\langle\Phi_i(\vec{x}_i)\Phi_j(\vec{x}_j)\rangle = \begin{cases}
        \frac{C_{ij}}{|\vec{x}_i - \vec{x}_j|^{2\Delta_i}} & \Delta_i = \Delta_j\\
        \,\,\,\quad0 & \Delta_i \neq \Delta_j
    \end{cases}
\end{equation}
where $\vec{b}$ is the parameter vector for SCT. Similarly, conformal covariance constrains the three-point function upto a normalization constant. However, the higher n-point functions are only constrained upto arbitrary functions of cross-ratios (like $U, V$ in (\ref{eq:fourpointfunc})). Therefore the three-point function and the relevant four-point function are given by the following. 
\begin{align}
    &\langle\Phi_i(\vec{x}_i)\Phi_j(\vec{x}_j)\Phi_k(\vec{x}_k)\rangle = \frac{C_{ijk}}{x_{ij}^{\Delta_i + \Delta_j - \Delta_k}x_{jk}^{\Delta_j + \Delta_k - \Delta_i}x_{ik}^{\Delta_i + \Delta_k - \Delta_j}}\label{eq:3pointconf}\\[0.5ex]
    &\langle\Phi_i(\vec{x}_i)\Phi_j(\vec{x}_j)\Phi_k(\vec{x}_k)\Phi_l(\vec{x}_l)\rangle = \left(\frac{x_{jl}^2}{x_{il}^2}\right)^{\frac{\Delta_{ij}}{2}}\left(\frac{x_{il}^2}{x_{ik}^2}\right)^{\frac{\Delta_{kl}}{2}}\frac{f(U, V)}{(x_{ij}^2)^{\frac{(\Delta_i + \Delta_j)}{2}}(x_{kl}^2)^{\frac{(\Delta_k + \Delta_l)}{2}}}\label{eq:fourpointfunc}\\[0.5ex]
    &\text{where}\,\,\,\,x_{ij} = |\vec{x}_i - \vec{x}_j|\,\,\,;\,\,\,\Delta_{ij} = \Delta_i - \Delta_j\,\,\,;\,\,\,U = \frac{x^2_{ij}x^2_{kl}}{x^2_{ik}x^2_{jl}}\,,\,\,V = \frac{x_{il}^2x_{jk}^2}{x_{ik}^2x_{jl}^2}\label{eq:crossratios}
\end{align}
The constrains on the aforementioned correlation functions or more generally the n-point correlation function are more formally dressed as solutions to the Conformal Ward identities, which are most easily derived in the Wick-rotated path integral representation. Denoting the n-point correlation function as $\langle X \rangle$, where $X = \Phi_1(\vec{x}_1)...\Phi_n(\vec{x}_n)$
\begin{equation}\label{eq:correlationpathintegral}
    \langle X \rangle = \frac{1}{Z}\int [D\Phi_i] \,\,X e^{-S[\Phi_i]}
\end{equation}
Consider a global symmetry of the action under infinitesimal $\delta_{\omega}$, characterized by the Lie algebra whose basis is $\{G_a\}$; $\Phi'(\vec{x}) = \Phi(\vec{x}) - i\omega_a G_a \Phi(\vec{x})$. Now if the symmetry is made local with local parameters, $\omega_a \rightarrow \omega_a(\vec{x})$, the action is no longer invariant and the first order variation can be written using the previously conserved current of the global symmetry $j_a^\mu$. Using this variation, and by performing the change of variables $\Phi \rightarrow \Phi'$
\begin{equation}
\begin{gathered}
    \langle X \rangle = \frac{1}{Z}\int [D\Phi_i] \,\,(X + \delta_{\omega} X)e^{-\left(S[\Phi_i] + \int d^d\vec{x} \,\,\partial_\mu j^\mu_a\omega_a(\vec{x})\right)}\\
    \text{where}\,\,\,\, \delta_\omega X = -i\int d^{d}\vec{x}\,\,\omega_a(\vec{x})\sum_{i = 1}^n\left[\Phi_1(\vec{x}_1)...G_a\Phi_i(\vec{x}_i)...\Phi_n(\vec{x}_n)\right]\delta(\vec{x} - \vec{x}_i)    
\end{gathered}
\end{equation}
where covariance of the integration measure $[D\Phi_i] = [D\Phi_i']$, and covariance of the path integral under change of variables, have been used. This gives the Ward identity for $j^{\mu}_a$
\begin{equation}\label{eq:conformalwardidentity}
    \frac{\partial}{\partial x^{\mu}}\langle j^\mu_a(\vec{x})\Phi_1(\vec{x}_1)...\Phi_n(\vec{x}_n) \rangle = -i\sum_{i = 1}^n \delta(\vec{x} - \vec{x}_i)\langle \Phi_1(\vec{x}_1)...G_a\Phi_i(\vec{x}_i)...\Phi_n(\vec{x}_n) \rangle
\end{equation}
Conformal Ward identities can then easily be written down using (\ref{eq:conformalwardidentity}), by substituting the generators and corresponding currents for the four conformal transformations. The resulting set of partial differential equations yield solutions of the form (\ref{eq:2pointconf}, \ref{eq:3pointconf}, \ref{eq:fourpointfunc}) for $n = 2, 3, 4$ respectively. It may also be noted that the operators inside correlation functions, that are `time' ordered, are most often chosen to be radially ordered on a plane, since Euclideanized spacetime allows for the freedom of choice with regards to the time direction. Prior to radial quantization, the theory quantized in Poincaré time can be viewed as being placed on the surface of an infinitely long cyclinder, with the flat and compact directions identified with time and space respectively. The definition of a conformal map from the cylinder ($\mathbb{R} \times S^{d - 1}$) to the plane ($\mathbb{R}^d$) then maps translations in Poincaré time to radial evolution, with the Dilation operator now taking the role of a Hamiltonian in the radially quantized CFT. The operators with definite values of the scaling dimension $\Delta$ are then the states in this Hilbert space, with this state-operator correspondence manifest under the aforementioned conformal map, where the $-\infty$ of Poincaré time (vacuum) is mapped to the origin of the complex plane. This implies that every state at a radial time `$r$' is determined by the radial evolution of a state obtained by the action of it's corresponding local operator at the origin on the vacuum; $\ket{\Phi} = \Phi(0)\ket{\text{vac}}$. Now that the fundamental nature of operators in CFTs has been established, there is another important concept called the Operator Product Expansion (OPE), which crucially holds at the level of operators ($\mathcal{V}$). The OPE states that the product of two local operators inserted at nearby points can be well approximated by a sum of composite local operators and their descendants at one of the points, as follows
\begin{equation}\label{eq:operatorproductexp}
   \lim_{\vec{x}_i\to\vec{x}_j}\mathcal{V}_i(\vec{x}_i)\mathcal{V}_j(\vec{x}_j)  = \sum_{k}C^k_{ij}(\vec{x}_i - \vec{x}_j, \partial_{\vec{x}_j})\mathcal{V}_k(\vec{x}_j)
\end{equation}
where it is understood that both the LHS and RHS expressions lie inside radially ordered correlation functions $\langle\cdot\rangle$. For the case of spin-0 scalar primaries, to be more in tune with the language used in the case of Witten diagrams, it is suggestive to write the OPE as
\begin{equation}\label{eq:scalaropes}
    \Phi_i(\vec{x}_i)\Phi_j(\vec{x}_j) \sim \sum_{\mathcal{V}}\lambda_{ij\mathcal{V}}\, C^{s_1...s_m}(\vec{x}_i - \vec{x}_j, \partial_{\vec{x}_j})\mathcal{V}_{s_1...s_m}(\vec{x}_j)
\end{equation}
where $s_1...s_m$ are the spin indices of a Spin-\textit{m} primary operator $\mathcal{V}$, $\lambda_{ij\mathcal{V}}$ are the OPE coefficients, and $\sim$ implies the limit mentioned in (\ref{eq:operatorproductexp}). These OPEs can now be used to extract more information about the arbitrary functions of cross-ratios in n-point functions. In  the interest of later relevance, let us focus only on four-point functions. Writing the OPEs for $\Phi_i(\vec{x}_i)\Phi_j(\vec{x}_j)$ and $\Phi_{k}(\vec{x}_k)\Phi_l(\vec{x}_l)$ using (\ref{eq:scalaropes}), the four-point function is then
\begin{equation*}
\begin{gathered}
\langle\Phi_i(\vec{x}_i)\Phi_j(\vec{x}_j)\Phi_k(\vec{x}_k)\Phi_l(\vec{x}_l)\rangle = \sum_{\mathcal{V}^{(1)}, \mathcal{V}^{(2)}} \lambda_{ij\mathcal{V}^{(1)}}\lambda_{kl\mathcal{V}^{(2)}}\,\mathcal{CB}_{\mathcal{V}^{(1)}\mathcal{V}^{(2)}}(\vec{x}_i, \vec{x}_j, \vec{x}_k, \vec{x}_l)\\
\mathcal{CB}_{\mathcal{V}^{(1)}\mathcal{V}^{(2)}}(\vec{x}_i, \vec{x}_j, \vec{x}_k, \vec{x}_l) = C^{s_1...\,s_m}(\vec{x}_{ij}, \partial_{\vec{x}_j})\, C^{t_1...t_n}(\vec{x}_{kl}, \partial_{\vec{x}_l}) \langle\mathcal{V}^{(1)}_{s_1...s_m}(\vec{x}_j)\mathcal{V}^{(2)}_{t_1...t_n}(\vec{x}_l)\rangle
\end{gathered}
\end{equation*}
where $\vec{x}_{ij} = \vec{x}_i - \vec{x}_j$. Now by using the fact that the two-point function in $\mathcal{CB}$ vanishes for $\mathcal{V}^{(1)}, \mathcal{V}^{(2)}$ belonging to different conformal families (\ref{eq:2pointconf}), $(1) = (2)$ and $m = n$.
\begin{equation}\label{eq:confblockdecomfour}
    \langle\Phi_i(\vec{x}_i)\Phi_j(\vec{x}_j)\Phi_k(\vec{x}_k)\Phi_l(\vec{x}_l)\rangle = \sum_{\mathcal{V}}\lambda_{ij\mathcal{V}}\lambda_{kl\mathcal{V}}\,\mathcal{CB}_{\mathcal{V}}(\vec{x}_i, \vec{x}_j, \vec{x}_k, \vec{x}_l)
\end{equation}
This $\mathcal{CB}_{\mathcal{V}}$, which receives contributions from the two-point functions of operator $\mathcal{V}$ and it's descendants is called a \textit{conformal block}. Now since the OPEs respect conformal covariance, this implies that $\mathcal{CB}_{\mathcal{V}}$ transforms under the conformal group in the same way as the four-point function on the LHS does. Therefore comparing with (\ref{eq:fourpointfunc}) then implies
\begin{align}
    &\mathcal{CB}_{\mathcal{V}}(\vec{x}_i, \vec{x}_j, \vec{x}_k, \vec{x}_l) = \left(\frac{x_{jl}^2}{x_{il}^2}\right)^{\frac{\Delta_{ij}}{2}}\left(\frac{x_{il}^2}{x_{ik}^2}\right)^{\frac{\Delta_{kl}}{2}}\frac{f_{\mathcal{V}}(U, V)}{(x_{ij}^2)^{\frac{(\Delta_i + \Delta_j)}{2}}(x_{kl}^2)^{\frac{(\Delta_k + \Delta_l)}{2}}}\label{eq:cbinvariantform}\\
    &\implies f(U, V) = \sum_{\mathcal{V}} \lambda_{ij\mathcal{V}}\lambda_{kl\mathcal{V}}\,f_{\mathcal{V}}(U, V)\label{eq:conformalpartialwavedecomposition}
\end{align}
The function $f_{\mathcal{V}}(U, V)$ is called a \textit{conformal partial wave}. In general, these partial waves depend on the scaling dimension $\Delta$ and spin of the operator $\mathcal{V}$, and also on the scaling dimensions $\Delta_i$ of the operators in the correlator. There is one more thing to address, which is the definition of $\lambda_{ij\mathcal{V}}$, and thereby the corresponding normalization in the OPE (\ref{eq:scalaropes}). Rewriting the OPE by separating the descendants ($\partial^n\mathcal{V}$) from the leading term
\begin{equation}\label{eq:OPEplusdescendants}
    \Phi_i(\vec{x}_i)\Phi_j(\vec{x}_j) \sim \sum_{\mathcal{V}}\lambda_{ij\mathcal{V}}\, C^{s_1...s_m}(\vec{x}_i - \vec{x}_j)\mathcal{V}_{s_1...s_m}(\vec{x}_j) + \text{Descendants}
\end{equation}
Acting with the Spin-$l$ Operator $\mathcal{V}_{t_1...t_m}(\vec{x}_k)$ from the right on the above, $\lambda_{ij\mathcal{V}}$ enters the definition via the correlator of the resulting product of operators on the LHS, as follows. These are the conformally covariant correlator expressions similar to the ones mentioned before, except that there is a Spin-\textit{l} operator now involved
\begin{equation}\label{eq:twopointspinspin}
    \langle \mathcal{V}_{s_1...s_l}(\vec{x}_j)\mathcal{V}_{t_1...t_l}(\vec{x}_k) \rangle = \frac{\prod_{m = 1}^l I_{s_mt_m}(\vec{x}_{jk})}{x_{jk}^{2\Delta}}\,\,\,\Bigg|\,\,\,I_{st}(\vec{x}_{jk}) \equiv \delta_{st} - 2\frac{(\vec{x}_{jk})_s(\vec{x}_{jk})_t}{x_{jk}^2}
\end{equation}
\begin{equation}\label{eq:threepointscalarspin}
    \begin{split}
        \langle \Phi_i(\vec{x}_i)\Phi_j(\vec{x}_j)\mathcal{V}_{t_1...t_l}(\vec{x}_k) \rangle = &\frac{\lambda_{ij\mathcal{V}}\prod_{m = 1}^lY_{t_m}}{(x_{ij}^2)^{\frac{1}{2}(\Delta_i + \Delta_j - \Delta + l)}(x_{jk}^2)^{\frac{1}{2}(\Delta_j + \Delta - \Delta_i - l)}(x_{ik}^2)^{\frac{1}{2}(\Delta_i + \Delta - \Delta_j - l)}}\\
        &\text{where}\,\,\,\,Y_t \equiv \frac{(\vec{x}_{ik})_t}{x_{ik}^2} - \frac{(\vec{x}_{jk})_t}{x_{jk}^2}
    \end{split}
\end{equation}
where $(\vec{x}_{ij})_t$ denotes the component of the said vector with index $t$. Now by finally taking the near-point limit of (\ref{eq:twopointspinspin}) and (\ref{eq:threepointscalarspin}), and comparing with (\ref{eq:scalaropes}), one obtains the appropriate normalization for the OPE (\ref{eq:scalaropes}) as follows
\begin{equation}\label{eq:normalizedopescalar}
    \Phi_i(\vec{x}_i)\Phi_j(\vec{x}_j) \sim \sum_{\mathcal{V}}\frac{\lambda_{ij\mathcal{V}}}{(x_{ij}^2)^{\frac{1}{2}(\Delta_i + \Delta_j - \Delta_{\mathcal{V}} + l)}}\left(\prod_{m = 1}^{l}(\vec{x}_{ij})^{s_m}\right)\mathcal{V}_{s_1...s_l}(\vec{x}_j) + \text{Descendants}
\end{equation}
Also, it is easy to see that for the descendant term with $n^{\text{th}}$ order derivative ($\partial^n \mathcal{V}$), there is a similar factor $(x_{ij}^2)^{\frac{1}{2}(\Delta_i + \Delta_j - \Delta + l) - n}$. As a final remark, the corresponding holographic Witten diagrams mentioned earlier in the chapter are usually difficult to evaluate in position space, thereby calling for a more convenient space of variables to be represented in. This space will be described in the next chapter along with some computations therein.

%% file: chapters/conceptual_model.tex
\chapter{Mellin Space and Four-point Functions}\label{chap:conceptual_model}
In the previous chapter, the components pertaining to the precise statement of $AdS/CFT$ correspondence were studied i.e., Witten diagrams on the gravity side and correlators on the CFT side. The CFT correlators, by virtue of their conformal covariance, were highly constrained to specific forms. However in the case of Witten diagrams, although the propagators and amplitudes for relevant diagrams were explicitly written in position space, these highly non-trivial integrals are in general hard to compute even at the tree level. The \textit{n}-point diagrams for $n \leq 3$ are easy to compute since they involve contributions from contact diagrams alone, but they are uninteresting since on the field theory side the corresponding $n$-point correlators are fixed upto a constant via conformal covariance. However from $n = 4$, the contributions from exchange and other more complicated diagrams, start to make the traditional procedure of computing in position space extremely lengthy and quite intractable. Infact, very few cases have actually been computed explicitly since the inception of these Witten diagrams, see for e.g. \cite{D_Hoker_1999}, \cite{Uruchurtu_2007}. Some simplifications have been discovered for certain types of multiplets in specific number of dimensions, such as the \textit{without really trying} method \cite{D_Hoker_1999_wrt}, where exchange diagrams are reduced to a finite sum of contact diagrams, but the issue of generality still looms at large. However recently, conformal bootstrap techniques using OPEs and conformal blocks, have been the source of inspiration for conjecturing more efficient ways of computing these diagrams. One such method resorts to the use of the \textit{Mellin space} formalism, first introduced by Mack \cite{mack2009dindependent}, which encodes the CFT data in an analytically elegant way. This chapter will mostly be devoted to computing the Mellin amplitude for the four-point functions of scalar operators dual to Witten diagrams in $AdS_{d+1}$, but first things first, let us start with the description of the Mellin space.

\section{Mellin Space and Corresponding Amplitudes}\label{sec:mellinspaceamplitudes}
In the previous chapter, it was mentioned near (\ref{eq:fourpointfunc}) that $n$-point correlators for $n \geq 4$ are constrained only upto arbitrary functions of cross-ratios. It was then shown for $n = 4$ that these functions can be decomposed into conformal partial waves. Although not shown explicitly, this can be extended to higher $n$ by recursively decomposing the correlators into conformal blocks. Taking this as a given, this section will then introduce the Mellin space for general $n$, for the sake of completeness. The \textit{n}-point correlator is
\begin{equation}
    G_{\Delta_1...\Delta_n}(\vec{x}_1,...,\vec{x}_n) = \langle 
\Phi_1(\vec{x}_1)...\Phi_n(\vec{x}_n) \rangle = \prod_{i < j}^n(x_{ij}^2)^{-\delta^0_{ij}}f(\varepsilon_r)\,\,\,\Bigg|\,\,\,\varepsilon_r \equiv \frac{x_{ij}^2 x_{kl}^2}{x_{il}^2 x_{kj}^2}
\end{equation}
where $\delta^{0}_{ij}$ are the exponents of the rotation invariant quantities ($x_{ij}^2$) in the correlator ansatz, which then obey the condition $\sum_{j \neq i}\delta^0_{ij} = \Delta_i$ by requiring dilation covariance of the correlator. Since $\delta^0_{ij}$ are symmetric and non-zero only for $i \neq j$, there are $\frac{n(n-1)}{2}$ of them that satisfy the aforementioned condition which inturn makes $n$ of them redundant, thereby yielding $\frac{n(n - 3)}{2}$ such independent weights. The set of variables $\varepsilon_r$ represent the cross-ratios, whose counting goes as follows
\begin{equation}
    \text{card}(\{\epsilon_r\}) = \begin{cases}
        \frac{n(n - 3)}{2} & n < d + 1\\
        nd - \frac{(d+1)(d+2)}{2} & n \geq d+1
    \end{cases} 
\end{equation}
The counting for $n \geq d + 1$ is due to the fact that the configuration space of $n$ points is $nd$-dimensional, while the dimension of the Wick-rotated conformal group $SO(d+1, 1)$ is $\frac{(d+1)(d+2)}{2}$. However for $n < d + 1$, the counting is obtained by adding the dimension of a non-trivial stability group that arises in this case, see \cite{Rastelli_2018}. This bodes well with the fact that there were two cross-ratios $U, V$ for $n = 4$. Also in this notation, for $n = 4$, it can be seen from (\ref{eq:fourpointfunc}) that; $\delta^0_{12} = \frac{\Delta_1 + \Delta_2}{2}, \delta^0_{13} = \frac{\Delta_3 - \Delta_1}{2}, \delta^0_{14} = \frac{\Delta_1 - \Delta_2 + \Delta_4 - \Delta_3}{2}, \delta^0_{23} = 0, \delta^0_{24} = \frac{\Delta_2 - \Delta_1}{2}, \delta^0_{34} = \frac{\Delta_3 + \Delta_4}{2}$. The idea of Mack \cite{mack2009dindependent} was then to promote the fixed weights $\delta^0_{ij}$ to $\delta_{ij}$, which are now variables that span the Mellin space. The motivation for this can partially be seen from the discussion above (\ref{eq:operatorproductexp}), where it was mentioned that the dilation generator takes the role of the Hamiltonian in radial quantization. Therefore going into the space of `energy' ($\Delta$) associated with the `particle' ($\mathcal{V}$) is similar to transforming the flat space scattering amplitude expressions into momentum space for convenience. The \textit{reduced Mellin amplitude} $M(\delta_{ij})$ is then defined via the inverse Mellin transform as follows
\begin{equation}\label{eq:reducedmellinamplitude}
\begin{split}
    &G^{conn.}_{\Delta_1...\Delta_n}(\vec{x}_1,...,\vec{x}_n) = \mathcal{N}\mathlarger{\int}_{-i\infty}^{i\infty} [d\delta_{ij}] M(\delta_{ij}) \prod_{i < j}(x_{ij}^2)^{-\delta_{ij}}\\
    &[d\delta_{ij}] = \frac{1}{(2\pi i)^{n(n - 3)/2}}d^{\frac{n(n - 3)}{2}}\delta_{ij}\,\,\,\Big|\,\,\,\delta_{ij} = \delta_{ji}\,,\,\, \sum_j \delta_{ij} = \Delta_i
\end{split}
\end{equation}
where $\mathcal{N}$ is a normalization constant. $G_{\Delta_1...\Delta_n}^{conn.}(\vec{x}_1,...,\vec{x}_n)$ corresponds to the connected part of the correlator i.e., recursively decomposing it into conformal blocks and hence decomposing $f(\varepsilon_r)$ into conformal partial waves, and considering only the connected diagrams in the resulting sum. This is most easily seen for $n = 4$ for e.g. where in (\ref{eq:confblockdecomfour}), the decomposition symbolically represents an exchange diagram where the operator $\mathcal{V}$ is exchanged. The disconnected diagrams that will have to be removed, then correspond to the exchange of unit operators $\mathcal{V} = \mathbb{1}$, where in the $s$-channel for e.g. it is two disconnected pieces $\Phi_1\Phi_2 \rightarrow nothing$ and $nothing \rightarrow \Phi_3\Phi_4$; These unsurprisingly arise from the presence of $O(1)$ terms in the OPE (\ref{eq:scalaropes}). The Mellin amplitude is singular and ill-defined for the disconnected diagram, which is the reason why it is removed from the definition. It may also be noted that the integration in (\ref{eq:reducedmellinamplitude}) is performed parallel to the imaginary axis as shown, so that all the poles of the integrand lie on one side of the integration axis. This method of contour integration is then particularly advantageous because, the CFT data coming from the OPEs of operators in the correlator is neatly encoded in the form of poles of $M(\delta_{ij})$, such that the rest of the integrand matches the OPE behavior by the use of the residue theorem. For e.g. consider the limit $\vec{x}_i \rightarrow \vec{x}_j$ (where $j = i+1$) in the $n$-point correlator, and use the normalized OPE $\Phi_i(\vec{x}_i)\Phi_j(\vec{x}_j)$ from (\ref{eq:normalizedopescalar})
\begin{flalign}
    \lim_{\vec{x}_i\to\vec{x}_j} G^{conn.}_{\Delta_1...\Delta_i\Delta_j...\Delta_n}(\vec{x}_1,...,\vec{x}_i,\vec{x}_j,...,\vec{x}_n) = \sum_{\mathcal{V}}\frac{\lambda_{ij\mathcal{V}}G^{l-conn.}_{\Delta_1...\Delta_{\mathcal{V}}...\Delta_n}(\vec{x}_1,...,\vec{x}_{i-1},\vec{x}_j,...\vec{x}_n)}{(x_{ij}^2)^{\frac{1}{2}(\Delta_i + \Delta_j - \Delta_{\mathcal{V}} + l)}} + G_D\nonumber
\end{flalign}
\vspace{-1.5em}
\begin{align}
    G^{l-conn.}_{\Delta_1...\Delta_{\mathcal{V}}...\Delta_n}(\vec{x}_1,...,\vec{x}_{i-1},\vec{x}_j,...\vec{x}_n) = \left(\prod_{k = 1}^l(\vec{x}_{ij})^{s_k}\right)\langle \Phi_1(\vec{x}_1)...\mathcal{V}_{s_1...s_l}(\vec{x}_j)...\Phi_n(\vec{x}_n) \rangle
\end{align}
where $G_D$ denotes correlators similar to $G^{l-conn.}$ that arise from descendants of $\mathcal{V}$. Now for the RHS of (\ref{eq:reducedmellinamplitude}) to reproduce the same $x_{ij}^2$ behavior as $\vec{x}_i \rightarrow \vec{x}_j$, the term $\prod_{a < b}(x_{ab}^2)^{-\delta_{ab}}$ has to pick up a residue at $\delta_{ij} =(\Delta_i + \Delta_j - \Delta_{\mathcal{V}} + l)/2$ for the exchange of operator $\mathcal{V}$, hence implying a pole in the Mellin amplitude at that value of $\delta_{ij}$. Consequently for the exchange of descendant $\partial^m \mathcal{V}$, the value of the pole is shifted by $m$. To summarize
\begin{equation}\label{eq:satellitepoles}
    \Phi_i\Phi_j \rightarrow \partial^m \mathcal{V} \rightarrow \prod_{k \neq i,j} \Phi_{k} \,\,\,\Bigg|\,\,\,\text{Poles in $M(\delta_{ab})$}\,\,:\,\, \delta_{ij} = \frac{\Delta_i + \Delta_j - \tau_{\mathcal{V}} - 2m}{2}\,;\,m \in \mathbb{Z}_{\geq 0} 
\end{equation}
where $\Phi_i\Phi_j \rightarrow \partial^m \mathcal{V} \rightarrow \prod_{k \neq i,j} \Phi_{k}$ denotes a tree level diagram where operator $\partial^m \mathcal{V}$ is exchanged between the two interaction vertices ($\Phi_i\Phi_j$) and ($\prod_{k \neq i,j} \Phi_{k}$), and $\tau_{\mathcal{V}}$ is the twist $\Delta_{\mathcal{V}} - l$. These poles are analogous to the resonance that occurs in flat space scattering amplitudes, at the mass value of the virtual particle that is exchanged. Although only the exchange of single-trace operators has been mentioned, there is nothing restricting $M(\delta_{ab})$ from acquiring poles due to exchange of multi-trace operators. This is not an issue in general, but in the large $N$ limit of CFTs all multi-trace operators can be written as a product of single-trace operators, and hence the information of multi-trace poles in $M(\delta_{ab})$ is redundant. Consider the following amplitude which Mack defines in \cite{mack2009dindependent} 
\begin{equation}\label{eq:mellinamplitude}
    \mathcal{M}(\delta_{ij}) \equiv \frac{M(\delta_{ij})}{\prod_{i < j}\Gamma(\delta_{ij})}
\end{equation}
\vspace{-0.2em}
This $\mathcal{M}(\delta_{ij})$ is the \textit{Mellin amplitude}, and will be more relevant for our use, since the large $N$ limit is what will ultimately be of interest for holographic supergravity. The constraints on $\delta_{ij}$ in (\ref{eq:reducedmellinamplitude}) are usually solved by introducing auxiliary momenta $k_i$ and corresponding Mellin space kinetic variables $s_{ij} = -(k_i + k_j)^2$, as follows
\begin{equation}\label{eq:auxiliarymomenta}
    \delta_{ij} = k_i \cdot k_j = \frac{\Delta_{i} + \Delta_j - s_{ij}}{2} \,\,\,\Big|\,\,\,\sum_{i = 1}^n k_i = 0\,,\,\,-k_i^2 = \Delta_i
\end{equation}
To see that $\mathcal{M}$ encodes only the single-trace poles, notice that the gamma function $\Gamma(\delta_{ij})$ has poles at $\delta_{ij} = -d\,\,|\,\, d \in \mathbb{Z}_{\geq 0}$. Correspondingly $s_{ij} = \Delta_i + \Delta_j + 2d$, which corresponds to double-trace operators of the form $\partial^r \Phi_1 \partial^{d - r} \Phi_2$. Therefore the product of gamma functions in (\ref{eq:mellinamplitude}) encodes multi-trace poles thereby leaving the Mellin amplitude $\mathcal{M}(\delta_{ij})$ with poles corresponding to single-trace exchanges alone. Now that the Mellin space formalism has been introduced for general $n$-point correlators in CFTs, it is also important to mention that these Mellin amplitudes are defined the same way for the position-space holographic amplitudes ($\mathcal{A}$) coming from Witten diagrams. Infact, as will be mentioned later, there is a conjectured relation between the Bulk S-matrix and the asymptotic Mellin amplitudes as given by Joao Penedones in \cite{Penedones_2011}. Moving on, let us now specialize again to the case relevant to us i.e. $n = 4$ and write down the Mellin transform explicitly. Introduce the kinetic variables using (\ref{eq:auxiliarymomenta}); $s \equiv s_{12} = s_{34} = -(k_1 + k_2)^2$, $t \equiv s_{13} = s_{24} = -(k_1 + k_3)^2$ and $u \equiv s_{14} = s_{23} = -(k_2 + k_3)^2$. The Mellin variables ($\delta_{ij}$) in terms of these $s_{ij}$ are
\begin{align}
    &\delta_{12} = \frac{\Delta_1 + \Delta_2 - s}{2}\,\,\,;\,\,\,\delta_{34} = \frac{\Delta_3 + \Delta_4 - s}{2}\label{eq:s-variables}\\
    &\delta_{13} = \frac{\Delta_1 + \Delta_3 - t}{2}\,\,\,\,;\,\,\,\delta_{24} = \frac{\Delta_2 + \Delta_4 - t}{2}\label{eq:t-variables}\\
    &\delta_{14} = \frac{\Delta_1 + \Delta_4 - u}{2}\,\,\,;\,\,\,\delta_{23} = \frac{\Delta_2 + \Delta_3 - u}{2}\label{eq:u-variables}
\end{align}
\vspace{-1em}
\begin{equation}
    \text{where}\,\,\,\,\sum_{j}\delta_{ij} = \Delta_i \implies s + t + u = \Delta_1 + \Delta_2 + \Delta_3 + \Delta_4
\end{equation}
With these definitions, the terms in the integrand of the Mellin transform are as follows
\begin{equation}\label{eq:mellinint1}
    \prod_{i < j}(x_{ij}^2)^{-\delta_{ij}} = \left(\frac{x_{24}^2}{x_{14}^2}\right)^{\frac{\Delta_{12}}{2}}\left(\frac{x_{14}^2}{x_{13}^2}\right)^{\frac{\Delta_{34}}{2}}\frac{U^{\frac{s}{2}}V^{\frac{u - \Delta_2 - \Delta_3}{2}}}{(x_{12}^2)^{\frac{\Delta_1 + \Delta_2}{2}}(x_{34}^2)^{\frac{\Delta_3 + \Delta_4}{2}}}
\end{equation}
where $U, V$ are the cross-ratios as defined in (\ref{eq:crossratios}). Now by substituting (\ref{eq:mellinamplitude}) and (\ref{eq:mellinint1}) into (\ref{eq:reducedmellinamplitude}), and comparing with (\ref{eq:fourpointfunc}; $i,j,k,l = 1,2,3,4$), the four-point Mellin transform 
\small
\begin{align}\label{eq:fourpointmellin}
    f^{conn.}(U, &V) = \frac{\mathcal{N}}{4\cdot (2\pi i)^2} \int_{-i\infty}^{i\infty} ds \,du\,\,U^{\frac{s}{2}}V^{\frac{u - \Delta_2 - \Delta_3}{2}} \mathcal{M}(s, u)\Gamma\left(\frac{\Delta_1 + \Delta_2 - s}{2}\right)\Gamma\left(\frac{\Delta_3 + \Delta_4 - s}{2}\right)\nonumber\\
    &\times \Gamma\left(\frac{\Delta_1 + \Delta_3 - t}{2}\right)\Gamma\left(\frac{\Delta_2 + \Delta_4 - t}{2}\right)\Gamma\left(\frac{\Delta_1 + \Delta_4 - u}{2}\right)\Gamma\left(\frac{\Delta_2 + \Delta_3 - u}{2}\right)
\end{align}
\normalsize
Now let us compute Mellin amplitudes for the Witten diagrams from the previous chapter.

\section{Mellin amplitudes for Witten diagrams}
Let us first start with the simplest diagram of them all i.e. the \textit{n}-point contact diagram, following the computation in \cite{Penedones_2011}. The amplitude is given by (\ref{eq:contactamplitude}) and the bulk-boundary propagator is given by (\ref{eq:scalarprop}). Define $h = d/2$ from now on for convenience
\begin{equation}
    \mathcal{A}^n_{contact}(\{\vec{x}_i\}) = \lambda_n \int \frac{d^{d+1}z}{z_0^{d+1}}\prod_{i = 1}^n G^{\Delta_i}_{b\partial}(z, \vec{x}_i)\,\,\,\Big|\,\,\,G^{\Delta}_{b\partial}(z, \vec{x}) = C_{\Delta}\left(\frac{z_0}{(z - \vec{x})^2}\right)^{\Delta}
\end{equation}
The constant $C_{\Delta} = \frac{\Gamma(\Delta)}{\pi^{h}\Gamma(\Delta - h)}$ is the normalization defined in association with the two point function (\ref{eq:2pointconf}) $\langle \Phi_i(\vec{x}_i)\Phi_j(\vec{x}_j) \rangle = C_{\Delta}/x_{ij}^{2\Delta}$. The bulk-boundary propagator can be rewritten using the integral representation of the gamma function as follows
\begin{equation}\label{eq:gammafunction}
\begin{gathered}
    \Gamma(\Delta) = \int_0^\infty dt\,t^{\Delta - 1}e^{-t} \implies G^{\Delta}_{b\partial}(z, \vec{x}) = \frac{C_{\Delta}}{\Gamma(\Delta)}\int_0^\infty \frac{dt}{t}\,t^{\Delta}e^{-t\frac{(z - \vec{x})^2}{z_0}}\\
    \implies \mathcal{A}^n_{contact}(\{\vec{x}_i\}) = \lambda_n\left(\prod_{i = 1}^nC_{\Delta_i}\right)D_{\Delta_1...\Delta_n}(\{\vec{x}_i\})
\end{gathered}
\end{equation}
As shown above, the amplitude in this new representation has a D-function $D_{\Delta_1...\Delta_n}(\{\vec{x}_i\})$, which is quite famous in the literature on traditional computations of Witten diagrams
\begin{equation}\label{eq:dfunction}
    D_{\Delta_1...\Delta_n}(\{\vec{x}_i\}) = \left(\prod_{i = 1}^n\frac{1}{\Gamma(\Delta_i)}\right)\int_0^\infty \frac{dt_1}{t_1}\,t_1^{\Delta_1}...\int_0^\infty \frac{dt_n}{t_n}\,t_n^{\Delta_n} \int \frac{d^{d+1}z}{z_0^{d+1}}\,e^{\,-\sum_{j = 1}^n t_j \frac{(z - \vec{x}_j)^2}{z_0}}
\end{equation}
Unpack the notation $(z - \vec{x}_j)^2 = z_0^2 + (\vec{z} - \vec{x}_j)^2$, the integral over $\vec{z}$ coordinates is a Gaussian 
\begin{align}
        &\int \frac{d^{d+1}z}{z_0^{d+1}} \,e^{\,-\sum_{j = 1}^n t_j \frac{(z - \vec{x}_j)^2}{z_0}} = \int \frac{dz_0}{z_0^{d+1}} e^{-\left(\sum_{k = 1}^n t_k\right)z_0}\int d^{d}\vec{z}\,\,e^{-\sum_{j = 1}^n \frac{t_j(\vec{z} - \vec{x}_j)^2}{z_0}}\label{eq:integraloverz}\\
        &\text{Gaussian}\,\,:\,\,\int d^{d}\vec{z}\,\,e^{-\sum_{j = 1}^n \frac{t_j(\vec{z} - \vec{x}_j)^2}{z_0}} = \frac{\pi^h z_0^h}{\left(\sum_{k = 1}^n t_k\right)^h}\,\text{exp}\left(\frac{\sum_{i,j = 1}^{n} t_it_j (\vec{x}_i - \vec{x}_j)^2}{2z_0 \sum_{k = 1}^n t_k}\right)\label{eq:gaussianintegral}
\end{align}
Use (\ref{eq:gaussianintegral}) in (\ref{eq:integraloverz}), and substitute it into (\ref{eq:dfunction}). Now first perform the change of variable $(z_0 \,\sum_{k = 1}^n t_k) \rightarrow z_0$, and then perform $t_i \rightarrow \sqrt{z_0}t_i$, it results in the following 
\begin{align}
    &D_{\Delta_1...\Delta_n}(\{\vec{x}_i\}) = \pi^h\Gamma\left(\frac{\sum_{k = 1}^n \Delta_k - d}{2}\right)\left(\prod_{i = 1}^n \frac{1}{\Gamma(\Delta_i)}\right)
    \mathcal{D}_{\Delta_1...\Delta_n}(\{\vec{x}_i\})\label{eq:dfunction2}\\
    &\text{where}\,\,\,\mathcal{D}_{\Delta_1...\Delta_n}(\{\vec{x}_i\}) = \int_0^\infty \frac{dt_1}{t_1}\,t_1^{\Delta_1}...\int_0^\infty \frac{dt_n}{t_n}\,t_n^{\Delta_n}e^{\,-\sum_{i < j}t_it_j(\vec{x}_i - \vec{x}_j)^2}
\end{align}
Note that $\Gamma(\frac{\sum_{k = 1}^n \Delta_k - d}{2})$ in (\ref{eq:dfunction2}) comes from the $z_0$ integral, which takes the integral representation of the gamma function as in (\ref{eq:gammafunction}). The integral $\mathcal{D}_{\Delta_1...\Delta_n}(\{\vec{x}_i\})$ can be rewritten in a more familiar form using the identity derived in \cite{Symanzik:1972wj}, which reads
\begin{equation}\label{eq:symanzikstarformula}
    \int_0^\infty \frac{dt_1}{t_1}\,t_1^{\Delta_1}...\int_0^\infty \frac{dt_n}{t_n}\,t_n^{\Delta_n}e^{\,-\sum_{i < j}t_it_jx_{ij}^2} = \frac{2^{-1}}{(2\pi i)^{n(n - 3)/2}}\int_{-i\infty}^{i\infty}d\delta_{ij} \prod_{i < j}^{n} \Gamma(\delta_{ij}) (x_{ij}^2)^{-\delta_{ij}}
\end{equation}
Therefore finally, using this identity in (\ref{eq:dfunction2}), and comparing (\ref{eq:gammafunction}) to the inverse Mellin transform (\ref{eq:reducedmellinamplitude}) while noting the Mellin amplitude as (\ref{eq:mellinamplitude}), we get
\begin{equation}\label{eq:mellincontactamplitude}
    \boxed{\mathcal{M}_{contact}^n(\delta_{ij}) = \lambda_n} \,\,\,\,\,\Bigg|\,\,\,\,\,\mathcal{N} = \frac{\pi^h}{2}\Gamma\left(\frac{\sum_{k = 1}^n \Delta_i - d}{2}\right)\prod_{i = 1}^n \frac{C_{\Delta_i}}{\Gamma(\Delta_i)}
\end{equation}
This result of constant Mellin amplitude for the contact vertex $\lambda_n \prod_{i = 1}^n \phi_i$, is reassuring of the analogy between Mellin representation for $AdS$ and momentum representation for flat space. Now moving onto the four-point scalar exchange ($\Delta$) diagram, the position space amplitude had been reduced to a sum of $D$-functions for the special case of $(\Delta_1 + \Delta_2 - \Delta)/2 \in \mathbb{N}$, in \cite{D_Hoker_1999}. Correspondingly, the Mellin amplitude was computed in \cite{Penedones_2011} via transforms of $D$-functions as seen before. However, this special case doesn't generalize, so to make the reader more equipped, the scalar exchange diagram will be computed using a different method of diagrammatic rules developed in (\cite{Fitzpatrick_2011}, \cite{Paulos_2011}, \cite{Fitzpatrick_2012}).
\vspace{-0.2em}
\subsection{Scalar Exchange using Diagrammatic rules}
Consider the scalar exchange diagram seen in the previous chapter. The exchanged scalar, characterized by a bulk-bulk propagator, was shown in section \ref{sec:scalarprop} to satisfy 
\begin{equation}\label{eq:bulkbulkeom}
    (\square_{AdS} - m^2) G_{bb}^{\Delta}(z, w) = \frac{\delta^{d+1}(z - w)}{\sqrt{g}}
\end{equation}
The key thing to note here is that the isometry group of the bulk Euclidean $AdS_{d+1}$ is $SO(d+1, 1)$. Therefore the Casimir constructed out of the generators of this group is mapped to the Laplacian on $AdS_{d+1}$ ($\square_{AdS}$). Now since $SO(d+1, 1)$ is also the conformal group of the CFT on the boundary, the action of the Casimir on the CFT correlator translates to the action of the Laplacian on the bulk-bulk propagator. Let the generators of $SO(d+1, 1)$ be $J_i$, then the Casimir $\frac{(\sum_{i} J_i)^2}{2}$ acts on the correlator $A$ as follows
\begin{equation}\label{eq:finitedifference}
    \left[\frac{1}{2}\left(\sum_i J_i\right)^2 - \Delta(d - \Delta)\right]A = A_0 \,\,\,\,\,\Bigg|\,\,\,\,\, \frac{1}{2}\left(\sum_i J_i\right)^2 \rightarrow -\square_{AdS}
\end{equation}
where $A_0$ is the correlator dual to the contact Witten diagram. This equation holds because action of the Laplacian collapses the bulk-bulk propagator to a point as seen in (\ref{eq:bulkbulkeom}), thereby converting the exchange diagram on the LHS ($A$) to the contact diagram on the RHS ($A_0$). This is the called the finite-difference equation, which when converted to Mellin space ($\mathcal{M}, \mathcal{M}_0$), takes an algebraic form as shown in \cite{Fitzpatrick_2011}, and can then be solved since $\mathcal{M}_0$ is known from (\ref{eq:mellincontactamplitude}). However to maintain generality with regards to potential use cases beyond the four point exchange diagram, the diagrammatic rules mentioned in the same paper and streamlined in \cite{Fitzpatrick_2012}, will now be described. Such diagrammatic rules are possible because (\ref{eq:finitedifference}) can be recursively generalized to arbitrary no. of three-point vertices and bulk-bulk propagators. Consider the following tree-level Witten diagram
\begin{figure}[h]
\centering
\includegraphics[width=\textwidth]{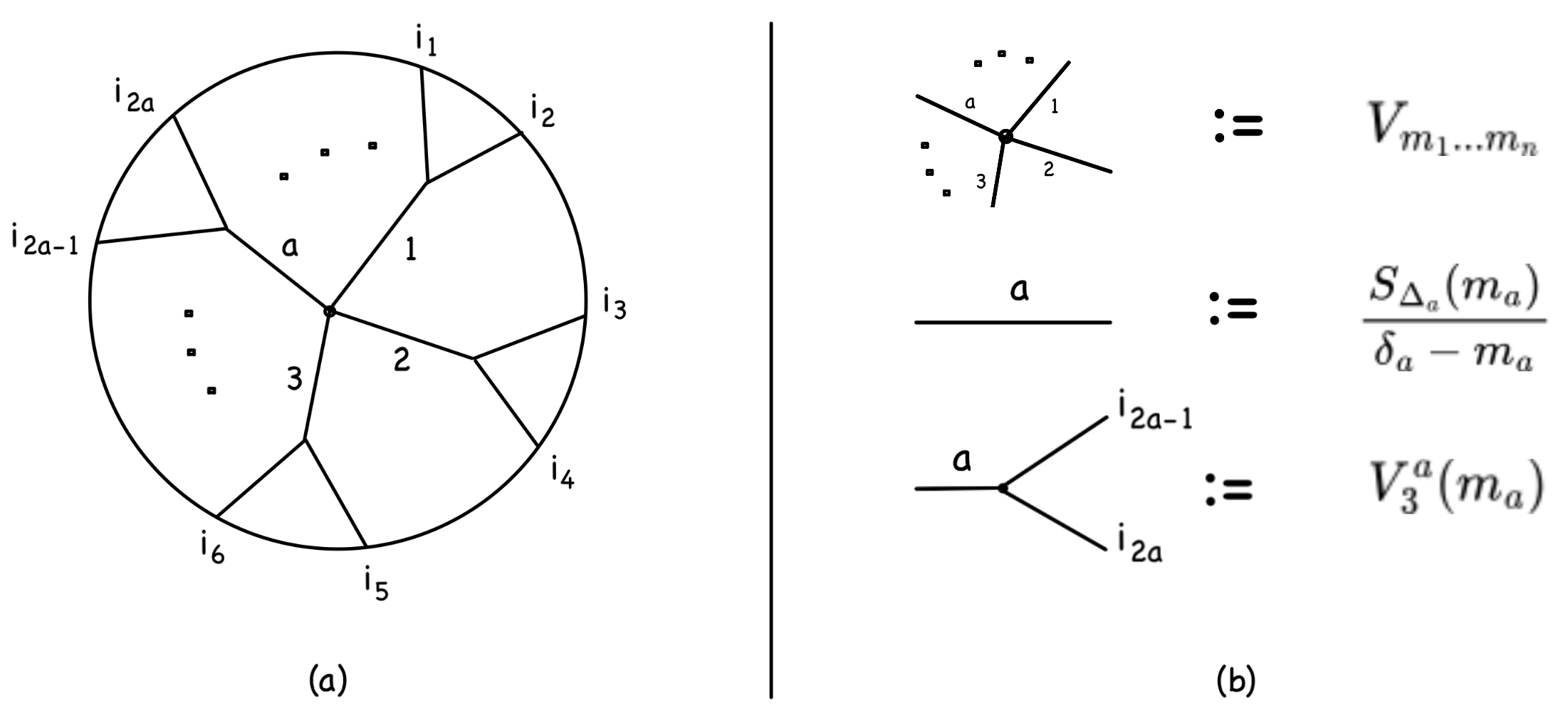}
\vspace{-1.75em}
\caption{(a) General tree-level scalar Witten diagram that connects $n$ 3-point vertices to an off-shell $n$-point vertex. (b) Diagrammatic rules for computing Mellin amplitude.}
\label{fig:wd_2}
\end{figure}
The external operators corresponding to an $n$-point correlator are inserted on the boundary at points $i_1,...,i_n$, that have scaling dimensions $\Delta_{i_1},...,\Delta_{i_n}$ respectively. The cubic interaction vertices are connected to the off-shell $n$-point vertex in the center via bulk-bulk propagators labelled by $1,...,a,...,n$, mediated by scalars of scaling dimensions $\Delta_1,...,\Delta_a,...\Delta_n$ respectively. Also, each bulk-bulk propagator is assigned a non-negative integer $m$ ($m_a$ for $a$), which classifies the infinite number of poles coming from the descendants of the scalar operator in the corresponding propagator (\ref{eq:satellitepoles}). Introduce a bit of notation, $2\delta_{a} = \Delta_{i_{2a - 1}} + \Delta_{i_{2a}} - \Delta_{a} - 2\delta_{2a-1, 2a} = 2\Delta_{i_{2a - 1}i_{2a},a} - 2\delta_{2a-1, 2a}$. Coming to the diagrammatic rules, the off-shell $n$-point vertex contributes a factor $V_{m_{1},...,m_n}$ which is
\small
\begin{align}
    &V_{m_1,...,m_n} = \lambda_n\left(\prod_{i = 1}^n \frac{(1 - h + \Delta_i)_{m_i}}{m_i!}\right) \cdot F_{A}^{(n)}\left(\Delta_{\Sigma} - h, \begin{Bmatrix}
        -m_1,...,-m_n \\ 1 + \Delta_1 - h,...,1 + \Delta_n - h
    \end{Bmatrix}; 1,...,1\right)\nonumber\\
    &\text{Lauricella}\,\,\,:\,\,\,F_A^{(n)}\left(g, \begin{Bmatrix}
        a_1,...a_n\\b_1,...b_n
    \end{Bmatrix}; x_1,...,x_n\right) = \sum_{k_1,...,k_n = 0}^\infty \left((g)_{k_1+...+k_n}\prod_{i = 1}^n\frac{(a_i)_{k_i}}{(b_i)_{k_i}}\frac{x_i^{k_i}}{k_i !}\right)
\end{align}
\normalsize
where $\Delta_{\Sigma} = \sum_{j = 1}^n \Delta_j /2$, $F_A^{(n)}$ is the Lauricella hypergeometric series, and $\lambda_n$ is the $n$-point coupling as seen before. Also note the notation of the Pochhammer symbol $(q)_i = \frac{\Gamma(q+i)}{\Gamma(q)}$. As mentioned in figure \ref{fig:wd_2} (b), each bulk-bulk propagator labelled $a$ then contributes a factor $\frac{S_{\Delta_a}(m_a)}{\delta_a - m_a}$, and each three-point vertex created by scalars $\Delta_{i_{2a - 1}}, \Delta_{i_{2a}}, \Delta_a$ contributes a factor $V_3^a(m_a)$. Given the three-point coupling $\lambda_{\Delta_{a}\Delta_{i_{2a-1}}\Delta_{i_{2a}}}$, the functions are as follows
\begin{align}\label{eq:factorspropvert}
    S_{\Delta_a}(m_a) = \frac{-2\pi^h \Gamma^2(\Delta_a - h + 1)m_a!}{\Gamma(\Delta_a + m_a - h + 1)} \,\,\,\,\,\Bigg|\,\,\,\,\,V_3^a(m_a) = \frac{\lambda_{\Delta_{a}\Delta_{i_{2a-1}}\Delta_{i_{2a}}}}{m_a!(\Delta_{{i_{2a - 1}i_{2a},a}})_{-m_a}}
\end{align}
By factoring in all these contributions and summing over the descendant poles $m_a$, one then obtains the Mellin amplitude for the tree-level scalar diagram in figure \ref{fig:wd_2} (a) as 
\begin{equation}\label{eq:mellinamplitudegeneraltree}
    \mathcal{M} = \sum_{m_1,...,m_n = 0}^{\infty}\kern -0.5em V_{m_1,...m_n} \prod_{a = 1}^n\frac{S_{\Delta_a}(m_a)V_3^a(m_a)}{\delta_a - m_a}
\end{equation}
Using these rules, the Mellin amplitude for the four-point scalar exchange is then very easy to compute. However, there is an ambiguity with regards to the ordering of the four scalar operators on the boundary, as to which three-point vertices are formed as a junction of which three scalar operators. This is where the advantage of the Mellin representation is in full display i.e. consider (\ref{eq:s-variables} to \ref{eq:u-variables}), these make it clear that there are three independent exchanges in the $s, t, u$ channels, as is evident in the multi-trace poles of the gamma functions in (\ref{eq:mellinint1}). The diagrams ($l = 0$) are as follows
\begin{figure}[h]
\centering
\includegraphics[width=\textwidth]{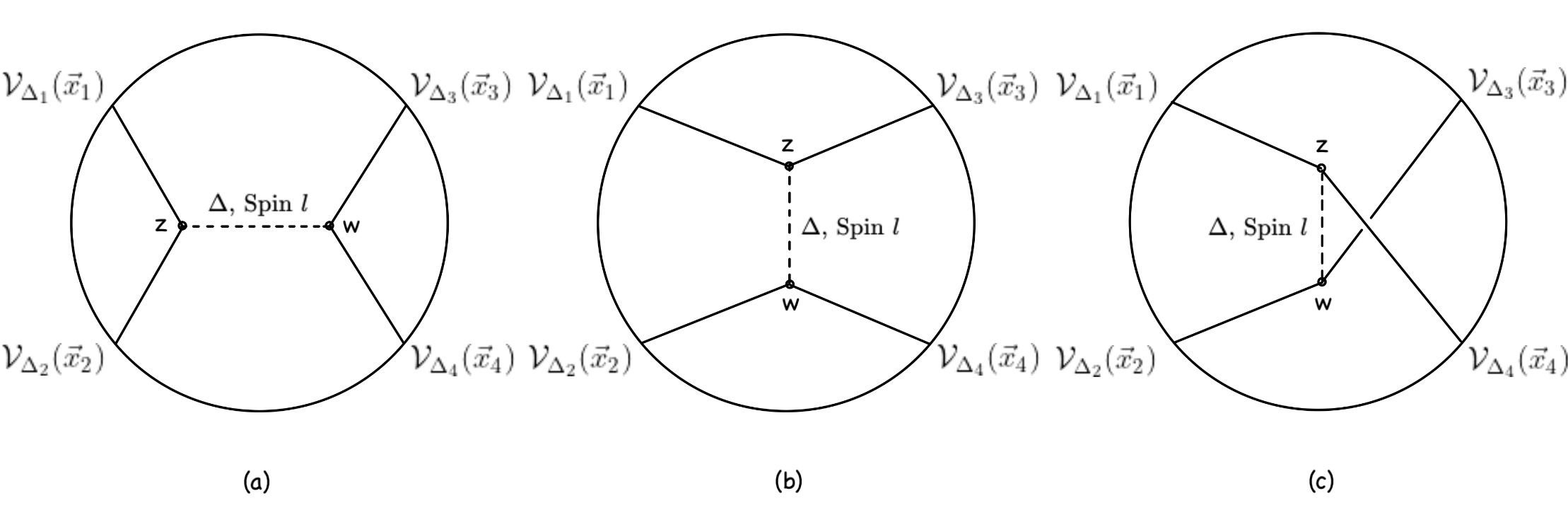}
\vspace{-1.5em}
\caption{Four-point spin-\textit{l} ($\Delta$) exchange in \textbf{(a)} $s$-channel, \textbf{(b)} $t$-channel, \textbf{(c)} $u$-channel}
\label{fig:wd_3}
\end{figure}

\noindent Let us compute the Mellin amplitude of $s$-channel exchange, for which the variable of interest is, $2\delta = \Delta_1 + \Delta_2 - \Delta - 2\delta_{12} = \Delta_3 + \Delta_4 - \Delta - 2\delta_{34} = s - \Delta$ (from \ref{eq:s-variables}). There is no off-shell vertex, the contributions are from the lone bulk-bulk propagator and the two 3-point vertices, which via the rules (\ref{eq:factorspropvert}) leads to the following Mellin amplitude
\begin{align}
    \mathcal{M}^{scalar}_{s-exch.} &= \sum_{m = 0}^\infty V_3(m; \Delta_1, \Delta_2) V_3(m; \Delta_3, \Delta_4) \frac{2S_{\Delta}(m)}{s - (2m + \Delta)}\label{eq:mellinamplitudescalarexchange}\\
    &= \sum_{m = 0}^\infty \frac{-4\pi^h\Gamma(\Delta - h + 1)\lambda_{\Delta\Delta_1\Delta_2}\lambda_{\Delta\Delta_3\Delta_4}}{m!(\Delta - h + 1)_m(\Delta_{12,b})_{-m}(\Delta_{34,b})_{-m}}\cdot\frac{1}{s - (2m + \Delta)}\nonumber
\end{align}
where $\Delta_{ij,b} = \Delta_i + \Delta_j - \Delta$. For $t$-channel : $s\rightarrow t\,; \,2 \leftrightarrow 3$. For $u$-channel : $s \rightarrow u\,; \,2\leftrightarrow 4$.

\subsection{Spin - \textit{l}\,\, Exchange using Conformal blocks}\label{sec:spinlexchange}
\begin{figure}[h]
\centering
\includegraphics[width=\textwidth]{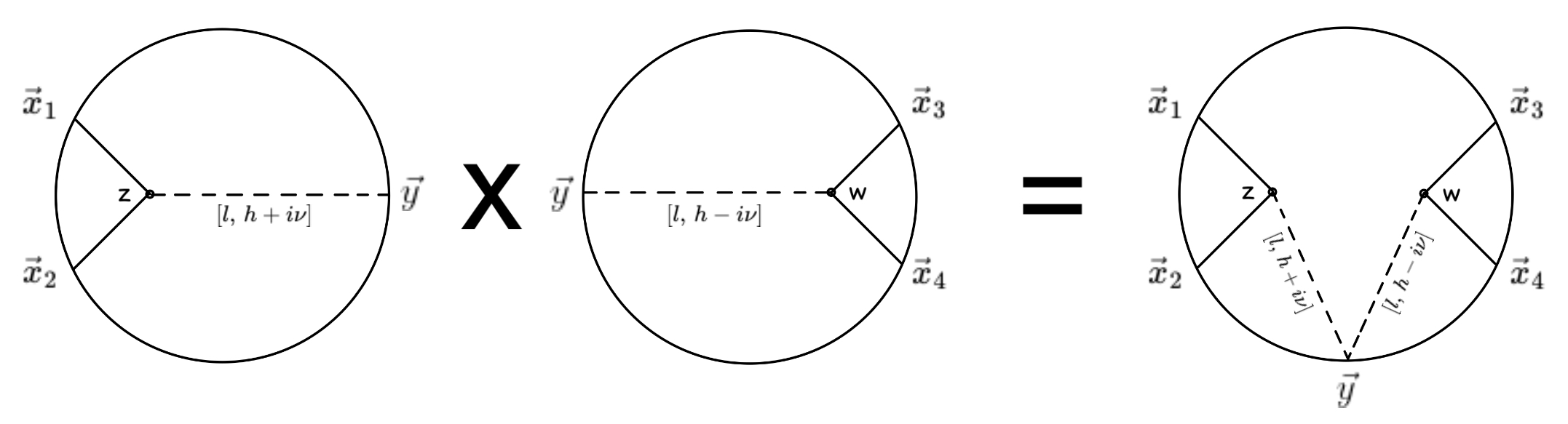}
\vspace{-1.5em}
\caption{Mellin amplitude dual to the block $\mathcal{CB}_{\nu, \,l}$, upon normalization (\ref{eq:3pointnormspectral}).}
\label{fig:wd_4}
\end{figure}

The Mellin space computations thus far, have been done purely on the Gravity side of the duality without requiring any result from the CFT i.e., D-functions for contact diagram and algebraic Mellin representation of E.O.M for scalar tree-level diagram. However for diagrams involving higher spin fields, it goes without mention that computations via these methods get significantly more complicated. Some special cases have been computed for e.g., Graviton exchange between minimally coupled massless scalars ($\Delta_i = d$) reduced to a sum of D-functions in \cite{D_Hoker_1999_2} and corresponding Mellin amplitude computed in \cite{Penedones_2011}. However, no such method exists yet for Graviton exchange in arbitrary dimension $d$ and arbitrary $\Delta_i$, let alone an algorithmic procedure for a general Spin-$l$ tree-level diagram. This therefore calls for a point of break-off from `testing' $AdS/CFT$ duality and move towards utilizing it by trusting it to hold, which is reasonable to expect at this point (atleast at the supergravity level). The answer to this call comes in the form of conformal blocks (see section \ref{sec:conformalcft}), which were discussed for position space amplitudes. Mack showed in \cite{mack2009dindependent} that a similar decomposition holds for Mellin amplitudes too as follows
\begin{equation}\label{eq:mellinpartialwave}
    \mathcal{M}(\delta_{ij}) = \sum_{l = 0}^{\infty} \int_{-\infty}^{\infty} d\nu \,b_{l}(\nu^2)\mathcal{M}_{\nu,\,l}(\delta_{ij})
\end{equation}
where $\nu$ is the parameter of the principal series $\chi = [l, h + i\nu]$ and it's dual $\bar{\chi} = [\bar{l}, h - \nu]$, representing a pair of operators with spins $l, \bar{l}$ and scaling dimensions $h + i\nu, h - i\nu$ respectively; $l = \bar{l}$ if external operators are scalar. This statement is less abstract in position space where for e.g., for four-point functions, substituting (\ref{eq:mellinpartialwave}) in (\ref{eq:fourpointmellin}) yields
\begin{equation}\label{eq:spectralpartialwave}
f^{conn.}(U, V) = \sum_{l = 0}^{\infty} \int_{-\infty}^{\infty} d\nu \, b_{l}(\nu^2)f^{conn.}_{\nu, \,l}(U, V)
\end{equation}
where $f^{conn.}_{\nu, \,l}(U, V)$ is obtained by replacing $\mathcal{M}(s, u)$ with $\mathcal{M}_{\nu,\,l}(s, u)$ in (\ref{eq:fourpointmellin}). To unpack this, first of all $f^{conn.}(U, V)$ has a partial wave decomposition weighted by $b_{l}(\nu^2)$ as seen above. These partial waves correspond to various physically exchanged operators of scaling dimensions $\Delta^b_i$, made continuous by the parameter $\nu$. Then this principal weight $b_{l}(\nu^2)$ is chosen to have poles at $\Delta_{i}^b$ i.e., $h+i\nu = \Delta_{i}^b$ and $h - i\nu = \Delta_{i}^b$, such that $f^{conn.}_{\nu,\,l}(U, V)$ decomposes into two further conformal partial waves as follows 
\begin{equation}\label{eq:normalizationpartialwave}
    f_{\nu,\,l}(U, V) = \kappa_{\nu,\,l}\cdot g_{h + i\nu,\,l}(U, V) + \kappa_{-\nu,\,l}\cdot g_{h - i\nu,\,l}(U, V)\,\,\,\textbf{;}\,\,\,\kappa_{\nu,\,l} = \frac{i\nu}{2\pi K_{h+i\nu,\,l}}
\end{equation}
\vspace{-1.5em}
\small
\begin{align}
    \langle \mathcal{V}^{s_1...s_l}_{h + i\nu}(\vec{x})\Phi_{\Delta_i}(\vec{x}_i)\Phi_{\Delta_j}(\vec{x}_j)\rangle = \int d^{d}\vec{y}\, \langle \mathcal{V}^{s_1...s_l}_{h+i\nu}(\vec{x})\mathcal{V}^{t_1...t_l}_{h+i\nu}(\vec{y}) \rangle\langle \mathcal{V}_{t_1...t_l}^{h - i\nu}(\vec{y})\Phi_{\Delta_i}(\vec{x}_i)\Phi_{\Delta_j}(\vec{x}_j) \rangle\label{eq:3pointnormspectral}
\end{align}
\normalsize
where the amputation identity in the second line (\ref{eq:3pointnormspectral}) is the choice of normalization for the three-point functions. Given this choice of normalization and the principal series representation described thus far, the Mellin amplitude dual to the conformal block $\mathcal{CB}_{\nu, \,l}(\vec{x}_1, \vec{x}_2, \vec{x}_3, \vec{x}_4)$ in the $s$-channel has a \textit{split} representation as shown in figure \ref{fig:wd_4} below
\vspace{-0.5em}

\noindent Therefore mathematically, this split representation splits the block $\mathcal{CB}_{\nu, \,l}$ into a product of two three-point functions integrated over an intermediate point. Via (\ref{eq:3pointnormspectral}), it reads
\begin{equation}
    \mathcal{CB}_{\nu, \,l}(\vec{x}_1, \vec{x}_2, \vec{x}_3, \vec{x}_4) \equiv \int d^{d}\vec{y}\,\, \langle \Phi_{\Delta_1}(\vec{x}_1)\Phi_{\Delta_2}(\vec{x}_2) \mathcal{V}_{h + i\nu}^{t_1...t_l} \rangle\,\langle \Phi_{\Delta_3}(\vec{x}_3)\Phi_{\Delta_4}(\vec{x}_4)\mathcal{V}^{h - i\nu}_{t_1...t_l} \rangle
\end{equation}
These three-point functions involving two scalars and a spin-$l$ operator are fixed upto a normalization constant as seen in (\ref{eq:threepointscalarspin}). Substituting it into the above then yields 
\small
\begin{equation*}
     \mathcal{CB}_{\nu, \,l}(\vec{x}_1, \vec{x}_2, \vec{x}_3, \vec{x}_4) = \frac{\lambda_{12(h + i\nu)}\lambda_{34(h - i\nu)}}{(x_{12}^2)^{\frac{1}{2}p_{12}(\nu)}(x_{34}^2)^{\frac{1}{2}p_{34}(-\nu)}}\int d^{d}\vec{y}\,\,\frac{(x_{10}^2)^{-\frac{1}{2}q_{12}(\nu)}(x_{30}^2)^{-\frac{1}{2}q_{34}(-\nu)}}{(x_{20}^2)^{\frac{1}{2}q_{21}(\nu)}(x_{40}^2)^{\frac{1}{2}q_{43}(-\nu)}}\langle f^{l}_{(12)},f_l^{(34)} \rangle
\end{equation*}
\normalsize
where $p_{ij}(\nu) = \Delta_i + \Delta_j - (h + i\nu) + l$ and $q_{ij}(\nu) = \Delta_i - \Delta_j + h + i\nu - l$. Also, $Y^{t}_{(ij)}$ in $\langle f^{l}_{(12)},f_l^{(34)} \rangle$ are as defined in (\ref{eq:threepointscalarspin}) with $k = 0$; where $\vec{x}_{i0} = \vec{x}_i - \vec{y}$. Expanding the $SO(d)$ invariant $\langle f^{l}_{(12)},f_l^{(34)} \rangle$ in the basis of spherical harmonics ($\mathcal{Y}_l^d(\text{cos}\,\theta)$) on a $(d-1)$-sphere
\small
\begin{flalign}
   \langle f^{l}_{(12)},f_l^{(34)} \rangle = \prod_{m = 1}^l Y^{t_m}_{(12)}Y_{t_m}^{(34)} = (|\vec{Y}_{(12)}||\vec{Y}_{(34)}|)^l \mathcal{Y}_l^d(\text{cos}\,\theta)\,\,\,\Bigg|\,\,\, \mathcal{Y}_l^d = \frac{2^{-l}l!}{(h - l)_l} C_{l}^{h - 1}(\text{cos}\,\theta)
\end{flalign}
\normalsize
where $\theta$ is the angle between $\vec{Y}_{(12)}$ and $\vec{Y}_{(34)}$, and $C_l^{h - 1}$ is the Gegenbauer polynomial which is the analogue of Legendre polynomial for spherical harmonics on a 2-sphere. Mack then generalized the Symanzik star formula (\ref{eq:symanzikstarformula}) to include these spherical harmonics
\begin{align}
    \int d^d{\vec{y}} \,\,\mathcal{Y}^d_l(\text{cos}\,\theta)\prod_{i = 1}^4 (x_{i0}^2)^{-\delta_{i}} = \frac{\pi^{\frac{d}{2}}}{(2\pi i)^2}\int_{-i\infty}^{i\infty} d \delta_{ij}\, \mathcal{P}_{l}&(\delta_{ij}) \Gamma\left(\delta_{12} - \frac{l}{2}\right) (x_{12}^2)^{-\delta_{12}} \Gamma\left(\delta_{34} - \frac{1}{2}\right)\nonumber\\&\times(x_{34}^2)^{-\delta_{34}} \tilde{\prod}\, \Gamma(\delta_{ij})(x_{ij}^2)^{-\delta_{ij}}\label{eq:generalizedsymanzik}
\end{align}
where $\tilde{\prod}$ is the product over $(ij) = (13), (14), (23), (24)$, and $\delta_i$ are such that $\sum_i \delta_i = d$. See Appendix 12.1, 12.2 in \cite{mack2009dindependent} for the derivation. Mack then rewrote the earlier integral expression for $\mathcal{CB}_{\nu, \,l}$ using (\ref{eq:generalizedsymanzik}), in a form that resembles the Mellin transform (\ref{eq:fourpointmellin}). This manipulation is the analogue of the one done for $D$-functions (\ref{eq:dfunction}), in which case the normal Symanzik star yielded a constant Mellin amplitude (\ref{eq:mellincontactamplitude}). However with the Gegenbauer polynomial now in the integrand, the generalized Symanzik star yields a non-trivial Mellin amplitude that involves Mack polynomials $\mathcal{P}_{\nu, \, l}(\delta_{ij})$. Following the normalization in \cite{Costa_2012}, the partial Mellin amplitude ($\mathcal{M}_{\nu, \, l}(\delta_{ij})$) derived by Mack reads
\begin{equation*}
    \mathcal{M}_{\nu, \, l} (t, s) = \mathcal{N}_{\nu, \,l}\,\omega_{\nu, \, l}(s) \mathcal{P}_{\nu, \, l}(t, s)\,\,\,\Bigg|\,\,\,\mathcal{N}_{\nu, \,l} = \frac{\Gamma\left[\frac{p_{12}(-\nu)}{2}\right]\Gamma\left[\frac{p_{34}(-\nu)}{2}\right]\Gamma\left[\frac{p_{12}(\nu)}{2}\right]\Gamma\left[\frac{p_{34}(\nu)}{2}\right]}{8\pi\Gamma(i\nu)\Gamma(-i\nu)}
\end{equation*}
\vspace{-0.5em}
\begin{equation}\label{eq:partialmellinamplitude}
    \text{where}\,\,\,\omega_{\nu, \, l}(s) = \frac{\Gamma\left(\frac{h + i\nu - l - s}{2}\right)\Gamma\left(\frac{h - i\nu - l - s}{2}\right)}{\Gamma\left(\frac{\Delta_1 + \Delta_2 -s}{2}\right)\Gamma\left(\frac{\Delta_3 + \Delta_4 - s}{2}\right)}
\end{equation}
where $s, t$ are as defined in (\ref{eq:s-variables}, \ref{eq:t-variables}). The Mack polynomials are given by the following
\begin{equation}\label{eq:mackpolynomials}
\begin{split}
    \mathcal{P}_{\nu, \, l}(t, s) = \sum_{r = 0}^{[l/2]} a_{l,\,r}&\frac{2^{l+2r}\left(\frac{h + i\nu - l - s}{2}\right)_r \left(\frac{h - i\nu - l -s}{2}\right)_r(l - 2r)!}{(h + i\nu - 1)_{l}(h - i\nu - 1)_l}\\
    &\quad\quad\quad\quad\times\sum_{\sum k_{ij} = l - 2r} (-1)^{k_{13} + k_{24}} \tilde{\prod_{(ij)}} \frac{(\delta_{ij})_{k_{ij}}}{k_{ij}!} \prod_{n = 1}^4 (\alpha_n)_{l - r - \sum_{j}k_{jn}}
\end{split}
\end{equation}
\vspace{-0.5em}
\begin{equation*}
    \begin{split}
        \alpha_1 = 1 - \frac{h + i\nu + l + \Delta_{12}}{2}\,,\,\,\,\alpha_2 = 1 - \frac{h + i\nu + l - \Delta_{12}}{2}\,, \,\,\, \alpha_3 = 1 - \frac{h - i\nu + l + \Delta_{34}}{2}
    \end{split}
\end{equation*}
\vspace{-0.5em}
\begin{equation*}
    \alpha_4 = 1 - \frac{h - i\nu + l - \Delta_{34}}{2}\,,\,\,\,a_{l,\,r} = (-1)^r\,\frac{l!\,(h + l - 1)_{-r}}{2^{2r}r! (l - 2r)!}
\end{equation*}
where note that the only non-vanishing integers $k_{ij} = k_{ji}$ for $s$-channel are $k_{13}, k_{14}, k_{23}, k_{24}$, which are the four possibilities that the product $\tilde{\prod}_{(ij)}$ enumerates over in (\ref{eq:mackpolynomials}). It is also important to make note of the poles in $\mathcal{M}_{\nu, \, l}(t, s)$, that arise from the negative integer arguments in Gamma functions, which are as follows for the variable $\nu$
\begin{alignat}{5}
        &(\ref{eq:partialmellinamplitude})\,\,:\,\, \Gamma\left(\frac{h \pm i\nu - l - s}{2}\right)\,\,\,\,&&\Big|\,\,\,\, \pm i\nu = h - l - s + 2m\,,\,\,&&&m = 0,1,2...\label{eq:relevantpoles}\\
        &(\ref{eq:partialmellinamplitude})\,\,:\,\,\Gamma\left[\frac{p_{12}(\pm \nu)}{2}\right]&&\Big|\,\,\,\,\pm i\nu = \Delta_{1} + \Delta_2 - h + l + 2m\,,\,\,\,&&&m = 0,1,2...\\
        &(\ref{eq:partialmellinamplitude})\,\,:\,\,\Gamma\left[\frac{p_{34}(\pm \nu)}{2}\right]&&\Big|\,\,\,\,\pm i\nu = \Delta_3 + \Delta_4 - h + l + 2m\,,\,\,\,&&&m = 0,1,2...\\[0.7ex]
        &(\ref{eq:mackpolynomials})\,\,:\,\,(h \pm i\nu - 1)_l&&\Big|\,\,\,\,\pm i\nu = h - 1 + l - q\,,\,\,&&&q = 1,2,...,l
\end{alignat}
With the partial Mellin amplitude $M_{\nu, \, l}$ and it's corresponding poles now in place, it is just a matter of choosing the right pole structure for the weight function $b_l(\nu^2)$ so that when the integration is done in (\ref{eq:mellinpartialwave}), $M(t, s)$ ends up having the expected poles. This then fixes the residue of the singular part of the Mellin amplitude in terms of the Mack Polynomials, as we will now see. Consider a physically exchanged operator of scaling dimension $\Delta$ and spin-$l$, then as mentioned in the discussion below (\ref{eq:spectralpartialwave}), $b_l(\nu^2)$ ought to have poles at $\pm i\nu = h - \Delta$. Given the normalization ($\kappa_{\nu,\,l}$) chosen for the partial wave decomposition (\ref{eq:normalizationpartialwave}), the normalized pole structure for $b_l(\nu^2)$ is then
\begin{equation}\label{eq:weightpolestructure}
    b_l(\nu^2) \approx \lambda_{12\mathcal{V}_{\Delta,\,l}}\lambda_{34\mathcal{V}_{\Delta,\,l}}\, \frac{K_{\Delta,\,l}}{\nu^2 + (\Delta - h)^2}
\end{equation}
where $\lambda_{12\mathcal{V}_{\Delta,\,l}}, \lambda_{34\mathcal{V}_{\Delta,\, l}}$ are the OPE coefficients as seen in (\ref{eq:conformalpartialwavedecomposition}, \ref{eq:normalizedopescalar}). Now coming to the expected pole structure of $\mathcal{M}(t, s)$, it was discussed near (\ref{eq:satellitepoles}) that it has poles at $\delta_{ij} = (\Delta_i + \Delta_j - \Delta + l - 2m)/2$, which for the $s$-channel exchange becomes $s = \Delta - l + 2m$. Mack infact argued that these simple poles are the only poles that correspond to single-trace exchanges, which are what $\mathcal{M}(t, s)$ entails by definition (\ref{eq:mellinamplitude}). Therefore the relevant poles to consider in $\mathcal{M}_{\nu, \, l}$ are (\ref{eq:relevantpoles}), which then justifies the pole structure (\ref{eq:weightpolestructure}) chosen for $b_l(\nu^2)$; Since two of the poles in (\ref{eq:relevantpoles}) coincide with the poles $\pm i\nu = (\Delta - h)$ for each $m$ as $s \rightarrow \Delta - l + 2m$, the integral diverges and therefore the two poles in (\ref{eq:weightpolestructure}) are enough to consider to compute the relevant residue for a given $m$. Actually $b_{l}(\nu^2)$ has other spurious poles that correspond to non-physical exchanged operators, which will then have to be accounted for. They will not be relevant to us but the interested reader can refer to Appendix A.5 of \cite{Costa_2012} for a discussion of the same. Back to the discussion at hand, substitute (\ref{eq:weightpolestructure}) into (\ref{eq:mellinpartialwave}), and use Cauchy Residue theorem to evaluate the residue of the singular part of $\mathcal{M}(t, s)$ (by also noting that $\mathcal{M}_{\nu,\,l} = \mathcal{M}_{-\nu,\, l}$).
\begin{equation}\label{eq:residuecomputation}
\begin{split}
    \mathcal{Q}_{l, m}(t) = \lim_{s\to \Delta - l + 2m} \,\text{Res}\left(\frac{K_{\Delta, \, l}}{\nu^2 + (\Delta - h)^2}\mathcal{M}_{\nu, \, l}(t, s)\right)\Bigg|_{\pm i\nu = \Delta - h} 
\end{split}
\end{equation}
The normalization factor $K_{\Delta, \,l}$ is chosen according to the normalization in \cite{Costa_2012} as follows 
\begin{equation}\label{eq:normalizationk}
    K_{\Delta, \,l} = \frac{\Gamma(\Delta + l)\Gamma(\Delta - h + 1)(\Delta - 1)_{l}}{4^{l - 1}\Gamma\left(\frac{\Delta + l \pm \Delta_{12}}{2}\right)\Gamma\left(\frac{\Delta + l \pm \Delta_{34}}{2}\right)\Gamma\left(\frac{\Delta_1 + \Delta_2 - h + l \pm (h - \Delta)}{2}\right)\Gamma\left(\frac{\Delta_3 + \Delta_4 - h + l \pm (h - \Delta)}{2}\right)} 
\end{equation}
where the notation $\Gamma(a \pm b) = \Gamma(a + b)\Gamma(a - b)$ is used for convenience. Now by substituting (\ref{eq:normalizationk}) and (\ref{eq:partialmellinamplitude}) into (\ref{eq:residuecomputation}), one then obtains the following residue 
\small
\begin{equation*}
    \mathcal{Q}_{l,\,m}(t) = \frac{-2.4^{-l}\,[\Gamma(\Delta + l)]\,(\Delta - 1)_l \times \mathcal{P}_{\,-i(\Delta - h),\,l}\,(t, s = \Delta - l + 2m)}{m!(\Delta - h + 1)_m\Gamma\left(\frac{\Delta + l \pm \Delta_{12}}{2}\right)\Gamma\left(\frac{\Delta + l \pm \Delta_{34}}{2}\right)\Gamma\left(\frac{\Delta_1 + \Delta_2 - \Delta + l - 2m}{2}\right)\Gamma\left(\frac{\Delta_3 + \Delta_4 - \Delta + l - 2m}{2}\right)}
\end{equation*}
\normalsize
Therefore finally, the Mellin amplitude $\mathcal{M}(t, s)$ for a spin-\textit{l} exchange in the $s$-channel is
\begin{equation}\label{eq:mellinamplitudespinexch}
    \boxed{\mathcal{M}_{s-exch.}^{l}(t, s) = \lambda_{12\mathcal{V}_{\Delta,\,l}}\lambda_{34\mathcal{V}_{\Delta,\,l}} \sum_{m = 0}^\infty \frac{\mathcal{Q}_{l,\,m}(t)}{s - (\Delta - l + 2m)} + \mathcal{R}_{\,l - 1}(t, s)}
\end{equation}
where $\mathcal{R}_{\,l - 1}(t, s)$ is the non-singular part of the Mellin amplitude which is a degree $l - 1$ polynomial, that arises from different interaction vertices in the contact diagram other than $\phi^4$. The $t$ and $u$ channel amplitudes are then obtained by performing the associated interchanges. Also note that for $l = 0$, to match the normalization for the scalar exchange, the OPE coefficients ($\lambda_{ij\mathcal{V}_{\Delta, 0}}$) in (\ref{eq:mellinamplitudespinexch}) and the coupling constants $\lambda_{\Delta\Delta_i\Delta_j}$ are related by
\begin{equation}
    \lambda_{\Delta_i\Delta_j\mathcal{V}_{\Delta, 0}} = \Gamma\left(\frac{\Delta_i + \Delta_j - \Delta}{2}\right)\Gamma\left(\frac{\Delta \pm \Delta_{ij}}{2}\right) \lambda_{\Delta\Delta_i\Delta_j}\,\,\,\Bigg|\,\,\,\mathcal{N} = \frac{2\pi^{h}\Gamma(\Delta - h + 1)}{\Gamma(\Delta)}
\end{equation}
where $\mathcal{N}$ is the factor from (\ref{eq:mellinamplitudescalarexchange}) absorbed into the normalization constant in (\ref{eq:reducedmellinamplitude}). With this normalization now sorted, we will use (\ref{eq:mellinamplitudespinexch}) for the scalar exchange as well from now on. After a long journey starting from the previous chapter, the requisite formalism has now finally been set up. Therefore in the following section, let us compute some examples of Mellin amplitudes tailored to our focus i.e. $AdS_4/CFT_3$.

\section{Examples in $AdS_4 / CFT_3$}\label{sec:ads4cft3examples}
In this section, we will compute the four-point function of the stress-tensor multiplet in the three dimensional CFT dual to M-theory on $AdS_4 \times S^7/\mathbb{Z}_k$. In the corresponding superconformal algebra i.e., $OSp(\mathcal{N} | 4)$, the stress tensor multiplet is a BPS multiplet, which is important to note as BPS multiplets are protected by supersymmetry. In general, a $p/q$-BPS multiplet is classified by it's highest weight state i.e. the $p/q$-BPS superconformal primary, which is annihilated by a $p/q$ fraction of the $\Delta$ raising $\mathcal{Q}$ supercharges (see section \ref{sec:ospn4algebra} for definitions). The stress tensor multiplet in particular consists of the conserved bosonic stress tensor ($T_{\mu\nu}$, $\Delta = d$), whose spatial integrals give rise to the three conformal generators $\mathcal{P}_\mu, \mathcal{D}$ and $\mathcal{K}_\mu$; It's conserved superpartner $\tilde{T}_{\mu\alpha}$ ($\Delta = d - \frac{1}{2}$), whose integrals correspondingly give rise to the Poincaré supercharges ($\mathcal{Q}$) and conformal supercharges ($\mathcal{S}$); The conserved $R$-symmetry current $j_\mu^{(R)}$ ($\Delta = d - 1$) which gives rise to the $R$-symmetry generators ($\mathcal{R}$) \cite{cordova2016multiplets}. The fact that the three conserved currents belong to the same multiplet can be seen from the commutation and anti-commutation relations in section \ref{sec:ospn4algebra} i.e. $\{\mathcal{Q}, \mathcal{Q}\} \sim \mathcal{P}$ (\ref{eq:cartanosp2}), $[\mathcal{Q}, \mathcal{P}] = 0$ and $[\mathcal{Q}, \mathcal{R}] \sim \mathcal{Q}$ (\ref{eq:cartanosp3}), which imply that upto total derivatives $\{\mathcal{Q}, \tilde{T}\} \sim T$, $[\mathcal{Q}, T] = 0$ and $[\mathcal{Q}, j^{(R)}] \sim \tilde{T}$ respectively. In addition to these three conserved currents, the remaining fields required to ensure the off-shell closure of supersymmetry can be found in for instance \cite{Bergshoeff_2010}.
\vspace{-0.5em}
\subsection{$k = 1, \mathcal{N} = 8$ Stress-tensor multiplet 4-point function}
The simplest case to consider is the Chern-Simons level $k = 1$, so that the compact part of the background in the gavity dual is simply $S^7$, thereby eliminating the possibility of additional constraints on the diagrams due to the orbifold $\mathbb{Z}_k$ action. This case also corresponds to the maximal $\mathcal{N} = 8$ supersymmetry, as mentioned in the previous chapters, with the superconformal group being $OSp(8 | 4)$ and the R-symmetry group being $SO(8)$. From a group theoretic analysis of simple roots and commutators, the stress tensor multiplet for this case is a half-BPS multiplet, whose superconformal primary is either $\mathcal{B}[0]_{1}^{(0,0,2,0)}$ or $\mathcal{B}[0]_{1}^{(0,0,0,2)}$ \cite{cordova2016multiplets}. The choice between the former or the latter is a matter of convention, and either one is defined in accordance with the $\mathcal{Q}$ supercharges transforming as $8_v = (1,0,0,0)$ under the R-symmetry. Refer to table \ref{eq:recombrulesn8} for a reminder on the notation; As mentioned below the same table, this is also of the form corresponding to an absolutely protected multiplet, which thereby ensures the reliability of any computations thus related, under superconformal deformations or interpolations to strong coupling. The R-symmetry Dynkin label of the scalar primary being $35_c = (0,0,2,0)$ is related to the fact that the stress-tensor multiplet is half-BPS i.e. As a more general statement, any superconformal algebra with the R-symmetry group being locally isomorphic to $SO(\mathcal{N})$ has half-BPS operators in the rank-$n$ traceless symmetric representation of SO($\mathcal{N}$) i.e., $\mathcal{V}_n^{\alpha_1...\alpha_n}$. The scaling dimension $\Delta$ of these operators is quantized via the rank $n$ as $\Delta = (h - 1)n$ \cite{Zhou_2018}. For convenience, these indices are removed by defining new operators upon contractions with the R-symmetry polarization vectors $Y_{\alpha_i}$ as follows
\begin{equation}\label{eq:polarizationrsymm}
    \Phi_\Delta(\vec{x}, \vec{Y}) = \Phi_\Delta^{\alpha_1...\alpha_n}(\vec{x})(\vec{Y})_{\alpha_1}...(\vec{Y})_{\alpha_n}\,\,\,\Big|\,\,\,(\vec{Y})_\alpha (\vec{Y})^\alpha = 0
\end{equation}
These newly defined operators are scalars under R-symmetry and therefore the correlators of these operators will have to be scalar covariant w.r.t R-symmetry. The ($\vec{x}$) dependent part of the correlators is $SO(\mathcal{N})_R$ invariant as position space is not acted upon by $SO(\mathcal{N})_R$ (no twisting). The $Y$-dependent terms however ought to be functions of $|\vec{Y}_i - \vec{Y}_j|^2$, and therefore of $\vec{Y}_i \cdot \vec{Y}_j = Y_{ij}$ (since $\vec{Y}_i \cdot \vec{Y}_i = 0$), in  order to be $SO(\mathcal{N})_R$ invariant. The precise functional form can then be fixed by considering the individual scaling of polarization vectors $\vec{Y}_i \rightarrow \lambda \vec{Y}_i$, under which the rank-$n$ operators (\ref{eq:polarizationrsymm}) scale as $\lambda^{n}$. Now by specializing to four-point functions (\ref{eq:fourpointfunc}), the first obvious step is to separate the scale invariant part from the scale covariant part, and define scale invariant cross-ratios ($\sigma, \tau$) similar to the dilation invariant conformal cross-ratios ($U, V$), as follows
\begin{equation}\label{eq:rsymmcrossratios}
    f(U, V) \rightarrow F(U, V; \sigma, \tau) \,\,\,\Bigg|\,\,\, \sigma = \frac{Y_{13}Y_{24}}{Y_{12}Y_{34}}\,,\,\,\tau = \frac{Y_{14}Y_{23}}{Y_{12}Y_{34}}
\end{equation}
The scale covariant part however takes the most general form $\prod_{i < j} (Y_{ij})^{\gamma_{ij}}$, with non-negative integers $\gamma_{ij}$. Since rank-\textit{n} operators scale as $\lambda^n$, $\gamma_{ij}$ are then constrained to obey the relation $\sum_{i \neq j} \gamma_{ij} = n_j$. This system of equations is obviously not a fully-determined system, and therefore leaves a freedom of choice with regards to the definition of the scale covariant part. The most convenient definition satisfying the constraint, that causes maximal simplification for the case of all operators with equal conformal weights ($\Delta_i$) is 
\begin{equation}\label{eq:rsymmfourpointfunc}
    \left\langle \prod_{i = 1}^4\Phi_i(\vec{x}_i, \vec{Y}_i)\right\rangle = \prod_{i < j} \left(\frac{Y_{ij}}{(x_{ij}^2)^{h - 1}}\right)^{\gamma_{ij}^0} \left(\frac{Y_{12}Y_{34}}{(x_{12}^2)^{h - 1}(x_{34}^2)^{h - 1}}\right)^{L} F(U, V; \sigma, \tau)
\end{equation}
\vspace{-0.1em}
where $\gamma^0_{ij}$ are the lower bounds on $\gamma_{ij}$ given the constraint they obey, which is most easily seen after the parametrization $2\gamma_{ij} = -a_{ij} + n_i + n_j$, with $a_{12} = a_{34},\, a_{13} = a_{24},\, a_{14} = a_{23}$ and $a_{ij} = a_{ji}$ (see section 2.1 of \cite{Zhou_2018}). These bounds and the weight $L$ are given by
\begin{alignat}{5}
    &\gamma_{12}^0 = \frac{n_1 + n_2 - n_3 - n_4}{2}\,\,\,\,\,;\,\,\,\,\,\gamma_{13}^0 = \frac{n_1 + n_3 - n_2 - n_4}{2}\,\,\,\,\,;\,\,\,\,\, \gamma^0_{34} = \gamma^0_{24} = 0\label{eq:boundsfourpoi}\\
    &\gamma^0_{14}, \,\gamma^0_{23} = \begin{cases}
        0,\, \frac{n_2 + n_3 - n_1 - n_4}{2} & n_1 + n_4 \leq n_2 + n_3\\
        \frac{n_1 + n_4 - n_2 - n_3}{2},\,0 & n_1 + n_4 > n_2 + n_3
    \end{cases}\,\,;\,\,L = \begin{cases}
         n_4 & n_1 + n_4 \leq n_2 + n_3\\
        \frac{n_2 + n_3 + n_4 - n_1}{2} & n_1 + n_4 > n_2 + n_3
    \end{cases}\nonumber
\end{alignat}
where $n_1 \geq n_2 \geq n_3 \geq n_4$ is assumed without loss of generality. 
Now returning to our case of the stress tensor multiplet in three dimensions ($2h = 3$), whose scalar primary is of dimension $\Delta = 1$; The four point function of these scalars ($n_1 = n_2 = n_3 = n_4 = 2$) is of the form (substitute \ref{eq:boundsfourpoi}) in (\ref{eq:rsymmfourpointfunc}, where $L = 2$ and $\gamma^0_{ij} = 0$)
\begin{equation}\label{eq:stresstensorfourpointcorrelator}
    \langle \Phi(\vec{x}_1, \vec{Y}_1)\,\Phi(\vec{x}_2, \vec{Y}_2)\,\Phi(\vec{x}_3, \vec{Y}_3)\,\Phi(\vec{x}_4, \vec{Y}_4) \rangle = \frac{Y_{12}^2Y_{34}^2}{x_{12}^2x_{34}^2} F(U, V; \sigma, \tau)
\end{equation}
Now that the R-symmetry invariant form of the correlator has been determined, it is time to decompose $F(U, V; \sigma, \tau)$ into superconformal partial waves and determine the operator content of the exchange channel, which is strongly constrained by group theory. 

\subsubsection*{Superconformal blocks and Operators in exchange channel}
Superconformal blocks are very similar to the conformal blocks introduced in section \ref{sec:conformalcft}, with the difference being that they are more constrained by superconformal invariance rather than just conformal invariance. To make this statement more rigorous, a few statements made previously will have to be revisited. For instance, it was claimed in the discussion above (\ref{eq:cbinvariantform}) that $\mathcal{CB}_{\mathcal{V}}$ transforms in the same way as the four-point function does due to conformal invariance of the OPEs. To see why this is the case, consider the conformal ward identities (\ref{eq:conformalwardidentity}) w.r.t to the four conformal generators $\mathcal{D}, \mathcal{P}, \mathcal{M}, \mathcal{K}$ (collectively called $\mathcal{CF}_{AB}$). Integrating over a spacetime surface that encloses all the points $\vec{x}_i$, assuming that $\langle j^\mu_a (\vec{x}) \Phi_1(\vec{x}_1)...\Phi_n(\vec{x}_n) \rangle$ on the LHS vanishes asymptotically, and sending the surface to infinity, one obtains 
\begin{equation}\label{eq:conformalwardinteg}
    \sum_{i = 1}^n \mathcal{CF}_{iAB} \langle \Phi_1(\vec{x}_1)...\Phi_n(\vec{x}_n) \rangle = 0
\end{equation}
where $\mathcal{CF}_{iAB}$ are the conformal generators w.r.t $\vec{x}_i$ acting on $\Phi_i(\vec{x}_i)$. Specializing to four-point functions, now act with the same differential operator on $\mathcal{CB}_{\mathcal{V}}(\vec{x}_1, \vec{x}_2, \vec{x}_3, \vec{x}_4)$ i.e. $\sum_{i = 1}^4 \mathcal{CF}_{iAB}\,\mathcal{CB}_{\mathcal{V}}$. The functions $C^{s_1...s_m}(\vec{x}_{ij})$ in the definition above (\ref{eq:confblockdecomfour}) are conformally invariant and therefore satisfy (\ref{eq:conformalwardinteg}), thereby resulting in the RHS as follows 
\begin{equation*}
    C^{s_1...s_l}(\vec{x}_{12})C^{t_1...t_l}(\vec{x}_{34}) (\mathcal{CF}_{2AB} + \mathcal{CF}_{4AB})\Bigl[\langle \mathcal{V}_{s_1...s_l}(\vec{x}_2)\mathcal{V}_{t_1...t_l}(\vec{x}_4)\rangle + \langle 
\text{Descendants} \rangle\Bigr]
\end{equation*}
\vspace{-0.1em}
This then vanishes as it is just $n = 2$ in (\ref{eq:conformalwardinteg}). Therefore conformal blocks satisify the same equation as the four-point correlator (\ref{eq:conformalwardinteg}) and hence have the same functional form. Now one can act with the Casimir of the conformal algebra as follows
\begin{align*}
    \mathcal{CF}^2_{(12)}\, \mathcal{CB}_{\mathcal{V}}(\vec{x}_1, \vec{x}_2, \vec{x}_3, \vec{x}_4) = C^{s_1...s_l}(\vec{x}_{12})C^{t_1...t_l}(\vec{x}_{34}) \Bigl[\Bigl\langle (\mathcal{CF}_{2AB}\,\mathcal{CF}^{AB}_2\,\mathcal{V}_{s_1...s_l}) \mathcal{V}_{t_1...t_l}
 \Bigr\rangle + ...\Bigr]\nonumber
\end{align*}
\vspace{-0.8em}
\begin{equation}
  \mathcal{CF}^2_{(12)} = \frac{1}{2}(\mathcal{CF}_{1AB} + \mathcal{CF}_{2AB})(\mathcal{CF}_{1}^{AB} + \mathcal{CF}_{2}^{AB})
\end{equation}
where $...$ represents descendants. Since the primary $\mathcal{V}$ lies in a representation of the conformal group, it has an eigen value w.r.t to the Casimir that classifies the representation i.e. $\mathcal{CF}_{2AB}\,\mathcal{CF}_2^{AB} \mathcal{V}_{s_1...s_l} = \Delta(\Delta - d) + l(l + d - 2) \mathcal{V}_{s_1...s_l}$. Therefore the contribution to the conformal block from the primary is given by an eigen value equation of the Casimir, which is then converted into a differential equation for the conformal partial wave using (\ref{eq:cbinvariantform}). This equation is usually solved in the embedding space $\mathbb{R}^{d+1, 1}$ which is naturally convenient to describe the action of the Euclidean conformal group $SO(d+1, 1)$, see \cite{Dolan_2004}. In this more formal language, the superconformal quantities are easy to distinguish from their conformal counterparts. The correlators satisfy superconformal ward identities, that are an extension to (\ref{eq:conformalwardinteg}) by considering $\mathcal{SCF}_{iAB}$, which now includes the supercharges $\mathcal{S}, \mathcal{Q}$ and R-symmetry generators $\mathcal{R}$ as well. Similar to the conformal case, the superconformal blocks ($\mathcal{SCB}_{\mathcal{V}}$) thereby have the same superconformal invariant form as the correlators, and now satisfy eigen value equations w.r.t the superconformal Casimir. Consequently, this leads to the differential equations for superconformal partial waves ($F_{\mathcal{V}}(U, V; \sigma, \tau)$) which can then in principle be solved (see \cite{Nirschl_2005}, \cite{Aprile_2023} for e.g.). Such calculations are quite involved and usually done by considering a specific superconformal algebra, and are beyond the scope of this thesis. We will only make use of results from such calculations when necessary rather than delving into details. Back to the case of interest, decomposing (\ref{eq:stresstensorfourpointcorrelator}) into superconformal blocks similar to (\ref{eq:confblockdecomfour})
\begin{equation}
    \langle \Phi(\vec{x}_1, \vec{Y}_1)\Phi(\vec{x}_2, \vec{Y}_2)\Phi(\vec{x}_3, \vec{Y}_3)\Phi(\vec{x}_4, \vec{Y}_4)\rangle = \sum_{\mathcal{V}\, \in \,osp(8|4)} \lambda_{\mathcal{V}}^2 \mathcal{SCB}_{\mathcal{V}}(\{\vec{x}_i, \vec{Y}_i\})\\
\end{equation}
\vspace{-0.8em}
\begin{equation*}
    \mathcal{SCB}_{\mathcal{V}}(\{\vec{x}_i, \vec{Y}_i\}) = \frac{Y_{12}^2Y_{34}^2}{x_{12}^2x_{34}^2} F_{\mathcal{V}}(U, V; \sigma, \tau) \implies F(U, V; \sigma, \tau) = \sum_{\mathcal{V}\, \in \,osp(8 | 4)}\lambda_{\mathcal{V}}^2 F_{\mathcal{V}}(U, V; \sigma, \tau)\nonumber
\end{equation*}
where $\lambda_{\mathcal{V}} = \lambda_{12\mathcal{V}} = \lambda_{34\mathcal{V}}$. The exchanged superconformal primaries $\mathcal{V}$ that appear in the OPEs of $\Phi(\vec{x}_1, \vec{Y}_1)\Phi(\vec{x}_2, \vec{Y}_2)$ and $\Phi(\vec{x}_3, \vec{Y}_3)\Phi(\vec{x}_4, \vec{Y}_4)$, are constrained by the superconformal algebra since the OPEs are superconformal invariant. The first and foremost is via R-symmetry which group theoretically constrains $\mathcal{V}$ to transform in a representation that is the tensor product of the $so(8)$ representations of $\Phi(\vec{x}_1, \vec{Y}_1)$ and $\Phi(\vec{x}_2, \vec{Y}_2)$ (\cite{yamatsu2020finitedimensional}, \cite{Ferrara_2002}).
\begin{equation}\label{eq:irrepsstresstensor}
    (0,0,m,0) \otimes (0,0,n,0) = \oplus_{a = 0}^n \oplus_{b = 0}^{n - a}\,(0, b, m + n - 2a - 2b, 0)
\end{equation}
\vspace{-1em}
\begin{equation*}
    m, n = 2 \implies \mathbf{35}_c \otimes \mathbf{35_c} = (\mathbf{1})_S \oplus (\mathbf{28})_A \oplus (\mathbf{35}_c)_S \oplus (\mathbf{294}_c)_S \oplus (\mathbf{300})_S \oplus (\mathbf{567}_c)_A
\end{equation*}
where the RHS shows the irreducible representations that constitute the tensor product of irreducible representations on the LHS. Also in the second line, note that the Dynkin labels have been converted to the dimension of representations they correspond to in $so(8)$. Therefore it is now suggestive to split the R-symmetry information of the conformal partial waves in the basis of these irreducible representations. To do so, first note that the cross-ratios $\sigma, \tau$ capture the R-symmetry information in $F(U, V; \sigma, \tau)$. Secondly, since $\Phi(\vec{x}_i, \vec{Y}_i) = \Phi_{\alpha_1\alpha_2}(\vec{Y}_i)^{\alpha_1}(\vec{Y}_i)^{\alpha_2}$, it is quadratic in $\vec{Y}_i$. Therefore the LHS of (\ref{eq:stresstensorfourpointcorrelator}) is quadratic in each $\vec{Y}_i$, and since the scale covariant factor on the RHS is quadratic in each $\vec{Y}_i$, $F(U, V; \sigma, \tau)$ ought to be quadratic in $\sigma, \tau$ to have the right dimensionality in each term. Thus the quadratic polynomial can conveniently be written in the basis of eigenfunctions of the $so(8)$ Casimir corresponding to the irreducible representations in (\ref{eq:irrepsstresstensor})
\begin{equation}\label{eq:decompconfblocksuconf}
    F_{\mathcal{V}}(U, V; \sigma, \tau) = \sum_{a = 0}^2 \sum_{b = 0}^a Y_{ab}(\sigma, \tau) \,G_{\mathcal{V}}^{ab}(U, V)
\end{equation}
where $Y_{ab}(\sigma, \tau)$ are the eigenfunctions corresponding to the irreducible representations $(0, a - b, 2b, 0)$; $a = 0, 1, 2$ and $b = 0,...,a$. This is a compact notation of the ones mentioned earlier i.e. $\mathbf{1} = (0,0,0,0), \,\mathbf{28} = (0,1,0,0),\, \mathbf{35}_c = (0,0,2,0),\, \mathbf{294}_c = (0,0,4,0), \,\mathbf{300} = (0,2,0,0)$ and $\mathbf{567}_c = (0,1,2,0)$. These eigenfunctions are given by the following \cite{Nirschl_2005}
\begin{align*}
    &Y_{00}(\sigma, \tau) = 1\,\,\boldsymbol{;}\,\,Y_{10}(\sigma, \tau) = \sigma - \tau\,\,\boldsymbol{;}\,\,Y_{11}(\sigma, \tau) = \sigma + \tau - \frac{1}{4}\,\,\boldsymbol{;}\,\,Y_{21}(\sigma, \tau) = \sigma^2 - \tau^2 - \frac{2}{5}(\sigma - \tau)\\
    &Y_{20}(\sigma, \tau) = \sigma^2 + \tau^2 - 2\sigma\tau - \frac{1}{3}(\sigma + \tau) + \frac{1}{21}\,\,\,\boldsymbol{;}\,\,\,Y_{22}(\sigma, \tau) = \sigma^2 + \tau^2 + 4\sigma\tau - \frac{2}{3}(\sigma + \tau) + \frac{1}{15}
\end{align*}
The next set of constraints on the superconformal primaries $\mathcal{V} = [l]^{(0, a-b, 2b, 0)}_\Delta$ are via the unitarity bounds. These can be read off from the $osp(8|4)$ table \ref{eq:recombrulesn8}, which are as follows
\begin{alignat}{7}
    &\mathbf{1} \,\,\,\,\,&&:\,\,\,\,\, \mathcal{L}[l]_{\Delta \,>\, l + 1}^{(0,0,0,0)}\,\,,\,\,\mathcal{A}_1[l \geq 1/2]_{l + 1}^{(0,0,0,0)}\,\,,\,\,\mathcal{A}_2[0]_{1}^{(0,0,0,0)}\,\,,\,\,\mathcal{B}_1[0]_0^{(0,0,0,0)}\\[0.5ex]
    &\mathbf{28} &&:\,\,\,\,\, \mathcal{L}[l]_{\Delta \,>\, l + 2}^{(0,1,0,0)} \,\,,\,\,\mathcal{A}_1[l \geq 1/2]_{l + 2}^{(0,1,0,0)} \,\,,\,\,\mathcal{A}_2[0]_{2}^{(0,1,0,0)} \,\,,\,\,\mathcal{B}_1[0]_{1}^{(0,1,0,0)}\\[0.5ex]
    &\mathbf{35}_c &&:\,\,\,\,\, \mathcal{L}[l]_{\Delta \,>\, l + 2}^{(0,0,2,0)}\,\,,\,\, \mathcal{A}_1[l \geq 1/2]_{l + 2}^{(0,0,2,0)}\,\,,\,\,\mathcal{A}_2[0]_{2}^{(0,0,2,0)}\,\,,\,\,\mathcal{B}_1[0]_{1}^{(0,0,2,0)}\\[0.5ex]
    &\mathbf{294}_c &&:\,\,\,\,\,\mathcal{L}[l]_{\Delta \,>\, l + 3}^{(0,0,4,0)}\,\,,\,\,\mathcal{A}_1[l \geq 1/2]_{l + 3}^{(0,0,4,0)}\,\,,\,\,\mathcal{A}_2[0]_{3}^{(0,0,4,0)}\,\,,\,\,\mathcal{B}_1[0]_{2}^{(0,0,4,0)}\\[0.5ex]
    &\mathbf{300} &&:\,\,\,\,\, \mathcal{L}[l]_{\Delta \,>\, l + 3}^{(0,2,0,0)}\,\,,\,\,\mathcal{A}_1[l \geq 1/2]_{j+3}^{(0,2,0,0)}\,\,,\,\,\mathcal{A}_2[0]_{3}^{(0,2,0,0)}\,\,,\,\,\mathcal{B}_1[0]_{2}^{(0,2,0,0)}\\[0.5ex]
    &\mathbf{567}_c &&:\,\,\,\,\, \mathcal{L}[l]_{\Delta \,>\, l + 3}^{(0,1,2,0)}\,\,,\,\,\mathcal{A}_1[l \geq 1/2]_{l + 3}^{(0,1,2,0)}\,\,,\,\,\mathcal{A}_2[0]_3^{(0,1,2,0)}\,\,,\,\,\mathcal{B}_1[0]_{2}^{(0,1,2,0)}  
\end{alignat}
Although there seems to be a lot of superconformal multiplets that appear in the OPE, a lot of these are not allowed, which leads us to the third set of constraints that come from two sources. The first source is the $S, A$ labels in (\ref{eq:irrepsstresstensor}) whose discussion has been postponed until now. They denote whether the representation lies in the  symmetric ($S$) or anti-symmetric ($A$) product of the two $35_c$'s i.e. under the exchange $\vec{Y}_1 \leftrightarrow \vec{Y}_2$ in the OPE $\Phi(\vec{x}_1, \vec{Y}_1)\Phi(\vec{x}_2, \vec{Y}_2)$. Since the two operators are identical, exchange $\vec{x}_1 \leftrightarrow \vec{x}_2$ in addition to the aforementioned exchange of polarization vectors corresponds to exchanging the operators, which will have to be symmetric as they are bosonic. However, $\vec{x}_1 \leftrightarrow \vec{x}_2$ introduces the parity factor $(-1)^{0}\cdot (-1)^{0}$ on the LHS and $(-1)^{l}$ on the RHS, due to the spin of the respective operators. Therefore for overall symmetry, the operators labelled $S$ ought to have even parity $(-1)^{l}$ i.e. even integer spin, and the operators labelled $A$ ought to have odd parity $(-1)^{l}$ i.e. odd integer spin. The second source of constraints come from the fact that these blocks have holographic duals; Consequently it was argued in \cite{Rastelli_2018} that, the exchanged operators will have to satisfy the twist selection rule $\Delta - l \leq \Delta_1 + \Delta_2$ ($\Delta - l \leq 2$) in order to have a finite amplitude for the three-point diagram. With these two constraints in place, the allowed superconformal blocks are then
\begin{align}
    &(\mathbf{1})_S \,\,\,\,\,\kern 0.05em\,\,\,:\,\,\,\Bigl\{\mathcal{L}[l]_{\Delta > l + 1}^{(0,0,0,0)}\,\,,\,\,\mathcal{A}[l \geq 1]_{l + 1}^{(0,0,0,0)}\,\,,\,\,\mathcal{A}_2[0]_1^{(0,0,0,0)} \Bigr\}\equiv \mathcal{L}[l]_{\Delta \geq l + 1}^{(0,0,0,0)}\,\,;\,\,l \in 2\,\mathbb{Z}_{\geq 0}\label{eq:allowedsuperconfblocks}\\[0.5ex]
    &(\mathbf{35}_c)_S \,\,\,:\,\,\,\left\{\mathcal{A}_1[l \geq 1]_{l + 2}^{(0,0,2,0)}\,\,,\,\,\mathcal{A}_2[0]_2^{(0,0,2,0)}\right\} \equiv \mathcal{A}_1[l]_{l+2}^{(0,0,2,0)}\,\,,\,\,\mathcal{B}_1[0]_1^{(0,0,2,0)}\,\,;\,\,l \in 2\,\mathbb{Z}_{\geq 0}\nonumber\\[0.5ex]
    &(\mathbf{28})_A \,\kern 0.1em\,\,\,:\,\,\, \mathcal{A}_1[l]_{l+2}^{(0,1,0,0)}\,\,,\,\,l \in 2\,\mathbb{Z}_{\geq 0} + 1\,\,\,\Big|\,\,\,(\mathbf{294}_c)_S \,\,\,:\,\,\,\mathcal{B}_1[0]_2^{(0,0,4,0)}\,\,\,\Big|\,\,\,(\mathbf{300})_S \,\,:\,\,\mathcal{B}_1[0]_2^{(0,2,0,0)}\nonumber
\end{align}
With all the possible superconformal blocks now determined, the next step is to apply the superconformal Ward identity to each block and thereby derive their functional forms. For four-point functions of half-BPS operators in any theory whose R-symmetry algebra is $so(\mathcal{N})$, the superconformal Ward identity was written down in \cite{Dolan_2004_2} for $3 \leq d \leq 6$
\begin{equation}\label{eq:superconformalwardidentity}
\begin{gathered}
    \Bigl[\chi \,\partial_{\chi} - (h - 1)\, \alpha\, \partial_\alpha\Bigr] F_{\mathcal{V}}(\chi, \chi'; \alpha, \alpha')\,\Big|_{\alpha = 1/\chi} = 0\\
    U = \chi\chi'\,\,,\,\,V = (1 - \chi)(1 - \chi')\,\,,\,\,\sigma = \alpha\alpha'\,\,,\,\,\tau = (1 - \alpha)(1 - \alpha')
\end{gathered}
\end{equation}
These conditions were solved for the superconformal blocks of interest in \cite{Chester_2014}, by using the strategy of decomposing each superconformal block into a linear combination of conformal blocks that belong to it, and then applying the identity (\ref{eq:superconformalwardidentity}). In terms of partial waves, this corresponds to further decomposing (\ref{eq:decompconfblocksuconf}) as follows
\begin{equation}\label{eq:suconftoconf}
    F_{\mathcal{V}}(U, V; \sigma, \tau) = \sum_{a = 0}^2 \sum_{b = 0}^a Y_{ab}(\sigma, \tau) \, \sum_{(\tilde{\Delta}, \tilde{l}) \in \mathcal{V}} A^{ab}_{\mathcal{V}\tilde{\Delta}\tilde{l}} f_{\tilde{\Delta}, \tilde{l}}(U, V)
\end{equation}
where $\tilde{\Delta}, \tilde{l}$ are the scaling dimension and spin of the conformal primary. Also note the slight change in notation from (\ref{eq:conformalpartialwavedecomposition}) i.e. $f_{\mathcal{V}}(U, V) \rightarrow f_{\tilde{\Delta}, \tilde{l}}(U, V)$, in order to not mix up the conformal primary ($\tilde{\Delta}, \tilde{l}$) with the superconformal primary ($\mathcal{V}$). Such a decomposition is also extremely convenient with regards to Mellin space computations, as we know the inverse Mellin transform for these conformal partial waves (\ref{eq:fourpointmellin}). Therefore the computation of Mellin amplitude of the exchange Witten diagrams dual to the concerned  four-point function is now algorithmic i.e. Consider each of the six allowed superconformal partial waves (\ref{eq:allowedsuperconfblocks}), decompose them into their corresponding conformal partial waves (\ref{eq:suconftoconf}), compute the Mellin amplitude corresponding to the exchange of each conformal primary (\ref{eq:mellinamplitudespinexch}), and finally add up all the contributions. It will be done in the next section.

\subsection{Mellin amplitudes of dual Witten diagrams}
To begin with, the connected part of the four-point correlator ($F^{conn.}(U, V;\sigma, \tau)$) is considered due to the definition of the Mellin transform. This four-point function is then dual to a sum of contact and exchange diagrams at tree-level as follows
\begin{align}
    &\mathcal{M}^{stress}_{tensor}(t, s; \sigma, \tau) = \mathcal{M}_{contact} + \mathcal{M}_{s-exch.} + \mathcal{M}_{t-exch.} + \mathcal{M}_{u-exch.}\label{eq:totalmellinstresstens}\\
    &\mathcal{M}_{s-exch.}(t, s; \sigma, \tau) = \sum_{i = 1}^6 \lambda_{(i)}^2\sum_{a = 0}^2 \sum_{b = 0}^a Y_{ab}(\sigma, \tau) \sum_{(\tilde{\Delta}, \tilde{l}) \in (i)} \tilde{A}^{ab}_{(i)\tilde{\Delta}\tilde{l}} \,\mathcal{M}^{(i)\tilde{\Delta}\tilde{l}}_{s-exch.}(t, s)\label{eq:Mellinexchangedecomp}
\end{align}
where $(i)$ denotes each of the six allowed superconformal blocks. Also, (\ref{eq:Mellinexchangedecomp}) is obtained by applying the inverse Mellin transform (\ref{eq:fourpointmellin}) on the decomposition (\ref{eq:suconftoconf}). It is helpful during later computations to first write down the residue $\mathcal{Q}_{l, m}(t)$ in (\ref{eq:mellinamplitudespinexch}) for $l = 0, 1, 2$, using (\ref{eq:mackpolynomials}) and $\Delta_1 = \Delta_2 = \Delta_3 = \Delta_4 = 1$. They are as follows
\begin{alignat}{3}
    &\mathcal{Q}_{0, m}(t) = &&F_{\Delta, 0}\label{eq:spin0residue}\\
    &\mathcal{Q}_{1, m}(t) = &&F_{\Delta, 1}\frac{2t - 5 + \Delta + 2m}{2}\label{eq:spin1residue}\\
    &\frac{\mathcal{Q}_{2, m}(t)}{F_{\Delta, 2}} = &&\frac{\Delta(\Delta - 3)}{(\Delta - 1)(\Delta - 2)}\left[\frac{(2 - t)^2 + (2 - u_2)^2}{8} + \frac{m(3 - 2\Delta - 2m)}{6}\right]\label{eq:spin2residue} \\
    &\,\,&&+ \frac{(4 - t)(2 - t) + (4 - u_2)(2 - u_2)}{8} - \frac{(2 - t)(2 - u)}{4}\left[\frac{\Delta}{\Delta - 2} + \frac{\Delta - 3}{\Delta - 1}\right]\nonumber\\
    &\,\,\kern 0.05em\text{where}\,\,\,\,&&F_{\Delta, l} = \frac{-2\cdot 4^{-l} [\Gamma(\Delta + l)] (\Delta - 1)_l}{m!\left(\Delta - \frac{1}{2}\right)_m\Gamma^2\left(\frac{\Delta + l}{2}\right)\Gamma^2\left(\frac{2 - \Delta + l - 2m}{2}\right)}\label{eq:prefactor}
\end{alignat}
where $u_2 = 6 - \Delta - 2m - t$ in (\ref{eq:spin2residue}). The analytic parts $R_{l - 1}(t, s)$ can be defined by a polynomial ansatz of degree $l - 1$ and moved into $\mathcal{M}_{contact}$.
\vspace{-0.2em}
\subsubsection*{Contribution from $(1) : \mathcal{B}_1[0]_1^{(0,0,2,0)}$} 
\noindent The coefficients $A^{ab}_{(1)\tilde{\Delta}\tilde{l}}$ in the partial wave decomposition (\ref{eq:suconftoconf}) as derived in \cite{Chester_2014} are
\begin{equation}\label{eq:decomp1coeff}
    A^{11}_{(1)10} = 1\,,\,\,A^{10}_{(1)21} = -1\,,\,\,A^{00}_{(1)32} = 1/4
\end{equation}
Therefore the $s$-exchange Mellin amplitude contribution from (1) block using (\ref{eq:Mellinexchangedecomp}) is 
\begin{equation}
    \mathcal{M}^{(1)}_{s-exch.}(t,s;\sigma,\tau) = \lambda_s^{\scaleobj{0.6}{(1)}} \left[(4\sigma + 4\tau - 1)\mathcal{M}^{(1)10}_{s-exch.} - 4(\sigma - \tau)\mathcal{M}^{(1)21}_{s-exch.} + \mathcal{M}^{(1)32}_{s-exch.}\right]
\end{equation}
where the OPE coefficients in (\ref{eq:mellinamplitudespinexch}) that are in the ratio of (\ref{eq:decomp1coeff}) i.e. $4\lambda^2_{1,0} = -4\lambda^2_{2, 1} = \lambda^2_{3, 2}$, have been absorbed into $\lambda_{(1)}^2$ in (\ref{eq:Mellinexchangedecomp}) and renamed as $\lambda_s^{(1)}$. Now by substituting (\ref{eq:spin0residue} - \ref{eq:spin2residue}) into (\ref{eq:mellinamplitudespinexch}) for $\Delta = 1, 2$ and 3 respectively, one obtains the individual Mellin amplitudes for the scalar\,, vector \,and \,graviton exchange, whose \,singular\, parts are then
\pagebreak
\begin{align}
    &\mathcal{M}^{(1)10}_{s-exch.} = \sum_{m = 0}^\infty \frac{-2}{\pi \,m!\,(1/2)_m \Gamma^2(\frac{1 - 2m}{2})} \cdot \frac{1}{s - (1 + 2m)}\label{eq:mellinamplitudescalarexchange}\\
    &\mathcal{M}^{(1)21}_{s-exch.} = \sum_{m = 0}^{\infty} \frac{-2}{\pi\, m!\,(3/2)_m\Gamma^2(\frac{1 - 2m}{2})}\cdot \frac{2t - 3 + 2m}{s - (1 + 2m)}\label{eq:mellinamplitudevectorexchange}\\
    &\mathcal{M}^{(1)32}_{s-exch.} = \sum_{m = 0}^{\infty}\frac{-4}{\pi\, m!\, (5/2)_m\Gamma^2(\frac{1-2m}{2})}\cdot\frac{8t^2 + 4m^2 + 16mt - 24t -24m + 19}{s - (1 + 2m)}\label{eq:mellinamplitudegravitonnexchange}
\end{align}
The amplitude for the contact diagram, which is now a constant (\ref{eq:mellincontactamplitude}) plus the non-singular parts of $(\ref{eq:mellinamplitudescalarexchange} - \ref{eq:mellinamplitudegravitonnexchange})$ is a degree 1 polynomial ($R_0(t, s) + R_1(t, s)$). Also from the earlier discussion above (\ref{eq:decompconfblocksuconf}), the amplitude is second order in $\sigma, \tau$, and therefore
\begin{equation}\label{eq:contactdiagramansatz}
    \mathcal{M}_{contact}^{(1)}(t, s;\sigma, \tau) = \sum_{\substack{0 \leq p, q \leq 2\\[0.2ex] 0 \leq p + q \leq 2}} \,\,\sum_{\substack{0 \leq m,n \leq 1\\[0.2ex] 0\leq m + n \leq 1}}\mu_{p,q;m,n}\, \sigma^p \tau^q s^m t^n
\end{equation}
Now consider the $t$-channel and $u$-channel exchanges. The $t$-channel exchange is obtained by $\vec{x}_2 \leftrightarrow \vec{x}_3$ and $\vec{Y}_2 \leftrightarrow \vec{Y}_3$, under which the correlator in (\ref{eq:stresstensorfourpointcorrelator}) becomes
\begin{equation}\label{eq:t-channelexchconstraints}
    \frac{Y_{12}^2Y_{34}^2}{x_{12}^2x_{34}^2}\rightarrow \frac{Y_{13}^2Y_{24}^2}{x_{13}^2x_{24}^2} = \sigma^2 \frac{Y_{12}^2Y_{34}^2}{x_{13}^2x_{24}^2}\,\,\,\,\Bigg|\,\,\,\,(\ref{eq:rsymmcrossratios})\,\,:\,\,\sigma, \tau \rightarrow \frac{1}{\sigma}, \frac{\tau}{\sigma}
\end{equation}
Now by the definition of Mellin space kinetic variables (\ref{eq:s-variables} - \ref{eq:t-variables}), $\vec{x}_1 \leftrightarrow \vec{x}_3$ corresponds to $s \leftrightarrow t$ in the Mellin amplitude. Therefore by considering this, (\ref{eq:t-channelexchconstraints}) then implies
\begin{align}
    &\mathcal{M}^{(1)}_{t-exch.}(t, s;\sigma, \tau) = \sigma^2 \mathcal{M}^{(1)}_{s-exch.}(s, t; 1/\sigma, \tau/\sigma)\label{eq:texchcrossing}\\
    &\mathcal{M}^{(1)}_{u-exch.}(t, s; \sigma, \tau) = \tau^2\mathcal{M}^{(1)}_{s-exch.}(t, u; \sigma/\tau, 1/\tau)\label{eq:uexchcrossing}
\end{align}
where the $u$-exchange (\ref{eq:uexchcrossing}) is also written down by similarly considering $\vec{x}_2 \leftrightarrow \vec{x}_4$ and $\vec{Y}_2 \leftrightarrow \vec{Y}_4$. However, the contact diagram via it's symmetry should remain invariant under $2 \leftrightarrow 3$, $2 \leftrightarrow 4$, which leads to the set of constraints on the ansatz (\ref{eq:contactdiagramansatz}) as follows
\begin{equation}\label{eq:contactsymmetryconstraints}
    \mathcal{M}_{contact}^{(1)}(t, s;\sigma, \tau) = \sigma^2\mathcal{M}_{contact}^{(1)}(s, t; 1/\sigma, \tau/\sigma) = \tau^2\mathcal{M}_{contact}^{(1)}(t, u; \sigma/\tau, 1/\tau) 
\end{equation}
\subsubsection*{Contribution from (2) : $\mathcal{B}_1[0]_2^{(0,0,4,0)}$}
\noindent The coefficients $A^{ab}_{(2)\tilde{\Delta}\tilde{l}}$ in the partial wave decomposition (\ref{eq:suconftoconf}) as derived in \cite{Chester_2014} are
\begin{equation*}
    A^{22}_{(2)20} = 1\,,\,\,A^{21}_{(2)31} = -\frac{4}{3}\,,\,\,A^{20}_{(2)40} = \frac{16}{45} \,,\,\,A^{11}_{(2)42} = \frac{256}{675}\,,\,\,A^{10}_{(2)51} = -\frac{128}{875} \,,\,\, A^{00}_{(2)60} = \frac{256}{18375}
\end{equation*}
As it can be seen, the twist $\tilde{\Delta} - \tilde{l} \geq 2$ for all the conformal blocks in the decomposition, and therefore don't contribute via the twist selection rule mentioned earlier. It should also be mentioned that, the twist selection rule ultimately applies to these conformal blocks, and thereby determines which ones contribute to the Mellin amplitudes. Therefore, the selection rule $\Delta - l \leq 2$ mentioned for the superconformal blocks earlier was because $\Delta - l > 2$ superconformal blocks have $\tilde{\Delta} - \tilde{l} \geq 2$ conformal blocks anyways, which don't contribute. So hopefully it is clear to the reader as to why the (= 2) case was included earlier, while it is excluded now. Moving on, as the reader can refer to (\cite{Chester_2014}), the story is the same for the superconformal blocks $(3): \mathcal{B}_1[0]_2^{(0,2,0,0)}, (4) : \mathcal{A}_1[l]_{l + 2}^{(0,0,2,0)}$ and $(5) : \mathcal{A}_1[l]^{(0,1,0,0)}_{l + 2}$ as well, which all have conformal blocks that violate the twist selection rule. However the long supermultiplet $(6) : \mathcal{L}[l]_{\Delta \geq l + 1}^{(0,0,0,0)}$ has contributing conformal blocks.
\subsubsection*{Contribution from (6) : $\mathcal{L}[l]_{\Delta \geq l + 1}^{(0,0,0,0)}$}
\noindent The coefficients corresponding to the contributing conformal blocks as per \cite{Chester_2014} are
\begin{align}
    &A^{00}_{(6)\Delta, \,l} \,\,\,\,\,\,\,\,\,\,\,= 1\label{eq:long0contr}\\[0.5ex]
    &A^{10}_{(6)\Delta + 1, \,l + 1} = \frac{-8(l + 1)(\Delta + l)}{(2l + 1)(\Delta + l + 1)}\\[0.5ex]
    &A^{11}_{(6)\Delta + 2, \,l + 2} = \frac{32(l+1)(l+2)(\Delta + l)(\Delta + l + 2)}{(4l^2 + 8l + 3)(\Delta + l + 1)(\Delta + l + 3)}\\[0.5ex]
    &A^{10}_{(6)\Delta + 3, \,l + 3} = \frac{-32(l + 1)(l + 2)(l + 3)\prod_{i = 0}^2(\Delta + l + 2i)}{(2l + 1)(2l + 3)(2l + 5)\prod_{j = 0}^2(\Delta + l + 2j + 1)}\label{eq:long3contr}\\[0.5ex]
    &A^{00}_{(6)\Delta + 4, \,l+4} = \frac{16(l + 1)(l + 2)(l + 3)(l + 4)\prod_{i = 0}^3 (\Delta + l + 2i)}{(2l + 1)(2l+3)(2l + 5)(2l + 7)\prod_{j = 0}^3(\Delta + l + 2j + 1)}\label{eq:long4contr}
\end{align}
Ofcourse as mentioned in the discussion above (\ref{eq:allowedsuperconfblocks}) earlier, $l \in 2\,\mathbb{Z}_{\geq 0}$ and also, only the case $l + 1 \leq \Delta < l + 2$ satisfies the twist selection rule in all the conformal blocks above. Let us assume the only fields present on the gravity side belong to the short supermultiplet $\mathcal{A}_2[0]^{(0,0,0,0)}_1 \in \mathcal{L}[l]_{\Delta \geq l+1}^{(0,0,0,0)}$ (as defined in (\ref{eq:allowedsuperconfblocks})). Therefore we would now also need to compute Spin-3 and Spin-4 exchange residues due to (\ref{eq:long3contr} - \ref{eq:long4contr}), which are
\begin{flalign}
    \frac{\mathcal{Q}_{3, m}(t)}{F_{\Delta, 3}} = \,&\frac{(t - 2)(t - 4)(t - 6)}{32} + \frac{3(t - 2)(u_3 - 2)(u_3 - 4)}{32}\left[\frac{\Delta - 4}{\Delta} + \frac{\Delta + 1}{\Delta - 3}\right] &&\\
    &+ \left[\frac{3(t - 4)(t - 2)^2}{32} - \frac{3m(2\Delta - 3 + 2m)(t - 2)}{40}\right]\frac{(\Delta + 1)(\Delta - 4)}{\Delta(\Delta - 3)} - t \leftrightarrow u_3\nonumber&&
\end{flalign}
\vspace{-1.3em}
\begingroup
\allowdisplaybreaks
\begin{align}
    \frac{\mathcal{Q}_{4, m}(t)}{F_{\Delta, 4}} = \,&\frac{(t - 2)(t - 4)(t - 6)(t - 8)}{128} + \frac{3(t - 2)^2(t - 4)^2}{128} \frac{\Delta(\Delta - 5)(\Delta - 3)(\Delta + 2)}{(\Delta - 4)(\Delta - 2)(\Delta - 1)(\Delta + 1)}\nonumber\\
    &+\frac{3(t - 2)(t - 4)(u_4 - 2)(u_4 - 4)}{128}\left[\frac{(\Delta - 3)(\Delta - 5)}{(\Delta - 1)(\Delta + 1)} + \frac{\Delta(\Delta + 2)}{(\Delta - 4)(\Delta - 2)}\right]\nonumber\\
    & + \frac{(t -2)^2(t - 4)(t - 6)}{32}\frac{(\Delta + 2)(\Delta - 5)}{(\Delta - 4)(\Delta + 1)} - \frac{3s_{4, 2}\,\Delta(\Delta + 2)(\Delta - 3)(\Delta - 5)}{35(\Delta - 1)(\Delta - 2)(\Delta - 4)(\Delta + 1)} \nonumber\\
    & + \frac{3s_{4, 1}(\Delta + 2)(\Delta - 5)}{28(\Delta - 4)(\Delta + 1)}\cdot\Biggl((t - 2)(t - 4) - (t - 2)(u_4 - 2)\left[\frac{\Delta - 3}{\Delta - 1} + \frac{\Delta}{\Delta - 2}\right]\nonumber
\end{align}
\endgroup
\vspace{-1.2em}
\begin{flalign}
     + (t - 2)^2 \frac{\Delta(\Delta - 3)}{(\Delta - 1)(\Delta - 2)}\Biggr) - \frac{(t - 2)\prod_{i = 1}^2(u_4 - 2i)}{32}\left[\frac{\Delta - 5}{\Delta + 1} + \frac{\Delta + 2}{\Delta - 4}\right]+ t \leftrightarrow u_4&&
\end{flalign}
where $s_{4, r} = \left(\frac{\Delta - 4 - s_4}{2}\right)_r\left(\frac{\Delta + 1 + s_4}{2}\right)_r$, and $s_l = \Delta - l + 2m, \,u_l = 4 + l - \Delta - 2m - t$. A lot of the terms in these seemingly daunting expressions vanish for the cases of interest to us i.e. $(\Delta; l) = (4, 5; 3, 4)$. Therefore the singular part of the Mellin amplitudes are
\begin{flalign}
    \mathcal{M}^{(6)43}_{s-exch.} = \sum_{m = 0}^\infty \frac{-12}{\pi\,m!\,(7/2)_m \Gamma^2(\frac{1 - 2m}{2})}\cdot \Biggl[\frac{32t^3 - 144t^2 + 238t + 96mt^2 - 288mt}{s - (1 + 2m)}& \nonumber &&\\
    + \frac{72m^2t + 8m^3 - 108m^2 + 238m - 141}{s - (1 + 2m)}&\Biggr]&&
\end{flalign}
\vspace{-1em}
\begin{align}
    \mathcal{M}^{(6)54}_{s-exch.} = \sum_{m = 0}^{\infty} \frac{-48}{\pi\,m!\,(9/2)_m \Gamma^2(\frac{1-2m}{2})}&\cdot \Biggl[\frac{128t^4 + 16m^4 - 768t^3 + 2016t^2 - 2592t + 512mt^3}{s - (1 + 2m)}\nonumber\\
    &+\frac{640m^2t^2 - 384m^3 - 2304mt^2 + 4032mt + 256m^3t}{s - (1 + 2m)}\nonumber\\
    &+ \frac{1624m^2 - 2592m  + 1321- 1920m^2t}{s - (1+2m)}\Biggr]
\end{align}
With regards to the $l = 0, 1, 2$ exchanges, $\mathcal{M}^{(6)10}_{s-exch.}, \mathcal{M}^{(6)21}_{s-exch.}, \mathcal{M}^{(6)32}_{s-exch.}$ are the same expressions as in (\ref{eq:mellinamplitudescalarexchange} - \ref{eq:mellinamplitudegravitonnexchange}). Also note that all these Mellin amplitudes are unique upto a purely spin dependent normalization factor, which the reader can introduce as necessary so that the Mellin space coefficients ($\tilde{A}$ in (\ref{eq:Mellinexchangedecomp})) lie in the same ratio as the position space coefficients ($A$ in (\ref{eq:decompconfblocksuconf})). With this normalization in mind, the $s$-exchange Mellin amplitude contribution from (6) block using (\ref{eq:Mellinexchangedecomp}) is the following
\begin{align}
    \mathcal{M}^{(6)}_{s-exch.}(t, s; \sigma, \tau) = \lambda_s^{(6)}\Bigl[\mathcal{M}^{(6)10}_{s-exch.} &-4(\sigma - \tau) \mathcal{M}^{(6)21}_{s-exch.} + 2(4\sigma + 4\tau - 1)\mathcal{M}^{(6)32}_{s-exch.}\nonumber\\
    &- 4(\sigma - \tau)\mathcal{M}^{(6)43}_{s-exch.} + \mathcal{M}^{(6)54}_{s-exch.}\Bigr]
\end{align}
Therefore the overall Mellin amplitude for the stress-tensor four point function is 
\begin{equation}
\begin{split}
    \mathcal{M}^{stress}_{tensor}(t, s; \sigma, \tau) = &\mathcal{M}_{contact} + \sum_{i = \{1, 6\}} \mathcal{M}_{s-exch.}^{(i)} + \mathcal{M}_{t-exch.}^{(i)} + \mathcal{M}_{u-exch.}^{(i)}\\
    \text{where}\,\,\,&\mathcal{M}_{contact} = \sum_{\substack{0 \leq p, q \leq 2\\[0.2ex] 0 \leq p + q \leq 2}} \,\,\sum_{\substack{0 \leq m,n \leq 3\\[0.2ex] 0\leq m + n \leq 3}}\mu_{p,q;m,n}\, \sigma^p \tau^q s^m t^n
\end{split}   
\end{equation}
The $t$-exchange and $u$-exchange amplitudes can be obtained as before using (\ref{eq:texchcrossing}, \ref{eq:uexchcrossing}). Also note that the contact amplitude is now a third degree polynomial in $s, t$ since the maximum degree analytic part is from the Spin-4 exchange.  Now the Mellin amplitude is all but determined, except for the coefficients $(\lambda^{(1)}_s, \lambda_s^{(6)})$ and $\mu_{p,q;m,n}$. The former can be determined by comparison with three-point functions, and the latter can be determined by using the superconformal Ward identity in Mellin space and the symmetry constraints (\ref{eq:contactsymmetryconstraints}). See \cite{Zhou_2018} for related computations.
\vspace{-0.45em}
\subsubsection*{Further Reading and Closing Remarks}
The $k = 1$, $\mathcal{N} = 8$ case was presented in quite some detail in the previous section, for the four point function of the stress-tensor multiplet superconformal primary. Similar superconformal block decomposition for the arbitrary $k$, $\mathcal{N} = 6$ case was derived in \cite{Binder_2021}, whose Mellin amplitudes can now easily be written with the fomalism described throughout the course of this chapter. With this sort of a procedural way of writing Mellin amplitudes for the Witten diagrams now in place, the attentive reader might have noticed a preference in utilizing the CFT data in order to study the dual gravitational quantities rather than the other way round. This bias started taking precedence from section \ref{sec:spinlexchange}, and would of course be a circular reasoning if one were to make claims that these computations stand as tests for the duality. So if not for tests, the reader may then question the concrete purpose that drives these computations of Witten diagrams, and one of the answers comes in the form of a conjecture put forth by Penedones in \cite{Penedones_2011}. He conjectured that the S-matrix of the bulk-theory (M-theory in this case) can be obtained by an integral transform of the Mellin amplitude in the flat space limit as follows
\begin{equation}\label{eq:penedonesconjecture}
\begin{split}
    &\mathcal{T}(p_i) = \frac{1}{\mathcal{N}}\lim_{R\to\infty} R^{2h - 3} \int_{-i\infty}^{i\infty} \frac{d\alpha}{2\pi i}\,\alpha^{h - \frac{1}{2}\sum \Delta_i} e^{\alpha} \mathcal{M}\left(\delta_{ij} = \frac{R^2}{2\alpha}p_i \cdot p_j\right)\\
    &\text{where}\,\,\,\mathcal{N} = \frac{1}{8\pi^h} \prod_{i = 1}^4\frac{1}{\sqrt{\Gamma(\Delta_i)\Gamma(\Delta_i - h + 1)}}
\end{split}
\end{equation}
where $R$ is the $AdS$-radius, $p_i$ are the momenta of particles, and the integration contour runs to the right of all the poles in the integrand in the $\alpha$ plane. The tree-level Witten diagram amplitudes computed in this chapter for the stress-tensor multiplet can then be plugged into (\ref{eq:penedonesconjecture}) to obtain the connected part of the four-graviton S-matrix in M-theory in the flat space limit. In \cite{Chester_2018}, this was shown to reproduce the $f_{R^4}$ term in small momentum expansion of the scattering amplitude at tree level obtained purely from 11D supergravity, which is constrained by the supersymmetry Ward identities as follows \cite{elvang2014scattering}
\begin{equation*}
    \mathcal{T} = \left(1 + l_{11}^6 f_{R^4}(s, t) + l_{11}^9 f_{1-loop}(s, t) + l_{11}^{12}f_{D^6R^4}(s, t) + l^{14}_{11} f_{D^8 R^4}(s, t) + ...\right)\mathcal{T}_{SG,\,tree}
\end{equation*}
where $\mathcal{T}_{SG, \,tree}$ is the tree-level scattering amplitude in 2-derivative supergravity, $l_{11}$ is the radius of the M-theory circle, and $f_{D^{2n}R^4}$ is a degree $n + 3$ symmetric polynomial. Correspondingly, the higher loop \cite{Penedones_2011} corrections to the stress-tensor multiplet tree-level amplitude, although not discussed in this thesis, were computed and matched in \cite{Alday_2022}.

%% file: chapters/additional_chapters.tex
\chapter*{\vspace*{-3.5em} Synthesis and Future Directions}\label{chap:additional_chapters}
\vspace{-1em}
The work in this thesis was aimed to be methodical, where each chapter slowly tried to establish one building block after the other in the overarching construction of the $AdS_4/CFT_3$ duality. Starting from the superspace construction of the ABJM theory in Chapter 1, to finding a host brane configuration in type IIB string theory in Chapter 2, and then lifting it to a pure geometry in M-theory in Chapter 3, the first half of the thesis laid the necessary groundwork for conjecturing the $AdS_4/CFT_3$ correspondence. With the correspondence statement and the more recent tests of it reviewed in Chapter 4, the second half of the thesis focused more on concrete computations within this duality; Starting with the position space description of Witten diagrams and CFT correlators in Chapter 5, and culminating in their computations in Mellin space in Chapter 6. Putting all of this together, it is remarkable that M-theory, a fundamental framework about which little is known beyond low-energy 11D supergravity and its dualities with string theories, reveals its profound depth when placed in an $AdS_4 \times S^7/\mathbb{Z}_k$ background, where it becomes equivalent to a quantum field theory with a well-defined Lagrangian description (ABJM). However, as it is the case with almost any work, it would be naive to claim that all possible directions have been explored and that there lie no open questions.

The example of the four-point function of the stress-tensor multiplet superconformal primary for $(k = 1, \mathcal{N} = 8)$, considered in section \ref{sec:ads4cft3examples}, had already been computed in \cite{Zhou_2018}. The corresponding information in the M-theory S-matrix had consequently been extracted in \cite{Chester_2018}. However, they had only considered the exchange of fields dual to operators belonging to the stress-tensor multiplet $\mathcal{B}_1[0]_1^{(0,0,2,0)}$. As part of this thesis, it has now been claimed that fields dual to the operators belonging to the short multiplet $\mathcal{A}_2[0]_1^{(0,0,0,0)}$ are also exchanged. The possible extension of this work then remains, in the form of computing contributions to the M-theory S-matrix from these additional amplitudes. More fundamentally, it was seen that two massive Spin-3 and Spin-4 fields are part of this newly considered exchange, which raises the question as to whether they arise due to yes-go theorems and Vasiliev theory (see \cite{article_vasiliev}) in compactified 11D supergravity, and whether they correspond to some interesting non-perturbative modes in M-theory. 

Another direction of research can be to extend the Mellin space formalism to Kerr-$AdS$ environments, with the obvious use case being the study of the different regions of a Kerr-$AdS_4$ Black hole and the interesting phenomena therein from an M-theory perspective, via the dual ABJM theory. To exhaust another possibility, it was mentioned in section \ref{sec:freeenergy} that for large $k$, M-theory on $AdS^4 \times S^7/\mathbb{Z}_k$ reduces to Type IIA on $AdS_4 \times \mathbb{CP}^3$; It is then also known that Type IIB theory on $AdS_5 \times S^5$ is dual to $\mathcal{N} = 4$ Super Yang-Mills theory (SYM) via $AdS_5/CFT_4$. So one can then possibly compute Mellin amplitudes in both cases, T-dualize on $AdS_4 \times S^5 \times S^1$ (symbolically) in the latter, and then analyze any topological effects that may arise from $S^1 \times S^5 \rightarrow \mathbb{CP}^3$, SYM $\rightarrow$ ABJM. In addition to these two, one can even possibly look for patterns in the d.o.f in $OSp(\mathcal{N} | 4)$ algebra to conjecture alternative descriptions of M-theory. The list of possibilities could go on, but hopefully it is already impressed upon the reader that the scope of future work possible, given the knowledge of the topics explored in this thesis, is vast and exciting.  

%% file: chapters/appendix.tex
\chapter*{\vspace{-3.5em}Appendix}\label{appendix}
% Following is required as above uses \chapter*{} (note the star). The start makes the chapter unnumbered, but also removes it from table of content. Former is desired, the latter is not:
\vspace{-1em}
\addcontentsline{toc}{chapter}{Appendix}
The bulk-bulk propagator for the massless Spin-2 particle was given in section \ref{subsec:massivespin2}, but the expression for the massive case is quite long and hence has been pushed to the Appendix. This was derived in the embedding formalism in \cite{Costa_2014} and is the following
\begin{equation*}
    G_{\mu\nu;\mu'\nu'}(u) = \sum_{i = 1}^5 A^{(i)}(u) T^{(i)}_{\mu\nu;\mu'\nu'}
\end{equation*}
where $T^{(i)}_{\mu\nu;\mu'\nu'}$ is the basis of five irreducible $SO(d, 1)$ tensors, given by the following 
\begin{align*}
    &T^{(1)}_{\mu\nu;\mu'\nu'} = g_{\mu\nu}\,g_{\mu'\nu'}\\
    &T^{(2)}_{\mu\nu;\mu'\nu'} = \partial_\mu u\, \partial_{\nu} u \,\partial_{\mu'} u \,\partial_{\nu'} u\\
    &T^{(3)}_{\mu\nu;\mu'\nu'} = \partial_\mu \partial_{\mu'} u \,\partial_{\nu}\partial_{\nu'} u + \partial_{\mu}\partial_{\nu'} u \,\partial_{\nu}\partial_{\mu'} u \\
    &T^{(4)}_{\mu\nu; \mu'\nu'} = \partial_{\mu'}u\,\partial_{\nu'}u\, g_{\mu\nu} + \partial_\mu u\, \partial_{\nu} u \,g_{\mu'\nu'}\\
    &T^{(5)}_{\mu\nu;\mu'\nu'} = \partial_{\mu}\partial_{\mu'} u\, \partial_{\nu} u\, \partial_{\nu'}u + \partial_{\nu}\partial_{\mu'}u \,\partial_{\mu}\partial_{\nu'}u + (\mu' \leftrightarrow \nu')
\end{align*}
The coefficient functions $A^{(i)}(u)$ are given in terms of the functions $g_k(u)$ as follows
\begin{align*}
    &A^{(4)}(u) = -\frac{1}{1+d} g_0(u) + \frac{(1+u)}{1+d}g_1(u) - \frac{u(u + 2)}{1 + d}g_2(u)\\
    &A^{(2)}(u) = g_2(u)\,\,, \,\,A^{(3)}(u) = \frac{1}{2}g_0(u)\,\,,\,\,A^{(5)}(u) = -\frac{1}{4}g_1(u)\\
    &A^{(1)}(u) = -\frac{1 + d - u(2 + u)}{(1 + d)^2}g_0(u) - \frac{u(1+u)(2 + u)}{(1 + d)^2}g_1(u) + \frac{u^2(2 + u)^2}{(1 + d)^2}g_2(u)\\
    &\text{where}\,\,\,g_k(u) = \sum_{i = k}^{2} (-1)^{i + k} \left(\frac{i!}{k!}\right)^2 \frac{1}{(i - k)!}f^{(k)}_i(u)
\end{align*}
where the notation $f^{(k)}_i(u) = \partial^k_u f_i(u)$. The recursive functions $f_i(u)$ are as follows
\begin{equation*}
    \begin{gathered}
        f_i = c_i \left[(d - 2i + 3)(df_{i - 1} + (1 + u)f^{(1)}_{i - 1}) + (4 - i)f_{i - 2}\right]\\
        f_0(u) = \frac{\Gamma(\Delta)}{2\pi^{\frac{d}{2}}\Gamma(\Delta + 1 - \frac{d}{2})}(2u)^{-\Delta} \,\tensor[_2]{F}{_1}\left(\Delta, \,\Delta + \frac{1 - d}{2}, \,2\Delta - d + 1,\,-\frac{2}{u}\right)\\
        c_i = -\frac{3 - i}{i(d + 2 - i)(\Delta + 1 - i)(d + 1 - \Delta - i)}
    \end{gathered}
\end{equation*}
where $\tensor[_2]{F}{_1}$ is the hypergeometric function mentioned in section \ref{sec:scalarprop}. These expressions were first derived in \cite{Naqvi_1999}. Also as mentioned in the main text, in the massless limit, three linear combinations of $T^{(3)}, T^{(4)}$ and $T^{(5)}$ correspond to diffeomorphims, thereby leaving the only physical components to be those of $T^{(1)}, T^{(2)}$, resulting in the expression (\ref{eq:gravitonpositionprop}).